\documentclass[useAMS,usenatbib,nofootinbib]{mn2e}

\usepackage[pdftex]{graphicx}

\usepackage{amsmath}
\usepackage{url}
\usepackage{natbib}
\usepackage{rotating}

\newcommand{\beq}{\begin{equation}}
\newcommand{\eeq}{\end{equation}}
\newcommand{\bdi}{\begin{displaymath}}
\newcommand{\edi}{\end{displaymath}}
\newcommand{\snr}{SNR}

\def\lsim{\mathrel{\lower2.5pt\vbox{\lineskip=0pt\baselineskip=0pt
           \hbox{$<$}\hbox{$\sim$}}}}

\def\gsim{\mathrel{\lower2.5pt\vbox{\lineskip=0pt\baselineskip=0pt
           \hbox{$>$}\hbox{$\sim$}}}}

\title[BLAST and LABOCA observations of the ECDF-S]{A joint analysis of
  BLAST 250--500\,\micron\ and LABOCA 870\,\micron\ observations in the
  Extended Chandra Deep Field South}

\author[Edward~L.~Chapin~et~al.]{
  \parbox[t]{\textwidth}{
    Edward~L.~Chapin$^{1}$\thanks{E-mail:~echapin@phas.ubc.ca},
    Scott~C.~Chapman$^{2}$,
    Kristen~E.~Coppin$^{3}$,
    Mark~J.~Devlin$^{4}$,
    James~S.~Dunlop$^{5}$,
    Thomas~R.~Greve$^{6,7}$,
    Mark~Halpern$^{1}$,
    Matthew~F.~Hasselfield$^{1}$,
    David~H.~Hughes$^{8}$,
    Rob~J.~Ivison$^{9,5}$,
    Gaelen~Marsden$^{1}$,
    Lorenzo~Moncelsi$^{10}$,
    Calvin~B.~Netterfield$^{11,12}$,
    Enzo~Pascale$^{10}$,
    Douglas~Scott$^{1}$,
    Ian~Smail$^{3}$,
    Marco~Viero$^{11}$,
    Fabian~Walter$^{6}$,
    Axel~Weiss$^{13}$,
    Paul~van~der~Werf$^{14}$
  }
  $^{1}$Dept. of Physics \& Astronomy, University of British Columbia,
  6224 Agricultural Road, Vancouver, B.C. V6T 1Z1, Canada\\
  $^{2}$Institute of Astronomy, University of Cambridge, Madingley Road,
  Cambridge CB3 0HA, UK \\
  $^{3}$Institute for Computational Cosmology, Durham University,
  South Road, Durham DH1 3LE, UK\\
  $^{4}$Department of Physics and Astronomy, University of Pennsylvania,
  209 South 33rd Street, Philadelphia, PA 19104, USA\\
  $^{5}$SUPA\footnote{Scottish Universities Physics Alliance} Institute for Astronomy, University of Edinburgh, Royal
  Observatory, Blackford Hill, Edinburgh, EH9 3HJ, UK\\
  $^{6}$Max-Planck Institute f\"ur Astronomie, K\"onigstuhl 17 D-69117,
  Heidelberg, Germany\\
  $^{7}$Dark Cosmology Centre, Niels Bohr Institute, University of Copenhagen,
  Juliane Maries Vej 30, DK-2100 Copenhagen, Denmark\\
  $^{8}$Instituto Nacional de Astrof\'isica, \'Optica y Electr\'onica
  (INAOE), Aptdo. Postal 51 y 216, Puebla, Mexico\\
  $^{9}$UK Astronomy Technology Centre, Royal Observatory, Blackford Hill,
  Edinburgh, EH9 3HJ, UK\\
  $^{10}$School of Physics \& Astronomy, Cardiff University, 5 The Parade,
  Cardiff, CF24 3AA, UK \\
  $^{11}$Department of Astronomy \& Astrophysics, University of Toronto,
  50 St. George Street Toronto, ON M5S 3H4, Canada\\
  $^{12}$Department of Physics, University of Toronto, 60 St. George Street,
  Toronto, ON M5S 1A7, Canada\\
  $^{13}$Max-Planck-Institut f\"ur Radioastronomie, Auf dem H\"ugel 69, Bonn,
  D-53121, Germany\\
  $^{14}$Leiden Observatory, Leiden University, PO Box 9513, NL-2300 RA Leiden,
  the Netherlands}
\begin{document}

\label{firstpage}

\maketitle

\begin{abstract}
  We present a joint analysis of the overlapping BLAST 250, 350,
  500\,\micron, and Large APEX Bolometer Camera 870\,\micron\
  observations (from the LESS survey) of the Extended Chandra Deep
  Field South. Out to $z \sim 3$, the BLAST filters sample near the
  peak wavelength of thermal far-infrared (FIR) emission from galaxies
  (rest-frame wavelengths $\sim 60$--200\,\micron), primarily produced
  by dust heated through absorption in star-forming clouds. However,
  identifying counterparts to individual BLAST peaks is very
  challenging, given the large beams (FWHM 36--60\,arcsec). In
  contrast, the ground-based 870\,\micron\ observations have a
  significantly smaller 19\,arcsec FWHM beam, and are sensitive to
  higher redshifts ($z\sim1$--5, and potentially beyond) due to the
  more favourable negative $K$-correction.  We use the LESS data, as
  well as deep {\em Spitzer} and VLA imaging, to identify 118
  individual sources that produce significant emission in the BLAST
  bands. We characterize the temperatures and FIR luminosities for a
  subset of 69 sources which have well-measured submm SEDs and
  redshift measurements out to $z\sim3$. For flux-limited sub-samples
  in each BLAST band, and a dust emissivity index $\beta=2.0$, we find
  a median temperature $T=30$\,K (all bands) as well as median
  redshifts: $z=1.1$ (interquartile range 0.2--1.9) for $S_{250} >
  40$\,mJy; $z=1.3$ (interquartile range 0.6--2.1) for $S_{350} >
  30$\,mJy; and $z=1.6$ (interquartile range 1.3--2.3) for $S_{500} >
  20$\,mJy. Taking into account the selection effects for our survey
  (a bias toward detecting lower-temperature galaxies), we find no
  evidence for evolution in the local FIR-temperature correlation out
  to $z\sim2.5$.  Comparing with star-forming galaxy SED templates,
  about 8\% of our sample appears to exhibit significant excesses in
  the radio and/or mid-IR, consistent with those sources harbouring an
  AGN.  Since our statistical approach differs from most previous
  studies of submm galaxies, we describe the following techniques in
  two appendices: our `matched filter' for identifying sources in the
  presence of point-source confusion; and our approach for identifying
  counterparts using likelihood ratios. This study is a direct
  precursor to future joint far-infrared/submm surveys, for which we
  outline a potential identification and SED measurement strategy.
\end{abstract}


\section{Introduction}
\label{sec:intro}

Observations in the submillimetre (submm) wavelength band (defined
here to be 200--1000\,\micron) are ideal for detecting light from
massive star-forming galaxies out to cosmological distances. It has
been known since the all-sky {\em Infrared Astronomical Satellite
  (IRAS)} survey of the 1980's that such sources contain significant
amounts of dust, so that the ultra-violet (UV) light of newly-formed
stars is absorbed by the galaxies' interstellar medium (ISM)
\citep{sanders1996}. The dust is typically heated to tens of Kelvin,
and most of the light is then thermally re-radiated at far-infrared
(FIR) wavelengths ($\sim$60--200\,\micron). In the submm, the thermal
spectral energy distribution (SED) drops off steeply, so that there is
a progressively stronger negative $K$-correction with increasing
observing wavelength. The correction is so strong that near
$\sim$1\,mm the observed flux density for a galaxy of fixed luminosity
is approximately constant from $1 \lsim z \lsim 10$ \citep{blain2002}.

Even though much of the submm band is obscured to ground-based
observations by atmospheric water vapour, a number of surveys over the
last decade have exploited transparency in several spectral windows to
successfully locate high-redshift ($z>1$) dusty star-forming galaxies
solely through their submm emission (submillimetre galaxies, or
SMGs). Their discovery was first made with the Submillimetre Common
User Bolometer Array \citep[SCUBA][]{holland1999} at 850\,\micron\
\citep[e.g.,][]{smail1997,hughes1998,barger1998,cowie2002,
  scott2002,borys2003,webb2003,coppin2006}. Several other instruments
confirmed their existence in the slightly more transparent
1.1--1.2\,mm band \citep[e.g.,][]{greve2004,laurent2005,scott2008,
  perera2008,austermann2010}. These ground-based surveys at
850--1200\,\micron\ typically cover $\ll 1$\,deg$^2$, and detect
several tens of sources per field. The typical angular resolution of
these surveys is in the range $\sim$9--20 arcsec full-width at
half-maximum (FWHM).  It is worth noting that observations at 350 and
450\,\micron\ have also been attempted from the ground
\citep[e.g.,][]{smail1997,hughes1998,fox2002,kovacs2006,khan2007,coppin2008}.
However, this wavelength range is much more difficult, due to
increased atmospheric opacity, so that these surveys have only
detected a handful of sources.

While the first generation surveys successfully demonstrated the
existence of these ultra-luminous infrared galaxies (ULIRGs) at
$z\sim1.5$--4 \citep[e.g][]{chapman2003b,aretxaga2003,chapman2005},
sample sizes have been modest (typically $\ll 100$ sources in a given
field). Most of what is known about SMGs is based on
cross-identifications with sources in higher-resolution data,
particularly in the radio \citep[primarily 1.4\,GHz Very Large Array
maps, e.g.,][]{smail2000,ivison2007} and in the mid-IR \citep[such as
24\,\micron\ {\em Spitzer} maps, e.g.,][]{ivison2004,pope2006}. While
these counterpart identification strategies could be biased toward
lower redshifts due to the {\em positive} $K$-corrections in the
radio/mid-IR, more observationally time-consuming mm-wavelength
interferometric observations
\citep[e.g.,][]{lutz2001,dannerbauer2004,iono2006,younger2007}
demonstrate reasonable correspondence with proposed radio/mid-IR
counterparts for a handful of sources. With accurate positions it is
then possible to identify optical counterparts, although they are
usually extremely faint due to obscuration by the same dust that makes
them bright in the submm--FIR, and the fact that stellar light from
the most distant objects gets red-shifted out of the optical bands
into the near-IR. Obviously, ground-based optical spectroscopy is even
more challenging given the difficulty in imaging the counterparts.

Recent observations by the 1.8-m Balloon-borne Large Aperture
Submillimeter Telescope (BLAST) at 250, 350, 500\,\micron\ \citep[a
pathfinder for {\em Herschel}/SPIRE,][]{pascale2008} toward the
Extended Chandra Deep Field South (ECDF-S) have provided the first
confusion-limited submm maps at these wavelengths which cover areas
larger than 1\,deg$^2$. These bands were chosen to bracket the peak
rest-frame FIR emission from the SMG population at
$z\sim1$--4. However, given the size of its primary mirror, the BLAST
diffraction-limited angular resolution of 36--60\,arcsec FWHM at
250--500\,\micron\ has made associations between submm emission peaks
and individual sources at other wavelengths considerably more
challenging than with the existing ground-based surveys at longer
wavelengths. For this reason many of the primary BLAST scientific
results to date have been derived from the statistics of brightness
fluctuations for entire maps, such as the number counts
\citep{devlin2009,patanchon2009}, contributions of known sources to
the Cosmic Infrared Background \citep[CIB][]{puget1996,fixsen1998} in
the BLAST bands \citep{devlin2009,marsden2009,pascale2009}, evolution
in the FIR--radio correlation \citep{ivison2009}, and the large-scale
clustering of infrared-bright galaxies \citep{viero2009}. We emphasize
that {\em none of these results depend on identifying individual submm
  sources in the BLAST maps}.

More traditional analyses of BLAST sources identified through peaks in
maps convolved with the point spread function (PSF) have also been
attempted \citep{dye2009,dunlop2010,ivison2009}. In general, it has
been a struggle to determine whether these peaks are produced
primarily by single galaxies, or blends of several faint sources,
necessitating either: conservative cuts in the signal-to-noise ratio
(\snr) to consider only the very brightest sources; or a careful
(though subjective) comparison of all the multi-wavelength data on a
case-by-case basis to decide whether single or multiple objects are
the likely source of the submm emission. In these earlier papers,
Poisson chance alignment `$P$' probabilities \citep{downes1986} have
been used to rank potential counterparts to the BLAST peaks from
external matching catalogues, showing that 10\% spurious threshold
probabilities must be adopted to obtain reasonable source statistics
(unlike the more conservative 5\% that is typical in the submm
community). These methods yield limited results for these wide-area
BLAST maps, despite the fact that the \snr\ of the individual peaks
rival those of most previous ground-based observations.

In this paper, driven by the apparent inadequacy of existing methods
for studying individual sources in the low-resolution BLAST maps, we
develop improved approaches for: (i) filtering confused maps to find
emission peaks that are more likely to be produced by individual (or
at least a small number of) sources; and (ii) identifying counterparts
to these peaks in external matching catalogues using Likelihood Ratios
(LR), a method which can incorporate more prior information than that
assumed in the calculation of $P$. In addition to deep BLAST
observations of the ECDF-S, we also make extensive use of the deepest
wide-area submm map at $\sim$1\,mm to date: the Large APEX Bolometer
Camera (LABOCA) ECDFS Submm Survey (LESS) at 870\,\micron\
\citep{weiss2009}, taken with the 12\,m APEX telescope
\citep{gusten2006}. This first detailed comparison between BLAST and
longer-wavelength ground-based submm data helps in two key
ways. First, the LABOCA beam has a 19\,arcsec FWHM (roughly half that
of the 250\,\micron\ BLAST beam), enabling us to ascertain {\em
  directly} whether some of the BLAST peaks resolve into multiple
submm counterparts. Second, like SCUBA, LABOCA \citep{siringo2009} is
more sensitive to $z>1$ sources than BLAST, and the most distant
sources \citep[$z>4$, e.g.,][]{coppin2009} are expected to be
LABOCA-detected BLAST-dropouts. Therefore this study will offer
superior constraints on the high-redshift submm galaxy population than
earlier BLAST studies.

Now that we have entered the era of {\em Herschel} surveys, we also
show that these techniques will be useful, despite the approximately
twofold improvement in angular resolution offered by SPIRE compared to
BLAST.  We explore this issue using simulations of SPIRE maps using
the smaller PSFs. While we find that the situation is certainly
improved for SPIRE, confusion will continue to seriously hamper the
interpretation of these new surveys.

The analysis is organized as follows. In Section~\ref{sec:submm} we
summarize our treatment of the submm data using our new `matched
filter' to identify individual peaks (full details are given in
Appendix~\ref{sec:matched}). We produce an external matching catalogue
in Section~\ref{sec:matchcat} combining 24\,\micron\ mid-IR and
1.4\,GHz radio priors to select sources from a deep {\em Spitzer} IRAC
near-IR catalogue. The LR identification technique is summarized in
Section~\ref{sec:lr_priors} (a full development, and calculation of
priors are given in Appendix~\ref{sec:lr}), and it is used to produce
a list of potential matches to the submm peaks in
Section~\ref{sec:potential}. Also, for cases where matches in the
catalogue could not be identified, we search for counterparts in the
higher-resolution LESS 870\,\micron\ peak catalogue.  At this stage we
have a collection of submm peaks, and a list of individual galaxies
that we believe produce these peaks -- in many cases blends of several
galaxies. To establish their submm flux densities we fit PSFs at all
of their locations simultaneously, in each of the submm maps, in
Section~\ref{sec:photometry}. The effects of confusion, missing
identifications, and clustering are explored in
Sections~\ref{sec:confusion}--\ref{sec:clustering} using simulations.
We derive redshifts for the proposed counterparts in
Section~\ref{sec:redshift}, and study the rest-frame properties of the
sample in Section~\ref{sec:restsed}, showing in particular how
confusion may have biased some of the earlier BLAST results. Finally,
in Section~\ref{sec:future} we simulate SPIRE data to demonstrate the
usefulness of our techniques for these new surveys.

\section{Data}
\label{sec:obs}

\subsection{Submillimetre Data}
\label{sec:submm}

\subsubsection{BLAST}
In 2006 BLAST conducted a two-tiered nested survey centred over the
Great Observatories Origins Deep Surveys South (GOODS-S): BLAST
GOODS-South Wide (BGS-Wide) over 10\,deg$^2$ to instrumental RMS
depths 36, 31 and 20\,mJy, at 250, 350 and 500\,\micron, respectively;
and BLAST GOODS-South Deep (BGS-Deep) over 0.9\,deg$^2$ to RMS depths
11, 9 and 6\,mJy. It is important to note that a significant
additional contribution to the noise in these maps is produced by
point source confusion, estimated to be 21, 17 and 15\,mJy in the
three bands \citep{marsden2009}. The ECDF-S is completely encompassed
by the BGS-Deep coverage. The BLAST maps were produced using SANEPIC
\citep{patanchon2008}, and were filtered to suppress residual noise on
scales larger than approximately 10\,arcmin (the array footprint). The
BLAST beams have FWHM 36, 42, and 60\,arcsec at 250, 350, and
500\,\micron. The maps and data reduction are discussed in detail in
\citet{devlin2009}. Details on instrument performance and calibration
are provided in \citet{pascale2008,truch2009}.

\subsubsection{LESS}
The LABOCA Survey of the Extended Chandra Deep Field South
\citep[LESS,][]{weiss2009} provides deep 870\,\micron\ data, with an
RMS better than 1.2\,mJy across the full $30\,\mathrm{arcmin} \times
30\,\mathrm{arcmin}$ ECDFS. A combination of time-domain filtering of
the raw bolometer data, as well as the suppression of residual noise
on scales $\gsim90$\,arcsec were incorporated as part of the reduction
procedure. Similar to the BLAST data, this map was then smoothed with
the 19\,arcsec FWHM diffraction-limited PSF to identify 126 point
sources in \citet{weiss2009} above a significance of $3.7\sigma$
(equivalent to a false detection rate of $<5$\%).

\subsubsection{Submm Peak Catalogues}
\label{sec:filter}

It is standard practice in the submm community to find sources in maps
by cross-correlating with the PSF (this strategy was used both in
earlier analyses of BLAST data and the LESS 870\,\micron\ map as noted
above). This operation is optimal for the case of an isolated point
source in a field of statistically un-correlated noise: the
cross-correlation gives the maximum-likelihood flux density of a point
source fit to every position in the map.\footnote{This procedure has
  long been understood in astronomical data analysis in other
  wavebands, e.g., \citet{stetson1987}.} However, sources are {\em
  not} isolated in the real submm maps under discussion. Their high
surface density, combined with the large beams, practically ensures
that every pixel in these maps has at least some contribution from
multiple overlapping sources. This ``confusion noise'',
$N_\mathrm{c}$, is independent of the approximately white instrumental
noise, $N_\mathrm{w}$. Confusion noise must be considered when asking
the question: what is the flux density of a {\em particular} point
source at an arbitrary location on the sky? One can think of
$N_\mathrm{c}$ as the distribution of flux densities in a map with no
instrumental noise source (i.e. a map of point sources smoothed by the
PSF), precisely the distribution that is modelled in a $P(D)$
analysis. As shown in \citet{patanchon2009} for BLAST, this
distribution is asymmetric, with a positive tail that converges to the
underlying differential counts distribution at large flux densities
(since brighter sources have a lower surface density, and therefore
stand out more against the confusion of fainter sources). For the
remainder of this paper, both $\sigma_\mathrm{w}$ and
$\sigma_\mathrm{c}$ will refer to the RMS of the respective noise
distributions, where the former is an output of the map-making
process, and the latter is estimated from simulated maps with no
instrumental noise added.

The previously published BLAST PSF-smoothed maps lie in a regime where
$\sigma_\mathrm{w}$ is about a factor of two smaller than confusion noise,
$\sigma_\mathrm{c}$. In contrast, for LESS the two noise components are
roughly equal.\footnote{For BLAST the ratios of confusion RMS to
  instrumental noise in the PSF-smoothed maps are 1.9, 1.9, and 2.4 at
  250, 350, and 500\,\micron. For the LESS PSF-smoothed map the ratio
  is 1.15.}  Since confusion is a source of noise that is {\em
  correlated on the scale of the PSF}, we have investigated a modified
filter for identifying peaks. If one imagines the extreme opposite
case of that used to motivate cross-correlation with the PSF -- that
there is no instrumental noise in the image whatsoever, and so the
only signal is the confused pattern of many overlapping sources,
clearly any additional smoothing will only make matters worse. In this
extreme case, one must clearly bin the data into a finely-sampled grid
and simply identify peaks without any filtering that might further
degrade the resolution --- or alternatively attempt to {\em
  de}-convolve the PSF.

We approach this problem by developing a `matched filter' that
maximizes the \snr\ of individual point sources in the presence of the
two noise components. The resulting filter in each band can be thought
of as an optimal balance between smoothing and de-convolving the PSF,
depending on the relative sizes of the white and confusion noise
components. We cross-correlate the raw submm maps with these new
filters, and produce source lists from peaks in the resulting \snr\
maps. A detailed description of our matched filter is given in
Appendix~\ref{sec:matched}, with Fig.~\ref{fig:matchedfilt} comparing
the matched filter with the PSF at 500\,\micron.

\begin{figure*}
\centering
\includegraphics[width=0.9\linewidth]{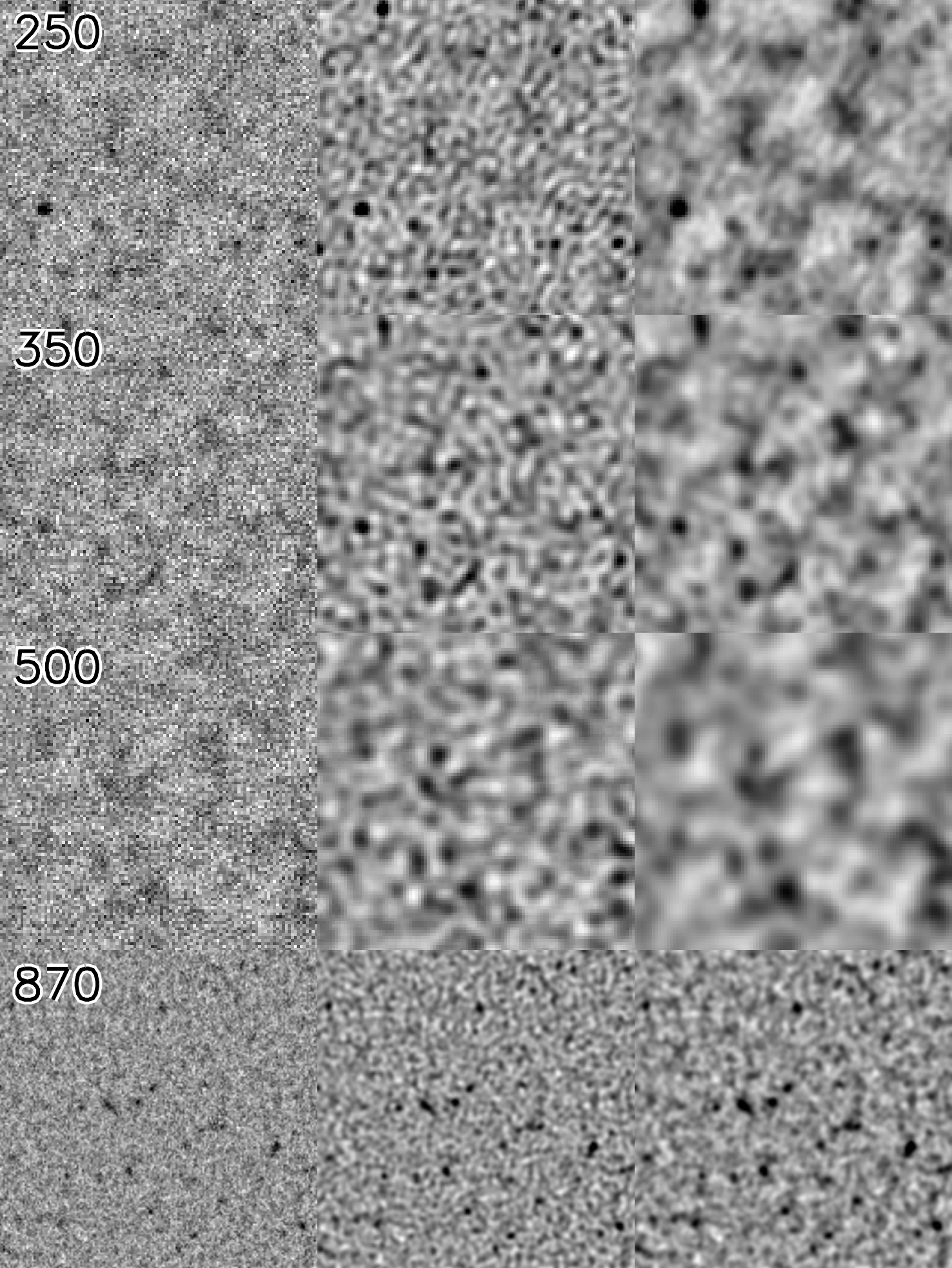}
\caption{Comparison of maps at all four wavelengths (rows), and
  smoothing scales in a $0.3\,\mathrm{deg} \times 0.3\,\mathrm{deg}$
  patch of the ECDF-S. The first column shows un-smoothed maps (noisy,
  but diffraction-limited resolution), the second column maps smoothed
  by the matched filter (greatly improved \snr\ at the expense of a
  slight degradation in the resolution), and the last column maps
  smoothed by the PSF (less improvement in the \snr, and resolution
  degraded by $\sqrt{2}$). The greyscales indicate the significance of
  point-source flux densities in the maps, ranging from $-3\,\sigma$
  (white) to $+4\,\sigma$ (black), considering both the instrumental
  and confusion noise contributions to each pixel (estimated from
  simulations -- see boldface values in Table~\ref{tab:matched}).}
\label{fig:mapcompare}
\end{figure*}

How much does this filter improve the \snr\ at each wavelength?  To
address this question we turn to simulations, drawing point source
flux densities from the best-fit number counts measured in each band
(from \citet{patanchon2009} and \citet{weiss2009} for BLAST and LESS
respectively), distributing them uniformly in maps, smoothing the maps
with the respective PSFs, adding instrumental white noise to match the
levels reported for the real observations, and finally subtracting the
map means.  We then compare the relative sizes of confusion
($\sigma_\mathrm{c}$), white ($\sigma_\mathrm{w}$) and combined
($\sigma_\mathrm{t} \equiv \sqrt{\sigma_\mathrm{c}^2 +
  \sigma_\mathrm{w}^2}$) noise components in: (i) the raw un-smoothed
maps; (ii) maps smoothed by the PSFs; and (iii) maps smoothed by the
matched filters. The results are summarized in
Table~\ref{tab:matched}, and in Fig.~\ref{fig:mapcompare} a sample
portion of the ECDF-S is compared in all four submm bands with no
filtering, matched filtering, and PSF filtering.

\begin{table}
  \setlength{\tabcolsep}{0.14cm}
  \caption{Comparison of confusion ($\sigma_\mathrm{c}$), white ($\sigma_\mathrm{w}$) and
    total ($\sigma_\mathrm{t}$) noise contributions in the raw submm maps, maps
    cross-correlated with the full PSF, and maps cross-correlated with the
    matched filter which compensates for confusion. Each quantity was estimated
    using simulations based on un-clustered realizations of sources drawn
    from the measured number counts distributions of \citet{patanchon2009} and
    \citet{weiss2009}, for BLAST and LESS respectively.
    All noise values are standard
    deviations in mJy. Note that our definition of confusion noise here is
    simply the RMS of a noise-free map containing point sources only. This
    calculation shows that
    white noise is more effectively suppressed by PSF
    filtering than matched filtering. However, the
    confusion noise is significantly larger in the PSF filtered maps than
    the match filtered maps. For this reason the {\em total} noise in the match
    filtered maps is smaller.}
  \vspace{0.2cm}
  \centering
  \begin{tabular}{c|ccc|ccc|ccc}
    \hline
    $\lambda$ &
    \multicolumn{3}{c}{Raw} &
    \multicolumn{3}{c}{PSF} &
    \multicolumn{3}{c}{Matched Filter} \\
    (\micron) & $\sigma_\mathrm{c}$ & $\sigma_\mathrm{w}$ & $\sigma_\mathrm{t}$&
    $\sigma_\mathrm{c}$ & $\sigma_\mathrm{w}$ & $\sigma_\mathrm{t}$&
    $\sigma_\mathrm{c}$ & $\sigma_\mathrm{w}$ &$\sigma_\mathrm{t}$\\
    \hline
250 & 14.5 & 31.2 &{\bf 34.4}&21.9 & 11.5 &{\bf 24.7}& 14.6 & 13.8 &{\bf 20.1}\\
350 & 12.5 & 28.5 &{\bf 31.1}&17.3 & 8.90 &{\bf 19.5}& 13.0 & 10.7 &{\bf 16.8}\\
500 & 11.6 & 27.4 &{\bf 29.8}&16.1 & 6.70 &{\bf 17.4}& 11.7 & 8.30 &{\bf 14.3}\\
870 & 0.70 & 2.00 &{\bf 2.12}&0.97 & 0.84 &{\bf 1.28}& 0.82 & 0.90 &{\bf 1.22}\\
    \hline
  \end{tabular}
  \label{tab:matched}
\end{table}

We find that the point-source sensitivities in the match-filtered
BLAST maps improve by $\sim$15--20\% over those reported for the PSF
smoothed maps in \citet{devlin2009} (comparing the bold-face
$\sigma_\mathrm{t}$ columns for the PSF and Matched Filter in
Table~\ref{tab:matched}). However, the increase in the LESS
point-source sensitivities, $\sim5$\%, is not as significant due to
the relatively smaller contribution of confusion.

Finally, we produce catalogues of submm peaks for which we will
attempt to identify the source(s) of submm emission. Given the
marginal improvement in the 870\,\micron\ map using the new filter,
and for the sake of simplicity, we use the full 870\,\micron\ LESS
catalogue from \citet{weiss2009} (down to a \snr\ of
3.7$\sigma$). However, we construct new 3.75$\sigma$ peak lists for
BLAST (relative to instrumental noise) from the match-filtered maps
(all submm catalogues are provided in
Appendix~\ref{sec:tables}).\footnote{The new BLAST GOODS-S
  match-filtered maps and peak lists are available at
  \url{http://blastexperiment.info/}.} These thresholds ensure that
the peaks are likely to be caused by submm emission rather than
spurious instrumental noise; such cuts typically result in
false-identification rates (the probability that the underlying flux
density within an instrument beam at that location is negative) of
order 5\% \citep[e.g.,][]{coppin2006,perera2008,weiss2009}. In total,
within the region of overlap, there are 64 peaks at 250\,\micron, 67
peaks at 350\,\micron, 55 peaks at 500\,\micron, and 81 peaks at
870\,\micron.  We emphasize, at this stage, that we do not know
whether they are produced primarily by one, or multiple overlapping
sources. We will will explore the properties of peaks selected this
way using simulations in Section~\ref{sec:sed}.

\subsection{Matching Catalogue}
\label{sec:matchcat}

\begin{figure*}
\centering
\includegraphics[width=\linewidth]{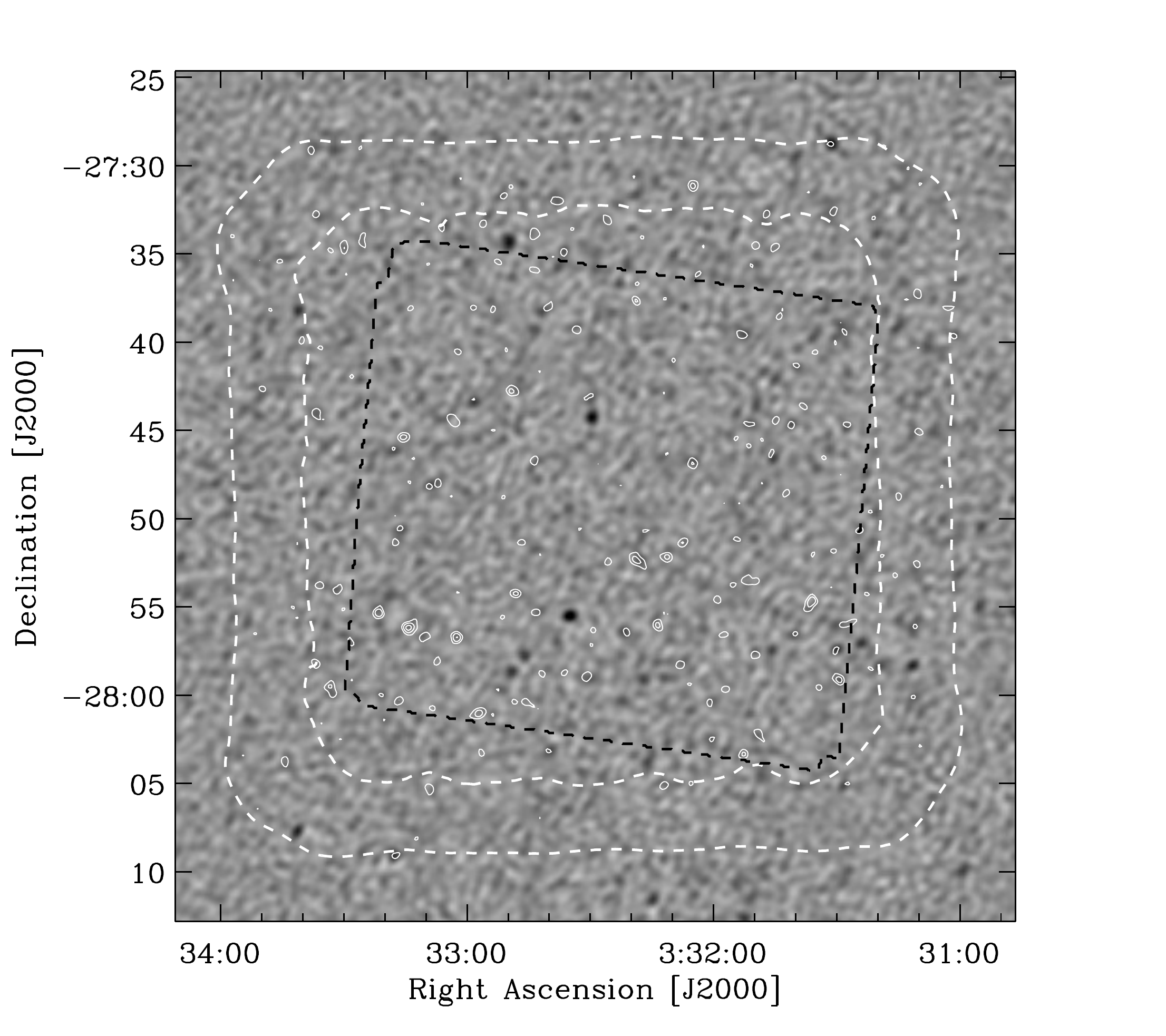}
\caption{Relative coverage of data sets in ECDF-S. The background
  greyscale image shows the BLAST 250\,\micron\ match-filtered \snr\
  map \citep[this field is completely encompassed by the BGS-deep
  region described in][]{devlin2009}, scaled between $-10\,\sigma$
  (white) and $+13\,\sigma$ (black). The solid white contours show the
  LESS \snr\ map at levels 3, 6 and 10\,$\sigma$. The white dashed
  lines show the LESS 1.3 and 2.2\,mJy instrumental noise
  contours. The dashed black line indicates the 708\,arcmin$^2$ region
  common to the VLA and FIDEL survey coverage within which we perform
  out counterpart search. The ECDF-S presently has the best (widest
  and deepest) submm coverage from 250--870\,\micron\ on the sky, with
  the mid/far-IR and radio data of matching quality required to
  identify counterparts. }
\label{fig:coverage}
\end{figure*}

The ideal matching catalogue for our lists of submm peaks would
contain {\em all} of the sources that emit significantly in the submm,
without having any spurious interlopers.  The starting point for our
catalogue is IRAC data from SIMPLE \citep[{\em Spitzer} IRAC/MUSYC
Public Legacy in ECDF-S,][]{gawiser2006}. This sample is approximately
flux-limited with $S_{3.6\mathrm{\micron}} \gsim 5$\,$\mu$Jy and
$S_{4.5\mathrm{\micron}} \gsim 5$\,$\mu$Jy (although there is
significant scatter at the faint end, presumably due to noise and
completeness effects). Anecdotally, such deep near-IR catalogues
appear to contain counterparts to some of the faintest and
highest-redshift submm sources detected in ground-based surveys
\citep[e.g.,][]{pope2006,chapin2009b,coppin2009}. However, the high
surface densities, in this case 48.3\,arcmin$^{-2}$ for the ECDF-S,
make it impossible to directly associate individual sources with the
much lower-resolution submm peaks. For this reason it is generally
necessary to use other priors to cull known spurious sources (such as
stars, or galaxies with little or no dust). Fortunately, in this field
there also exist deep 24\,\micron\ {\em Spitzer} and 1.4\,GHz VLA
maps. These two wavelength regimes have been used considerably in the
past to identify submm sources as they are sensitive to the presence
of dust and star-formation activity, respectively. We check for
emission in these additional data sets to reduce the 33962 sources
from SIMPLE covering the 703\,arcmin$^2$ region of overlap
(Fig.~\ref{fig:coverage}), to a total of 9216 entries, a surface
density of 13.1\,arcmin$^{-2}$. There are 8833 sources detected at
24\,\micron, 1659 at 1.4\,GHz, and 1276 at both 24\,\micron\ and
1.4\,GHz.

In Sections~\ref{sec:fidelcat} and \ref{sec:radiocat} we describe our
treatment of the mid-IR and radio data, respectively. Then, in
addition to the previous plausibility arguments, we use stacking to
check directly whether our combined catalogue reproduces the diffuse
measurements of the CIB, and hence determine how complete it is in the
four submm bands under consideration in Section~\ref{sec:stacks}.

\subsubsection{FIDEL}
\label{sec:fidelcat}
A deep Multiband Imaging Photometer for Spitzer (MIPS) 24 and
70\,\micron\ catalogue was produced by \citet{magnelli2009}, combining
{\em Spitzer} data from GOODS-S with the Far-Infrared Deep
Extragalactic Legacy survey (FIDEL, P.I. Mark Dickinson). The PSFs
have FWHM 5.5\,arcsec and 16\,arcsec, in each band respectively. The
catalogue uses SIMPLE as a positional prior enabling de-blending of
sources down to separations as small as 0.5 times the MIPS FWHM. This
catalogue is the same that was used for other recent BLAST studies in
ECDF-S \citep{devlin2009,dye2009,marsden2009,pascale2009}, and is
consistent with producing the entire CIB across the BLAST bands. This
catalogue is also used in the LESS identification paper (Biggs et
al. submitted), although it is known that such catalogues do not
reproduce the entire CIB at wavelengths approaching $\sim1$\,mm
\citep{wang2006,pope2007,marsden2009}.

\subsubsection{Radio}
\label{sec:radiocat}

The VLA 1.4\,GHz data are from the survey of \citet{miller2008}.
Given the low elevation of this southern field, and observing from the
VLA in the north, the synthesized beam is significantly elongated with
dimensions $2.8\,\mathrm{arcsec} \times 1.7\,\mathrm{arcsec}$.  We use
a new radio catalogue produced by Biggs et al. (submitted) that was
developed to identify counterparts to LESS sources. An initial
catalogue of 3\,$\sigma$ peaks in the map is produced, and then
Gaussians are fit at each of those positions, allowing the sizes to
vary. Since this fit is particularly noisy at the faint end, we cull
sources with integrated flux densities that are less significant than
2\,$\sigma$. Finally, we only include sources that lie within 2\,arcsec
of the IRAC positions. This strategy enables us to go significantly
deeper than the published 7\,$\sigma$ catalogue from \citet{miller2008},
or the 5\,$\sigma$ catalogue from \citet{dye2009}, at the expense of
missing a handful of brighter radio sources that do not appear to have
IRAC associations. Since Monte Carlo simulations based on the submm
data and matching catalogue are used in Section~\ref{sec:potential} to
calculate the probability that individual counterparts are real, any
spurious radio sources near the detection threshold will simply reduce
the identification efficiency (see also
Fig.~\ref{fig:frhist}). Finally, we note that Biggs et al. (submitted)
also find identifications for LESS sources using a combination of the
radio, MIPS, and IRAC catalogues, although the IRAC data are not
explicitly used as a prior for the radio positions.

\subsubsection{Stacks}
\label{sec:stacks}

\begin{table*}
  \caption{Surface brightnesses resulting from stacks on different catalogues as
    compared to absolute measurements of the total CIB
    \citep[values shown in brackets are from][]{fixsen1998}, in order of
    decreasing
    surface density (quoted values in arcmin$^{-2}$): SIMPLE is the entire IRAC
    catalogue; FIDEL+VLA includes sources from SIMPLE that exhibit 24\,\micron\
    emission, and additional VLA 1.4\,GHz sources that have a significance
    $>2\sigma$ within 2\,arcsec of a SIMPLE position; FIDEL is the
    24\,\micron\ catalogue based on SIMPLE from \citet{magnelli2009}; and VLA
    is the subset of SIMPLE sources with 1.4\,GHz emission that are {\em not}
    members of FIDEL. The units of the surface brightness measurements are
    nW\,m$^{-2}$\,sr$^{-1}$. The FIDEL+VLA catalogue (indicated in boldface) is
    used for matching throughout this paper.}
  \vspace{0.2cm}
  \centering
  \begin{tabular}{c|ccccc}
    \hline
    Catalogue & Surface & 250\,\micron\ & 350\,\micron\ & 500\,\micron\ & 870\,\micron \\
              & Density & ($10.4\pm2.3$)& ($5.4\pm1.6$) & ($2.4\pm0.6$) & ($0.47\pm0.1$)\\
    \hline
    SIMPLE & 48.3 & $4.33\pm0.87$ & $3.71\pm0.51$ & 1.75$\pm$0.30 & $0.13\pm0.01$ \\
    {\bf FIDEL+VLA} & {\bf 13.1} & $\mathbf{8.25\pm0.46}$ & $\mathbf{4.60\pm0.27}$ &
    $\mathbf{2.09\pm0.16}$ & $\mathbf{0.17\pm0.01}$\\
    FIDEL & 12.6 & $8.16\pm0.45$ & $4.56\pm0.26$ & $2.02\pm0.16$ & $0.17\pm0.01$ \\
    VLA only & 0.54 & $0.35\pm0.10$ & $0.16\pm0.06$ & $0.15\pm0.03$ & $0.004\pm0.001$ \\
    \hline
  \end{tabular}
  \label{tab:stacks}
\end{table*}

We use the method of \citet{marsden2009} to measure the submm surface
brightnesses of the complete SIMPLE catalogue, as well as the
24\,\micron\ (FIDEL), 1.4\,GHz (VLA), and combined 24\,\micron\ and
1.4\,GHz (FIDEL+VLA) subsets of SIMPLE in
Table~\ref{tab:stacks}. These values are compared to the absolute
measurements of \citet{fixsen1998}.

First, we find that the stack on the {\em complete} (and very
high-surface density) SIMPLE catalogue results in lower values than
the lower-surface density FIDEL and FIDEL+VLA catalogues. Even though
we know that this IRAC catalogue contains sources that are not strong
submm emitters (e.g., stars), following the arguments in Section~3.1
of \citet{marsden2009} we would not expect these extraneous sources to
bias the result unless they were somehow correlated on the sky (they
are simply an additional source of noise). Therefore, we believe that
this peculiar result simply demonstrates the presence of an
un-characterized systematic in the SIMPLE spatial distribution (i.e.,
they are anti-correlated with submm emission).  We note that
\citet{marsden2009} also found unexpected behaviour when stacking on a
high-surface density optical catalogue.

Second, we find that stacks on the FIDEL catalogue yield results
consistent with those quoted in \citet{marsden2009} (our values are
slightly lower, as expected, because we have not corrected for
completeness in FIDEL at faint flux densities as they did). Within
the uncertainties, this catalogue is consistent with reproducing the
entire CIB at 250--500\,\micron, but recovers less than half of the
CIB at 870\,\micron, consistent with the previous stacking
measurements noted earlier.

Finally, we see that the addition of 1.4\,GHz detected, but
24\,\micron\ un-detected sources (``VLA only''), to FIDEL
systematically increases the value of the stack slightly; i.e. the
stack on FIDEL+VLA is greater than the stack on FIDEL by itself. We
use this fact as our primary justification for including the
additional faint radio sources.

While this catalogue arguably contains the majority of the sources
that produce significant submm flux in the BLAST bands, the catalogue
is certainly missing a significant portion of the 870\,\micron\
emitters. Unfortunately, we do not know whether these missing sources
are, on average, the same sources that produce the 870\,\micron\
peaks, or fainter sources that do not typically contribute to the
brighter peaks. We note that the LESS identification paper, Biggs et
al. submitted, also fails to identify counterparts for a number of the
870\,\micron\ sources using similar matching data.  We will attempt to
address the impact of this shortcoming in Section~\ref{sec:missing}.

\section{Cross-identifications}
\label{sec:id}

Given the poor positional uncertainties inherent to the current
generation of submillimetre waveband surveys (typically several
arcsec), there are usually many potential optical counterparts for
each SMG. Thus it has usually been necessary to search for
identifications in lower surface density catalogues at radio and
mid-IR wavelengths. Also, as mentioned in Section~\ref{sec:matchcat},
emission in these two wavebands are expected to be physically
correlated with the submm--FIR emission; there is not a similarly
strong correlation for optically-selected sources. The method usually
adopted is to estimate `$P$' Poisson chance alignment probabilities
\citep{downes1986} in order to exclude the least likely candidates
(although this does not provide a probability that a given source `is
{\em the} counterpart'). This calculation only uses the source counts
of the matching catalogue and an empirically derived maximum search
radius. The expected distribution of offsets for true matches is {\em
  not} used, except perhaps to set the search radius.

In this paper we take a different approach, using a `likelihood ratio'
(LR) formalism. The basic idea attempts to answer the following
question: given a potential counterpart to the submm peak, what is the
relative likelihood that it could be a real counterpart given its
measured properties (e.g., radial offset, flux density, colour etc.),
versus the probability that it is a chance interloper (given the
background source counts as a function of measured properties)?
Versions of this technique have been used in a variety of contexts
\citep[e.g.,][]{sutherland1992,mann1997,rutledge2000}. Clearly the LR
can explicitly use information such as the expected positional
uncertainties, whereas the $P$ calculation does not (although the
maximum search radius implicitly incorporates some of this
information). It should be noted that, in the past, colour-based
priors have been used to cull matching catalogues, and hence reduce
the surface density of spurious sources, before finding counterparts
using $P$ statistics \citep[e.g][Biggs et
al. submitted]{pope2006,yun2008}. By itself the LR can still only be
used to {\em rank} potential counterparts (similar to the problem with
$P$ statistics), but we attempt to establish both the false
identification rates, and identification completeness rates for given
absolute LR thresholds in each band.

It appears that the reason the LR formalism has not been used for
submm identification work in the past is due to its reliance on prior
information which historically has been extremely difficult to
estimate \citep[see discussion in][]{serjeant2003,clements2004}. With
individual surveys covering $\ll$1\,deg$^2$ and having typically fewer
than several tens of peaks per field, and very low \snr, the precise
positional uncertainty distribution is unknown. However, the BLAST and
LESS data, combined with the deep radio and {\em Spitzer} mid-IR data
covering the ECDF-S, enable estimates of priors (such as the radial
offset distribution of counterparts) with sufficient precision to
produce useful results. Forthcoming {\em Herschel}, SCUBA-2, and LMT
surveys will have better angular resolution, depth, and cover
substantially larger areas, so that the methods employed here will
also be fruitful (however, note that these future surveys will depend
on radio and mid/near-IR data of comparable area and depth for
identifying counterparts, and they do not presently exist). In
Section~\ref{sec:lr_priors} we summarize the LR method and the priors
that we have developed. In Section~\ref{sec:potential} we then use
this method to identify a list of potential counterparts from our
matching catalogue to the submm peaks.

\subsection{Likelihood Ratios and priors}
\label{sec:lr_priors}

As described in Section~\ref{sec:obs} we have produced a matching
catalogue based on sources from SIMPLE (IRAC) that exhibit either
mid-infrared ({\em Spitzer} MIPS photometry from FIDEL), or VLA
1.4\,GHz emission. For the purpose of identifying counterparts we have
focussed on three features of this catalogue in addition to positions:
24\,\micron\ and 1.4\,GHz flux densities (when available) --- both of
which have commonly been used in the calculation of $P$ values; and
the $c\equiv[3.6]-[4.5]$ IRAC colour, which is sensitive to redshift
\citep[e.g.,][]{simpson1999,sawicki2002,pope2006,yun2008,devlin2009}
as it traces the peak of the rest-frame stellar bump.

We fully develop the LR formulae and priors in Appendix~\ref{sec:lr},
but provide the main results here. Given the flux densities $S$ (at
24\,\micron\ and 1.4\,GHz in this case), the IRAC colour $c$, and the
distance $r$ to the $j$th matching catalogue source from the $i$th
submm peak for which we are searching for a counterpart, we calculate
the LR:
\begin{equation}
  L_{i,j} =
  \frac{q(S_j,c_j) e^{-r_{i,j}^2 / 2\sigma^2}}{2\pi \sigma^2 \rho(S_j,c_j)}.
\label{eq:lr}
\end{equation}
Here $\sigma$ characterizes a radially-symmetric Gaussian positional
uncertainty, $q(S,c)$ is the prior distribution for flux densities and
colours of matches to submm peaks, and $\rho(S,c)$ is the background
source distribution.

All of the priors, $\sigma$, $q(S,c)$ and $\rho(S,c)$, have been
estimated directly from the data by counting sources in the matching
catalogue as a function of each property around submm peak positions
(of order $\sim$60 in each submm band), and comparing to the counts
for the entire matching catalogue over the full survey area. We find
that sources in the matching catalogue near submm peaks (i.e.,
potential counterparts): have radial offset distributions that are
proportional to the instrumental PSF sizes, as expected; tend to have
24\,\micron\ and 1.4\,GHz flux densities that are brighter than
typical sources in the catalogue; and have redder $[3.6]-[4.6]$ IRAC
colours than typical sources in the catalogue, particularly at 500 and
870\,\micron. We also find that, on average, there are {\em multiple}
extra sources from the matching catalogue near each submm peak, an
excess $E$, of: 3.2 at 250\,\micron; 3.4 at 350\,\micron; 3.7 at
500\,\micron; and 2.2 at 870\,\micron. This result demonstrates that
the submm data are highly confused, but clustering in the matching
catalogue may also have an impact.

There could be submm-faint matching catalogue members that cluster
around a smaller number of submm-bright sources which produce the
observed submm peaks. However, we note that $E$ is strongly correlated
with the PSF size; if this clustering scenario were the dominant
effect we would expect the same integrated excess regardless of beam
size. We will further explore the potential impact of clustering in
Section~\ref{sec:clustering}.

We believe that our excess counting procedure yields a high-\snr\
measurement of the positional offset distribution, since we need only
bin measurements for the submm peaks (in each band) along one
coordinate, $r$, which we then fit with a simple one-parameter model,
$f(r) =
(r/\sigma_\mathrm{r}^2)e^{-r^2/2\sigma_\mathrm{r}^2}$. However, there
is an implicit assumption that the sources (both in the submm and
matching catalogues) are spatially un-clustered. In
Table~\ref{tab:priors} we summarize the matching catalogue selection
function, (described in Section~\ref{sec:matchcat}), as well as the
basic effect of the priors used in the LR calculation.

\begin{table}
  \caption{Summary of the matching catalogue selection function and priors for
    Likelihood Ratios. Flux cuts for the matching catalogue are only
    approximate (see Section~\ref{sec:matchcat}). No hard cutoffs are used as
    priors, there are only weights applied as a function of the distance
    between the submm peak and proposed counterparts, the 24\,\micron\ and
    1.4\,GHz flux densities, and $[3.6]-[4.5]$ near-IR colours (see
    Appendix~\ref{sec:lr} and Figs.~\ref{fig:r_excess}-\ref{fig:chist})}
  \vspace{0.2cm}
  \centering
  \begin{tabular}{c|l}
    \hline
    Catalogue selection & Notes \\
    \hline
    $S_{24\micron} \gsim 13\,\mu$Jy      & {\em Spitzer}/MIPS, FIDEL \\
                                        & (uses SIMPLE IRAC positions as a \\
                                        & prior, see below) \\
    & \\
    \multicolumn{2}{c}{{\em or}} \\
    & \\
    $S_{1.4\mathrm{GHz}} \gsim 20\,\mu$Jy & VLA \\
    \multicolumn{2}{c}{{\em and}} \\
    $S_{3.6\micron} \gsim 5\,\mu$Jy       & {\em Spitzer}/IRAC, SIMPLE \\
    $S_{4.5\micron} \gsim 5\,\mu$Jy       & '' \\
    \hline
    Prior &  \\
    \hline
    $f(r)$                   & Favour nearby counterparts \\
    $q(S_{24\micron)}$         & Favour brighter $S_{24\micron}$ \\
    $q(S_{1.4\mathrm{GHz}})$   & Favour brighter $S_{1.4\mathrm{GHz}}$ \\
    $q(c)$                   & Favour redder $[3.6]-[4.5]$ colour \\
    \hline

  \end{tabular}
  \label{tab:priors}
\end{table}

Ideally we would bin $q(S,c)$ and $\rho(S,c)$ along all three axes
simultaneously (two flux densities and a colour), but in practice this
is not feasible given the numbers of submm peaks that we have to work
with. We therefore handle the priors {\em independently}, e.g.,
estimating $q(S,c) \simeq q(S_{24}) q(S_\mathrm{r}) q(c)$, and
$\rho(S,c) \simeq \rho(S_{24}) \rho(S_\mathrm{r}) \rho(c)$. This
assumption certainly introduces a bias, since in practice these
properties are correlated \citep[for example, see Fig.~7 in][showing
the correlation between 24\,\micron\ and 1.4\,GHz flux densities for
counterparts to bright BLAST sources]{dye2009}. These correlations are
simply a reflection of the fact that submm galaxies have a particular
range of SED {\em shapes} (including the radio and near-IR discussed
here). We have not attempted to measure these shapes, but an
alternative method might consider a range of plausible,
physically-motivated SEDs as part of the identification process
\citep[e.g.][]{roseboom2009}. Instead, we have chosen to compensate
for the bias by using Monte Carlo simulations to estimate a threshold
in the LR that produces a false identification rate of 10\%.

We establish the appropriate level by choosing 10,000 random positions
in the field, and calculating the LR of matching catalogue sources
around those positions. In each submm band we choose a threshold in LR
that rejects 90\% of the matched sources from this random sample. For
convenience, we then calculate normalized LRs for counterparts to
submm peaks in each band by dividing the raw LR from Eq.~\ref{eq:lr}
by these thresholds. We therefore only consider potential counterparts
to submm peaks those objects for which the normalized LR is greater
than 1. While this choice of normalization is arbitrary, the relative
LRs for potential counterparts are meaningful (i.e., a LR of 2
indicates double the relative likelihood that it is real compared to a
LR of 1). We also estimate that a proposed counterpart with a
normalized LR of 1 is about three times more likely to be real than
spurious.

Finally, note that the LR only gives the relative chance that a
proposed counterpart is real. We have {\em not} attempted to derive an
absolute reliability, $R$, the probability that a proposed counterpart
is correct, as in \citet{sutherland1992}, since their formulation
assumes that there is a {\em single} counterpart to each peak (see
Appendix~\ref{sec:lr}). Instead we rely on our threshold in LR to
ensure our spurious fraction of 10\% for the ensemble of proposed
counterparts.

In summary, while our simplifying assumption that the flux densities
and colour are independent of one another is incorrect, we have
established a cut on LR that will restrict the number of false
positive identifications in the matching catalogue to around 10\%
(with respect to the number of submm peaks). This idea of setting a
threshold to reject unrelated sources is similar to adopting a cut on
$P$, although here we have included more prior information to improve
the efficiency.  However, we warn the reader that the calibration of
the LR and $P$ are both tied to our assumption that clustering in the
matching catalogue around submm positions is a negligible effect.

\subsection{Potential counterparts}
\label{sec:potential}

We search for counterparts to all of the submm peaks independently in
all four submm bands out to a search radius of 60\,arcsec. In practice
we could use a much larger search radius, but, since the expected
radial probability density, $f(r)$, rolls off to $\sim0$ before this
radius, even at 500\,\micron, there is no difference in the list of
counterparts with normalized $\mathrm{LR}>1$ if it is increased. We
identify the following numbers of potential counterparts in each band:
52 at 250\,\micron; 50 at 350\,\micron; 31 at 500\,\micron; and 66 at
850\,\micron.

As noted in the previous section, we expect to encounter several
counterparts, on average, for each submm peak. However, with the
threshold LR chosen, we only detect a fraction of the total number of
counterparts expected: 22\% at 250\,\micron, 19\% at 350\,\micron,
13\% at 500\,\micron, and 32\% at 870\,\micron. It is probably the
case that this low identification rate is simply due to the quality of
the submm data.

For comparison, we have also identified potential counterparts using
$P$ statistics based on the 24\,\micron\ and 1.4\,GHz flux
densities. We counted the numbers of IDs with $P<0.1$ using both of
these sub-catalogues around random positions (as a control, similar to
the method employed to normalize the LR), and around submm positions
\citep[similar to Fig.~3 in][]{chapin2009b}. We discovered that, while
the radio catalogue was well-behaved (a 10\% false ID rate was
obtained for random positions), we obtained too many false IDs with
the 24\,\micron\ catalogue. This result is probably demonstrating that
the 24\,\micron\ catalogue is slightly clustered. We therefore tuned
the cut on $P_{24}$ to 0.08 to obtain the desired 10\% rate using
random positions. The search radii that we used were
1.5\,$\sigma_\mathrm{r}$ (the single parameter in the radial offset
distributions); these search radii are roughly comparable to those
adopted in previous studies of BLAST peaks
\citep{dye2009,ivison2009,dunlop2010}, and encompass approximately
68\% of the true counterparts inferred from the excess counting
statistics. Using $P_{24}<0.08$ we would find 45, 31, 12 and 33
potential counterparts, and using $P_\mathrm{r}<0.10$ we would find
51, 35, 17 and 44, at 250, 350, 500, and 870\,\micron,
respectively. Clearly there is a significant (although modest)
improvement using our LR calculation over $P$ statistics, particularly
at the longer wavelengths where the $[3.6] - [4.5]$ IRAC colour is a
good discriminator of redshift.

As our goal is to study the properties of BLAST-selected peaks, we
identify all of the unique sources from our matching catalogue that
potentially produce the observed 250, 350 and 500\,\micron\
emission. However, we only consider matches to the LESS peaks that lie
within a 2\,$\sigma$ search radius of {\em any} 3.75$\sigma$ BLAST
peak (combining both the BLAST and LABOCA positional uncertainties in
quadrature). With this search radius we expect to find 95\% of the
real matches, and given the surface density of the LESS catalogue, we
will have a 10\% spurious ID rate (the same as that adopted for the
cut on LR). In total, 42 out of the 81 LESS peaks that land within the
survey region are associated with BLAST sources, and 36 of them are
identified in the matching catalogue using LRs. The remaining 6 LESS
peaks are still included as potential matches for BLAST peaks,
although they may themselves also be blends of multiple galaxies
within the LABOCA beam, and none of them have associated radio, mid-
or near-IR flux densities with which to conduct further analysis.

The end result of our matching procedure is a list of 118 unique
sources that are believed to contribute to the submm peaks in all four
bands (including the 6 that are simply LESS 870\,\micron\ sources with
no matches at other wavelengths). The coordinates of these sources,
and the submm peaks to which they were matched, are given in
Table~\ref{tab:match}. Postage stamps showing the locations of the
matches in relation to the submm positions are shown in
Fig.~\ref{fig:stamps1}. For each ID, in addition to the LR, we also
provide $P$ values for matches to the 24\,\micron\ and 1.4\,GHz radio
catalogues, using search radii $1.5\sigma_\mathrm{r}$ (see
Table~\ref{tab:pos}).

Note that the third columns in the lists of submm peaks,
Tables~\ref{tab:250cat}--\ref{tab:870cat}, give references to the
individual identifications from the matching catalogue in
Table~\ref{tab:match} (again, within a 2\,$\sigma$ search
radius). Since these sources may have been matched to submm peaks in
{\em any} band, the total numbers of counterparts listed here exceed
the numbers of matches found independently in each band. In other
words, we can now see how the simultaneous observations in the four
submm bands have helped one another: 77\%, 72\%, 76\%, and 86\% of the
250, 350, 500, and 870\,\micron\ peaks have at least one potential
counterpart identified. In many cases the same sources appear in
several bands, and the higher-resolution observations have enabled
counterpart identifications where the lower-resolution observations
failed. However, counting the total number of proposed counterparts,
we find 64, 65, 59, and 51 sources at 250, 350, 500, and
870\,\micron\footnote{Note that at 870\,\micron\ 57 matches are
  indicated, but 6 of those are the LESS sources themselves, leaving
  51. We have only searched for counterparts to the 42 LESS peaks that
  appear to be associated with BLAST peaks, so the average
  counterparts per peak is actually 51/42=1.21}. These numbers still
fall far short of the total numbers expected, especially in the BLAST
bands. We will attempt to quantify the impact of the missing matches
in Section~\ref{sec:missing}.

\section{Submillimetre Spectral Energy Distributions}
\label{sec:sed}

In this section we re-measure the submm flux densities from the
combined list (i.e., selected in {\em any} BLAST band) at the
positions of their proposed counterparts in the raw submm maps. Using
simulations, we explore the impact of confusion, missing sources, and
clustering on these measurements. Finally, we fit these observed-frame
SEDs with simple isothermal models, and measure the observed number
counts in our catalogue as a function of limiting flux
density. Together, these calculations allow us to explore bias and
completeness effects for our sample.

\subsection{Submm photometry}
\label{sec:photometry}

Under the assumption that the potential counterparts identified in the
previous section produce the observed submm emission (and are not
simply other galaxies clustered around the submm peaks), we return to
the four submm maps and perform simultaneous fits of point sources at
all 118 locations to measure their flux densities. This procedure is
expected to reduce the Eddington-like bias, or {\em flux-boosting},
inherent to low-\snr\ submm surveys \citep{coppin2005} in two
ways. First, since peaks are initially selected in three different
bands, the component of bias introduced by instrumental noise is
reduced (as it is independent in each map) --- submm peaks
preferentially detected on positive noise excursions in one map will
not necessarily also land on positive noise excursions in other
maps. Second, since we allow for the possibility of {\em multiple}
counterparts to each submm peak (a hypothesis confirmed by the excess
counting statistics described in Section~\ref{sec:pos}), the
simultaneous fit can, to some extent, de-blend some of the brighter
confused sources (confusion itself also contributes to Eddington bias,
and unlike the instrumental noise, is correlated between the submm
bands).

The fit is performed by modeling the emission of the counterparts as
the submm PSFs scaled by their unknown submm flux densities $S_i$ at
the locations from the matching catalogue. Under the assumption that
the instrumental noise in our submm maps is un-correlated from one map
pixel to the next (a reasonable assumption for the raw, un-smoothed
maps on the scale of the PSF), there is a simple maximum-likelihood
solution for the $S_i$ that takes into account the correlations that
arise in cases where multiple sources overlap within a PSF footprint
--- we follow the derivation provided in Appendix~A of
\citet{scott2002}. This solution only uses the maps, instrumental
noise estimates, and source positions. No preferential weight is given
to counterparts with larger LRs, so the noise in the answer only
depends on how well the map is fit using our simple parameterization.

A downfall of this approach is that we ignore the additional component
of confusion noise from un-identified sources that is correlated on
the scale of the PSF. We therefore estimate the total noise by adding
the confusion noise for the simulated raw maps from
Table~\ref{tab:matched} to the variances for each source flux density,
$\sigma_i^2 = \mathrm{Cov}(S_i,S_i)$. This operation should give good
estimates for isolated sources, but we warn that it produces an
under-estimate of the variances for the most confused sources. In
Section~\ref{sec:missing} we will test the validity of our estimated
uncertainties using simulated data sets.

For isolated sources, the recovered flux density is identical to that
obtained from the PSF-smoothed map at the location of its counterpart,
and its value is un-correlated with the measured flux densities for
all other sources in the map. However, for blended sources, the total
flux density in the map is divided among the multiple counterparts,
and there are non-negligible covariances $\mathrm{Cov}(S_i,S_j)$ for
all sources $i,j$ that lie roughly within a FWHM of each other. We
therefore evaluate the full expression for the covariances between
measured flux densities, i.e., using the off-diagonals of Eq.~A11 in
\citet{scott2002}.

The individual observed-frame submm SEDs based on these measurements
are given in Table~\ref{tab:stuff}. We also display the submm SEDs in
Fig.~\ref{fig:seds}, along with the 1.4\,GHz, MIPS 24 and 70\,\micron\
flux densities (when available), and the photometry from the 3.6, 4.5,
5.8 and 8.0\,\micron\ IRAC bands (all sources) from the matching
catalogue.

\subsection{Confusion}
\label{sec:confusion}

Our method assumes that the proposed counterparts to the submm peaks
comprise {\em all} of the galaxies that contribute significant submm
emission. However, we have made two fairly arbitrary choices: we
select only peaks that have a significance of 3.75\,$\sigma$ over the
instrumental noise levels; and we only consider sources with a 10\%
threshold false association rate from the LR analysis. To assess the
impact these choices have on the measured flux densities and
completeness, we have generated simulated maps drawing sources from
the measured number counts in the BLAST bands from
\citet{patanchon2009}, and then adding appropriately scaled white
noise to mimic the estimated instrumental noise levels.

First, we identify individual peaks in the simulated BLAST maps above
the same 3.75$\sigma$ \snr\ threshold as for the real data using the
same matched filter. Given the sizes of the BLAST beams, and the
surface density of the sources, {\em every} location in the filtered
maps has a contribution from multiple submm galaxies (even if they are
extremely faint). For each 3.75$\sigma$ peak we therefore identify
{\em the single} source that makes the {\em largest} contribution to
the observed flux density from the input catalogue at that location
--- considering the PSFs in each band, input source brightnesses and
their distances from the peaks in the filtered maps. In this way a
faint source will only be identified provided that it is very close to
the submm peak in question, and exceeds the brightnesses of the tails
of all the more distant sources in the catalogue.  We then re-run this
procedure on 100 independent realizations of the maps at each
wavelength to fully characterize the scatter in the results. Each time
we randomly select a different differential counts distribution from
the actual Markov Chains produced from the the $P(D)$ fits in
\citet{patanchon2009}.

We find that these brightest sources statistically contribute
fractions $0.73^{+0.57}_{-0.57}$, $0.57^{+0.56}_{-0.22}$, and
$0.58^{+0.65}_{-0.25}$ (means and 95\% confidence intervals) of the
peak flux densities in the filtered maps at 250, 350 and
500\,\micron. Note that the fraction can be greater than one since the
simulation is noisy (the source may have landed on a large positive
noise excursion). This test shows us that, using a 3.75$\sigma$ cut,
submm peaks are usually a significant blend of two or more individual
sources, although there is an incredibly large scatter; a peak may
have {\em many contributors}, but it is also true, in some cases, that
a peak is dominated by a single bright source. This result is broadly
consistent with a similar set of simulations used by
\citet{moncelsi2010} to correct BLAST peak flux density biases. Also,
we have noted that the radial distribution of the brightest sources
identified for each submm peak (not shown) broadly resemble the radial
distributions estimated for the LR analysis (Eq.~\ref{eq:pos}).

There are numerous obvious examples of blends of sources from the
matching catalogue in the real submm maps.  We have flagged 42/118 of
the most extreme cases with the letter `C' in Fig.~\ref{fig:seds}
indicating that they are confused to the point that the submm
photometry cannot be used reliably, particularly in the BLAST
channels. For example, sources 2, 3, 4 and 5 comprise one of the most
confused regions of BLAST emission in the entire ECDF-S, as can
clearly be seen in the postage stamps (Fig.~\ref{fig:stamps1}). By
comparison, the superior LESS resolution can nearly resolve the entire
feature into a string of individual peaks.  The inferred flux
densities at 250--500\,\micron\ therefore have strong {\em
  anti-correlations}, since the emission from those four sources must
sum to the total integrated flux density of the feature. In fact, the
maximum-likelihood solution we have adopted can even allow negative
values, a problem which occurs in a number of the most confused
examples.  The low \snr\ of the submm maps, combined with confusion
from fainter submm sources, and the close proximity of the
counterparts, has resulted in flux densities with drastically
under-estimated error bars in cases such as sources 2--5. In contrast,
sources 61 and 62 are an example where the joint-fit at the positions
of two nearby counterparts has recovered plausible flux densities in
all the submm bands \citep[this is a low-redshift interacting pair
first discussed in][]{dunlop2010}. In this case, the BLAST \snr\ is
much higher, and the two potential counterparts have sufficient
separation to disentangle them.

Next, we investigate the completeness by counting the number of
sources above different flux limits in the input catalogue that are
recovered in the match-filtered source list (again, considering only
the single brightest submm galaxies that contribute to the observed
brightness). The recovered percentages are: 50\% above 30\,mJy, and
90\% above 60\,mJy at 250\,\micron; 50\% above 15\,mJy, and 90\% above
45\,mJy at 350\,\micron; and 50\% above 10\,mJy, and 90\% above
25\,mJy at 500\,\micron.

\subsection{Missing counterparts}
\label{sec:missing}

Another significant problem that we face is the issue of missing
source matches to the submm peaks. As mentioned in
Section~\ref{sec:stacks}, the matching catalogue is probably missing a
significant fraction of the 870\,\micron\ emitters, and possibly a
smaller fraction of the 250--500\,\micron\ emitters. More importantly,
in Section~\ref{sec:potential} we were not able to identify a large
portion of the counterparts to the submm peaks expected in the
matching catalogue. Therefore, the measured flux densities for the
sources that {\em were} correctly identified are noisier, and perhaps
biased, due to these missing sources in the fitting procedure. We
attempted to account for this noise in Section~\ref{sec:photometry} by
adding the RMS of simulated, instrumental noise-free maps in
quadrature to the noise returned from the fitting procedure. Here we
will use simulations to estimate how biased and noisy this procedure
is.

We use the simulated maps from the previous section that were
generated using realizations of sources drawn from the measured counts
distributions for the real maps, and assigning them random
positions. We now also simulate 870\,\micron\ maps using the best-fit
counts as reported in \citet{weiss2009}. For each realization we add a
random 20\% uncertainty to the total number of sources drawn from the
distribution to approximate the error indicated in the faintest bin of
their cumulative catalogue-based counts.

To approximate the source identification procedure, we first produce
3.75$\sigma$ peak lists, and then identify the brightest sources from
the input catalogue that contributed to each of the peaks. We include
as matches only those sources from the input catalogue that contribute
more than a threshold fraction of the observed peak, chosen to produce
the same average number of matches per peak as we obtained using the
real maps and matching catalogue. In this way we associate a range of
matches from the input catalogue (with known flux densities) to each
observed peak, with a similar surface density of matches as for the
real data. However, we stress that this is only a {\em plausible}
simulation, since there is no guarantee that the matches proposed for
the real data are in fact the brightest contributors as is the case
for this simulation. A significantly more complicated simulation could
be undertaken in which we: generate galaxies with full
radio--submm--IR SEDs (that are consistent with the true surface
densities of sources in each band); create a matching catalogue (i.e.,
simulating radio, mid- and near-IR catalogues with realistic noise);
repeat the process of estimating priors; and finally use LRs to
propose matches for each peak. However, we felt that such a simulation
was beyond the scope of this paper, and opted instead for the simpler
approach that captures most of the necessary ingredients.

With simulated maps, and lists of proposed matches to each peak, we
repeat the maximum-likelihood fitting operation of
Section~\ref{sec:photometry}, in all of the submm bands. Since we know
the true flux densities for each of the matched sources,
$S_\mathrm{t}$, we are able to directly probe the scatter of the
observed flux densities, $R \equiv (S_\mathrm{t} -
S_\mathrm{o})/\sigma_\mathrm{o}$, where $S_\mathrm{o}$ is the inferred
flux density, and $\sigma_\mathrm{o}$ its uncertainty derived from the
fitting process (which accounts for instrumental noise and overlap
with other nearby sources), and then adding the additional confusion
noise for the simulated raw maps from Table~\ref{tab:matched} in
quadrature. The expectation is that this distribution has a mean of
zero, and a standard deviation of one. Combining the results for all
100 simulations in each band, we measure the following mean values and
standard deviations for $R$: $0.17\pm1.23$ at 250\,\micron;
$0.19\pm1.20$ at 350\,\micron; $0.25\pm1.20$ at 500\,\micron; and
$-0.05\pm0.97$ at 870\,\micron.

These results suggest that there is only a small upward bias in the
measured flux densities across the BLAST bands, of order
$\sigma_\mathrm{o}/4$, and negligible bias at 870\,\micron. In
absolute terms, we can scale this bias to approximate flux density
units by multiplying by the mean values of $\sigma_\mathrm{o}$ in each
band: $+3.2$\,mJy at 250\,\micron; $+3.2$\,mJy at 350\,\micron;
$+3.5$\,mJy at 500\,\micron; and $-0.05$\,mJy at 870\,\micron. We also
find that our inclusion of the confusion noise only under-estimates
the noise in the BLAST bands by at most $\sim$23\% (as discussed in
Section~\ref{sec:photometry} our estimate is a lower-limit on the
noise for any particular source).

Finally, as noted at the end of Section~\ref{sec:stacks}, our matching
catalogue is potentially very incomplete at 870\,\micron. However, the
results of our 870\,\micron\ simulation suggest that the missing
sources are so faint that they do not contribute significantly to the
submm peaks we are analyzing. Again, for this to be true, the sources
that {\em are} in the matching catalogue need to account for the bulk
of the brightest 870\,\micron\ emitters in the sky.

\subsection{The impact of clustering}
\label{sec:clustering}

In addition to the confusion arising from a uniformly distributed
population of submm emitters (i.e., chance superpositions of objects
at different redshifts, as in Section~\ref{sec:confusion}), the submm
emitters themselves, and/or the matching catalogue, could also be
clustered. There are three distinct cases that would affect the
results in this paper worth considering:

\begin{enumerate}

\item The low-resolution submm peak could be resolved into multiple
  components at approximately the same redshift. This is plausible,
  since it is known that about 10\% of SCUBA sources are associated
  with double radio sources
  \citep[e.g.,][]{ivison2002,chapman2005,pope2006,ivison2007}, and
  \citet{vaisanen2010} show that many of their 180\,\micron-selected
  sources are blends of multiple galaxies at $z<0.3$. We also note
  that a significant clustering signal on angular scales $<1$\,arcmin
  has been measured for the LESS catalogue in excess of the Poisson
  expectation \citep{weiss2009}.

\item The submm peaks could instead be dominated primarily by single
  matching catalogue sources with a lower surface density, in which
  case the extra sources counted in the radial excess plots may be
  spatially correlated, but are not otherwise directly associated with
  the submm emitter (e.g., galaxies with lower star-formation rates in
  the same structures).

\item Massive foreground structures could enhance the brightnesses of
  background galaxies through lensing. This scenario would have a
  similar effect to the previous one; there could be a number of
  additional foreground galaxies near the positions of submm peaks,
  even though they do not themselves contribute significantly to the
  submm flux.

\end{enumerate}

The impact of these clustering scenarios is not easy to assess
accurately with simulations because it is simply unknown how {\em all}
galaxy populations cluster throughout the history of the Universe;
indeed, this is one of the major outstanding questions in modern
Cosmology. While there are a number of ways to simulate clustered
galaxy populations, such as the Halo model \citep[][]{mo1996} used to
fit BLAST data in \citet{viero2009}, or more complicated
semi-analytical models \citep[e.g.][]{baugh2005}, it is beyond the
scope of this paper to test the full range of models that are
plausible. On the other hand, there is clear evidence that some
clustering is required to explain both the BLAST and LESS submm
maps. Since we have argued that the bulk of the submm emission is
produced by sources in the matching catalogue, a reasonable approach
is to incorporate these real, and clustered positions in our simulated
maps to see what impact they have.

We repeat the simulations of the previous sections, drawing sources
from the measured counts distributions, but now randomly assigning
them positions from the real matching catalogue. With this limitation,
we can only draw the same number of sources as in the real matching
catalogue, so we restrict ourselves to a flux-limited sample that
results in the same surface density. We then produce maps, and
identify 3.75$\sigma$ peaks as before.

To assess the relative impact of clustering, we also produce
simulations with the same surface density of input sources, but using
uniformly distributed positions. In both cases we measure the radial
excess counts in the matching catalogue around peaks, with respect to
the entire catalogue (as in Figure~\ref{fig:r_excess}). We find that
the clustered simulations yield systematically larger excesses per
submm peak, by factors 1.07, 1.02, 1.02, and 1.24 at 250, 350, 500,
and 870\,\micron, respectively. The trend is for this excess to be
larger for the {\em highest-resolution} measurements, which might be
expected given the large 2-point correlation measurements from
\citet{weiss2009} growing towards scales $<1$\,arcmin. However,
considering the scatter in the 100 simulations, we also find that the
uncertainties in the excess measurements are 21\%, 14\%, 19\%, and
16\% in the four bands. In other words, when testing the hypothesis
that the difference observed for the clustered simulations are
significant compared to the un-clustered simulations, only the excess
at 870\,\micron\ is marginally significant (a 1.5$\sigma$ outlier).

However, we warn that our simulation is only testing a particularly
weak form of clustering. It is probable that subsets of our matching
catalogue have significantly different angular clustering signatures
when compared to the catalogue as a whole, and it may be possible to
identify them through their redshifts, colours, and brightnesses. If
such populations are correlated more, or less strongly with the submm
emitters, their may be additional significant biases in our LR
approach -- effectively cases~(ii) and (iii) described above. We note
that \citet{chary2010} investigate the impact of just such an effect
on stacking analyses in this field, although the results depend
substantially on their model for the redshift evolution of submm
galaxies and their multi-wavelength SEDs.

We note additional support for the hypothesis that clustering has a
negligible impact in Section~\ref{sec:potential}: the excess matching
catalogue counts around submm positions are correlated with the
angular resolution of the submm maps. The same integrated excess would
be measured, regardless of the PSF shape, if the submm emission were
produced by single galaxies.

As a final word on this subject, what would the effect be on our
analysis if there {\em were} significant clustering in the matching
catalogue around the submm peaks?  In this case the radial excess
distributions (Fig.~\ref{fig:r_excess}), and our measurements of
$q(S,c)$ (Figs.~\ref{fig:f24hist}--\ref{fig:chist}) would trace the
properties of the clustered objects (e.g., spatial extent, colours and
brightnesses), rather than the properties of submm sources, and we
would end up with many more false positives than the target 10\%
rate. While we do not believe this is the most likely scenario based
on the arguments made in this section, and the fact that the SEDs for
the counterparts seem to match models for star-forming galaxies, only
higher-resolution studies will be able to settle this issue
unambiguously.

\subsection{Isothermal SED models and number counts}

\begin{figure}
\centering
\includegraphics[width=\linewidth]{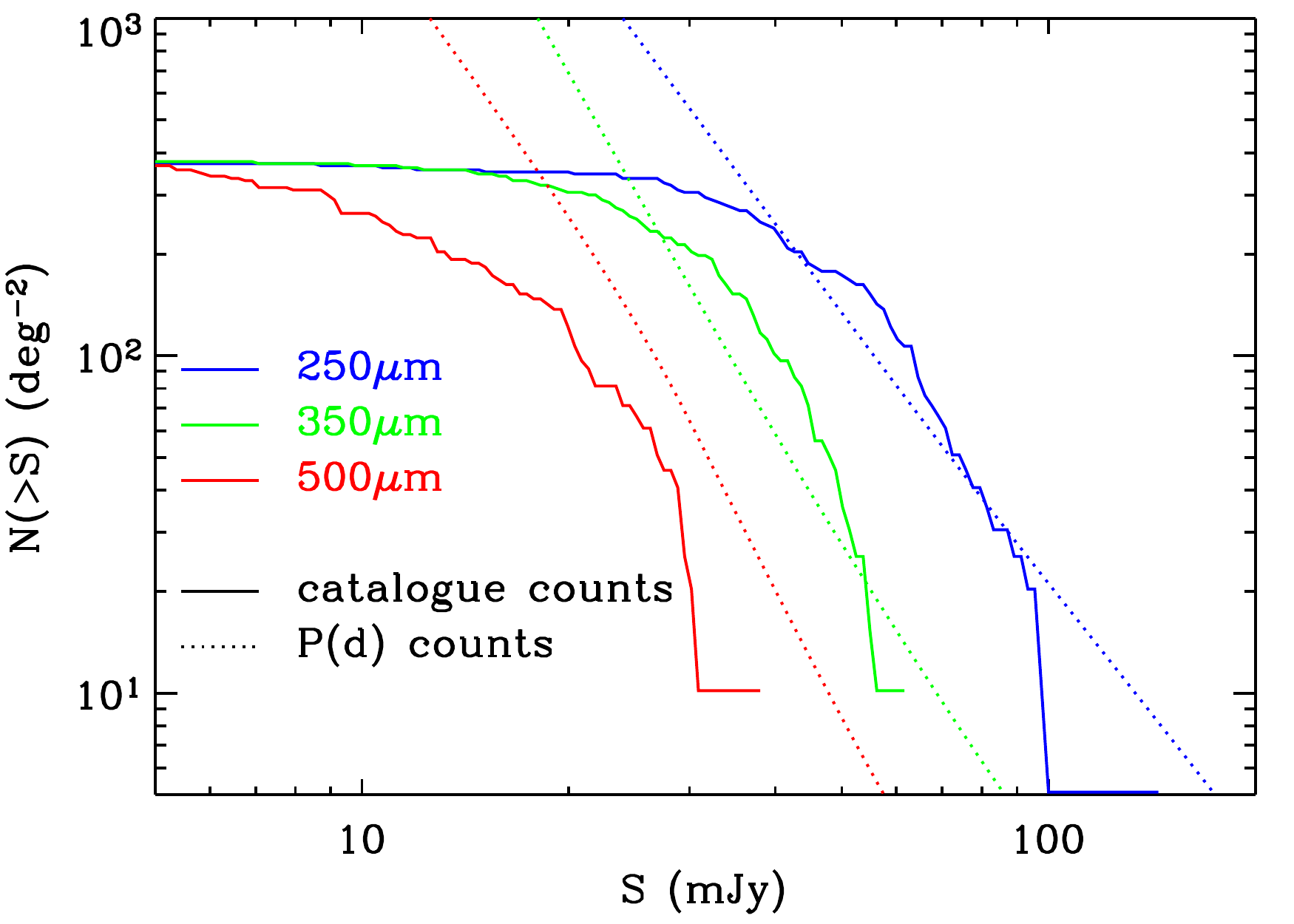}
\caption{The integral number counts for our sample (solid lines),
  compared with the total counts inferred from the $P(D)$ analysis of
  \citet{patanchon2009} (dotted lines), at 250 (blue), 350 (green)
  and 500\,\micron\ (red).}
\label{fig:counts}
\end{figure}

We fit optically-thin isothermal modified blackbody functions
(modified blackbodies henceforth), $S_\nu \propto \nu^{\beta}
B_\nu(T)$, to the new submm photometry, and use Monte Carlo
simulations to characterize the uncertainties \citep[as described in
Section~4 of][]{chapin2008}. Following the work of \citet{wiebe2009},
who examined the detailed spatially-resolved SEDs of several nearby
resolved galaxies observed with BLAST, as well as an earlier study by
\citet{klaas2001} who combined ground-based submm photometry with
FIR measurements of ULIRGs, we choose to model the submm emission
using an emissivity index of $\beta=2.0$. The only free parameters are
the amplitudes and temperatures (see Table~\ref{tab:stuff} and
Fig.~\ref{fig:seds}).

To test the completeness of our catalogue, and also to gauge the
degree to which our procedure has dealt with flux boosting in the
submm peak catalogues, we compare the integral source counts from our
sample (summing the number of sources in the catalogue above a given
flux density limit, and dividing by the survey area of
708\,arcmin$^2$) to the total counts inferred from $P(D)$ analysis in
\citet{patanchon2009}. Rather than using the submm photometry
directly, we use the flux densities from the fitted SED models
evaluated in each band. The results of this comparison are shown in
Fig.~\ref{fig:counts}. In the 250 and 350\,\micron\ channels the
catalogue counts slightly exceed the $P(D)$ counts above approximately
30 and 40\,mJy, respectively. While this excess shows there is still
some influence from boosting, its effect has been drastically reduced
when compared with the individual flux-limited BLAST catalogues
\citep[see Fig.~11 in][]{patanchon2009}. Below these levels the
catalogue is clearly incomplete at 250 and 350\,\micron\ as the counts
rapidly flatten. At 500\,\micron\ the catalogue counts have a similar
qualitative shape, but lie below the $P(D)$ counts at all flux
densities; the completeness is about 66\% above 20\,mJy. This result
is expected given the more limited success we have had in identifying
counterparts at 500\,\micron\ (see Table~\ref{tab:pos}). Note that
these approximate completeness estimates are consistent with the
simulations described at the end of Section~\ref{sec:confusion}.

\section{Discussion}
\label{sec:discussion}

\subsection{Redshifts}
\label{sec:redshift}

Many of the proposed counterparts have either optical spectroscopic or
photometric redshifts in previously published catalogues
\citep{wolf2004,wolf2008,grazian2006,brammer2008,
  rowan-robinson2008,taylor2009}. For those counterparts that do not,
the IRAC colours may be used as a crude redshift estimator. We use the
redshift catalogue from \citet{pascale2009} which combines the various
photometric and spectroscopic redshfts in the literature with the
BLAST redshift survey of \citet{eales2009}, and then we add additional
redshifts identified in more recent BLAST follow-up studies
\citep{ivison2009,dunlop2010,casey2010}. In Fig.~\ref{fig:seds}, the
redshifts are indicated with `s', `p' or `i', indicating
spectroscopic, optical photometric, or IRAC-based photometric redshift
measurements, respectively.  In total, there are 76/118 sources with
usable submm photometry. Of those, 69 have redshift estimates: 23 are
optical spectroscopic redshifts; 35 are optical photometric redshifts;
and 11 are IRAC photometric redshifts.  The full list of redshifts
that we have adopted is given in Table~\ref{tab:stuff}. We warn that
the IRAC-based photometric redshifts are highly uncertain on an
object-by-object basis, and are biased low for the higher-redshift
($z>2$) sources \citep[see Fig.~4 in][]{pascale2009}.

\begin{figure}
\centering
\includegraphics[width=\linewidth]{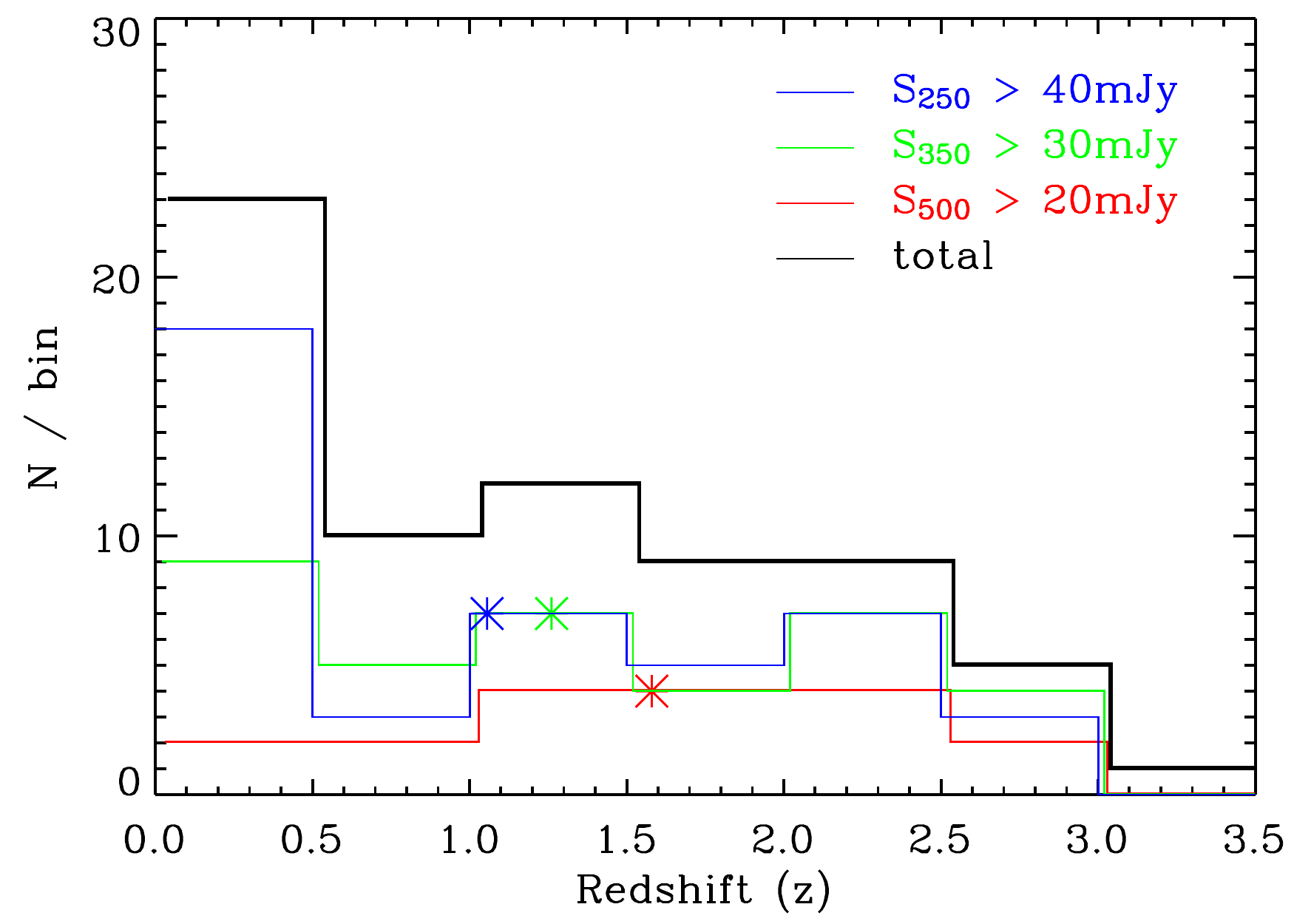}
\caption{The redshift distribution for the 69 non-confused sources
  with redshift estimates. Also shown are the redshift distributions
  for flux-limited sub-samples, $S_{250}>40$\,mJy, $S_{350}>30$\,mJy,
  and $S_{500}>20$\,mJy, chosen to correspond approximately to where
  the counts in our catalogues flatten significantly compared to the
  total population -- see Fig.~\ref{fig:counts}. The asterisks
  indicate the medians of these sub-samples.}
\label{fig:zhist}
\end{figure}

In Fig.~\ref{fig:zhist} we show the redshift distribution for these 69
sources, as well as sub-samples using flux density-limits in each of
the BLAST bands corresponding roughly to the flux densities at which
the counts begin to turn over significantly in Fig.~\ref{fig:counts}
(as a rough proxy for the point at which completeness begins to
drop). The median of the entire distribution is $z=1.1$ with an
interquartile range 0.3--1.9. Note that if we exclude the the
IRAC-based photometric redshifts the median of the entire sample
increases slightly to $z=1.3$.

This redshift distribution is qualitatively similar to the deep
250\,\micron\ survey of \citet{dunlop2010} in GOODS-S at the center of
the ECDF-S, but shows a significantly greater tail of sources beyond
$z=2$ compared to the shallower survey of \citet{dye2009}. This latter
discrepancy is probably due to a combination of increased depth in our
submm catalogue, and better completeness in the high-redshift
counterpart identifications.

The flux-limited distributions clearly show a trend from low to high
redshift with increasing wavelength in Fig.~\ref{fig:zhist}: a median
$z=1.1$ with an interquartile range 0.2--1.9 at 250\,\micron; a median
$z=1.3$ with an interquartile range 0.6--2.1 at 350\,\micron; and a
median $z=1.6$ with an interquartile range 1.3--2.3 at
500\,\micron. This trend is consistent with the results from BLAST
stacking analyses \citep{devlin2009,marsden2009,pascale2009} which
show that the CIB is produced by higher-redshift galaxies with
increasing wavelength.

With redshift estimates for our proposed counterparts, we are also
able to check whether clusters of sources contributing to single submm
peaks lie at a single redshift, or at a range of redshifts. If, as we
have asserted, most of our submm peaks are chance superpositions of
un-related galaxies, we would expect most clumps of proposed
counterparts to lie at different redshifts. On the other hand,
clusters of proposed counterparts at the same redshift would be
consistent with the clustering scenarios described in
Section~\ref{sec:clustering}.

There are cases of what appear to be groups of proposed counterparts
to single peaks at different redshifts. See, for example, sources 2--5
in Fig.~\ref{fig:stamps1}. In this particular case there also appears
to be a fifth significant source of emission in most of the submm maps
that is not identified, this being west and slightly south of the main
clump, perhaps coincident with a faint radio source that lies within
the saturated source apparent in the 3.6\,\micron\ map. The clump of
sources 13--15 is a similar example. Sources 52 and 53 form a more
well-separated example, with the former peaking in the 250\,\micron\
map, the latter in the 500\,\micron\ map, and with a double peak in
the 350\,\micron\ map. The two optical photometric redshifts appear to
be significantly different (0.1 and 0.6, respectively). Some of these
sources could also be examples of foreground galaxies lensing
background sources, although this effect is more difficult to quantify
without accurate spectroscopic redshifts for most of the sample, and
lensing models for each case.

There are also examples of peaks that could plausibly be interacting
pairs at the same redshift (case~(i) from
Section~\ref{sec:confusion}), more along the lines of radio-doubles
detected in SCUBA surveys. The clearest example is the low-redshift
interacting pair of sources 61 and 62 \citep[also noted
in][]{dunlop2010}. Some more possible examples are: 74 and 75; 76 and
77; 83 and 85; and 113 and 114.

\subsection{Rest-frame SEDs}
\label{sec:restsed}

\begin{figure}
\centering
\includegraphics[width=\linewidth]{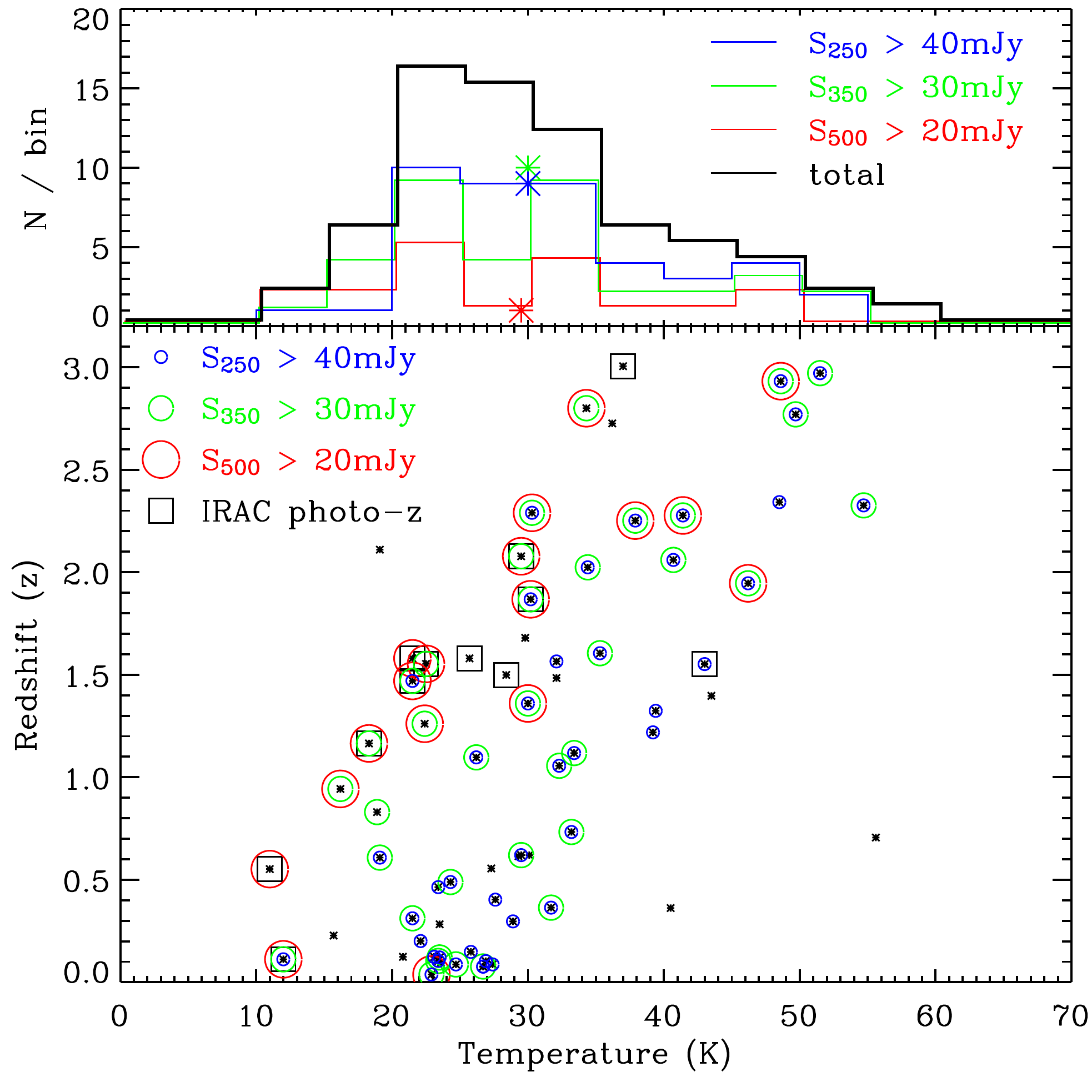}
\caption{The dust temperature distribution, assuming $S_\nu
  \propto\nu^{2.0}B_\nu(T)$, for the 69 non-confused sources with
  redshift estimates. The bottom panel shows redshifts plotted against
  temperature, and the top panel collapses the redshift axis into a
  histogram.  Similar to Fig.~\ref{fig:zhist} we indicate the
  distributions for flux-limited sub-samples in each BLAST band, and
  indicate the medians with asterisks. We indicate the sources with
  IRAC-based photometric redshifts, as they are highly uncertain, and
  may be biased low. If these sources are at higher redshifts, they
  will move up and right in the lower panel. There is a negligible
  trend in rest-frame temperature for sources selected in different
  BLAST bands.}
\label{fig:thist}
\end{figure}

The SEDs for all 118 sources are shown in Fig.~\ref{fig:seds}. For the
69 sources with redshift estimates and useful photometry, we convert
the observed temperatures from the modified blackbody fits of
Section~\ref{sec:sed} to rest-frame temperatures, allowing us to probe
the cold dust SEDs of the sample. The distributions of these
temperatures for the entire sample, and flux-limited sub-samples in
each BLAST band are shown in Fig.~\ref{fig:thist}. The total
distribution has a median $T=29$\,K and interquartile range
23--36\,K. For reference, re-running the analysis with $\beta=1.5$
increases the temperatures by about 5\,K.

There is almost no variation in the temperatures of the different
flux-limited sub-samples: a median $T=30$\,K with an interquartile
range 25--39\,K at 250\,\micron; a median $T=30$\,K with an
interquartile range 23--35\,K at 350\,\micron; and a median $T=30$\,K
with an interquartile range 22--38\,K at 500\,\micron. The reason for
this can be seen in the bottom panel of Fig.~\ref{fig:thist}: even
though the longer-wavelength channels tend to be biased to selecting
higher-redshift (and hence more luminous and warmer) galaxies, the
extra galaxies picked up at low redshift (and hence lower luminosity)
by the short-wavelength channels are biased toward warmer temperatures
in those volumes. We have also checked for biases in the total
temperature distribution due to the IRAC-based photometric redshifts,
but do not find a significant difference for the $z>1.5$ galaxy
temperatures when sources with these photometric redshifts are
excluded.

We infer cold-dust temperatures that are generally {\em warmer} than
the mean temperature of 23.4\,K and 1\,$\sigma$ width of 5.2\,K (also
assuming $\beta=2.0$) reported for the robust, but shallower, BLAST
sample in \citet{dye2009}. While this discrepancy may partly be due to
the relative depths of the samples, there is also strong evidence that
our treatment of confusion and the inclusion of LABOCA 870\,\micron\
data have resulted in less-biased temperatures.

The 870\,\micron\ data place a firm constraint on the Rayleigh-Jeans
part of the SED. Since the longer-wavelength BLAST data are more
confused (and hence more biased), there is a systematic preference for
shallower spectra, or cooler apparent temperatures. However, the
higher-resolution LESS data are less prone to this problem, and
therefore greatly improve the fits (if a smooth SED model such as the
modified blackbody is assumed). Sources 6, 23, 25, 34, 36, 44, 48, 67,
104, and 118 in Fig.~\ref{fig:seds} are more extreme examples that
show this effect clearly. In all these cases, the SED model is a
reasonable fit to the 250, 350, and 870\,\micron\ data points, but the
500\,\micron\ data lie above the model.

Restricting ourselves to the the 26 galaxies at $z\ge1.5$ with usable
submm photometry we find a median temperature of 36\,K. This is in
good agreement with Fig.~9 from \citet{dunlop2010} which shows that
their $S_{250}/S_{870}$ data are consistent with a $\beta=1.5$,
$T=40$\,K modified blackbody at these same redshifts (remembering that
if we had chosen $\beta=1.5$ instead of $2.0$ our temperatures would
be larger by about 5\,K).

We also fit the star-forming galaxy SED templates of \citet{dale2001}
to the submm and FIR photometry (870--70\,\micron). Since these
provide a fairly restricted range of SED shapes, they are only used to
illustrate several basic features in the radio and mid/near-IR SEDs of
our sample with respect to the rest-frame FIR peak (i.e., they should
not be considered accurate fits beyond the 870--70\,\micron\
data). First, they give an indication of the location of the
$\sim1.6$\,\micron\ stellar bump that is redshifted into the IRAC
bands for many of our sources (i.e., if the wavelength of the bump in
the model corresponds to a local maximum in the IRAC data the redshift
estimate is plausible). Also, since these are templates for
star-forming galaxies, sources from our sample with large excesses in
either the mid-IR or radio may harbour AGN. Of the 69 sources with
usable submm photometry and redshifts, 6 clear examples (8\%) seem to
exhibit such excesses: source 6 (mid-IR excess); source 34 (radio
excess); 49 (radio and mid-IR excess); 51 (mid-IR excess); and 52
(radio excess); and 67 (mid-IR excess). For the remaining 42 submm
confused sources, we fit the \citet{dale2001} SED templates to all of
non-confused data (VLA and {\em Spitzer}) to illustrate what plausible
values in the submm might be. Finally, we note that some of the
sources with IRAC-based photometric redshifts have particularly poor
SED fits in the near-IR when compared to these templates, such as
sources 33, 46, 50, and 117, as well as particularly cool inferred
rest-frame temperatures ($<20$\,K). It is quite likely that these
photometric redshifts are biased low, resulting in under-estimates of
the inferred temperatures, and luminosities.

\begin{figure}
\centering
\includegraphics[width=\linewidth]{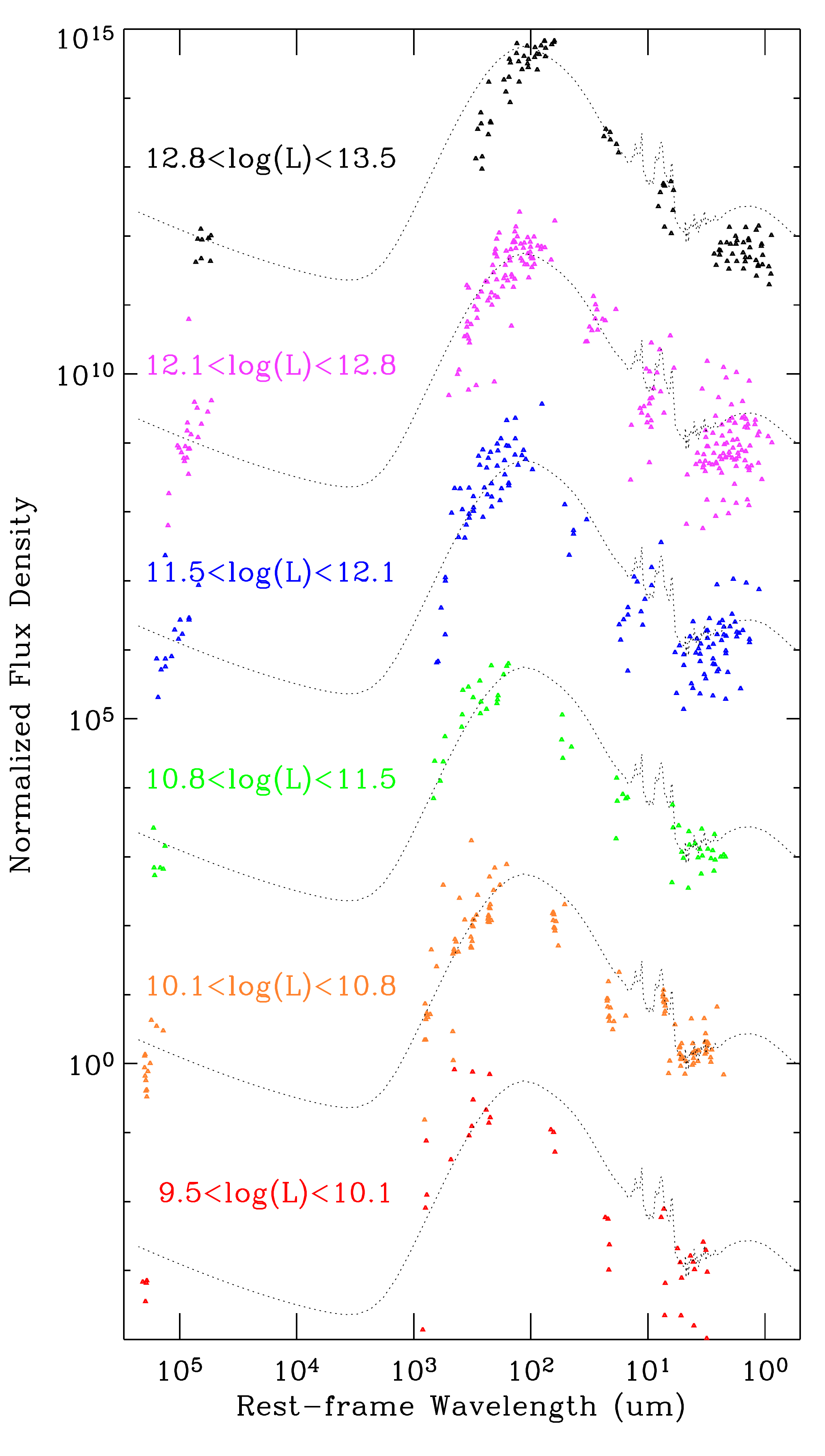}
\caption{SEDs of the 69 galaxies with non-confused submm photometry
  and redshift estimates, normalized by the total infrared (TIR)
  10--1000\,\micron\ luminosities of their modified blackbody
  fits. The SEDs are plotted in bins of $L_\mathrm{TIR}$ and offset by
  multiples of $10^3$ for clarity. The SEDs of individual galaxies are
  shown in Fig.~\ref{fig:seds}. A single $L_\mathrm{TIR}$-normalized
  SED template from \citet{dale2001}, for which
  $\log_{10}(S_{60}/S_{100}) = -0.2$, is shown as a dotted black line
  for reference in each bin. This comparison highlights several
  features: more luminous sources peak at shorter wavelengths (they
  are warmer); the radio-FIR correlation (the normalized model passes
  through the radio data irrespective of the FIR peak wavelength); the
  spread in the near/mid-IR SEDs (largest in the bins spanning $11.5 <
  \log(L) < 12.8$); and an increase in the size of the FIR peak with
  respect to stellar light (the 1.6\,\micron\ bump) with increasing
  luminosity.}
\label{fig:stacksed}
\end{figure}

Next, we normalize each of the full radio--submm--IR SEDs by their
total IR (TIR) 10--1000\,\micron\ luminosities using the modified
blackbody fits. The sample is then plotted in bins of $L_\mathrm{TIR}$
in Fig.~\ref{fig:stacksed}. This strategy minimizes scatter in the
region of the SEDs probed by the submm photometry, and enables us to
examine trends in the relative radio and mid/near-IR scatter as a
function of luminosity. For reference, the same TIR-normalized SED
from \citet{dale2001} (with $\log_{10}(S_{60}/S_{100}) = -0.2$) is
also plotted over each luminosity range to highlight differences. In
the first instance this figure demonstrates that the vast majority of
our proposed multi-wavelength counterparts are plausible: while there
are a handful of significant outliers, the relative intensities of the
radio and IR emission to the rest-frame FIR peak show reasonable
consistency with each other, and with the models. It is clear that the
radio--FIR correlation holds for most of the sample since the
reference FIR-normalized star-forming galaxy SED has a radio spectrum
that passes through most of the VLA data in each luminosity bin,
regardless of the wavelength of the FIR peak \citep[consistent with
the findings of][]{ivison2009}. On the other hand, the mid/near-IR
SEDs exhibit significantly larger scatter; particularly in the range
$11.5<\log(L)<12.8$ (a factor of $\sim100$ peak-to-peak). Most of the
sources mentioned previously with mid-IR excesses land in the two
luminosity bins that span this range. While a correlation between AGN
activity and FIR luminosity has been observed before
\citep[e.g.][]{takeuchi2003,takeuchi2004b,valiante2009}, it is curious
that our most luminous $12.8<\log(L)<13.5$ bin exhibits a more compact
spread in SED properties consistent with pure star-formation. This
result may not be significant, however, since this last luminosity bin
has only 10 sources (whereas the next two fainter bins have 26 and 20
sources respectively).  Fig.~\ref{fig:stacksed} also shows a strong
correlation between luminosity and temperature: the least luminous
sources peak at wavelengths $\lambda > 100$\,\micron, while the most
luminous sources peak at wavelengths $\lambda <
100$\,\micron. Finally, this plot shows that the more luminous sources
have a greater ratio of FIR luminosity to starlight (the 1.6\,\micron\
bump); the lowest-luminosity sources have SEDs more closely resembling
normal star-forming galaxies than ULIRGs.

\begin{figure*}
\centering
\includegraphics[width=0.9\linewidth]{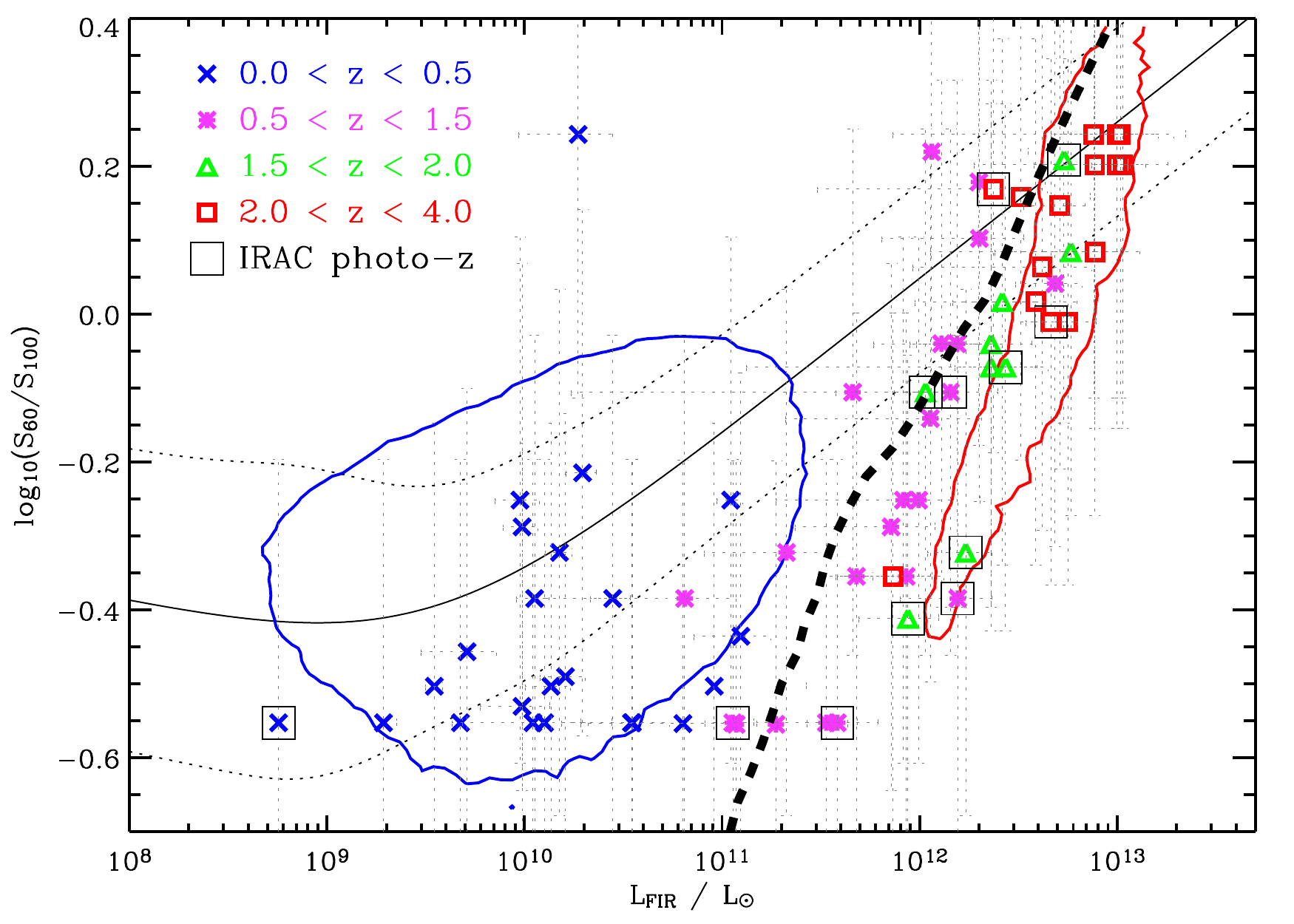}
\caption{Rest-frame $\log_{10}(S_{60}/S_{100})$ colours vs. integrated
  42.5--122.5\,\micron\ far-infrared (FIR) luminosities inferred from
  fits of \citet{dale2001} SED templates to the submm and FIR
  photometry for our sample. The solid black line shows the local
  correlation derived from {\em IRAS} galaxies in \citet{chapin2009},
  with the dotted black line its 68\% confidence interval. The symbols
  are data points from our survey, divided into sources at $z<0.5$,
  $0.5<z<1.5$, $1.5<z<2.0$, and $2.0<z<4.02$ (including the
  highest-redshift source in the sample). We indicate the sources with
  IRAC-based photometric redshifts as they are highly uncertain, and
  may be biased low. If these sources are at higher redshifts, they
  will move up and right in this plot.  For comparison, we generated a
  realization of sources from a non-evolving FIR colour-luminosity
  distribution \citep{chapin2009}, extrapolating to the BLAST
  wavelengths using the \citet{dale2001} SED templates, and applying
  flux density cuts 40, 30, and 20\,mJy at 250, 350, and 500\,\micron\
  respectively. Sub-samples were then made for the $0.0<z<0.5$ and
  $2.0<z<4.0$ redshifts bins, and their 90\% confidence intervals are
  shown as solid blue and red contours respectively. For further
  illustration we also show the BLAST selection function at $z=0.5$
  with a thick black dashed line. The reasonable agreement between the
  model contours and the sample demonstrates that the generally cooler
  (more negative) colours for our sample can be explained by selection
  effects, rather than evolution in the colour-luminosity correlation
  at high redshift.}
\label{fig:L-C}
\end{figure*}

To explore the correlation between luminosity and temperature more
fully, and the potential for its redshift evolution, we compare our
sample with the local Universe. We use as our reference the
distribution in integrated 42.5--122.5\,\micron\ FIR luminosity, and
$C\equiv \log_{10}(S_{60}/S_{100})$ colour as measured by
\citet{chapin2009} based on {\em IRAS} data \citep[an updated version
of the analysis in][]{chapman2003}.  At redshift $\sim2.5$ the entire
BLAST 250--500\,\micron\ bandpass closely matches the
42.5--122.5\,\micron\ coverage from the {\em IRAS} 60 and
100\,\micron\ channels \citep[see Fig.~6 in][]{chapin2009}, so that
tests for evolution at that redshift have only a minimal dependence on
the SED model used to fit the data. However, at lower redshifts, the
BLAST bandpass samples significantly longer wavelengths in the
rest-frame. Since the modified blackbody models used to fit the submm
data fall-off much faster in the mid/far-IR than for real galaxies
(since the shorter wavelengths sample warmer and/or optically-thick
dust), we have fit the submm and FIR data for our sample using the
more realistic star-forming galaxy SED templates of
\citet{dale2001}. In Fig.~\ref{fig:L-C} we plot the colours and
luminosities for our sample using these SED fits. We colour-code the
sources by redshift bin: $z<0.5$ (blue crosses); $0.5<z<1.5$ (grey
asterisks); $1.5<z<2.0$ (green triangles); and $2<z<42$ (red
squares). Note that there is some quantization along the vertical axis
for the sample since there are only 64 SED templates in the library,
with colours ranging from about $-0.55$ to $0.26$. Since this range is
not quite as broad as that observed for the real sources, there is
also some clipping (mostly problematic for the coolest sources at the
bottom of the plot which are simply assigned the template SED with the
most negative colour).  We then compare this distribution with the
local-Universe measurement of \citet{chapin2009}, including the
1$\sigma$ envelope (solid and dotted black lines respectively).

We find that the temperatures for our sample lie systematically toward
cooler (more negative) values than in the local Universe. At low
redshifts, the lower temperatures are expected, since the submm bands
sample the Rayleigh-Jeans tail of the thermal emission and are
therefore biased toward the detection of cooler sources for a given
flux-density--limit. Similarly, observations on the shorter-wavelength
side of the thermal emission peak (such as with {\em IRAS}) are biased
toward the detection of warmer sources --- an effect which was
accounted for in \citet{chapin2009} in order to infer the properties
of the entire population. Selection effects alone may therefore
explain much of the discrepancy in temperatures seen here, and for
850\,\micron\ \citep[e.g.,][]{kovacs2006,coppin2008} selected galaxy
populations \citep[see also the discussion in Section~4.3
of][]{dye2009}.

To test this theory, we have used a simple simulation to explore the
effects of our selection function on the temperatures and luminosities
of galaxies in different redshift slices. We take the local FIR
colour-luminosity distribution, $\Phi(L,C)$, from \citet{chapin2009}
and assume that it does not evolve with redshift. We then calculate
the total number of objects as a function of $L$ and $C$ in a redshift
slice $z_0<z<z_1$ as $\int_{z_0}^{z_1} \Phi(L,C) (dV/dz) dz$, where
$dV/dz$ is the differential volume element, and then use the
\citet{dale2001} SED templates to extrapolate to the BLAST
wavelengths, and hence estimate observed flux densities. While the
relative numbers of galaxies in different redshift slices will be
incorrect, since it is known that significant luminosity and/or
density evolution in the local luminosity function is required to fit
observed submm number counts
\citep[e.g.,][]{rowan-robinson2001,scott2002,lagache2003,valiante2009,
  rowan-robinson2009,wilman2010}, this calculation should give a good
idea of the selection function {\em within} a redshift slice (provided
that evolution across the slice is negligible).  We then apply flux
density cuts of 40, 30 and 20\,mJy at 250, 350, and 500\,\micron\
respectively to mimic our observational selection function. We show
90\% confidence intervals for these model distributions in the $z<0.5$
and $2.0<z<4.0$ redshift bins with solid blue and red contours in
Fig.~\ref{fig:L-C}, showing that they extend significantly below the
rest-frame colour-luminosity correlation. To illustrate how this
selection happens, we show the effect of the BLAST flux limits in the
colour-luminosity plane at a single fixed redshift $z=0.5$ as a thick
dashed line (sources at that redshift should only be observable to the
right of the limit). The detected galaxies tend to pile-up against
this limit, since there are many more galaxies at fainter luminosities
than at brighter luminosities in a given volume. If this limit were a
vertical line in the plot, we would then expect the observed
luminosities and colours to cluster around the rest-frame correlation
(solid and dotted lines). However, since the limit is {\em inclined},
a dis-proportionate number of objects are detected toward the
lower-left (lower luminosities, and cooler temperatures).  Despite the
simplicity of the model, the large noise in the observed sample, and
potential for incorrect redshifts, the general trends in the data are
clearly reproduced. This result suggests that there is no significant
evolution in the correlation between FIR luminosity and temperature at
high redshifts, at least for the most luminous ($L_\mathrm{FIR} \gsim
10^{12}$\,L$_\odot$) galaxies probed by our sample at $z\gsim1$.

While the lack of evolution in our sample appears to be at odds with
the evolution observed in the {\em stacked} SEDs with redshift
described in \citet{pascale2009}, we note that our sample contains
only ULIRGs at $z>1.5$, whereas the stacked SEDs are dominated by
significantly fainter galaxies. It is possible that `normal'
star-forming galaxies are generally warmer in the early Universe than
in the present-day, while massive starburst galaxies do not differ
appreciably. Another possible explanation is simply that the
completeness to less luminous (and presumably cooler) galaxies drops
off faster at high redshift than it does for the more luminous (and
warmer) galaxies in \citet{pascale2009} --- resulting in a bias to
warmer stacked temperatures.

Our finding may also have interesting consequences for recent studies
which indicate that SMGs are generally more extended and cooler than
local ULIRGs. These claims have been made based on a variety of
observations, including resolved MERLIN radio morphologies
\citep[e.g.][]{chapman2004}, {\em Spitzer} MIPS and IRAC SEDs
\citep[e.g.][]{hainline2009}, and {\em Spitzer} IRS spectra
\citep[e.g.][]{menendez2009}. Such observations point to a systematic
difference in the physical conditions of major-mergers which produce
SMGs \citep[e.g., see resolved molecular gas observations and
discussion in][]{tacconi2008}, compared to ULIRGs in the local
Universe. We note that \citet{chanial2007} found a positive FIR
luminosity-size correlation with a similar form to the
temperature-luminosity correlation that we have used here. It may
therefore be the case that the low-temperature bias for submm samples
has picked out galaxies that are also more extended.

\subsection{Predictions for {\em Herschel}/SPIRE surveys}
\label{sec:future}

\begin{figure*}
\centering
\includegraphics[width=\linewidth]{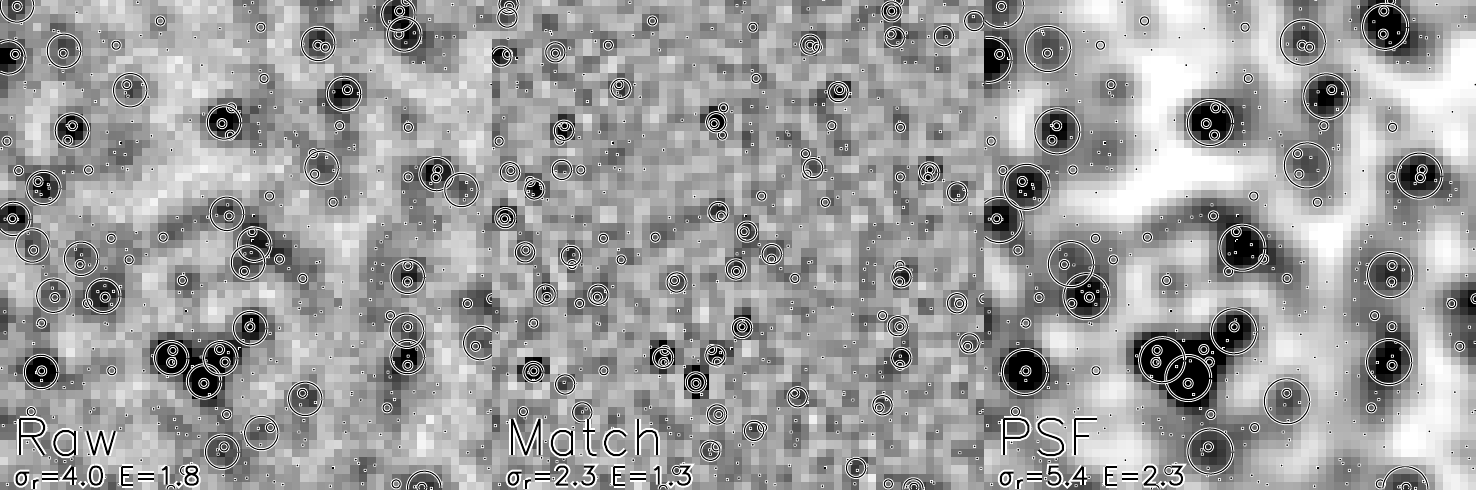}
\caption{A simulated $0.1\,\mathrm{deg} \times 0.1\,\mathrm{deg}$
  250\,\micron\ SPIRE observation, with a depth similar to the real
  {\em Herschel} Science Demonstration Phase GOODS-N data taken as a
  part of {\em HerMES}. The palette is scaled from $-15$\,mJy (white)
  to $+20$\,mJy (black). The panels show, from left to right, the raw,
  match-filtered, and PSF-filtered maps. The large circles indicate
  submm peaks with flux densities $S_{250} > 10$\,mJy, and their radii
  are $3\,\sigma_\mathrm{r}$, where $\sigma_r$ is estimated as in
  Appendix~\ref{sec:pos} using a matching catalogue consisting of
  input sources with $S_{250} > 4$\,mJy. The small circles and dots
  indicate input sources with flux densities $S_{250}>10$\,mJy, and
  $4\,\mathrm{mJy} < S_{250} < 10\,\mathrm{mJy}$, respectively. The
  match-filtered image is significantly less confused than the raw and
  PSF-filtered maps, but in many cases the submm peaks are still
  clearly blends of multiple sources.}
\label{fig:spiresmooth}
\end{figure*}

Will the techniques described in this paper be useful for the new
generation of {\em Herschel}/SPIRE surveys? To address this question,
we have again turned to simulations, creating realizations of sources
with flux densities drawn from the BLAST $P(D)$ counts. These sources
are assigned uniformly-distributed random positions covering the same
area as the ECDF-S observations described in this paper, and then
smoothed with the smaller 18, 25, and 36\,arcsec FWHM Gaussians to
approximate the SPIRE beams. We have chosen a pixel size of 6\,arcsec,
and added Gaussian noise with RMS 1.8, 2.3, and 5.0\,mJy at 250, 350,
and 500\,\micron, respectively. We find that the RMS due purely to
point source confusion is 7.0, 6.8, and 6.6\,mJy at 250, 350, and
500\,\micron\, respectively. This simulation is a reasonable
approximation of the SPIRE Science Demonstration Phase observations of
GOODS-N, taken as part of the {\em Herschel} Multi-tiered
Extra-galactic Survey (HerMES). Initial number counts were shown to be
consistent with the BLAST results in \citet{oliver2010}, and our
instrumental and confusion noise estimates are close to those reported
in \citet{nguyen2010}. A $0.1\,\mathrm{deg} \times 0.1\,\mathrm{deg}$
subset of the simulated 250\,\micron\ data are shown in the left panel
of Fig.~\ref{fig:spiresmooth}.

Next, we produce match-filtered maps, following the prescription of
Appendix~\ref{sec:matched}. The total noise (combining white and
confusion noise) is estimated to be 5.8, 5.6, and 5.7\,mJy at 250,
350, and 500\,mJy, respectively. We compare the 250\,\micron\
match-filtered map with the raw and PSF-filtered maps in
Fig.~\ref{fig:spiresmooth}. In all three cases we indicate submm peaks
with brightnesses $S_{250} > 10$\,mJy (which are $>5\,\sigma$
detections with respect to instrumental noise), as well as the
locations of the known input sources to the simulation. Clearly these
data have a \snr\ (with respect to instrumental noise) significantly
higher than the BLAST 250\,\micron\ data; therefore the improvement in
the match-filtered results over the raw and PSF-filtered maps is more
pronounced than for the BLAST data shown in
Fig.~\ref{fig:mapcompare}.

We draw attention to the fact that the SPIRE peaks are obviously
blends of multiple significant sources in most cases. To quantify the
confusion we have repeated the radial excess counts following
Appendix~\ref{sec:pos}, using the known input source positions as the
matching catalogue. At 250\,\micron, which we emphasize is the {\em
  least} confused SPIRE band, we find: $\sigma_r=4.0$\,arcsec (radial
positional uncertainty) and $E=1.8$ (average number of excess sources
contributing to peak) in the raw maps; $\sigma_r=2.3$\,arcsec and
$E=1.3$ in the match-filtered maps; and $\sigma_r=5.4$\,arcsec and
$E=2.3$ in the PSF-filtered maps. In other words, peaks in the raw
250\,\micron\ maps are often blends of about two galaxies, while the
matched-filter is able to significantly de-convolve the beam and
improve the positional uncertainties. This example also clearly
demonstrates that convolution with the full PSF is {\em the wrong}
approach to take with deep SPIRE data.

We have repeated this procedure for the longer wavelengths (also using
a limiting flux density of 10\,mJy for the peak catalogues), and find
for the match-filtered maps: $\sigma_r=4.1$\,arcsec and $E=1.8$ at
350\,\micron; and $\sigma_r=6.8$\,arcsec and $E=2.5$ for the
match-filtered maps at 500\,\micron. Therefore, confusion is a
significant issue, even in the match-filtered maps, at the longer
wavelengths that are better suited to detecting the highest-redshift
sources.

In summary, deep SPIRE data are considerably confused at levels that
lie between the LABOCA and BLAST data, as expected given the relative
beam sizes. Matched-filtering has an even greater potential to
de-blend and improve the positional uncertainty for SPIRE than for
BLAST due to the improved \snr. However, SPIRE peak catalogues will
still generally represent blends of multiple sources. It will
therefore be necessary to identify multiple potential counterparts to
each peak, and we advocate performing a simultaneous fit to the data
as we have done to the BLAST data in order to produce the least-biased
photometry. Finally, while not simulated here, the considerably larger
catalogues of SPIRE peaks will enable estimates of priors for
Likelihood Ratios (as in Section~\ref{sec:id}) with considerably
smaller statistical errors. In fact, LRs have already been use to
propose identifications for SPIRE sources in
\citet{smith2010}. However, they have calculated counterpart
reliabilities following \citet{sutherland1992}, which we suspect may
be biased due to the impact of confusion (see discussion in
Appendix~\ref{sec:lr}).

\section{Conclusions}
\label{sec:conclusions}

We have performed a deep multi-wavelength study of individual peaks
from the BLAST 250, 350, and 500\,\micron\ survey within the Extended
Chandra Deep Field South (ECDF-S). By comparing the BLAST data with
LABOCA ECDFS Submm Survey (LESS) maps at 870\,\micron\ we have been
able to greatly improve the positional uncertainties and
longer-wavelength (Rayleigh-Jeans tail) SEDs for our sample. Compared
to the earlier BLAST studies of \citet{dye2009}, \citet{ivison2009}
and \citet{dunlop2010}, our methodology for identifying counterparts
and measuring flux densities differ in several key respects:

\begin{itemize}

\item It is recognized that peaks in the submm maps (particularly the
  BLAST bands) are generally blends of several unrelated sources. It
  is therefore important to search for {\em multiple} counterparts to
  each peak in a matching catalogue, whereas earlier BLAST
  catalogue-based studies tended to focus on {\em single}
  counterparts.

\item We use a new `matched filter' that compensates for source
  confusion when searching for submm sources. For the BLAST bands, in
  which the contribution to the total noise by source confusion is
  roughly a factor of two larger than the instrumental noise, there is
  a significant improvement in \snr\ of approximately 15--20\%. In
  contrast, there is only a minimal improvement of about 5\% for the
  LESS data since the confusion and instrumental noise components are
  approximately the same.

\item Identifications in our combined radio/IR matching catalogue have
  been made using Likelihood Ratios that incorporate more prior
  information than traditional `$P$' statistics. With a threshold
  spurious ID rate set at 10\%, we find 52, 50, and 31 matches to 64,
  67, and 55 submm peak positions at 250, 350, and 500\,\micron,
  respectively. This is a significant improvement compared to the
  match rates obtained using 10\% cuts on $P$ giving 45, 31, and 12
  matches in the 24\,\micron\ catalogue, and 51, 35, and 17 matches in
  1.4\,GHz catalogue.  Most of the gain is at 500\,\micron\ where the
  beam is so large that any additional information to help the process
  clearly makes a large difference. Combining the identifications made
  independently in each band, approximately 75\% of the BLAST selected
  peaks have at least one potential counterpart.

\item We obtain submm photometry for the 118 unique identifications in
  the matching catalogue by performing a simultaneous fit of the submm
  PSFs at the precise ID locations in each of the four original
  (un-smoothed) submm maps. This procedure results in useful
  measurements for 76 sources; the remaining 42 sources are too close
  to one another to reliably disentangle the flux densities produced
  by each of the contributing objects. In order to make further
  progress in such cases, prior information for the SEDs will likely
  be needed \citep[see, for example, the work of][]{roseboom2009}.

\item Our procedure for identifying counterparts and measuring their
  flux densities has compensated for much of the `flux-boosting'
  present in the original flux-limited submm peak lists. Comparing the
  number counts of our final catalogue in the BLAST bands with the
  $P(D)$ counts from \citet{patanchon2009} we find good agreement
  above 40\,mJy at 250\,\micron, 30\,mJy at 350\,\micron\, and with
  somewhat lower completeness reaching a maximum of $\sim66$\% above
  20\,mJy at 500\,\micron.

\end{itemize}

We then proceeded to identify redshifts for the counterparts, and
hence probe the rest-frame properties of our sample:

\begin{itemize}

\item Of the 76 sources with usable submm photometry, 69 counterparts
  have redshifts: 23 are optical spectroscopic redshifts; 35 are
  optical photometric redshifts; and 11 are IRAC photometric
  redshifts. The median of the entire distribution is $z=1.1$ with an
  interquartile range 0.3--1.9. Restricting ourselves to sub-samples
  above the flux-density limits mentioned previously shows a clear
  trend in increasing redshift with observed wavelength: a median
  $z=1.1$ with an interquartile range 0.2--1.9 at 250\,\micron; a
  median $z=1.3$ with an interquartile range 0.6--2.1 at 350\,\micron;
  and a median $z=1.6$ with an interquartile range 1.3--2.3 at
  500\,\micron. In general, there is a higher-redshift tail for our
  sample than in the earlier study of \citet{dye2009} which we believe
  is due to the fainter flux densities probed.

\item Fitting modified blackbody SEDs of the form $S_\nu \propto
  \nu^{2.0}B_\nu(T)$ to the 69 sources with useable submm photometry
  and redshifts, we establish rest-frame cold-dust temperatures. The
  total distribution has a median $T=29$\,K and interquartile range
  23--36\,K. There is almost no variation in the temperatures of the
  different flux-limited sub-samples: a median $T=30$\,K with an
  interquartile range 25--39\,K at 250\,\micron; a median $T=30$\,K
  with an interquartile range 23--35\,K at 350\,\micron; and a median
  $T=30$\,K with an interquartile range 22--38\,K at
  500\,\micron. These temperatures are systematically warmer by about
  7\,K than those inferred in \citet{dye2009}. While this discrepancy
  may be partly caused by the relative depths of the samples, we
  believe it is mostly due to improved compensation for confusion
  (which reduces the estimated flux densities at longer BLAST
  wavelengths more than at shorter wavelengths), and the inclusion of
  higher-resolution 870\,\micron\ data from LESS.

\item The primary improvement over earlier catalogue-based BLAST
  studies is the characterization of submm SEDs for the fainter, but
  higher-redshift ($z\gsim1$) and more-luminous galaxies. Even though
  the earlier analyses typically considered single counterparts to the
  BLAST peaks, these counterparts {\em are} probably significant submm
  emitters (although the measured flux densities were typically high,
  especially in the most confused 500\,\micron\ band). Therefore the
  measured properties of the optical counterparts (approximate
  redshift distributions, galaxy types etc.), and even the FIR
  temperatures and luminosities of the brighter, lower-redshift
  galaxies are generally correct.

\item We have also fit star-forming galaxy SED templates from the
  library of \citet{dale2001} (which span the radio--near-IR) to the
  submm and FIR photometry of our sample. Generally speaking there is
  good correspondence between these models and the photometry in the
  radio and near/mid-IR, confirming that most of the emission is
  probably powered by star-formation. However, about 8\% of the sample
  exhibits a significant excess in either (or both) of these
  wavelength regimes, which could indicate AGN activity. We note that
  these features are restricted to sources with luminosities primarily
  in the range $11.5<\log(L_\mathrm{FIR})<12.8$, near the bright end
  of our sample.

\item We compare the distribution of luminosity and cold dust
  temperature in our sample with the local-Universe measurement of
  \citet{chapin2009}. While the submm colours of our sources appear
  systematically cooler at all luminosities compared to the local
  distribution, we have determined that the distribution is consistent
  with selection effects in our survey. We therefore find no evidence
  for evolution in the temperature-luminosity correlation out to
  $z\sim2.5$.

\item Finally, we investigate the utility of the methods described in
  this paper for the new generation of {\em Herschel}/SPIRE
  surveys. Using a simple simulation we show that these new surveys
  are significantly confused, despite a factor of $\sim$2 improvement
  in angular resolution over BLAST. We find that the matched-filter
  will yield a {\em greater} improvement in the detection of point
  sources than for BLAST due to the significantly lower instrumental
  noise of SPIRE. However, even with these improvements, peaks in
  SPIRE maps will be confused, and we advocate the use of Likelihood
  Ratios to identify the expected multiple counterparts.

\end{itemize}

\section{Acknowledgements}

BLAST acknowledges the support of NASA through grant numbers
NAG5-12785, NAG5-13301, and NNGO-6GI11G, the NSF Office of Polar
Programs, the Canadian Space Agency, the Natural Sciences and
Engineering Research Council (NSERC) of Canada, and STFC.  This paper
contains data obtained by APEX, which is operated by the
Max-Planck-Institut f\"ur Radioastronomie, the European Southern
Observatory, and the Onsala Space Observatory. LESS is a joint MPI-ESO
collaboration and the data were obtained during the following ESO
observing runs: 078.F-9028(A), 079.F-9500(A), 080.A-3023(A), and
081.F-9500(A). This work is based in part on observations made with
the Spitzer Space Telescope, which is operated by the Jet Propulsion
Laboratory, California Institute of Technology under a contract with
NASA. IRS and KEKC acknowledge support from STFC.  JSD acknowledges
the support of the Royal Society through a Wolfson Research Merit
Award.  EC thanks John Peacock and Seb Oliver for useful discussions
regarding Likelihood Ratios.

\bibliographystyle{mn2e}
\bibliography{mn-jour,refs}

\begin{thebibliography}{}

\bibitem[\protect\citeauthoryear{{Aretxaga}, {Hughes}, {Chapin}, {Gazta{\~
  n}aga}, {Dunlop} \& {Ivison}}{{Aretxaga} et~al.}{2003}]{aretxaga2003}
{Aretxaga} I.,  {Hughes} D.~H.,  {Chapin} E.~L.,  {Gazta{\~ n}aga} E.,
  {Dunlop} J.~S.,    {Ivison} R.~J.,  2003, \mnras, 342, 759

\bibitem[\protect\citeauthoryear{{Austermann}, {Dunlop}, {Perera}, {Scott},
  {Wilson}, {Aretxaga}, {Hughes}, {Almaini}, {Chapin}, {Chapman}, {Cirasuolo},
  {Clements}, {Coppin} \& {Dunne}}{{Austermann} et~al.}{2010}]{austermann2010}
{Austermann} J.~E.,  {Dunlop} J.~S.,  {Perera} T.~A.,  {Scott} K.~S.,  {Wilson}
  G.~W.,  {Aretxaga} I.,  {Hughes} D.~H.,  {Almaini} O.,  {Chapin} E.~L.,
  {Chapman} S.~C.,  {Cirasuolo} M.,  {Clements} D.~L.,  {Coppin} K.~E.~K.,
  {Dunne} L.,  2010, \mnras, 401, 160

\bibitem[\protect\citeauthoryear{{Barger}, {Cowie}, {Sanders}, {Fulton},
  {Taniguchi}, {Sato}, {Kawara} \& {Okuda}}{{Barger} et~al.}{1998}]{barger1998}
{Barger} A.~J.,  {Cowie} L.~L.,  {Sanders} D.~B.,  {Fulton} E.,  {Taniguchi}
  Y.,  {Sato} Y.,  {Kawara} K.,    {Okuda} H.,  1998, \nat, 394, 248

\bibitem[\protect\citeauthoryear{{Barnard}, {Vielva}, {Pierce-Price}, {Blain},
  {Barreiro}, {Richer} \& {Qualtrough}}{{Barnard} et~al.}{2004}]{barnard2004}
{Barnard} V.~E.,  {Vielva} P.,  {Pierce-Price} D.~P.~I.,  {Blain} A.~W.,
  {Barreiro} R.~B.,  {Richer} J.~S.,    {Qualtrough} C.,  2004, \mnras, 352,
  961

\bibitem[\protect\citeauthoryear{{Barreiro}, {Sanz}, {Herranz} \&
  {Mart{\'{\i}}nez-Gonz{\'a}lez}}{{Barreiro} et~al.}{2003}]{barreiro2003}
{Barreiro} R.~B.,  {Sanz} J.~L.,  {Herranz} D.,
  {Mart{\'{\i}}nez-Gonz{\'a}lez} E.,  2003, \mnras, 342, 119

\bibitem[\protect\citeauthoryear{{Baugh}, {Lacey}, {Frenk}, {Granato}, {Silva},
  {Bressan}, {Benson} \& {Cole}}{{Baugh} et~al.}{2005}]{baugh2005}
{Baugh} C.~M.,  {Lacey} C.~G.,  {Frenk} C.~S.,  {Granato} G.~L.,  {Silva} L.,
  {Bressan} A.,  {Benson} A.~J.,    {Cole} S.,  2005, \mnras, 356, 1191

\bibitem[\protect\citeauthoryear{{Blain}, {Smail}, {Ivison}, {Kneib} \&
  {Frayer}}{{Blain} et~al.}{2002}]{blain2002}
{Blain} A.~W.,  {Smail} I.,  {Ivison} R.~J.,  {Kneib} J.-P.,    {Frayer} D.~T.,
   2002, PhR, 369, 111

\bibitem[\protect\citeauthoryear{{Borys}, {Chapman}, {Halpern} \&
  {Scott}}{{Borys} et~al.}{2003}]{borys2003}
{Borys} C.,  {Chapman} S.,  {Halpern} M.,    {Scott} D.,  2003, \mnras, 344,
  385

\bibitem[\protect\citeauthoryear{{Brammer}, {van Dokkum} \& {Coppi}}{{Brammer}
  et~al.}{2008}]{brammer2008}
{Brammer} G.~B.,  {van Dokkum} P.~G.,    {Coppi} P.,  2008, \apj, 686, 1503

\bibitem[\protect\citeauthoryear{{Casey}, {Chapman}, {Smail},
  {Alaghband-Zadeh}, {Bothwell} \& {Swinbank}}{{Casey}
  et~al.}{2010}]{casey2010}
{Casey} C.~M.,  {Chapman} S.~C.,  {Smail} I.,  {Alaghband-Zadeh} S.,
  {Bothwell} M.~S.,    {Swinbank} A.~M.,  2010, ArXiv e-prints

\bibitem[\protect\citeauthoryear{{Chanial}, {Flores}, {Guiderdoni}, {Elbaz},
  {Hammer} \& {Vigroux}}{{Chanial} et~al.}{2007}]{chanial2007}
{Chanial} P.,  {Flores} H.,  {Guiderdoni} B.,  {Elbaz} D.,  {Hammer} F.,
  {Vigroux} L.,  2007, \aap, 462, 81

\bibitem[\protect\citeauthoryear{{Chapin}, {Ade}, {Bock}, {Brunt}, {Devlin},
  {Dicker}, {Griffin}, {Gundersen}, {Halpern}, {Hargrave} \& {Hughes}}{{Chapin}
  et~al.}{2008}]{chapin2008}
{Chapin} E.~L.,  {Ade} P.~A.~R.,  {Bock} J.~J.,  {Brunt} C.,  {Devlin} M.~J.,
  {Dicker} S.,  {Griffin} M.,  {Gundersen} J.~O.,  {Halpern} M.,  {Hargrave}
  P.~C.,    {Hughes} D.~H.,  2008, \apj, 681, 428

\bibitem[\protect\citeauthoryear{{Chapin}, {Hughes} \& {Aretxaga}}{{Chapin}
  et~al.}{2009}]{chapin2009}
{Chapin} E.~L.,  {Hughes} D.~H.,    {Aretxaga} I.,  2009, \mnras, 393, 653

\bibitem[\protect\citeauthoryear{{Chapin}, {Pope}, {Scott}, {Aretxaga},
  {Austermann}, {Chary}, {Coppin}, {Halpern}, {Hughes}, {Lowenthal},
  {Morrison}, {Perera}, {Scott}, {Wilson} \& {Yun}}{{Chapin}
  et~al.}{2009}]{chapin2009b}
{Chapin} E.~L.,  {Pope} A.,  {Scott} D.,  {Aretxaga} I.,  {Austermann} J.~E.,
  {Chary} R.,  {Coppin} K.,  {Halpern} M.,  {Hughes} D.~H.,  {Lowenthal} J.~D.,
   {Morrison} G.~E.,  {Perera} T.~A.,  {Scott} K.~S.,  {Wilson} G.~W.,    {Yun}
  M.~S.,  2009, \mnras, 398, 1793

\bibitem[\protect\citeauthoryear{{Chapman}, {Blain}, {Ivison} \&
  {Smail}}{{Chapman} et~al.}{2003}]{chapman2003b}
{Chapman} S.~C.,  {Blain} A.~W.,  {Ivison} R.~J.,    {Smail} I.~R.,  2003,
  \nat, 422, 695

\bibitem[\protect\citeauthoryear{{Chapman}, {Blain}, {Smail} \&
  {Ivison}}{{Chapman} et~al.}{2005}]{chapman2005}
{Chapman} S.~C.,  {Blain} A.~W.,  {Smail} I.,    {Ivison} R.~J.,  2005, \apj,
  622, 772

\bibitem[\protect\citeauthoryear{{Chapman}, {Helou}, {Lewis} \&
  {Dale}}{{Chapman} et~al.}{2003}]{chapman2003}
{Chapman} S.~C.,  {Helou} G.,  {Lewis} G.~F.,    {Dale} D.~A.,  2003, \apj,
  588, 186

\bibitem[\protect\citeauthoryear{{Chapman}, {Smail}, {Windhorst}, {Muxlow} \&
  {Ivison}}{{Chapman} et~al.}{2004}]{chapman2004}
{Chapman} S.~C.,  {Smail} I.,  {Windhorst} R.,  {Muxlow} T.,    {Ivison} R.~J.,
   2004, \apj, 611, 732

\bibitem[\protect\citeauthoryear{{Chary} \& {Pope}}{{Chary} \&
  {Pope}}{2010}]{chary2010}
{Chary} R.,  {Pope} A.,  2010, ArXiv e-prints

\bibitem[\protect\citeauthoryear{{Clements}, {Eales}, {Wojciechowski}, {Webb},
  {Lilly}, {Dunne}, {Ivison}, {McCracken}, {Yun}, {James}, {Brodwin}, {Le
  F{\`e}vre} \& {Gear}}{{Clements} et~al.}{2004}]{clements2004}
{Clements} D.,  {Eales} S.,  {Wojciechowski} K.,  {Webb} T.,  {Lilly} S.,
  {Dunne} L.,  {Ivison} R.,  {McCracken} H.,  {Yun} M.,  {James} A.,  {Brodwin}
  M.,  {Le F{\`e}vre} O.,    {Gear} W.,  2004, \mnras, 351, 447

\bibitem[\protect\citeauthoryear{{Coppin}, {Chapin}, {Mortier}, {Scott},
  {Borys}, {Dunlop}, {Halpern}, {Hughes}, {Pope}, {Scott}, {Serjeant}, {Wagg}
  \& {Alexander}}{{Coppin} et~al.}{2006}]{coppin2006}
{Coppin} K.,  {Chapin} E.~L.,  {Mortier} A.~M.~J.,  {Scott} S.~E.,  {Borys} C.,
   {Dunlop} J.~S.,  {Halpern} M.,  {Hughes} D.~H.,  {Pope} A.,  {Scott} D.,
  {Serjeant} S.,  {Wagg} J.,    {Alexander} D.~M.,  2006, \mnras, 372, 1621

\bibitem[\protect\citeauthoryear{{Coppin}, {Halpern}, {Scott}, {Borys} \&
  {Chapman}}{{Coppin} et~al.}{2005}]{coppin2005}
{Coppin} K.,  {Halpern} M.,  {Scott} D.,  {Borys} C.,    {Chapman} S.,  2005,
  \mnras, 357, 1022

\bibitem[\protect\citeauthoryear{{Coppin}, {Halpern}, {Scott}, {Borys},
  {Dunlop}, {Dunne}, {Ivison}, {Wagg}, {Aretxaga}, {Battistelli}, {Benson},
  {Blain}, {Chapman}, {Clements}, {Dye}, {Farrah}, {Hughes} \&
  {Jenness}}{{Coppin} et~al.}{2008}]{coppin2008}
{Coppin} K.,  {Halpern} M.,  {Scott} D.,  {Borys} C.,  {Dunlop} J.,  {Dunne}
  L.,  {Ivison} R.,  {Wagg} J.,  {Aretxaga} I.,  {Battistelli} E.,  {Benson}
  A.,  {Blain} A.,  {Chapman} S.,  {Clements} D.,  {Dye} S.,  {Farrah} D.,
  {Hughes} D.,    {Jenness} T.,  2008, \mnras, 384, 1597

\bibitem[\protect\citeauthoryear{{Coppin}, {Smail}, {Alexander}, {Weiss},
  {Walter}, {Swinbank}, {Greve}, {Kovacs}, {De Breuck}, {Dickinson}, {Ibar},
  {Ivison}, {Reddy}, {Spinrad}, {Stern}, {Brandt}, {Chapman} \&
  {Dannerbauer}}{{Coppin} et~al.}{2009}]{coppin2009}
{Coppin} K.~E.~K.,  {Smail} I.,  {Alexander} D.~M.,  {Weiss} A.,  {Walter} F.,
  {Swinbank} A.~M.,  {Greve} T.~R.,  {Kovacs} A.,  {De Breuck} C.,  {Dickinson}
  M.,  {Ibar} E.,  {Ivison} R.~J.,  {Reddy} N.,  {Spinrad} H.,  {Stern} D.,
  {Brandt} W.~N.,  {Chapman} S.~C.,    {Dannerbauer} H.,  2009, \mnras, 395,
  1905

\bibitem[\protect\citeauthoryear{{Cowie}, {Barger} \& {Kneib}}{{Cowie}
  et~al.}{2002}]{cowie2002}
{Cowie} L.~L.,  {Barger} A.~J.,    {Kneib} J.-P.,  2002, \aj, 123, 2197

\bibitem[\protect\citeauthoryear{{Dale}, {Helou}, {Contursi}, {Silbermann} \&
  {Kolhatkar}}{{Dale} et~al.}{2001}]{dale2001}
{Dale} D.~A.,  {Helou} G.,  {Contursi} A.,  {Silbermann} N.~A.,    {Kolhatkar}
  S.,  2001, \apj, 549, 215

\bibitem[\protect\citeauthoryear{{Dannerbauer}, {Lehnert}, {Lutz}, {Tacconi},
  {Bertoldi}, {Carilli}, {Genzel} \& {Menten}}{{Dannerbauer}
  et~al.}{2004}]{dannerbauer2004}
{Dannerbauer} H.,  {Lehnert} M.~D.,  {Lutz} D.,  {Tacconi} L.,  {Bertoldi} F.,
  {Carilli} C.,  {Genzel} R.,    {Menten} K.~M.,  2004, \apj, 606, 664

\bibitem[\protect\citeauthoryear{{Devlin}, {Ade}, {Aretxaga}, {Bock}, {Chapin},
  {Griffin}, {Gundersen}, {Halpern}, {Hargrave}, {Hughes}, {Klein}, {Marsden},
  {Martin}, {Mauskopf}, {Moncelsi} \& {Netterfield}}{{Devlin}
  et~al.}{2009}]{devlin2009}
{Devlin} M.~J.,  {Ade} P.~A.~R.,  {Aretxaga} I.,  {Bock} J.~J.,  {Chapin}
  E.~L.,  {Griffin} M.,  {Gundersen} J.~O.,  {Halpern} M.,  {Hargrave} P.~C.,
  {Hughes} D.~H.,  {Klein} J.,  {Marsden} G.,  {Martin} P.~G.,  {Mauskopf} P.,
  {Moncelsi} L.,    {Netterfield} C.~B.,  2009, \nat, 458, 737

\bibitem[\protect\citeauthoryear{{Downes}, {Peacock}, {Savage} \&
  {Carrie}}{{Downes} et~al.}{1986}]{downes1986}
{Downes} A.~J.~B.,  {Peacock} J.~A.,  {Savage} A.,    {Carrie} D.~R.,  1986,
  \mnras, 218, 31

\bibitem[\protect\citeauthoryear{{Dunlop}, {Ade}, {Bock}, {Chapin},
  {Cirasuolo}, {Coppin}, {Devlin}, {Griffin}, {Greve} \& {Gundersen}}{{Dunlop}
  et~al.}{2010}]{dunlop2010}
{Dunlop} J.~S.,  {Ade} P.~A.~R.,  {Bock} J.~J.,  {Chapin} E.~L.,  {Cirasuolo}
  M.,  {Coppin} K.~E.~K.,  {Devlin} M.~J.,  {Griffin} M.,  {Greve} T.~R.,
  {Gundersen} J.~O.,  2010, \mnras, pp 1354--+

\bibitem[\protect\citeauthoryear{{Dye}, {Ade}, {Bock}, {Chapin}, {Devlin},
  {Dunlop}, {Eales}, {Griffin}, {Gundersen}, {Halpern}, {Hargrave}, {Hughes} \&
  {Klein}}{{Dye} et~al.}{2009}]{dye2009}
{Dye} S.,  {Ade} P.~A.~R.,  {Bock} J.~J.,  {Chapin} E.~L.,  {Devlin} M.~J.,
  {Dunlop} J.~S.,  {Eales} S.~A.,  {Griffin} M.,  {Gundersen} J.~O.,  {Halpern}
  M.,  {Hargrave} P.~C.,  {Hughes} D.~H.,    {Klein} J.,  2009, \apj, 703, 285

\bibitem[\protect\citeauthoryear{{Eales}, {Chapin}, {Devlin}, {Dye}, {Halpern},
  {Hughes}, {Marsden}, {Mauskopf}, {Moncelsi}, {Netterfield}, {Pascale},
  {Patanchon}, {Raymond}, {Rex}, {Scott}, {Semisch}, {Siana}, {Truch} \&
  {Viero}}{{Eales} et~al.}{2009}]{eales2009}
{Eales} S.,  {Chapin} E.~L.,  {Devlin} M.~J.,  {Dye} S.,  {Halpern} M.,
  {Hughes} D.~H.,  {Marsden} G.,  {Mauskopf} P.,  {Moncelsi} L.,  {Netterfield}
  C.~B.,  {Pascale} E.,  {Patanchon} G.,  {Raymond} G.,  {Rex} M.,  {Scott} D.,
   {Semisch} C.,  {Siana} B.,  {Truch} M.~D.~P.,    {Viero} M.~P.,  2009, \apj,
  707, 1779

\bibitem[\protect\citeauthoryear{{Fixsen}, {Dwek}, {Mather}, {Bennett} \&
  {Shafer}}{{Fixsen} et~al.}{1998}]{fixsen1998}
{Fixsen} D.~J.,  {Dwek} E.,  {Mather} J.~C.,  {Bennett} C.~L.,    {Shafer}
  R.~A.,  1998, \apj, 508, 123

\bibitem[\protect\citeauthoryear{{Fox}, {Efstathiou}, {Rowan-Robinson},
  {Dunlop}, {Scott}, {Serjeant}, {Mann}, {Oliver}, {Ivison}, {Blain},
  {Almaini}, {Hughes}, {Willott}, {Longair}, {Lawrence} \& {Peacock}}{{Fox}
  et~al.}{2002}]{fox2002}
{Fox} M.~J.,  {Efstathiou} A.,  {Rowan-Robinson} M.,  {Dunlop} J.~S.,  {Scott}
  S.,  {Serjeant} S.,  {Mann} R.~G.,  {Oliver} S.,  {Ivison} R.~J.,  {Blain}
  A.,  {Almaini} O.,  {Hughes} D.,  {Willott} C.~J.,  {Longair} M.,  {Lawrence}
  A.,    {Peacock} J.~A.,  2002, \mnras, 331, 839

\bibitem[\protect\citeauthoryear{{Gawiser}, {van Dokkum}, {Herrera}, {Maza},
  {Castander}, {Infante}, {Lira}, {Quadri}, {Toner}, {Treister}, {Urry},
  {Altmann}, {Assef} \& {Christlein}}{{Gawiser} et~al.}{2006}]{gawiser2006}
{Gawiser} E.,  {van Dokkum} P.~G.,  {Herrera} D.,  {Maza} J.,  {Castander}
  F.~J.,  {Infante} L.,  {Lira} P.,  {Quadri} R.,  {Toner} R.,  {Treister} E.,
  {Urry} C.~M.,  {Altmann} M.,  {Assef} R.,    {Christlein} D.,  2006, \apjs,
  162, 1

\bibitem[\protect\citeauthoryear{{Grazian}, {Fontana}, {de Santis}, {Nonino},
  {Salimbeni}, {Giallongo}, {Cristiani}, {Gallozzi} \& {Vanzella}}{{Grazian}
  et~al.}{2006}]{grazian2006}
{Grazian} A.,  {Fontana} A.,  {de Santis} C.,  {Nonino} M.,  {Salimbeni} S.,
  {Giallongo} E.,  {Cristiani} S.,  {Gallozzi} S.,    {Vanzella} E.,  2006,
  \aap, 449, 951

\bibitem[\protect\citeauthoryear{{Greve}, {Ivison}, {Bertoldi}, {Stevens},
  {Dunlop}, {Lutz} \& {Carilli}}{{Greve} et~al.}{2004}]{greve2004}
{Greve} T.~R.,  {Ivison} R.~J.,  {Bertoldi} F.,  {Stevens} J.~A.,  {Dunlop}
  J.~S.,  {Lutz} D.,    {Carilli} C.~L.,  2004, \mnras, 354, 779

\bibitem[\protect\citeauthoryear{{G{\"u}sten}, {Nyman}, {Schilke}, {Menten},
  {Cesarsky} \& {Booth}}{{G{\"u}sten} et~al.}{2006}]{gusten2006}
{G{\"u}sten} R.,  {Nyman} L.~{\AA}.,  {Schilke} P.,  {Menten} K.,  {Cesarsky}
  C.,    {Booth} R.,  2006, \aap, 454, L13

\bibitem[\protect\citeauthoryear{{Hainline}, {Blain}, {Smail}, {Frayer},
  {Chapman}, {Ivison} \& {Alexander}}{{Hainline} et~al.}{2009}]{hainline2009}
{Hainline} L.~J.,  {Blain} A.~W.,  {Smail} I.,  {Frayer} D.~T.,  {Chapman}
  S.~C.,  {Ivison} R.~J.,    {Alexander} D.~M.,  2009, \apj, 699, 1610

\bibitem[\protect\citeauthoryear{{Holland}, {Robson}, {Gear}, {Cunningham},
  {Lightfoot}, {Jenness}, {Ivison}, {Stevens}, {Ade}, {Griffin}, {Duncan},
  {Murphy} \& {Naylor}}{{Holland} et~al.}{1999}]{holland1999}
{Holland} W.~S.,  {Robson} E.~I.,  {Gear} W.~K.,  {Cunningham} C.~R.,
  {Lightfoot} J.~F.,  {Jenness} T.,  {Ivison} R.~J.,  {Stevens} J.~A.,  {Ade}
  P.~A.~R.,  {Griffin} M.~J.,  {Duncan} W.~D.,  {Murphy} J.~A.,    {Naylor}
  D.~A.,  1999, \mnras, 303, 659

\bibitem[\protect\citeauthoryear{{Hughes}, {Serjeant}, {Dunlop},
  {Rowan-Robinson}, {Blain}, {Mann}, {Ivison}, {Peacock}, {Efstathiou}, {Gear},
  {Oliver}, {Lawrence}, {Longair}, {Goldschmidt} \& {Jenness}}{{Hughes}
  et~al.}{1998}]{hughes1998}
{Hughes} D.~H.,  {Serjeant} S.,  {Dunlop} J.,  {Rowan-Robinson} M.,  {Blain}
  A.,  {Mann} R.~G.,  {Ivison} R.,  {Peacock} J.,  {Efstathiou} A.,  {Gear} W.,
   {Oliver} S.,  {Lawrence} A.,  {Longair} M.,  {Goldschmidt} P.,    {Jenness}
  T.,  1998, \nat, 394, 241

\bibitem[\protect\citeauthoryear{{Iono}, {Peck}, {Pope}, {Borys}, {Scott},
  {Wilner}, {Gurwell}, {Ho}, {Yun}, {Matsushita}, {Petitpas}, {Dunlop},
  {Elvis}, {Blain} \& {Le Floc'h}}{{Iono} et~al.}{2006}]{iono2006}
{Iono} D.,  {Peck} A.~B.,  {Pope} A.,  {Borys} C.,  {Scott} D.,  {Wilner}
  D.~J.,  {Gurwell} M.,  {Ho} P.~T.~P.,  {Yun} M.~S.,  {Matsushita} S.,
  {Petitpas} G.~R.,  {Dunlop} J.~S.,  {Elvis} M.,  {Blain} A.,    {Le Floc'h}
  E.,  2006, \apjl, 640, L1

\bibitem[\protect\citeauthoryear{{Ivison}, {Alexander}, {Biggs}, {Brandt},
  {Chapin}, {Coppin}, {Devlin}, {Dickinson}, {Dunlop}, {Dye}, {Eales}, {Frayer}
  \& {Halpern}}{{Ivison} et~al.}{2010}]{ivison2009}
{Ivison} R.~J.,  {Alexander} D.~M.,  {Biggs} A.~D.,  {Brandt} W.~N.,  {Chapin}
  E.~L.,  {Coppin} K.~E.~K.,  {Devlin} M.~J.,  {Dickinson} M.,  {Dunlop} J.,
  {Dye} S.,  {Eales} S.~A.,  {Frayer} D.~T.,    {Halpern} M.,  2010, \mnras,
  402, 245

\bibitem[\protect\citeauthoryear{{Ivison}, {Greve}, {Dunlop}, {Peacock},
  {Egami}, {Smail}, {Ibar}, {van Kampen}, {Aretxaga}, {Babbedge}, {Biggs},
  {Blain}, {Chapman} \& {Clements}}{{Ivison} et~al.}{2007}]{ivison2007}
{Ivison} R.~J.,  {Greve} T.~R.,  {Dunlop} J.~S.,  {Peacock} J.~A.,  {Egami} E.,
   {Smail} I.,  {Ibar} E.,  {van Kampen} E.,  {Aretxaga} I.,  {Babbedge} T.,
  {Biggs} A.~D.,  {Blain} A.~W.,  {Chapman} S.~C.,    {Clements} D.~L.,  2007,
  \mnras, 380, 199

\bibitem[\protect\citeauthoryear{{Ivison}, {Greve}, {Serjeant}, {Bertoldi},
  {Egami}, {Mortier}, {Alonso-Herrero}, {Barmby}, {Bei}, {Dole}, {Engelbracht},
  {Fazio} \& {Frayer}}{{Ivison} et~al.}{2004}]{ivison2004}
{Ivison} R.~J.,  {Greve} T.~R.,  {Serjeant} S.,  {Bertoldi} F.,  {Egami} E.,
  {Mortier} A.~M.~J.,  {Alonso-Herrero} A.,  {Barmby} P.,  {Bei} L.,  {Dole}
  H.,  {Engelbracht} C.~W.,  {Fazio} G.~G.,    {Frayer} D.~T.,  2004, \apjs,
  154, 124

\bibitem[\protect\citeauthoryear{{Ivison}, {Greve}, {Smail}, {Dunlop}, {Roche},
  {Scott}, {Page}, {Stevens}, {Almaini}, {Blain}, {Willott}, {Fox}, {Gilbank},
  {Serjeant} \& {Hughes}}{{Ivison} et~al.}{2002}]{ivison2002}
{Ivison} R.~J.,  {Greve} T.~R.,  {Smail} I.,  {Dunlop} J.~S.,  {Roche} N.~D.,
  {Scott} S.~E.,  {Page} M.~J.,  {Stevens} J.~A.,  {Almaini} O.,  {Blain}
  A.~W.,  {Willott} C.~J.,  {Fox} M.~J.,  {Gilbank} D.~G.,  {Serjeant} S.,
  {Hughes} D.~H.,  2002, \mnras, 337, 1

\bibitem[\protect\citeauthoryear{{Khan}, {Shafer}, {Serjeant}, {Willner},
  {Pearson}, {Benford}, {Staguhn}, {Moseley}, {Sumner}, {Ashby}, {Borys},
  {Chanial}, {Clements}, {Dowell}, {Dwek}, {Fazio}, {Kov{\'a}cs}, {Le Floc'h}
  \& {Silverberg}}{{Khan} et~al.}{2007}]{khan2007}
{Khan} S.~A.,  {Shafer} R.~A.,  {Serjeant} S.,  {Willner} S.~P.,  {Pearson}
  C.~P.,  {Benford} D.~J.,  {Staguhn} J.~G.,  {Moseley} S.~H.,  {Sumner} T.~J.,
   {Ashby} M.~L.~N.,  {Borys} C.~K.,  {Chanial} P.,  {Clements} D.~L.,
  {Dowell} C.~D.,  {Dwek} E.,  {Fazio} G.~G.,  {Kov{\'a}cs} A.,  {Le Floc'h}
  E.,    {Silverberg} R.~F.,  2007, \apj, 665, 973

\bibitem[\protect\citeauthoryear{{Klaas}, {Haas}, {M{\" u}ller}, {Chini},
  {Schulz}, {Coulson}, {Hippelein}, {Wilke}, {Albrecht} \& {Lemke}}{{Klaas}
  et~al.}{2001}]{klaas2001}
{Klaas} U.,  {Haas} M.,  {M{\" u}ller} S.~A.~H.,  {Chini} R.,  {Schulz} B.,
  {Coulson} I.,  {Hippelein} H.,  {Wilke} K.,  {Albrecht} M.,    {Lemke} D.,
  2001, \aap, 379, 823

\bibitem[\protect\citeauthoryear{{Kov{\'a}cs}, {Chapman}, {Dowell}, {Blain},
  {Ivison}, {Smail} \& {Phillips}}{{Kov{\'a}cs} et~al.}{2006}]{kovacs2006}
{Kov{\'a}cs} A.,  {Chapman} S.~C.,  {Dowell} C.~D.,  {Blain} A.~W.,  {Ivison}
  R.~J.,  {Smail} I.,    {Phillips} T.~G.,  2006, \apj, 650, 592

\bibitem[\protect\citeauthoryear{{Lagache}, {Dole} \& {Puget}}{{Lagache}
  et~al.}{2003}]{lagache2003}
{Lagache} G.,  {Dole} H.,    {Puget} J.-L.,  2003, \mnras, 338, 555

\bibitem[\protect\citeauthoryear{{Laurent}, {Aguirre}, {Glenn}, {Ade}, {Bock},
  {Edgington}, {Goldin}, {Golwala}, {Haig}, {Lange}, {Maloney}, {Mauskopf},
  {Nguyen}, {Rossinot}, {Sayers} \& {Stover}}{{Laurent}
  et~al.}{2005}]{laurent2005}
{Laurent} G.~T.,  {Aguirre} J.~E.,  {Glenn} J.,  {Ade} P.~A.~R.,  {Bock} J.~J.,
   {Edgington} S.~F.,  {Goldin} A.,  {Golwala} S.~R.,  {Haig} D.,  {Lange}
  A.~E.,  {Maloney} P.~R.,  {Mauskopf} P.~D.,  {Nguyen} H.,  {Rossinot} P.,
  {Sayers} J.,    {Stover} P.,  2005, \apj, 623, 742

\bibitem[\protect\citeauthoryear{{Lutz}, {Dunlop}, {Almaini}, {Andreani},
  {Blain}, {Efstathiou}, {Fox}, {Genzel}, {Hasinger} \& {Hughes}}{{Lutz}
  et~al.}{2001}]{lutz2001}
{Lutz} D.,  {Dunlop} J.~S.,  {Almaini} O.,  {Andreani} P.,  {Blain} A.,
  {Efstathiou} A.,  {Fox} M.,  {Genzel} R.,  {Hasinger} G.,    {Hughes} D.,
  2001, \aap, 378, 70

\bibitem[\protect\citeauthoryear{{Magnelli}, {Elbaz}, {Chary}, {Dickinson}, {Le
  Borgne}, {Frayer} \& {Willmer}}{{Magnelli} et~al.}{2009}]{magnelli2009}
{Magnelli} B.,  {Elbaz} D.,  {Chary} R.~R.,  {Dickinson} M.,  {Le Borgne} D.,
  {Frayer} D.~T.,    {Willmer} C.~N.~A.,  2009, \aap, 496, 57

\bibitem[\protect\citeauthoryear{{Mann}, {Oliver}, {Serjeant},
  {Rowan-Robinson}, {Baker}, {Eaton}, {Efstathiou}, {Goldschmidt}, {Mobasher}
  \& {Sumner}}{{Mann} et~al.}{1997}]{mann1997}
{Mann} R.~G.,  {Oliver} S.~J.,  {Serjeant} S.~B.~G.,  {Rowan-Robinson} M.,
  {Baker} A.,  {Eaton} N.,  {Efstathiou} A.,  {Goldschmidt} P.,  {Mobasher} B.,
     {Sumner} T.~J.,  1997, \mnras, 289, 482

\bibitem[\protect\citeauthoryear{{Marsden}, {Ade}, {Bock}, {Chapin}, {Devlin},
  {Dicker}, {Griffin}, {Gundersen}, {Halpern}, {Hargrave}, {Hughes}, {Klein},
  {Mauskopf}, {Magnelli}, {Moncelsi}, {Netterfield} \& {Ngo}}{{Marsden}
  et~al.}{2009}]{marsden2009}
{Marsden} G.,  {Ade} P.~A.~R.,  {Bock} J.~J.,  {Chapin} E.~L.,  {Devlin} M.~J.,
   {Dicker} S.~R.,  {Griffin} M.,  {Gundersen} J.~O.,  {Halpern} M.,
  {Hargrave} P.~C.,  {Hughes} D.~H.,  {Klein} J.,  {Mauskopf} P.,  {Magnelli}
  B.,  {Moncelsi} L.,  {Netterfield} C.~B.,    {Ngo} H.,  2009, \apj, 707, 1729

\bibitem[\protect\citeauthoryear{{Men{\'e}ndez-Delmestre}, {Blain}, {Smail},
  {Alexander}, {Chapman}, {Armus}, {Frayer}, {Ivison} \&
  {Teplitz}}{{Men{\'e}ndez-Delmestre} et~al.}{2009}]{menendez2009}
{Men{\'e}ndez-Delmestre} K.,  {Blain} A.~W.,  {Smail} I.,  {Alexander} D.~M.,
  {Chapman} S.~C.,  {Armus} L.,  {Frayer} D.,  {Ivison} R.~J.,    {Teplitz} H.,
   2009, \apj, 699, 667

\bibitem[\protect\citeauthoryear{{Miller}, {Fomalont}, {Kellermann},
  {Mainieri}, {Norman}, {Padovani}, {Rosati} \& {Tozzi}}{{Miller}
  et~al.}{2008}]{miller2008}
{Miller} N.~A.,  {Fomalont} E.~B.,  {Kellermann} K.~I.,  {Mainieri} V.,
  {Norman} C.,  {Padovani} P.,  {Rosati} P.,    {Tozzi} P.,  2008, \apjs, 179,
  114

\bibitem[\protect\citeauthoryear{{Mo} \& {White}}{{Mo} \&
  {White}}{1996}]{mo1996}
{Mo} H.~J.,  {White} S.~D.~M.,  1996, \mnras, 282, 347

\bibitem[\protect\citeauthoryear{{Moncelsi}, {Ade}, {Chapin}, {Cortese},
  {Devlin}, {Dye}, {Eales}, {Griffin}, {Halpern}, {Hargrave}, {Marsden},
  {Mauskopf}, {Netterfield}, {Pascale}, {Scott}, {Truch}, {Tucker}, {Viero} \&
  {Wiebe}}{{Moncelsi} et~al.}{2010}]{moncelsi2010}
{Moncelsi} L.,  {Ade} P.~A.~R.,  {Chapin} E.~L.,  {Cortese} L.,  {Devlin}
  M.~J.,  {Dye} S.,  {Eales} S.,  {Griffin} M.,  {Halpern} M.,  {Hargrave}
  P.~C.,  {Marsden} G.,  {Mauskopf} P.,  {Netterfield} C.~B.,  {Pascale} E.,
  {Scott} D.,  {Truch} M.~D.~P.,  {Tucker} C.,  {Viero} M.,    {Wiebe} D.,
  2010, ArXiv e-prints

\bibitem[\protect\citeauthoryear{{Nguyen}, {Schulz}, {Levenson}, {Amblard},
  {Arumugam}, {Aussel}, {Babbedge}, {Blain}, {Bock}, {Boselli}, {Buat},
  {Castro-Rodriguez}, {Cava}, {Chanial} \& {Chapin}}{{Nguyen}
  et~al.}{2010}]{nguyen2010}
{Nguyen} H.~T.,  {Schulz} B.,  {Levenson} L.,  {Amblard} A.,  {Arumugam} V.,
  {Aussel} H.,  {Babbedge} T.,  {Blain} A.,  {Bock} J.,  {Boselli} A.,  {Buat}
  V.,  {Castro-Rodriguez} N.,  {Cava} A.,  {Chanial} P.,    {Chapin} E.,  2010,
  \aap, 518, L5+

\bibitem[\protect\citeauthoryear{{Oliver}, {Wang}, {Smith}, {Altieri},
  {Amblard}, {Arumugam}, {Auld}, {Aussel}, {Babbedge}, {Blain}, {Bock},
  {Boselli}, {Buat} \& {Burgarella}}{{Oliver} et~al.}{2010}]{oliver2010}
{Oliver} S.~J.,  {Wang} L.,  {Smith} A.~J.,  {Altieri} B.,  {Amblard} A.,
  {Arumugam} V.,  {Auld} R.,  {Aussel} H.,  {Babbedge} T.,  {Blain} A.,  {Bock}
  J.,  {Boselli} A.,  {Buat} V.,    {Burgarella} D.,  2010, \aap, 518, L21+

\bibitem[\protect\citeauthoryear{{Pascale}, {Ade}, {Bock}, {Chapin}, {Chung},
  {Devlin}, {Dicker}, {Griffin}, {Gundersen}, {Halpern}, {Hargrave}, {Hughes},
  {Klein} \& {MacTavish}}{{Pascale} et~al.}{2008}]{pascale2008}
{Pascale} E.,  {Ade} P.~A.~R.,  {Bock} J.~J.,  {Chapin} E.~L.,  {Chung} J.,
  {Devlin} M.~J.,  {Dicker} S.,  {Griffin} M.,  {Gundersen} J.~O.,  {Halpern}
  M.,  {Hargrave} P.~C.,  {Hughes} D.~H.,  {Klein} J.,    {MacTavish} C.~J.,
  2008, \apj, 681, 400

\bibitem[\protect\citeauthoryear{{Pascale}, {Ade}, {Bock}, {Chapin}, {Devlin},
  {Dye}, {Eales}, {Griffin}, {Gundersen}, {Halpern}, {Hargrave}, {Hughes} \&
  {Klein}}{{Pascale} et~al.}{2009}]{pascale2009}
{Pascale} E.,  {Ade} P.~A.~R.,  {Bock} J.~J.,  {Chapin} E.~L.,  {Devlin} M.~J.,
   {Dye} S.,  {Eales} S.~A.,  {Griffin} M.,  {Gundersen} J.~O.,  {Halpern} M.,
  {Hargrave} P.~C.,  {Hughes} D.~H.,    {Klein} J.,  2009, \apj, 707, 1740

\bibitem[\protect\citeauthoryear{{Patanchon}, {Ade}, {Bock}, {Chapin},
  {Devlin}, {Dicker}, {Griffin}, {Gundersen}, {Halpern}, {Hargrave}, {Hughes},
  {Klein}, {Marsden} \& {Martin}}{{Patanchon} et~al.}{2008}]{patanchon2008}
{Patanchon} G.,  {Ade} P.~A.~R.,  {Bock} J.~J.,  {Chapin} E.~L.,  {Devlin}
  M.~J.,  {Dicker} S.,  {Griffin} M.,  {Gundersen} J.~O.,  {Halpern} M.,
  {Hargrave} P.~C.,  {Hughes} D.~H.,  {Klein} J.,  {Marsden} G.,    {Martin}
  P.~G.,  2008, \apj, 681, 708

\bibitem[\protect\citeauthoryear{{Patanchon}, {Ade}, {Bock}, {Chapin},
  {Devlin}, {Dicker}, {Griffin}, {Gundersen}, {Halpern}, {Hargrave}, {Hughes},
  {Klein}, {Marsden} \& {Mauskopf}}{{Patanchon} et~al.}{2009}]{patanchon2009}
{Patanchon} G.,  {Ade} P.~A.~R.,  {Bock} J.~J.,  {Chapin} E.~L.,  {Devlin}
  M.~J.,  {Dicker} S.~R.,  {Griffin} M.,  {Gundersen} J.~O.,  {Halpern} M.,
  {Hargrave} P.~C.,  {Hughes} D.~H.,  {Klein} J.,  {Marsden} G.,    {Mauskopf}
  P.,  2009, \apj, 707, 1750

\bibitem[\protect\citeauthoryear{{Perera}, {Chapin}, {Austermann}, {Scott},
  {Wilson}, {Halpern}, {Pope}, {Scott}, {Yun}, {Lowenthal}, {Morrison},
  {Aretxaga}, {Bock}, {Coppin}, {Crowe}, {Frey}, {Hughes}, {Kang}, {Kim} \&
  {Mauskopf}}{{Perera} et~al.}{2008}]{perera2008}
{Perera} T.~A.,  {Chapin} E.~L.,  {Austermann} J.~E.,  {Scott} K.~S.,  {Wilson}
  G.~W.,  {Halpern} M.,  {Pope} A.,  {Scott} D.,  {Yun} M.~S.,  {Lowenthal}
  J.~D.,  {Morrison} G.,  {Aretxaga} I.,  {Bock} J.~J.,  {Coppin} K.,  {Crowe}
  M.,  {Frey} L.,  {Hughes} D.~H.,  {Kang} Y.,  {Kim} S.,    {Mauskopf} P.~D.,
  2008, \mnras, 391, 1227

\bibitem[\protect\citeauthoryear{{Pope}, {Scott}, {Dickinson}, {Chary},
  {Morrison}, {Borys}, {Sajina}, {Alexander}, {Daddi}, {Frayer}, {MacDonald} \&
  {Stern}}{{Pope} et~al.}{2006}]{pope2006}
{Pope} A.,  {Scott} D.,  {Dickinson} M.,  {Chary} R.-R.,  {Morrison} G.,
  {Borys} C.,  {Sajina} A.,  {Alexander} D.~M.,  {Daddi} E.,  {Frayer} D.,
  {MacDonald} E.,    {Stern} D.,  2006, \mnras, 370, 1185

\bibitem[\protect\citeauthoryear{{Pope}}{{Pope}}{2007}]{pope2007}
{Pope} E.~A.,  2007, PhD thesis, The University of British Columbia (Canada

\bibitem[\protect\citeauthoryear{{Puget}, {Abergel}, {Bernard}, {Boulanger},
  {Burton}, {Desert} \& {Hartmann}}{{Puget} et~al.}{1996}]{puget1996}
{Puget} J.,  {Abergel} A.,  {Bernard} J.,  {Boulanger} F.,  {Burton} W.~B.,
  {Desert} F.,    {Hartmann} D.,  1996, \aap, 308, L5+

\bibitem[\protect\citeauthoryear{{Roseboom}, {Oliver}, {Parkinson} \&
  {Vaccari}}{{Roseboom} et~al.}{2009}]{roseboom2009}
{Roseboom} I.~G.,  {Oliver} S.,  {Parkinson} D.,    {Vaccari} M.,  2009,
  \mnras, 400, 1062

\bibitem[\protect\citeauthoryear{{Rowan-Robinson}}{{Rowan-Robinson}}{2001}]{ro%
wan-robinson2001}
{Rowan-Robinson} M.,  2001, \apj, 549, 745

\bibitem[\protect\citeauthoryear{{Rowan-Robinson}}{{Rowan-Robinson}}{2009}]{ro%
wan-robinson2009}
{Rowan-Robinson} M.,  2009, \mnras, 394, 117

\bibitem[\protect\citeauthoryear{{Rowan-Robinson}, {Babbedge}, {Oliver},
  {Trichas}, {Berta}, {Lonsdale}, {Smith}, {Shupe}, {Surace} \&
  {Arnouts}}{{Rowan-Robinson} et~al.}{2008}]{rowan-robinson2008}
{Rowan-Robinson} M.,  {Babbedge} T.,  {Oliver} S.,  {Trichas} M.,  {Berta} S.,
  {Lonsdale} C.,  {Smith} G.,  {Shupe} D.,  {Surace} J.,    {Arnouts} S.,
  2008, \mnras, 386, 697

\bibitem[\protect\citeauthoryear{{Rutledge}, {Brunner}, {Prince} \&
  {Lonsdale}}{{Rutledge} et~al.}{2000}]{rutledge2000}
{Rutledge} R.~E.,  {Brunner} R.~J.,  {Prince} T.~A.,    {Lonsdale} C.,  2000,
  \apjs, 131, 335

\bibitem[\protect\citeauthoryear{{Sanders} \& {Mirabel}}{{Sanders} \&
  {Mirabel}}{1996}]{sanders1996}
{Sanders} D.~B.,  {Mirabel} I.~F.,  1996, \araa, 34, 749

\bibitem[\protect\citeauthoryear{{Sawicki}}{{Sawicki}}{2002}]{sawicki2002}
{Sawicki} M.,  2002, \aj, 124, 3050

\bibitem[\protect\citeauthoryear{{Scott}, {Austermann}, {Perera}, {Wilson},
  {Aretxaga}, {Bock}, {Hughes}, {Kang}, {Kim}, {Mauskopf}, {Sanders},
  {Scoville} \& {Yun}}{{Scott} et~al.}{2008}]{scott2008}
{Scott} K.~S.,  {Austermann} J.~E.,  {Perera} T.~A.,  {Wilson} G.~W.,
  {Aretxaga} I.,  {Bock} J.~J.,  {Hughes} D.~H.,  {Kang} Y.,  {Kim} S.,
  {Mauskopf} P.~D.,  {Sanders} D.~B.,  {Scoville} N.,    {Yun} M.~S.,  2008,
  \mnras, 385, 2225

\bibitem[\protect\citeauthoryear{{Scott}, {Fox}, {Dunlop}, {Serjeant},
  {Peacock}, {Ivison}, {Oliver}, {Mann}, {Lawrence}, {Efstathiou},
  {Rowan-Robinson}, {Hughes}, {Archibald}, {Blain} \& {Longair}}{{Scott}
  et~al.}{2002}]{scott2002}
{Scott} S.~E.,  {Fox} M.~J.,  {Dunlop} J.~S.,  {Serjeant} S.,  {Peacock} J.~A.,
   {Ivison} R.~J.,  {Oliver} S.,  {Mann} R.~G.,  {Lawrence} A.,  {Efstathiou}
  A.,  {Rowan-Robinson} M.,  {Hughes} D.~H.,  {Archibald} E.~N.,  {Blain} A.,
   {Longair} M.,  2002, \mnras, 331, 817

\bibitem[\protect\citeauthoryear{{Serjeant}, {Dunlop}, {Mann},
  {Rowan-Robinson}, {Hughes}, {Efstathiou}, {Blain}, {Fox}, {Ivison},
  {Jenness}, {Lawrence}, {Longair}, {Oliver} \& {Peacock}}{{Serjeant}
  et~al.}{2003}]{serjeant2003}
{Serjeant} S.,  {Dunlop} J.~S.,  {Mann} R.~G.,  {Rowan-Robinson} M.,  {Hughes}
  D.,  {Efstathiou} A.,  {Blain} A.,  {Fox} M.,  {Ivison} R.~J.,  {Jenness} T.,
   {Lawrence} A.,  {Longair} M.,  {Oliver} S.,    {Peacock} J.~A.,  2003,
  \mnras, 344, 887

\bibitem[\protect\citeauthoryear{{Simpson} \& {Eisenhardt}}{{Simpson} \&
  {Eisenhardt}}{1999}]{simpson1999}
{Simpson} C.,  {Eisenhardt} P.,  1999, PASP, 111, 691

\bibitem[\protect\citeauthoryear{{Siringo}, {Kreysa}, {Kov{\'a}cs}, {Schuller},
  {Wei{\ss}}, {Esch}, {Gem{\"u}nd}, {Jethava}, {Lundershausen}, {Colin},
  {G{\"u}sten}, {Menten}, {Beelen}, {Bertoldi}, {Beeman} \& {Haller}}{{Siringo}
  et~al.}{2009}]{siringo2009}
{Siringo} G.,  {Kreysa} E.,  {Kov{\'a}cs} A.,  {Schuller} F.,  {Wei{\ss}} A.,
  {Esch} W.,  {Gem{\"u}nd} H.,  {Jethava} N.,  {Lundershausen} G.,  {Colin} A.,
   {G{\"u}sten} R.,  {Menten} K.~M.,  {Beelen} A.,  {Bertoldi} F.,  {Beeman}
  J.~W.,    {Haller} E.~E.,  2009, \aap, 497, 945

\bibitem[\protect\citeauthoryear{{Smail}, {Ivison} \& {Blain}}{{Smail}
  et~al.}{1997}]{smail1997}
{Smail} I.,  {Ivison} R.~J.,    {Blain} A.~W.,  1997, \apjl, 490, L5+

\bibitem[\protect\citeauthoryear{{Smail}, {Ivison}, {Owen}, {Blain} \&
  {Kneib}}{{Smail} et~al.}{2000}]{smail2000}
{Smail} I.,  {Ivison} R.~J.,  {Owen} F.~N.,  {Blain} A.~W.,    {Kneib} J.,
  2000, \apj, 528, 612

\bibitem[\protect\citeauthoryear{{Smith}, {Dunne}, {Maddox}, {Eales},
  {Bonfield}, {Jarvis}, {Sutherland}, {Fleuren} \& {Rigby}}{{Smith}
  et~al.}{2010}]{smith2010}
{Smith} D.~J.~B.,  {Dunne} L.,  {Maddox} S.~J.,  {Eales} S.,  {Bonfield} D.~G.,
   {Jarvis} M.~J.,  {Sutherland} W.,  {Fleuren} S.,    {Rigby} E.~E.,  2010,
  ArXiv e-prints

\bibitem[\protect\citeauthoryear{{Stetson}}{{Stetson}}{1987}]{stetson1987}
{Stetson} P.~B.,  1987, \pasp, 99, 191

\bibitem[\protect\citeauthoryear{{Sutherland} \& {Saunders}}{{Sutherland} \&
  {Saunders}}{1992}]{sutherland1992}
{Sutherland} W.,  {Saunders} W.,  1992, \mnras, 259, 413

\bibitem[\protect\citeauthoryear{{Tacconi}, {Genzel}, {Smail}, {Neri},
  {Chapman}, {Ivison}, {Blain}, {Cox}, {Omont}, {Bertoldi} \&
  {Greve}}{{Tacconi} et~al.}{2008}]{tacconi2008}
{Tacconi} L.~J.,  {Genzel} R.,  {Smail} I.,  {Neri} R.,  {Chapman} S.~C.,
  {Ivison} R.~J.,  {Blain} A.,  {Cox} P.,  {Omont} A.,  {Bertoldi} F.,
  {Greve} T.,  2008, \apj, 680, 246

\bibitem[\protect\citeauthoryear{{Takeuchi}, {Yoshikawa} \& {Ishii}}{{Takeuchi}
  et~al.}{2003}]{takeuchi2003}
{Takeuchi} T.~T.,  {Yoshikawa} K.,    {Ishii} T.~T.,  2003, \apjl, 587, L89

\bibitem[\protect\citeauthoryear{{Takeuchi}, {Yoshikawa} \& {Ishii}}{{Takeuchi}
  et~al.}{2004}]{takeuchi2004b}
{Takeuchi} T.~T.,  {Yoshikawa} K.,    {Ishii} T.~T.,  2004, \apjl, 606, L171

\bibitem[\protect\citeauthoryear{{Taylor}, {Franx}, {van Dokkum}, {Quadri},
  {Gawiser}, {Bell}, {Barrientos}, {Blanc}, {Castander}, {Damen},
  {Gonzalez-Perez}, {Hall}, {Herrera} \& {Hildebrandt}}{{Taylor}
  et~al.}{2009}]{taylor2009}
{Taylor} E.~N.,  {Franx} M.,  {van Dokkum} P.~G.,  {Quadri} R.~F.,  {Gawiser}
  E.,  {Bell} E.~F.,  {Barrientos} L.~F.,  {Blanc} G.~A.,  {Castander} F.~J.,
  {Damen} M.,  {Gonzalez-Perez} V.,  {Hall} P.~B.,  {Herrera} D.,
  {Hildebrandt} H.,  2009, \apjs, 183, 295

\bibitem[\protect\citeauthoryear{{Tegmark} \& {de Oliveira-Costa}}{{Tegmark} \&
  {de Oliveira-Costa}}{1998}]{tegmark1998}
{Tegmark} M.,  {de Oliveira-Costa} A.,  1998, \apjl, 500, L83+

\bibitem[\protect\citeauthoryear{{Truch}, {Ade}, {Bock}, {Chapin}, {Devlin},
  {Dicker}, {Griffin}, {Gundersen}, {Halpern}, {Hargrave}, {Hughes}, {Klein} \&
  {Marsden}}{{Truch} et~al.}{2009}]{truch2009}
{Truch} M.~D.~P.,  {Ade} P.~A.~R.,  {Bock} J.~J.,  {Chapin} E.~L.,  {Devlin}
  M.~J.,  {Dicker} S.~R.,  {Griffin} M.,  {Gundersen} J.~O.,  {Halpern} M.,
  {Hargrave} P.~C.,  {Hughes} D.~H.,  {Klein} J.,    {Marsden} G.,  2009, \apj,
  707, 1723

\bibitem[\protect\citeauthoryear{{V{\"a}is{\"a}nen}, {Kotilainen}, {Juvela},
  {Mattila}, {Efstathiou} \& {Kahanp{\"a}{\"a}}}{{V{\"a}is{\"a}nen}
  et~al.}{2010}]{vaisanen2010}
{V{\"a}is{\"a}nen} P.,  {Kotilainen} J.~K.,  {Juvela} M.,  {Mattila} K.,
  {Efstathiou} A.,    {Kahanp{\"a}{\"a}} J.,  2010, \mnras, 401, 1587

\bibitem[\protect\citeauthoryear{{Valiante}, {Lutz}, {Sturm}, {Genzel} \&
  {Chapin}}{{Valiante} et~al.}{2009}]{valiante2009}
{Valiante} E.,  {Lutz} D.,  {Sturm} E.,  {Genzel} R.,    {Chapin} E.~L.,  2009,
  \apj, 701, 1814

\bibitem[\protect\citeauthoryear{{Viero}, {Ade}, {Bock}, {Chapin}, {Devlin},
  {Griffin}, {Gundersen}, {Halpern}, {Hargrave}, {Hughes}, {Klein}, {MacTavish}
  \& {Marsden}}{{Viero} et~al.}{2009}]{viero2009}
{Viero} M.~P.,  {Ade} P.~A.~R.,  {Bock} J.~J.,  {Chapin} E.~L.,  {Devlin}
  M.~J.,  {Griffin} M.,  {Gundersen} J.~O.,  {Halpern} M.,  {Hargrave} P.~C.,
  {Hughes} D.~H.,  {Klein} J.,  {MacTavish} C.~J.,    {Marsden} G.,  2009,
  \apj, 707, 1766

\bibitem[\protect\citeauthoryear{{Vio}, {Andreani} \& {Wamsteker}}{{Vio}
  et~al.}{2004}]{vio2004}
{Vio} R.,  {Andreani} P.,    {Wamsteker} W.,  2004, \aap, 414, 17

\bibitem[\protect\citeauthoryear{{Wang}, {Cowie} \& {Barger}}{{Wang}
  et~al.}{2006}]{wang2006}
{Wang} W.,  {Cowie} L.~L.,    {Barger} A.~J.,  2006, \apj, 647, 74

\bibitem[\protect\citeauthoryear{{Webb}, {Eales}, {Lilly}, {Clements}, {Dunne},
  {Gear}, {Ivison}, {Flores} \& {Yun}}{{Webb} et~al.}{2003}]{webb2003}
{Webb} T.~M.,  {Eales} S.~A.,  {Lilly} S.~J.,  {Clements} D.~L.,  {Dunne} L.,
  {Gear} W.~K.,  {Ivison} R.~J.,  {Flores} H.,    {Yun} M.,  2003, \apj, 587,
  41

\bibitem[\protect\citeauthoryear{{Weiss}, {Kovacs}, {Coppin}, {Greve},
  {Walter}, {Smail}, {Dunlop}, {Knudsen}, {Alexander}, {Bertoldi}, {Brandt},
  {Chapman} \& {Cox}}{{Weiss} et~al.}{2009}]{weiss2009}
{Weiss} A.,  {Kovacs} A.,  {Coppin} K.,  {Greve} T.~R.,  {Walter} F.,  {Smail}
  I.,  {Dunlop} J.~S.,  {Knudsen} K.~K.,  {Alexander} D.~M.,  {Bertoldi} F.,
  {Brandt} W.~N.,  {Chapman} S.~C.,    {Cox} P.,  2009, ArXiv e-prints

\bibitem[\protect\citeauthoryear{{Wiebe}, {Ade}, {Bock}, {Chapin}, {Devlin},
  {Dicker}, {Griffin}, {Gundersen}, {Halpern}, {Hargrave}, {Hughes}, {Klein},
  {Marsden} \& {Martin}}{{Wiebe} et~al.}{2009}]{wiebe2009}
{Wiebe} D.~V.,  {Ade} P.~A.~R.,  {Bock} J.~J.,  {Chapin} E.~L.,  {Devlin}
  M.~J.,  {Dicker} S.,  {Griffin} M.,  {Gundersen} J.~O.,  {Halpern} M.,
  {Hargrave} P.~C.,  {Hughes} D.~H.,  {Klein} J.,  {Marsden} G.,    {Martin}
  P.~G.,  2009, \apj, 707, 1809

\bibitem[\protect\citeauthoryear{{Wilman}, {Jarvis}, {Mauch}, {Rawlings} \&
  {Hickey}}{{Wilman} et~al.}{2010}]{wilman2010}
{Wilman} R.~J.,  {Jarvis} M.~J.,  {Mauch} T.,  {Rawlings} S.,    {Hickey} S.,
  2010, ArXiv e-prints

\bibitem[\protect\citeauthoryear{{Wilson}, {Austermann}, {Perera}, {Scott},
  {Ade}, {Bock}, {Glenn}, {Golwala}, {Kim}, {Kang}, {Lydon}, {Mauskopf},
  {Predmore}, {Roberts}, {Souccar} \& {Yun}}{{Wilson}
  et~al.}{2008}]{wilson2008}
{Wilson} G.~W.,  {Austermann} J.~E.,  {Perera} T.~A.,  {Scott} K.~S.,  {Ade}
  P.~A.~R.,  {Bock} J.~J.,  {Glenn} J.,  {Golwala} S.~R.,  {Kim} S.,  {Kang}
  Y.,  {Lydon} D.,  {Mauskopf} P.~D.,  {Predmore} C.~R.,  {Roberts} C.~M.,
  {Souccar} K.,    {Yun} M.~S.,  2008, \mnras, 386, 807

\bibitem[\protect\citeauthoryear{{Wolf}, {Hildebrandt}, {Taylor} \&
  {Meisenheimer}}{{Wolf} et~al.}{2008}]{wolf2008}
{Wolf} C.,  {Hildebrandt} H.,  {Taylor} E.~N.,    {Meisenheimer} K.,  2008,
  \aap, 492, 933

\bibitem[\protect\citeauthoryear{{Wolf}, {Meisenheimer}, {Kleinheinrich},
  {Borch}, {Dye}, {Gray}, {Wisotzki}, {Bell}, {Rix}, {Cimatti}, {Hasinger} \&
  {Szokoly}}{{Wolf} et~al.}{2004}]{wolf2004}
{Wolf} C.,  {Meisenheimer} K.,  {Kleinheinrich} M.,  {Borch} A.,  {Dye} S.,
  {Gray} M.,  {Wisotzki} L.,  {Bell} E.~F.,  {Rix} H.,  {Cimatti} A.,
  {Hasinger} G.,    {Szokoly} G.,  2004, \aap, 421, 913

\bibitem[\protect\citeauthoryear{{Younger}, {Fazio}, {Huang}, {Yun}, {Wilson},
  {Ashby}, {Gurwell}, {Lai}, {Peck}, {Petitpas}, {Wilner}, {Iono}, {Kohno} \&
  {Kawabe}}{{Younger} et~al.}{2007}]{younger2007}
{Younger} J.~D.,  {Fazio} G.~G.,  {Huang} J.,  {Yun} M.~S.,  {Wilson} G.~W.,
  {Ashby} M.~L.~N.,  {Gurwell} M.~A.,  {Lai} K.,  {Peck} A.~B.,  {Petitpas}
  G.~R.,  {Wilner} D.~J.,  {Iono} D.,  {Kohno} K.,    {Kawabe} R.,  2007, \apj,
  671, 1531

\bibitem[\protect\citeauthoryear{{Yun}, {Aretxaga}, {Ashby}, {Austermann},
  {Fazio}, {Giavalisco}, {Huang}, {Hughes}, {Kim}, {Lowenthal}, {Perera},
  {Scott}, {Wilson} \& {Younger}}{{Yun} et~al.}{2008}]{yun2008}
{Yun} M.~S.,  {Aretxaga} I.,  {Ashby} M.~L.~N.,  {Austermann} J.,  {Fazio}
  G.~G.,  {Giavalisco} M.,  {Huang} J.-S.,  {Hughes} D.~H.,  {Kim} S.,
  {Lowenthal} J.~D.,  {Perera} T.,  {Scott} K.,  {Wilson} G.,    {Younger}
  J.~D.,  2008, \mnras, 389, 333

\end{thebibliography}

\appendix
\section[]{Matched Filter}
\label{sec:matched}

Here we describe a filter for identifying point sources in maps
containing significant contributions of both instrumental noise and
confusion due to blending of other point sources \citep[see e.g.][for
related studies of point sources in maps of the Cosmic Microwave
Background]{tegmark1998,barreiro2003,vio2004}. We formulate the
problem as follows: we wish to find the `matched filter', $F$, that
maximizes the \snr\ one would obtain when cross-correlating $F$ with
the signal of interest, $S$ -- in our case, a point source whose shape
is identical to the PSF -- in the presence of noise, $N$. We proceed
by expressing the total \snr\ resulting from this operation, $\phi$,
in Fourier space,
\begin{equation}
\phi \equiv \frac{\mathrm{Signal}}{\mathrm{Noise}} =
\frac{ \sum_k \hat{F}^T_k\hat{S}_k }
{ \left( \sum_k | \hat{F}^T_k \hat{N}_k |^2 \right)^{1/2}}.
\end{equation}
In this expression the carets denote discrete Fourier transforms, and
the index $k$ runs over all elements in the spatial frequency
domain. The superscript `$T$' indicates that we use the transpose of
$F$ in the expression.\footnote{The cross-correlation of the filter
  with the signal is equivalent to convolving the signal with the
  complex conjugate transpose of the filter. Since we are dealing with
  real-valued signals this operations reduces to a transpose. The
  convolution is then a simple product in Fourier space between the
  two transformed quantities.}  Taking the partial derivatives of
$\phi$ with respect to each mode $j$ in the filter, $\hat{F}^T_j$, and
setting them to 0, one can solve for the filter that maximizes the
\snr:
\begin{align}
  0 & = \frac{\partial \phi}{\partial \hat{F}^T_j} \\
  & = \frac{\hat{S}_j}
  {\left( \sum_k | \hat{F}^T_k \hat{N}_k |^2 \right)^{1/2}} -
  \hat{F}^T_j |\hat{N}_j|^2
  \left( \frac{\sum_k \hat{F}^T_k \hat{S}_k}
    {\left( \sum_k | \hat{F}^T_k \hat{N}_k |^2 \right)^{3/2}} \right)
  \nonumber \\
  & \Rightarrow \hat{F}^T_j = \frac{\hat{S}_j}{|\hat{N}_j|^2}
  \left( \frac{\sum_k | \hat{F}^T_k \hat{N}_k |^2}
    {\sum_k \hat{F}^T_k \hat{S}_k} \right) \propto
  \frac{\hat{S}_j}{|\hat{N}_j|^2}.
\label{eq:matched}
\end{align}
In other words, in Fourier space the matched filter is simply the PSF
weighted by the inverse noise variance at each spatial frequency. In
the case of an isolated point source in a field of white noise, this
expression results in the PSF, since $\hat{N}$ is constant. This is
the well-known result \citep[e.g.][]{stetson1987} that the best way to
find point sources in a noisy map is to convolve with the
PSF. However, this procedure is only optimal if the noise is white.
For our submm maps we consider two components of noise: instrumental
white noise, $N_\mathrm{w}$, and the confusion noise caused by other
point sources, $N_\mathrm{c}$.

For the case at hand we estimate $N_\mathrm{w}$ (constant at all
spatial frequencies) assuming typical noise values at the centres of
each submm ECDF-S map: 32, 29, 27, and 2\,mJy\footnote{These are the
  noise values in the raw {\em un-smoothed} maps at each wavelength.}
at 250, 350, 500, and 870\,\micron, respectively
\citep{devlin2009,weiss2009}. For the confusion noise, $N_\mathrm{c}$,
we use the counts inferred using $P(D)$ analyses from
\citet{patanchon2009} and \citet{weiss2009}, and knowledge of the
PSF. In the absence of noise (and assuming no spatial clustering of
sources), the sky can be thought of as a random superposition of PSF
shapes with different amplitudes (i.e., $\delta$ functions smoothed by
the PSF and scaled by the point source flux densities). We therefore
approximate the Fourier transform of the map using a white noise
distribution\footnote{In practice, even if the angular locations of
  galaxies are un-clustered, the $\delta$ function map power spectrum
  is not perfectly white because the flux densities of galaxies are
  drawn from the number counts rather than a Gaussian distribution.}
multiplied by the Fourier transform of the transpose of the PSF (by
the convolution theorem); in other words the power spectrum of
confusion noise rolls off in the same way as the PSF, but with a
different normalization. To determine how the power spectrum of the
PSF should be scaled we simulate maps in each band as described above,
and use the ratios of the standard deviations of these noise
realizations to the standard deviations of the PSFs.

\begin{figure*}
\centering
\includegraphics[width=0.49\linewidth]{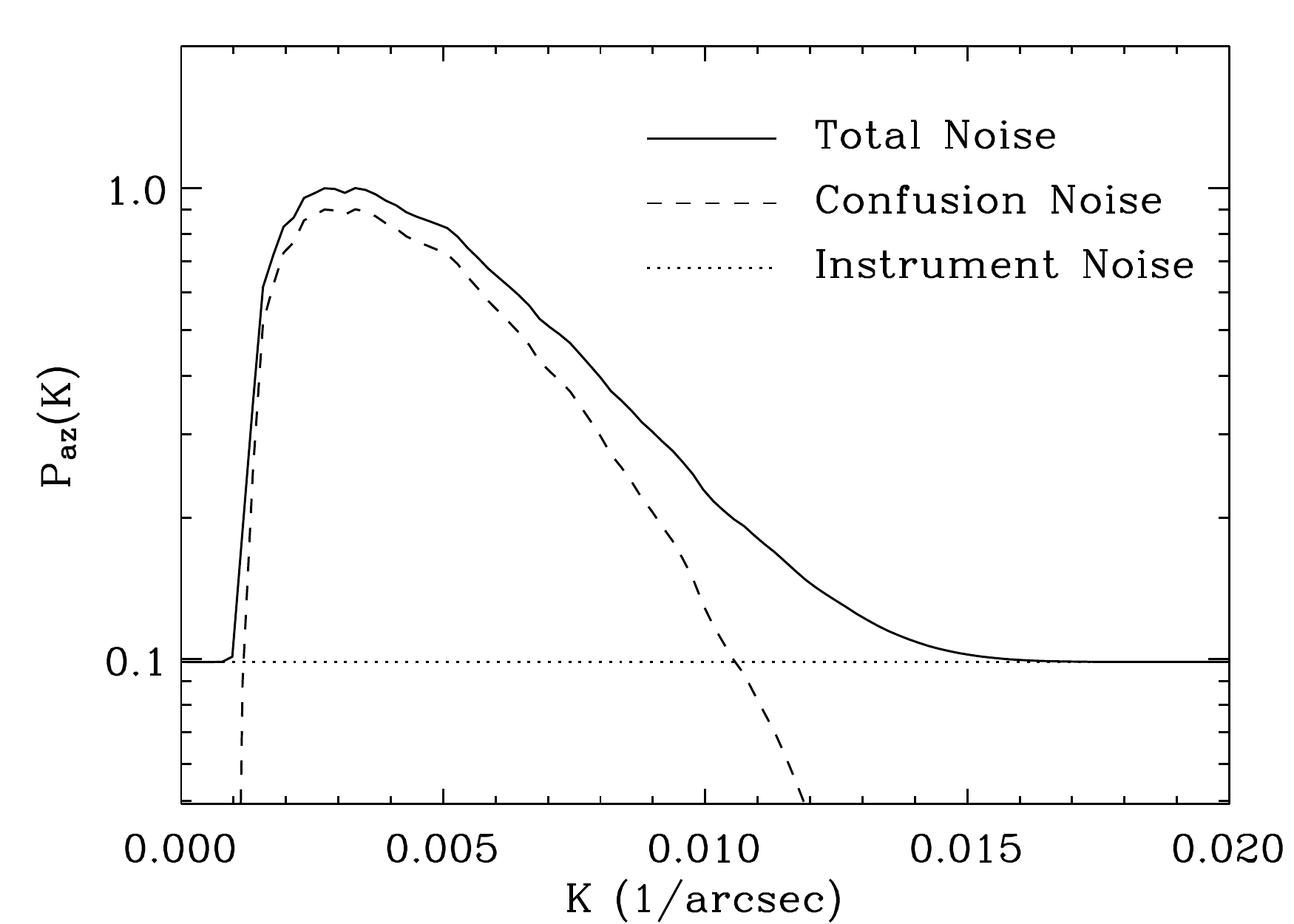}
\includegraphics[width=0.49\linewidth]{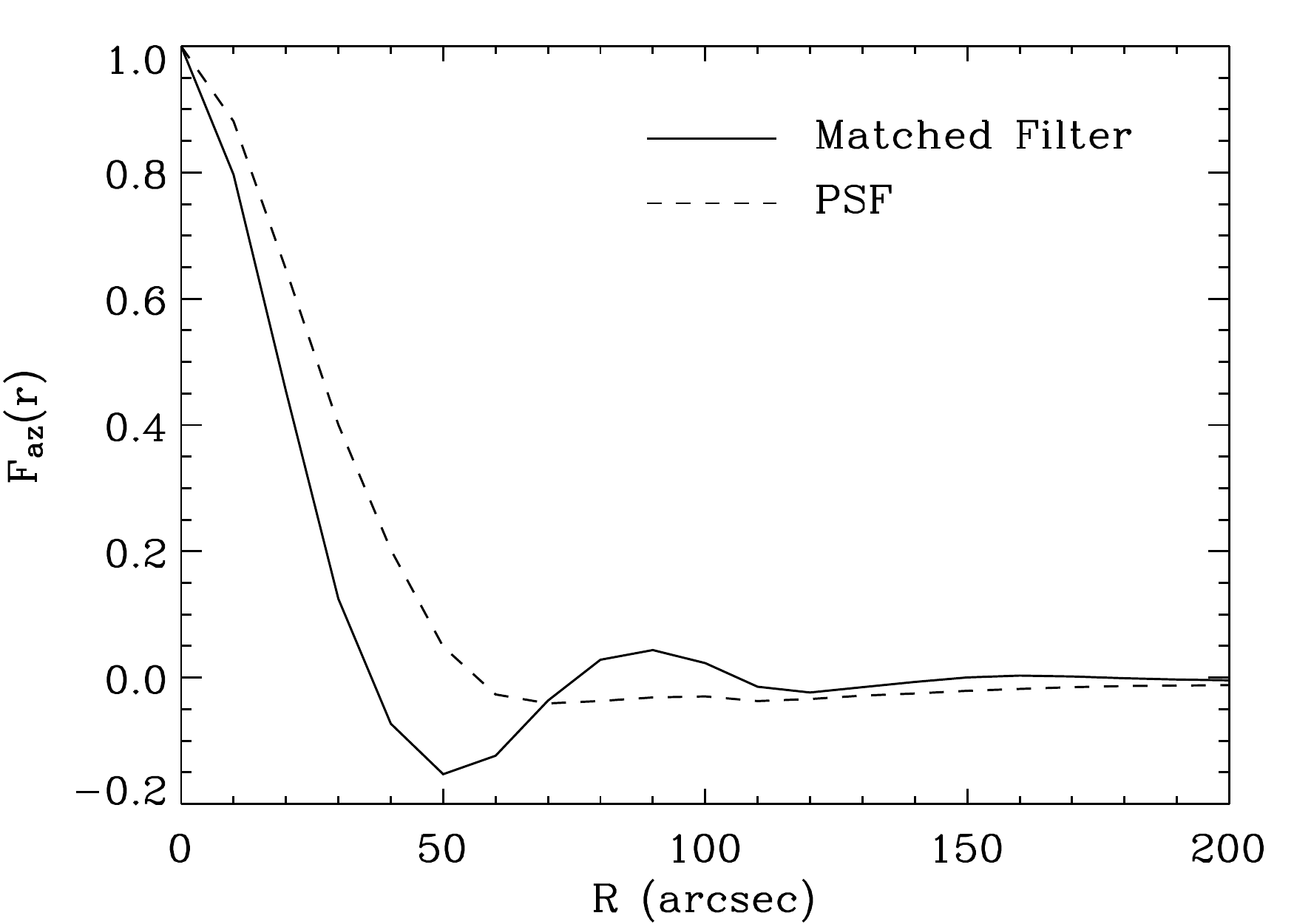}
\caption{{\em left:} The peak-normalized total azimuthally averaged
  500\,\micron\ angular noise power spectrum (solid line) and its two
  components: instrumental (white) noise (dotted line), and confusion
  from point sources (dashed line). The latter function has the same
  shape as the PSF, since point sources are modelled as a spatially
  unclustered collection of point sources smoothed by the PSF. The
  roll-off at $K < 0.003$ in the PSF is a result of the spatial
  whitening filter that has been applied to the BLAST maps
  \citep{devlin2009}. {\em right:} The azimuthally averaged
  500\,\micron\ matched filter (solid line) compared with the PSF
  (dashed line), now in angular rather than frequency space. The
  matched filter is obtained by dividing the power spectrum of the PSF
  (same shape as the dashed line in the left-hand plot) by the noise
  power spectrum (solid line in the left-hand plot), and taking the
  inverse Fourier Transform. For the noise sources considered here,
  this procedure augments high-frequency components of the PSF,
  resulting in a narrower central profile, while introducing ringing
  at angular scales $\gsim50$\,arcsec.}
\label{fig:matchedfilt}
\end{figure*}

As an example of this procedure, the estimated noise power spectrum at
500\,\micron, and resulting matched filter in real space (both
azimuthally averaged) are shown in Fig.~\ref{fig:matchedfilt}. One
sees that the matched filter is somewhat narrower than the PSF filter
(which has been used in most previous SMG studies) and has ringing at
larger angular scales, giving it similar properties to the `Mexican
hat' kernel \citep[see for example discussion and references
within][]{barnard2004}. In this particular case, the first large
negative ring effectively removes a local baseline estimated at radii
$\sim50$\,arcsec, and hence corrects the flux density slightly for
other nearby blended sources. The matched filter is also qualitatively
similar to the filter used to detect sources in deep AzTEC maps
\citep[e.g.,][]{scott2008,perera2008}. However, in those cases only
sources of atmospheric and instrumental noise are considered, since
the spatial noise power spectrum is estimated using `jackknife maps'
(produced by differencing alternating portions of the data), which
explicitly removes the contribution of astronomical sources. We
emphasize that here we are using such a filter to optimally extract
sources in a {\em confused} background. Note also that in a language
familiar in other fields, our filter is essentially a Wiener filter
which includes source confusion explicitly as a noise term.

In addition to the calculation we describe here, we also convolved
simulated maps (containing both point sources and instrumental noise)
with Gaussian filters of different widths, and determined the FWHMs
that optimized the \snr\ of point sources. These simple tests gave
results comparable to the more detailed calculation above, although
the \snr\ was slightly lower due to the lack of ringing to compensate
for the noise at larger angular scales.

It should be noted that in the development of our filter we have
assumed that the instrumental noise is constant across the map. Of
course, in general this is not true, and one could in principle derive
a different filter at each position. For the BLAST maps under
discussion, a separate filter should certainly be used to detect
sources in BGS-Wide and BGS-Deep, but for the present study the noise
was close to uniform across the ECDF-S, and hence we used a single
filter for the whole map.

As a consequence of Eq.~\ref{eq:matched}, the peak \snr\ for an
isolated point source in a field of white noise (for which $\hat{N}_k$
is constant) is obtained by convolution with a Gaussian of the same
FWHM as the instrumental PSF. In the opposite extreme, for a map of
confused point sources with no instrumental noise, the optimal filter
is the inverse of the PSF in Fourier space, i.e., the map is {\em
  de}-convolved by the beam. For real data in which instrumental noise
and confusion are both present, the degree to which the map is
actually de-convolved depends on the relative amplitudes of the two
noise components (see left panel of Fig.~\ref{fig:matchedfilt}).  The
source confusion depends on the number counts (and therefore the SEDs
and redshift distribution) of galaxies in the observed band, and beam
size, and the instrumental noise depends on the detector sensitivity
and integration time. For the data described here, these two
contributions are comparable, since the submm surveys of the ECDF-S
were designed to be {\em just} confusion-limited. The bottom-line is
that in the BLAST bands, considering both noise terms, convolving with
kernels that are {\em smaller} than the instrumental PSFs can give
approximately 15--20\% improvement in the \snr\ --- quite significant
considering the majority of the sources under discussion have
instrumental \snr\ of about 4$\sigma$ (see Table~\ref{tab:matched})!
The improvement is less impressive in the LESS 870\,\micron\ image,
since the instrumental noise dominates the RMS resulting from
confusion.  Most previous studies of SMGs in the regime where
confusion is important have extracted sources using the PSF, and our
results show that they have smoothed away some of the point source
information in the maps.

\section[]{Likelihood Ratios}
\label{sec:lr}

Here we describe our formulation of Likelihood Ratios (LR) as a means
for identifying counterparts to submm peaks. The matching catalogue
for which we estimate priors is described in
Section~\ref{sec:matchcat}.  To begin, we make two basic assumptions:
(i) that the radial offset of a potential ID is uncorrelated with its
other properties; and (ii) that the submm flux density is uncorrelated
with its other properties. The first assumption is fairly standard and
uncontroversial, although the latter could be considered
problematic. For example, the brightest peaks at 250\,\micron\ are
known to be low-luminosity {\em IRAS} galaxies, whereas the fainter
sources are thought to be a mixture of both low- and high-redshift
galaxies. However, for now we take the practical route and choose not
to make any distinction based on these properties so that we can
estimate priors with reasonable \snr\ given the data. We then express
the differential density {\em of true counterparts} to submm peaks as
a function of their offsets, $r$, flux densities, $S$ (both at
24\,\micron\ and 1.4\,GHz), and colour $c$ (IRAC $[3.6]-[4.5]$):
\begin{equation}
  n_\mathrm{c}(S,c,r) dS\,dc\,dr = q(S,c) f(r) dS\,dc\,dr,
\end{equation}
where $q(S,c)$ is the distribution of counterpart flux densities and
colours (in the matching catalogue), and $f(r)$ is the positional
probability distribution as a function of radial offset $r$. We assume
a symmetric Gaussian probability distribution as a function of
orthogonal positional coordinates which results in the Rayleigh radial
probability distribution,
\begin{equation}
f(r) = \frac{r}{\sigma_\mathrm{r}^2} e^{-r^2/2\sigma_\mathrm{r}^2}.
\label{eq:pos}
\end{equation}
The normalization is chosen so that $\int_0^\infty f(r) dr = 1$.
Following \citet{sutherland1992} $q(S,c)$ is normalized so that it
integrates to {\em the average expected number of counterparts per
  submm peak}. However, while those authors assumed that this number
is in the range 0--1 (i.e., the emission is produced by a {\em single}
source, which may or may not be present in the matching catalogue), we
consider cases in which the number of counterparts may be greater than
this.

We can also estimate the differential density of background sources as
a function of these same properties
\begin{equation}
  n_\mathrm{b}(S,c,r) dS\,dc\,dr = 2\pi r\rho(S,c) dS\,dc\,dr,
\end{equation}
where $\rho(S,c)$ is the surface density of background sources as a
function of $S$ and $c$, and multiplying by $2\pi r$ converts this
quantity to the infinitesimal number density of sources at a distance
$r$ from the submm peak.

Therefore, given $S,c$ and $r$ for the $j$th candidate counterpart to
the $i$th submm peak, the relative number of expected true
counterparts to background sources is the LR:
\begin{equation}
  L_{i,j} = \frac{n_\mathrm{c}}{n_\mathrm{b}} =
  \frac{q(S_j,c_j) e^{-r_{i,j}^2 / 2\sigma_\mathrm{r}^2}}
  {2\pi \sigma_\mathrm{r}^2 \rho(S_j,c_j)}.
\end{equation}
Candidates with large values of $L_{i,j}$ are more likely to be
associated with the submm emission.

Note that while the LR contains all of the information about the
potential identification, it is not itself a probability
distribution. In the case that only a single source in the matching
catalogue is believed to correspond to the submm peak, and where
$q(S,c)$, $\rho(S,c)$, and $\sigma$ are all known precisely, it is
possible to calculate the reliability $R$, the probability that the
candidate is {\em the unique counterpart}, following
\citet{sutherland1992}. Using this technique on a case for which there
are multiple strong candidate matches, for example, the reliability is
divided among them in such a way that their sum does not exceed 1.
For the case at hand, since we consider the possibility of multiple
real counterparts to a single submm peak, and since our estimates of
the priors are ultimately quite noisy, Eq.~5 from
\citet{sutherland1992} does not apply. Instead we will take the
approach of using Monte Carlo simulations to establish a threshold LR
that will provide candidates with a desired false identification rate.

\subsection{Background Counts and Positional Uncertainties}
\label{sec:pos}

We estimate priors directly from the data themselves. First,
$\rho(S,c)$, the background number counts, are determined by binning
the entire matching catalogue (Section~\ref{sec:matchcat}) as a
function of the two flux densities and colour, and dividing by the
survey area. This should give a good estimate for the average
properties of the background sources since the entire catalogue
contains 9216 entries, whereas there are only of order $\sim60$ submm
peaks in each band.

The positional uncertainty parameter, $\sigma_\mathrm{r}$, is more
difficult to estimate from the data. We proceed by counting the number
of catalogue sources around submm peaks, and checking for
statistically significant excesses compared to the background counts
predicted by $\rho(S,c)$, as a function of search radius. The
expectation, if each submm peak were produced by a single source in
the matching catalogue, is that the cumulative excess would grow from
0 at a search radius of 0, and converge to 1 at large radii.

\begin{figure*}
\centering
\includegraphics[width=0.49\linewidth]{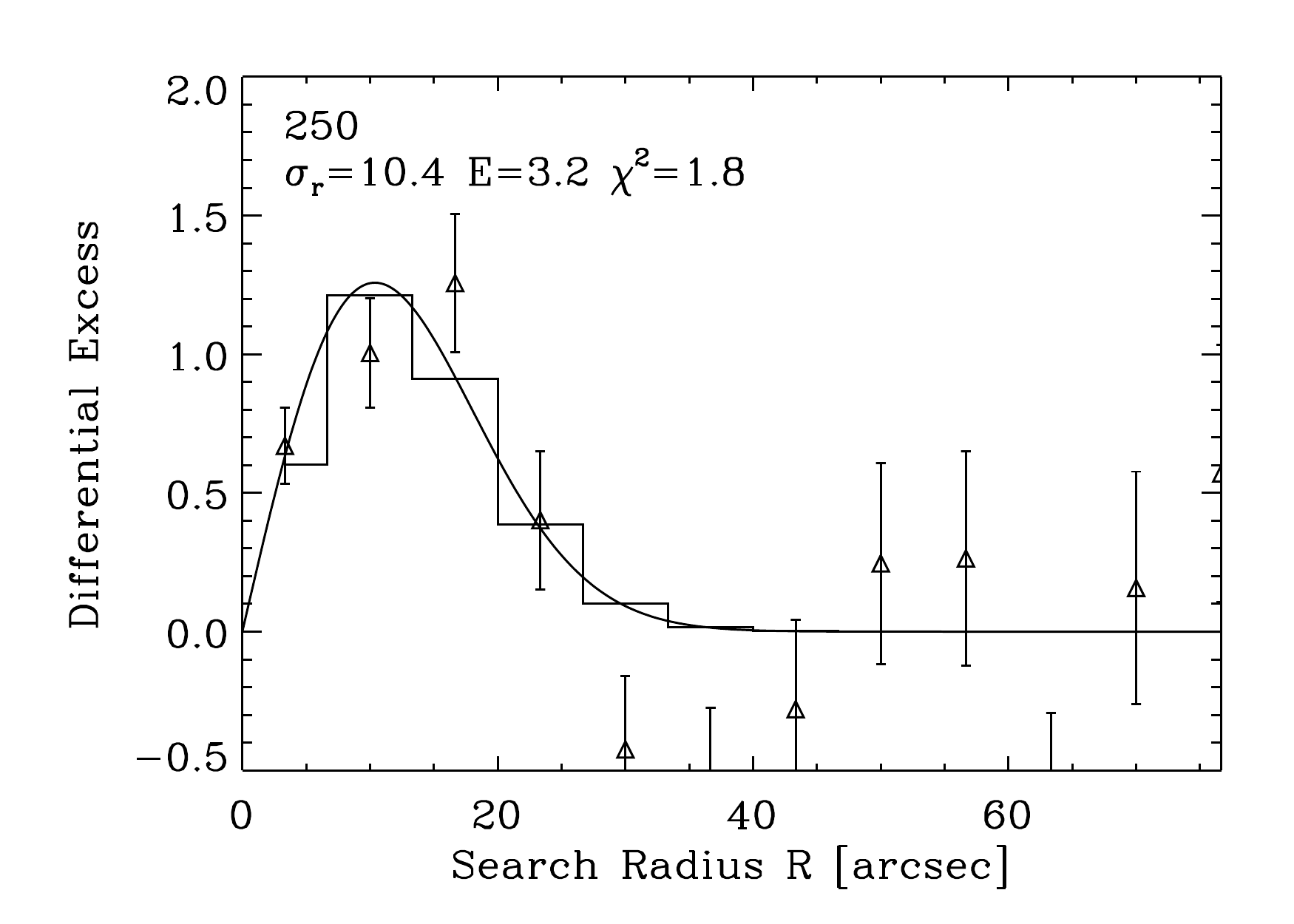}
\includegraphics[width=0.49\linewidth]{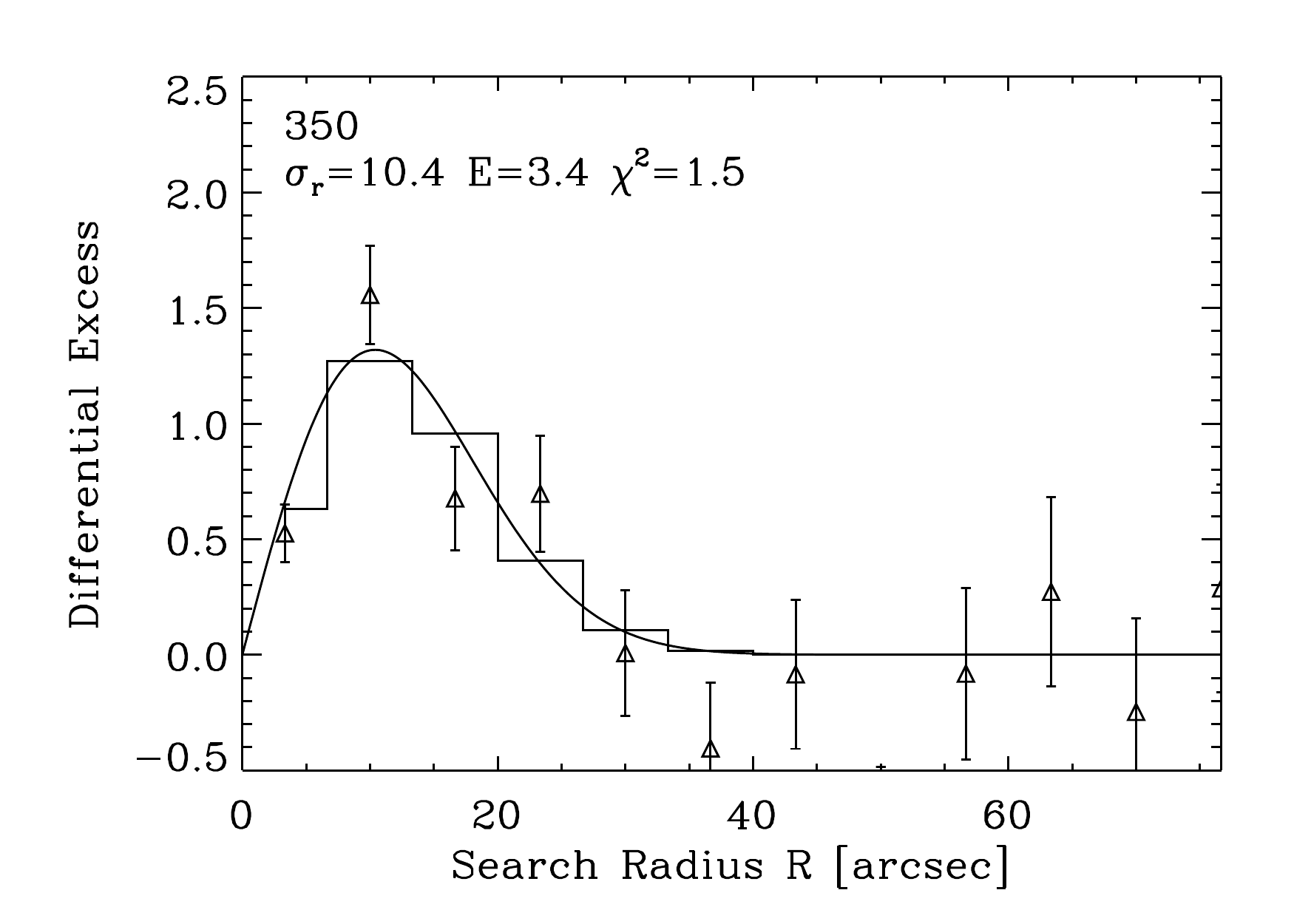}
\includegraphics[width=0.49\linewidth]{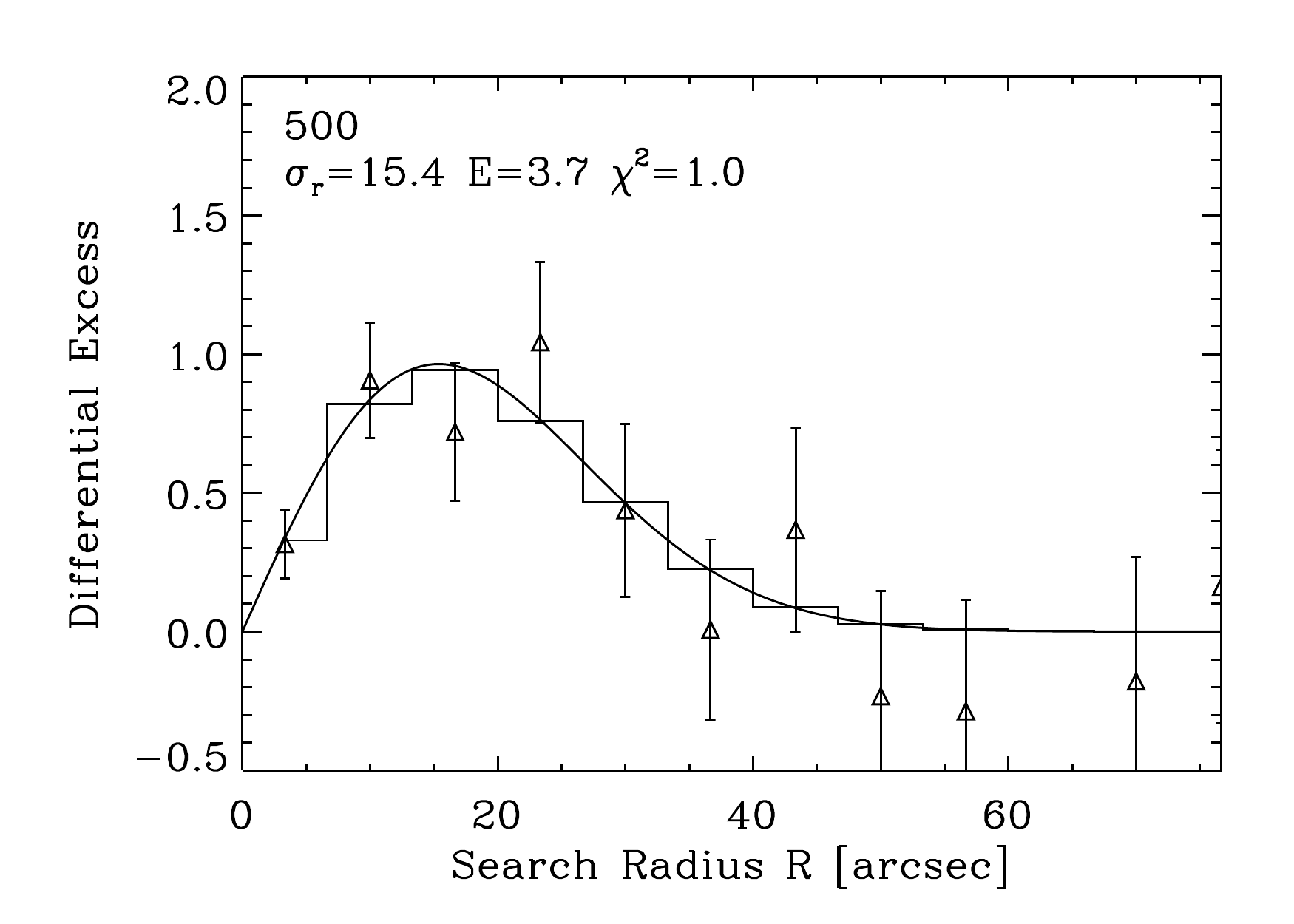}
\includegraphics[width=0.49\linewidth]{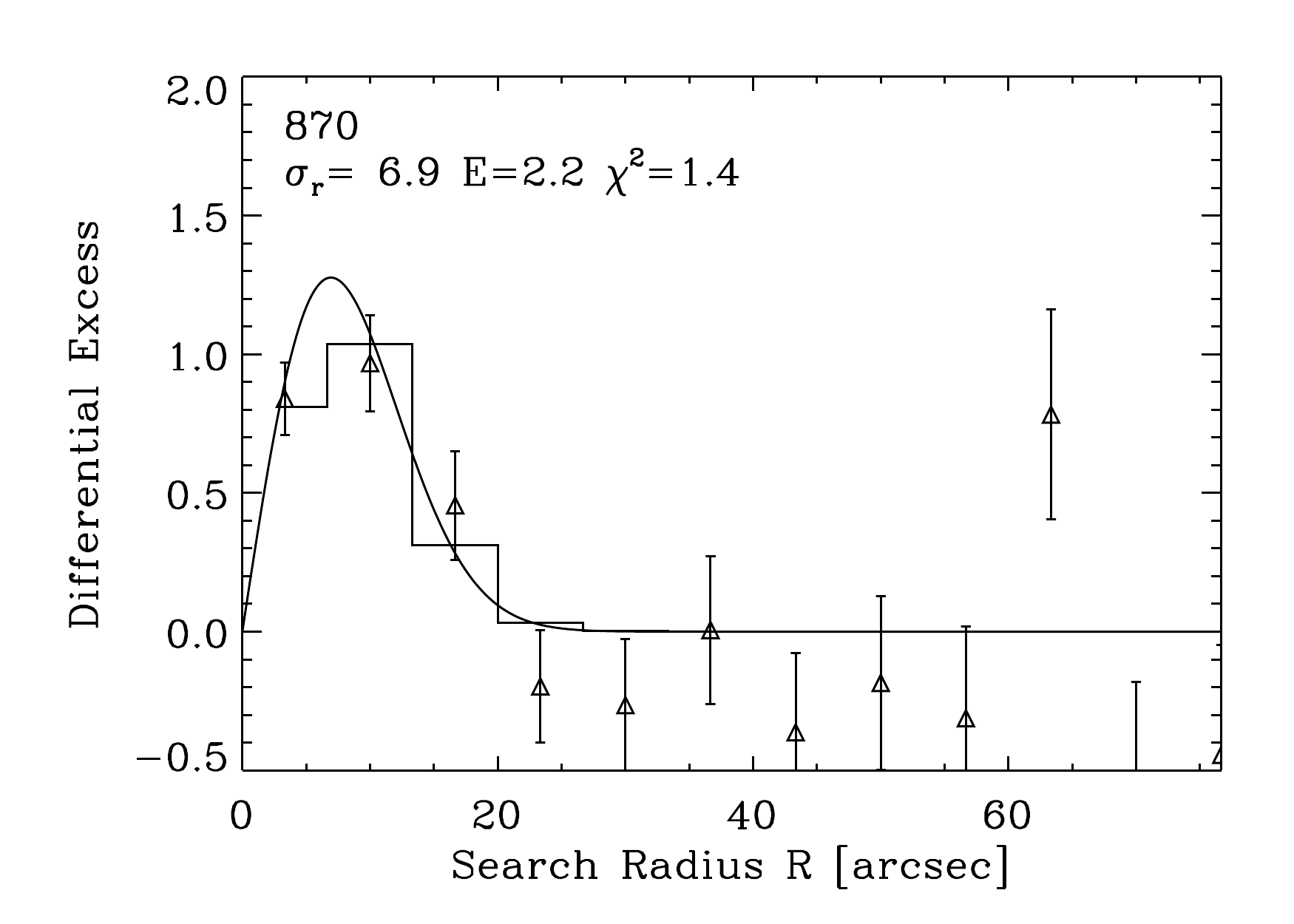}
\caption{Excess source counts in annuli around submm peaks compared to
  the expected background as a function of search radius. Poisson
  uncertainties are plotted. The solid lines are fits of a Gaussian
  radial probability density function $p(r) = (Er/\sigma_\mathrm{r})
  \exp (-r^2/2\sigma_\mathrm{r}^2)$, where the scale factor $E$ is the
  total excess counts encountered on average (fit values are indicated
  in each panel). This function is fit to the differential excess
  counts (since the uncertainties are uncorrelated). The histograms in
  the differential plots show the smooth model integrated across each
  bin (giving the actual predicted model values used to calculate
  $\chi_\mathrm{r}^2$).}
\label{fig:r_excess}
\end{figure*}

We show the differential excess distribution (calculated in annuli of
different sizes) for each submm band in
Fig.~\ref{fig:r_excess}. Uncertainties for each data point are
estimated from Poisson counting statistics (considering both the
number of objects in the full catalogue for the uncertainties in the
background counts, and the number of objects in the measurement
annulus). We fit the Rayleigh radial offset distribution to the excess
counts, multiplying the single-parameter expression for $f(r)$ in
Eq.~\ref{eq:pos} by a second parameter, $E$, giving the integrated
excess number of sources around submm peaks (note that $E = \int \int
q(S,c)\,dS\,dc$). The smooth models are shown as solid lines in the
differential excess plots. In addition, the solid histograms show the
models integrated across the same bins as the data, giving the model
predictions used to calculate $\chi^2$. Clearly, $E$ is significantly
larger than 1 in each submm band, demonstrating that the submm peaks
from our survey are typically blends of several sources in the
matching catalogue. One could use this approach to determine a
criterion which might be called `counterpart confusion' -- for a given
source map and catalogue (in a different waveband) one could conclude
that the source identification process will be relatively
straightforward if $E<2$ (say), but significantly complicated by
confusion if $E \ge 2$.

\subsection{Flux density and colour priors}

\begin{figure*}
\centering
\includegraphics[width=0.49\linewidth]{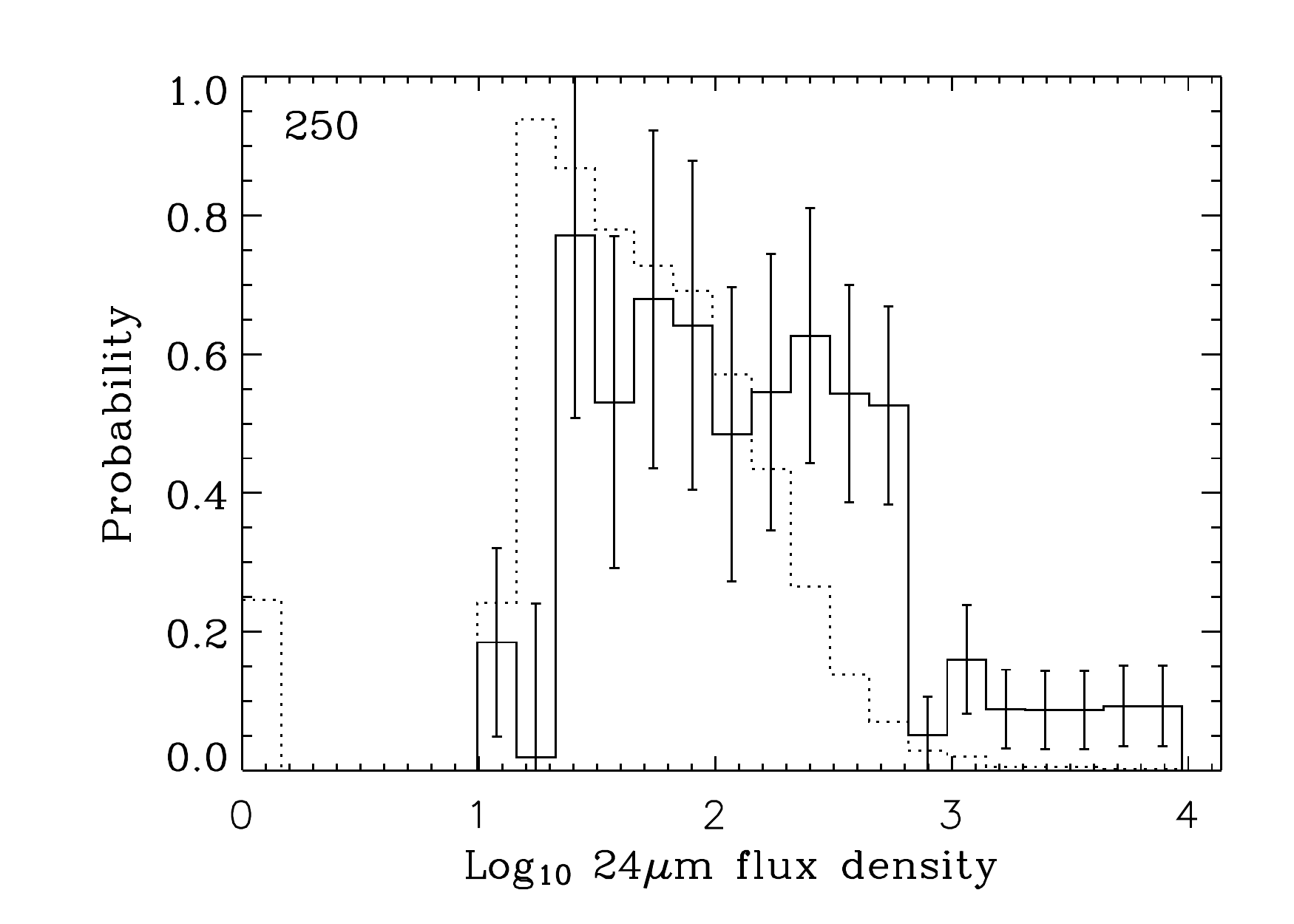}
\includegraphics[width=0.49\linewidth]{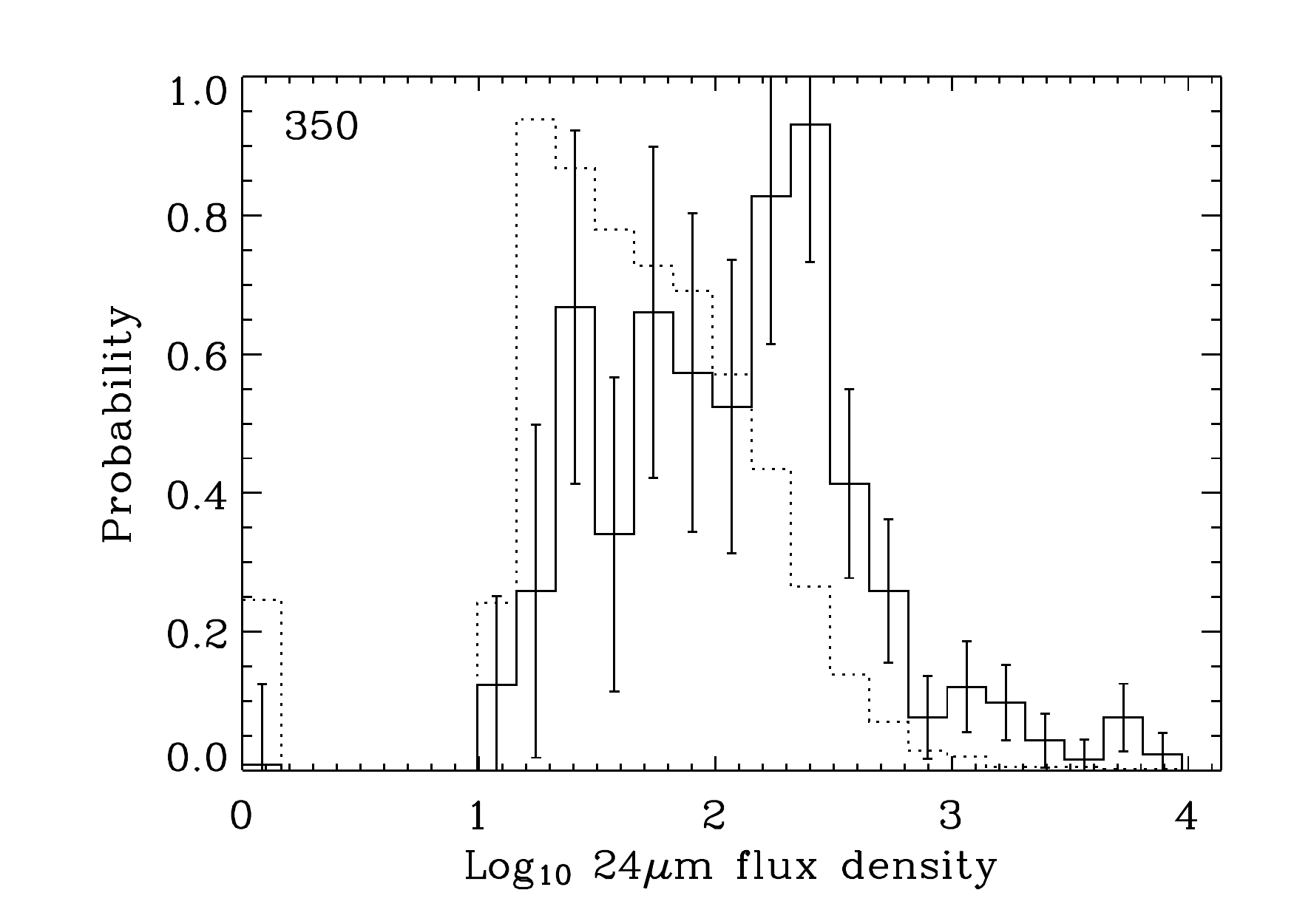}
\includegraphics[width=0.49\linewidth]{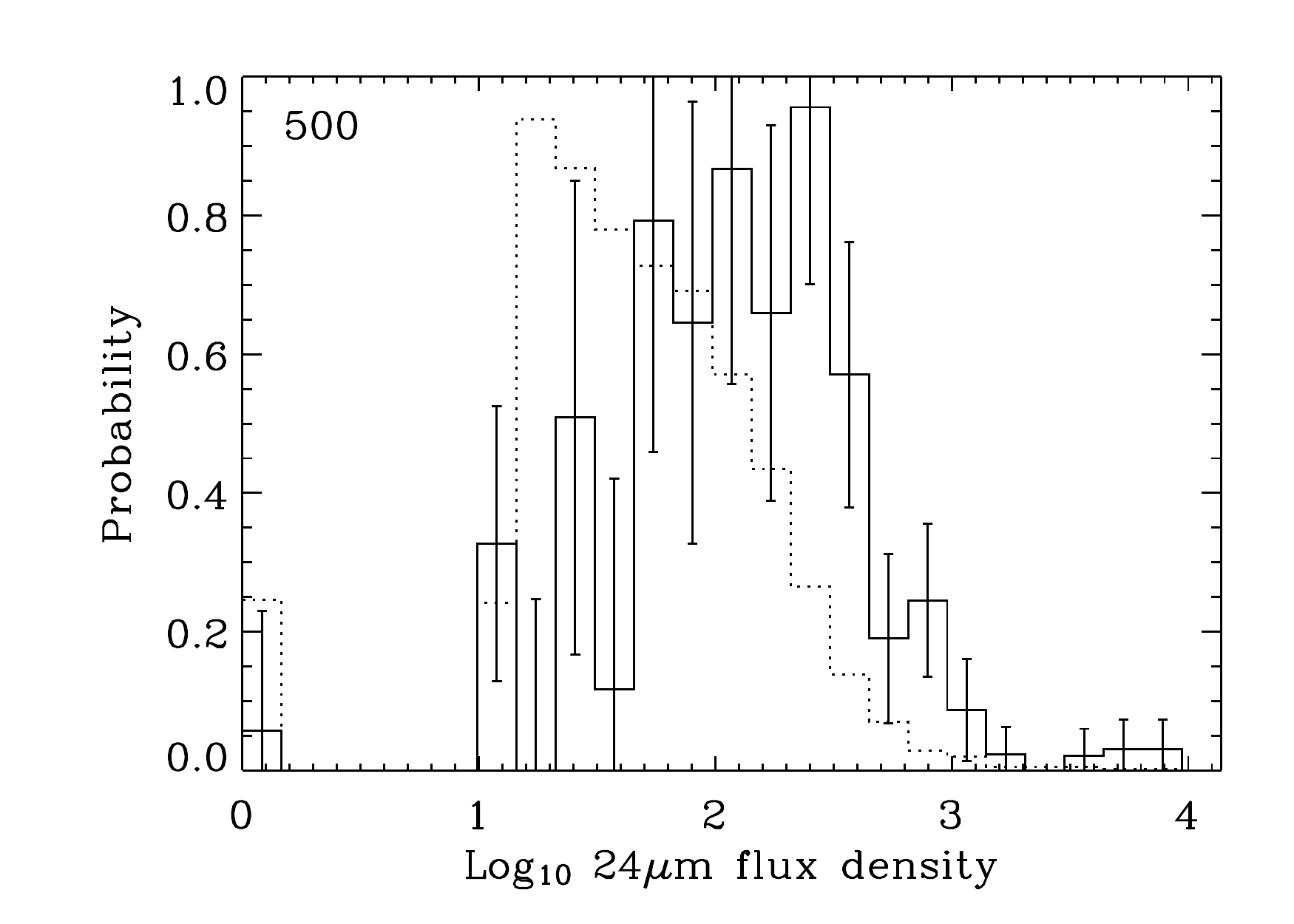}
\includegraphics[width=0.49\linewidth]{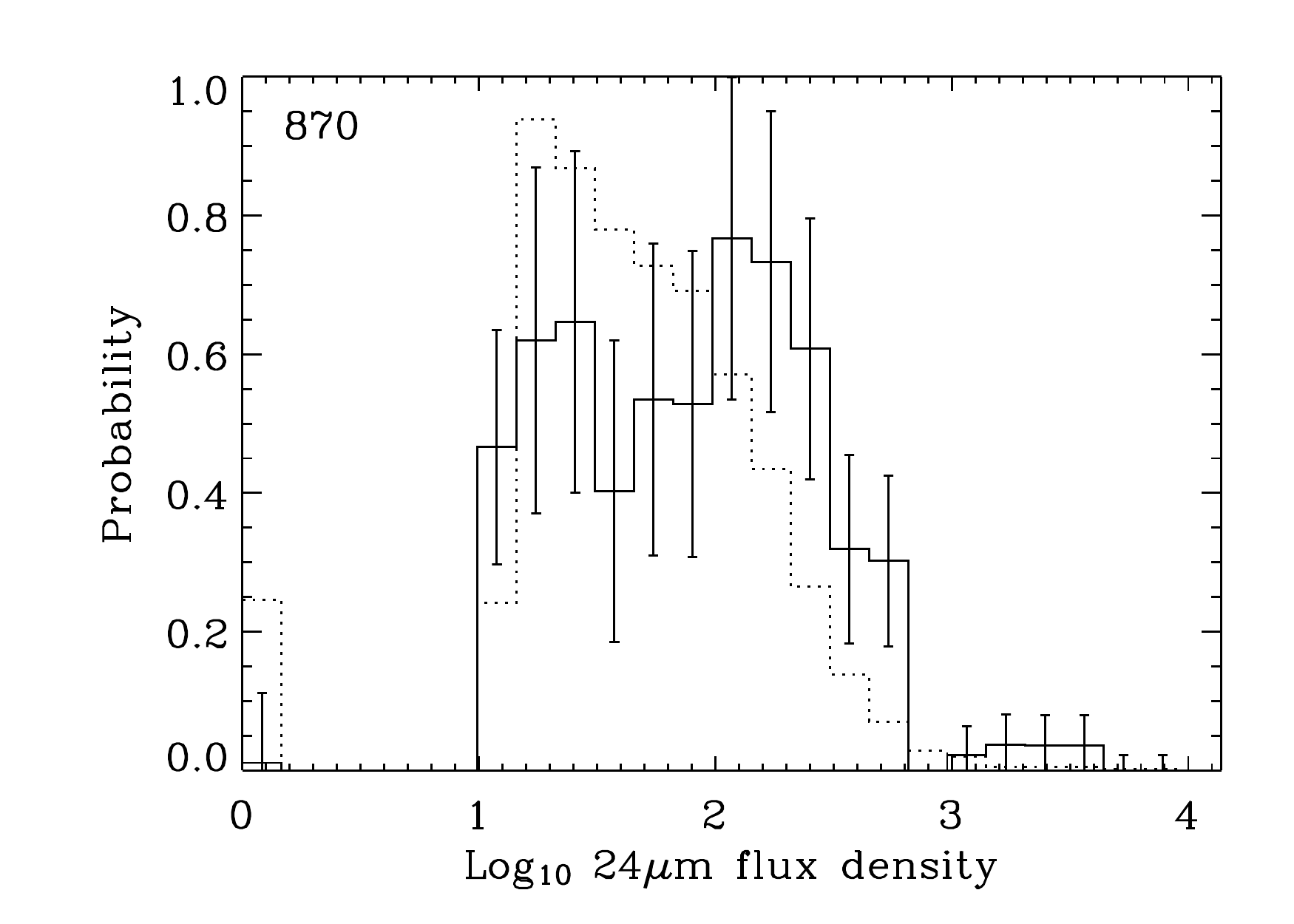}
\caption{MIPS 24\,\micron\ flux density distributions of excess
  sources around submm peaks in all four bands (solid histogram with
  Poisson uncertainties), our estimate of $p(S_{24})$, compared to the
  general population (dotted histogram), our estimate of
  $\rho(S_{24})$. The bin at 0 contains all radio sources that are not
  present in the FIDEL 24\,\micron\ catalogue. Galaxies selected near
  submm positions are on average brighter at 24\,\micron\ than
  galaxies selected randomly, and this is true for all four bands.}
\label{fig:f24hist}
\end{figure*}

\begin{figure*}
\centering
\includegraphics[width=0.49\linewidth]{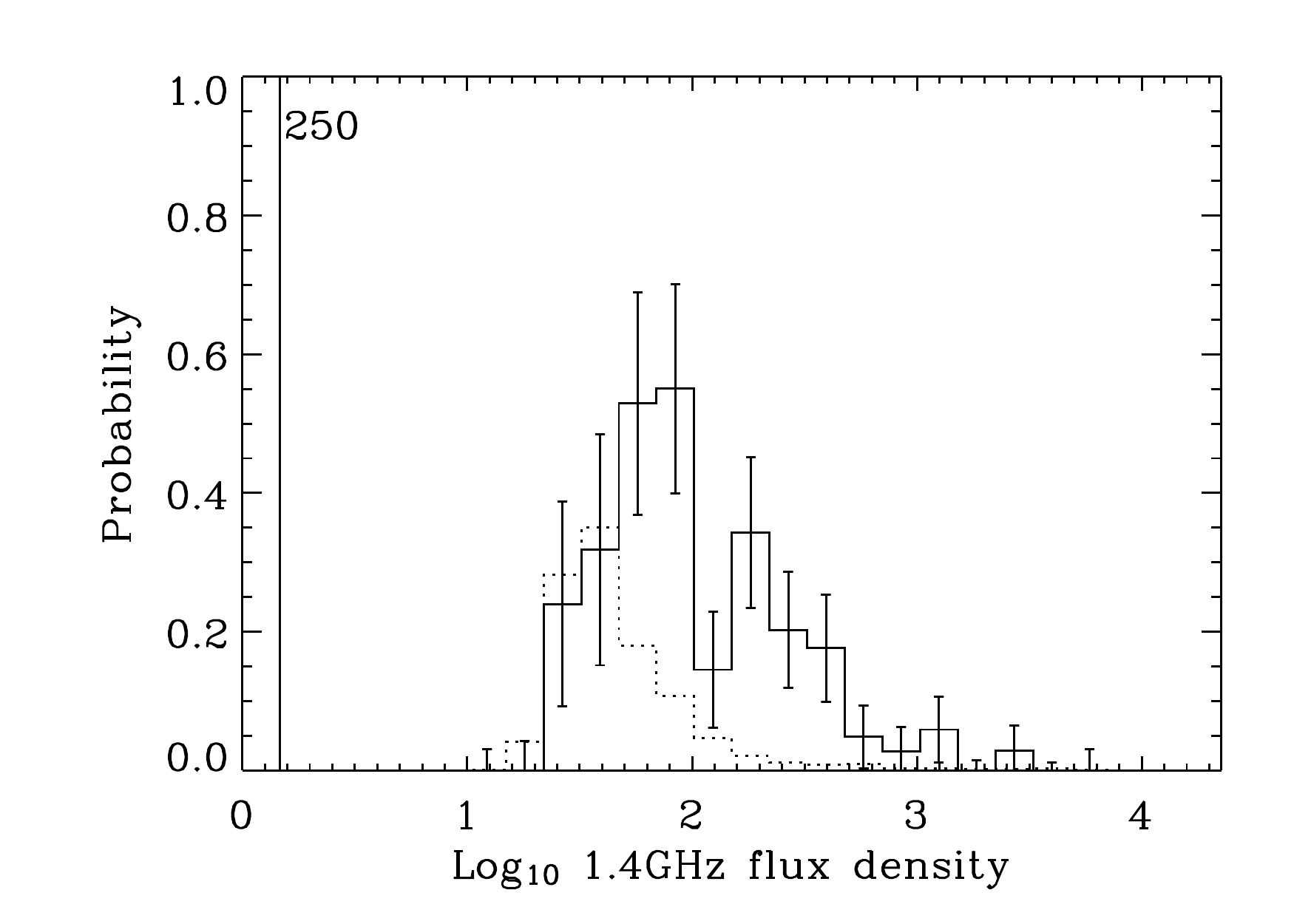}
\includegraphics[width=0.49\linewidth]{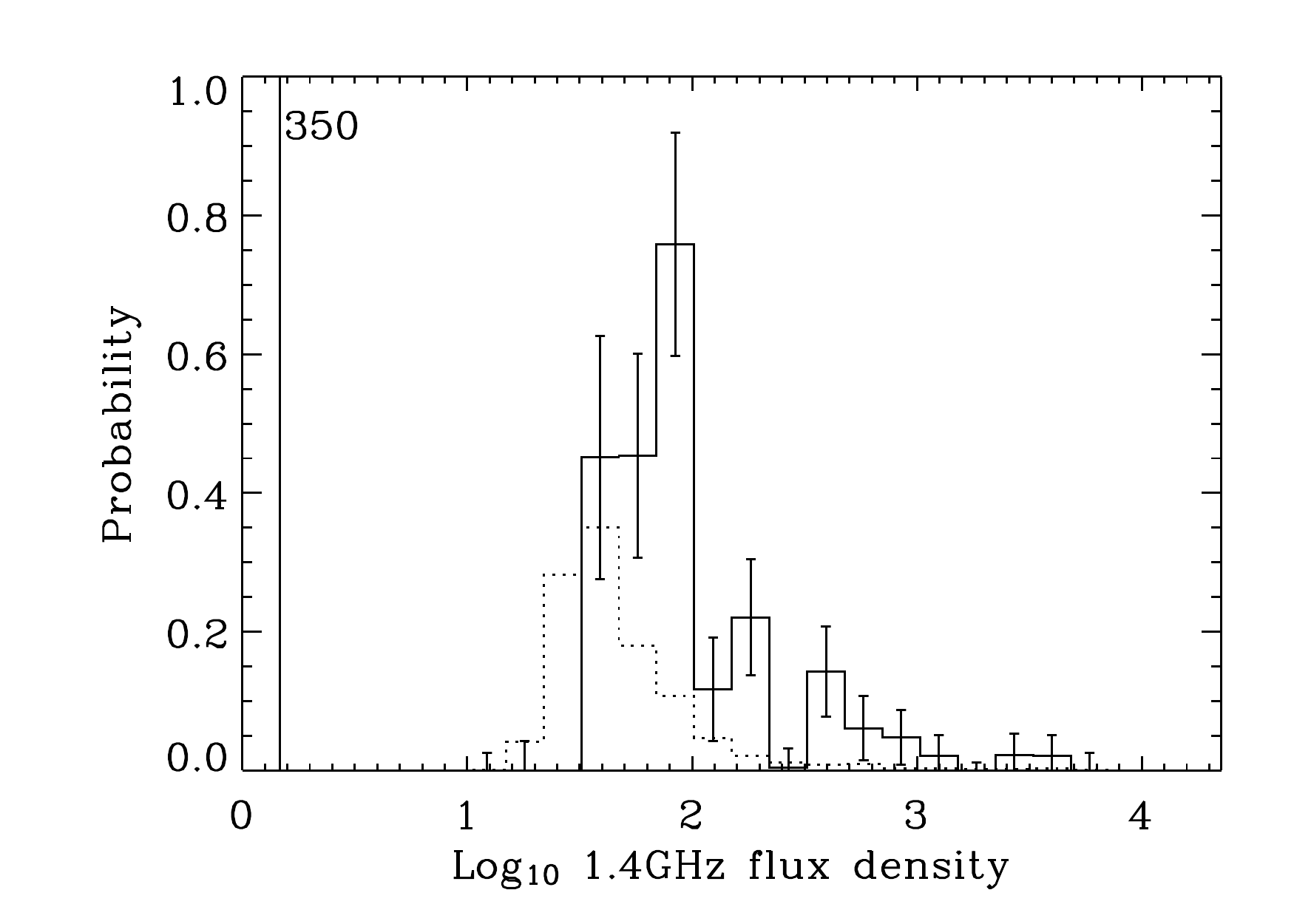}
\includegraphics[width=0.49\linewidth]{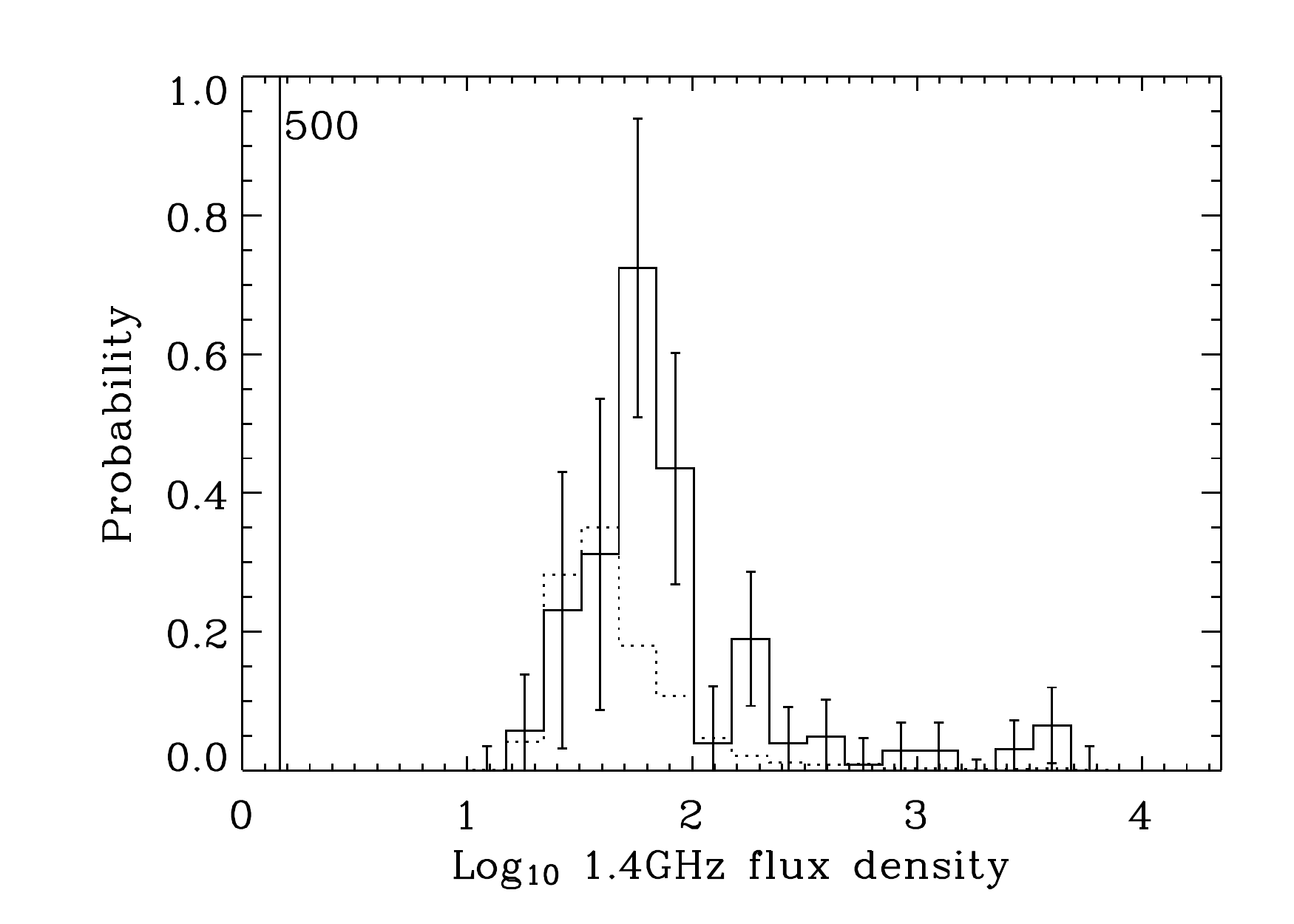}
\includegraphics[width=0.49\linewidth]{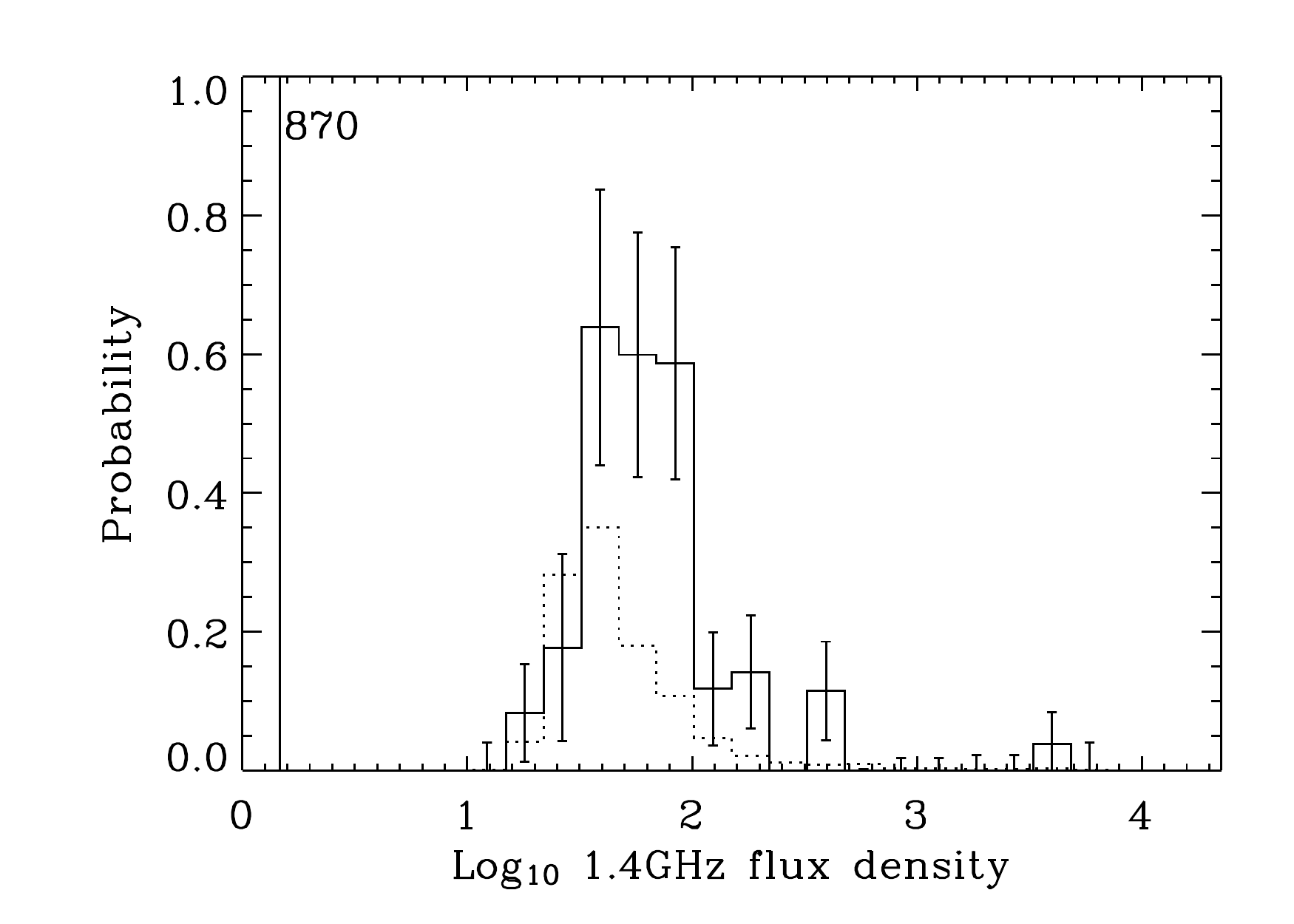}
\caption{VLA 1.4\,GHz flux density distributions of excess sources
  around submm peaks in all four bands (solid histogram with Poisson
  uncertainties), our estimate of $p(S_{1.4})$, compared to the
  general population (dotted histogram), our estimate of
  $\rho(S_{1.4})$. The bin at 0 contains all 24\,\micron\ FIDEL sources
  that do not exhibit significant radio flux. Galaxies selected near
  submm positions are on average brighter at 1.4\,GHz than galaxies
  selected randomly, and this is true for all four bands. We also note
  that many more of the radio sources with 24\,\micron\ emission are
  found near submm positions than random radio sources. Many of these
  radio sources without 24\,\micron\ emission could be spurious given
  the low significance cut on the catalogue that has been used (see
  Section~\ref{sec:radiocat}). Our prior estimation procedure will
  give much greater weight to potential radio identifications that
  also exhibit 24\,\micron\ emission.}
\label{fig:frhist}
\end{figure*}

Next we estimate the joint distribution of flux density and colour for
counterparts, $q(S,c)$. Since we assume that these properties are
un-correlated with their distance from the submm centroid, we first
identify the search radius within which the \snr\ of the excess counts
is highest, $r_\mathrm{s}$. As this radius is increased from 0, the
background counts grow as $r_\mathrm{s}^2$, but the counts around
submm positions grow even faster as the counterparts are included. The
ratio of the difference (the excess), compared to the Poisson
uncertainty in the difference, therefore increases until the number of
new counterparts drops significantly. We find that for our matching
catalogue and submm peaks lists the maximum excess \snr\ are achieved
at 23, 23, 35 and 15\,arcsec for 250, 350, 500 and 870\,\micron,
respectively. Using these search radii, we then compare the normalized
histogram of 24\,\micron\ and 1.4\,GHz flux densities for the entire
matching catalogues with the normalized histogram of excess sources
around submm peaks, $p(S_{24})$ and $p(S_\mathrm{r})$, in
Figs.~\ref{fig:f24hist} and \ref{fig:frhist}. This operation shows
that the extra sources around submm peaks tend to be {\em brighter} at
24\,\micron\ and 1.4\,GHz on average than the entire populations in
the BLAST bands.


\begin{figure*}
\centering
\includegraphics[width=0.49\linewidth]{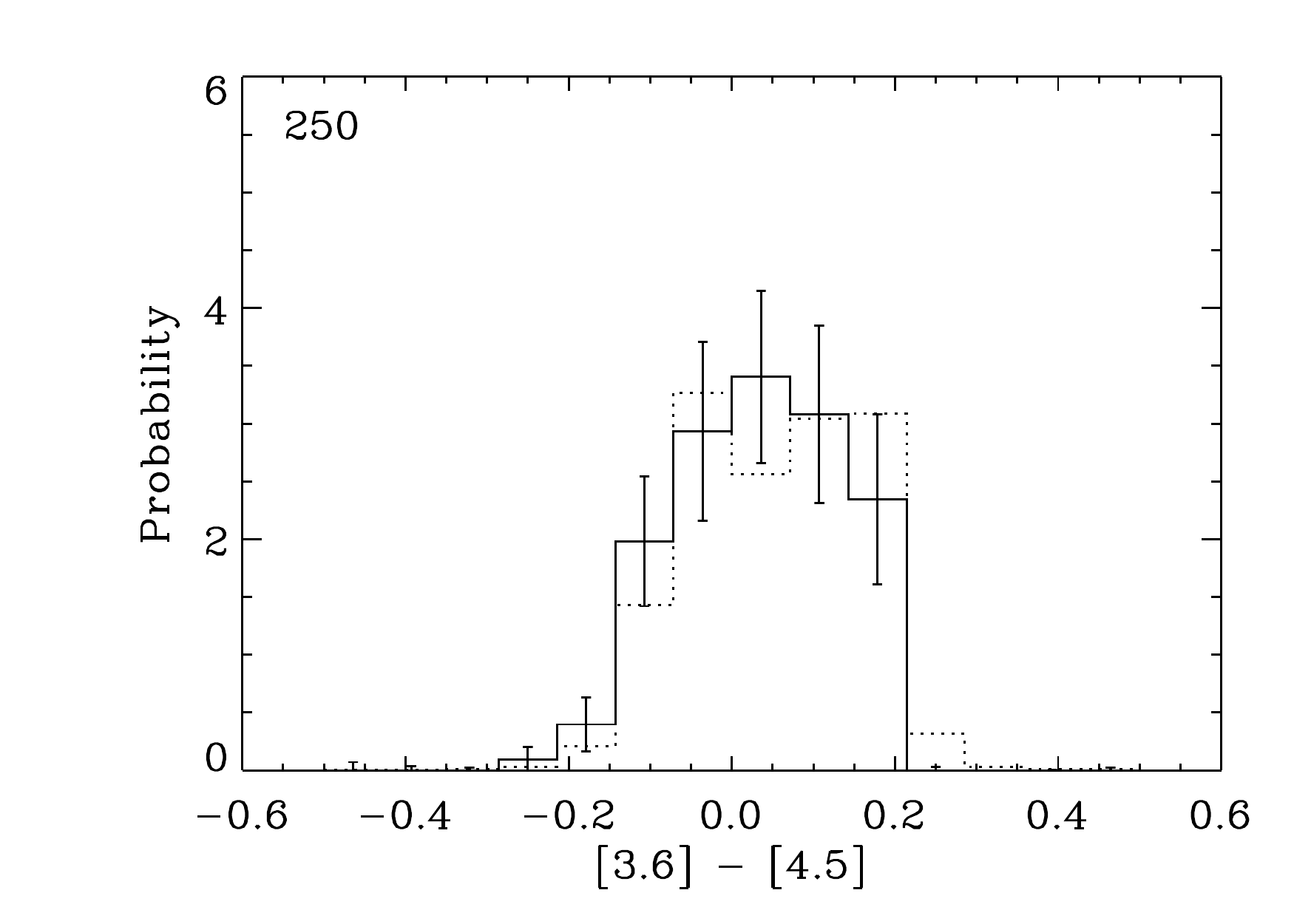}
\includegraphics[width=0.49\linewidth]{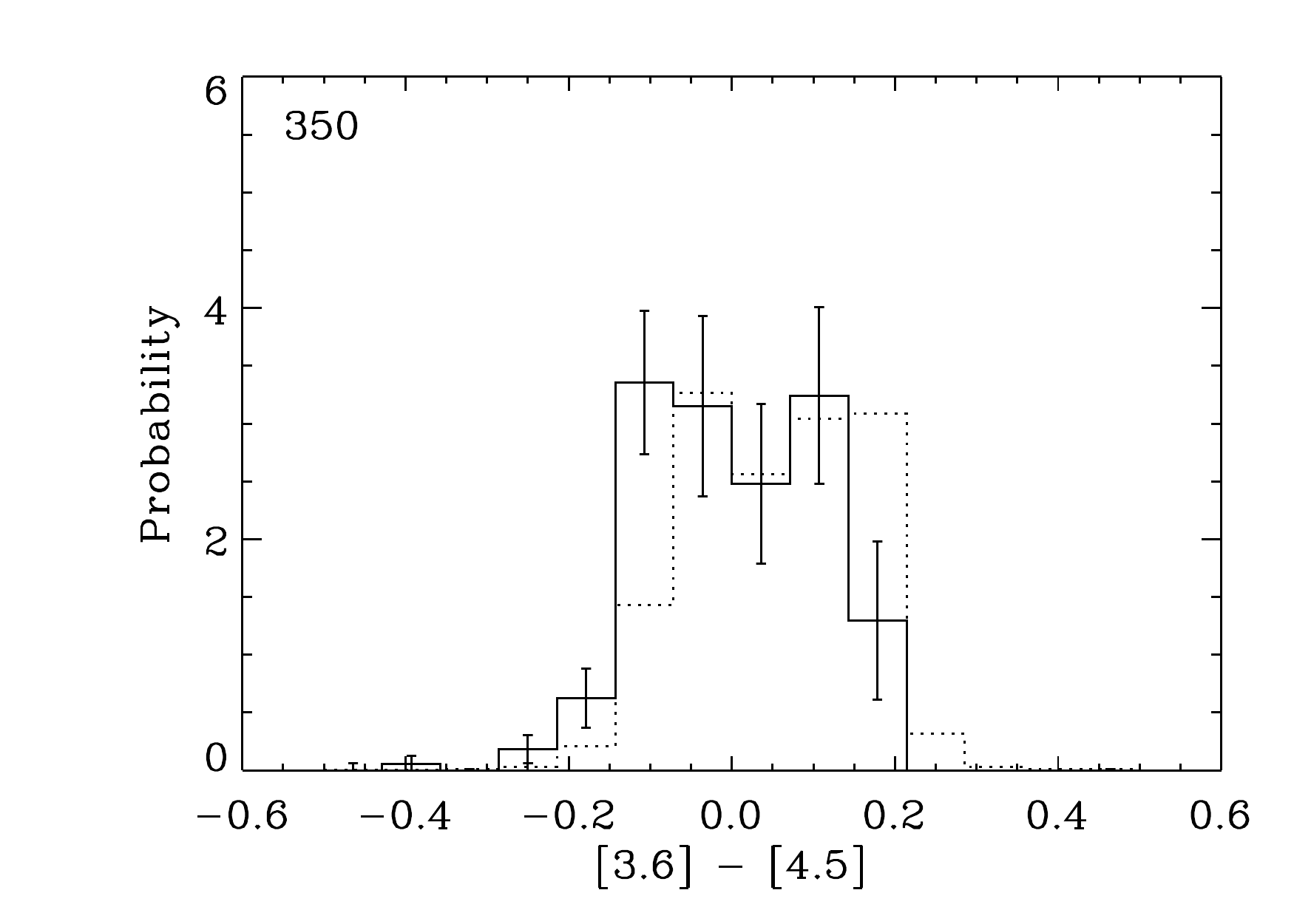}
\includegraphics[width=0.49\linewidth]{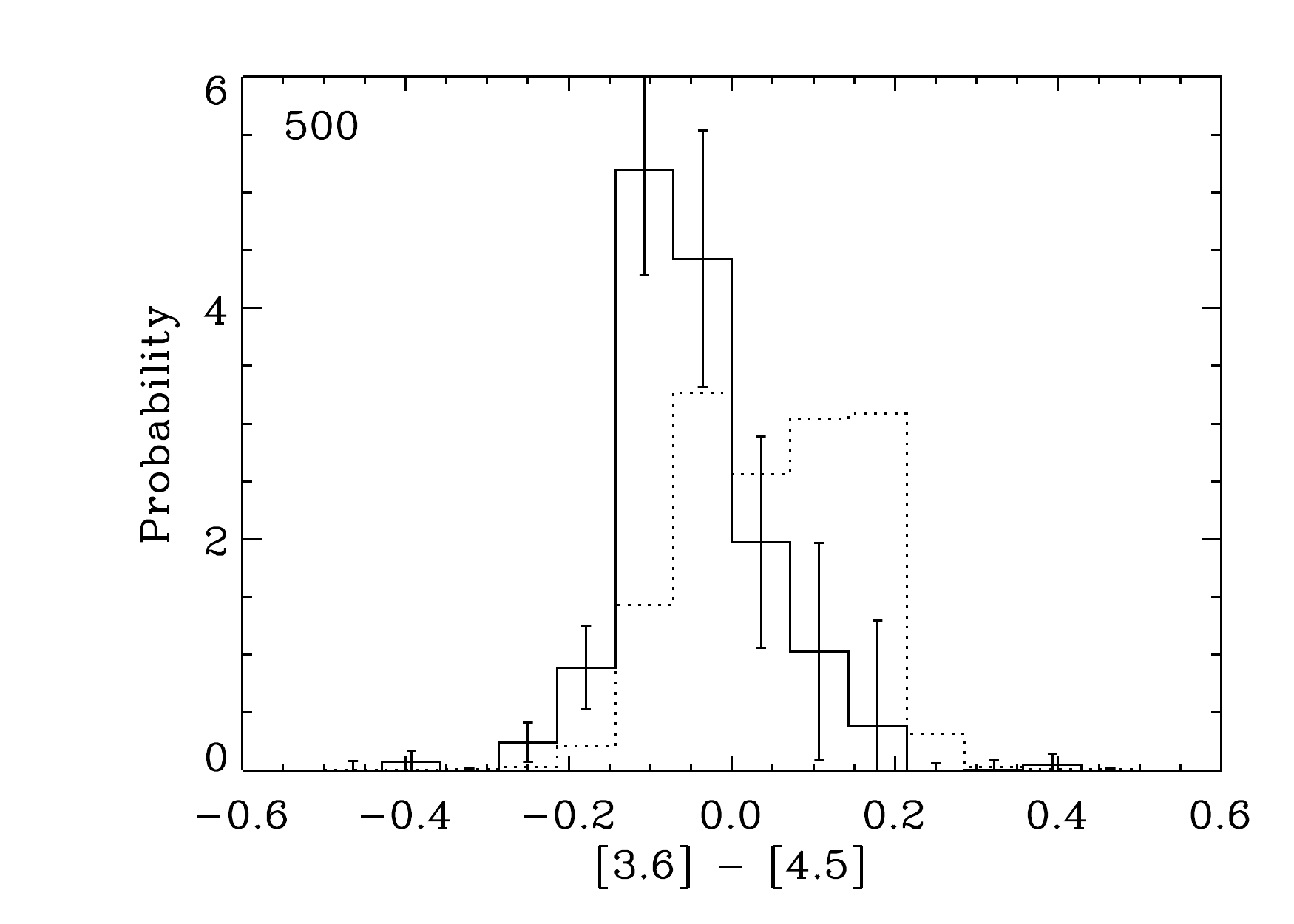}
\includegraphics[width=0.49\linewidth]{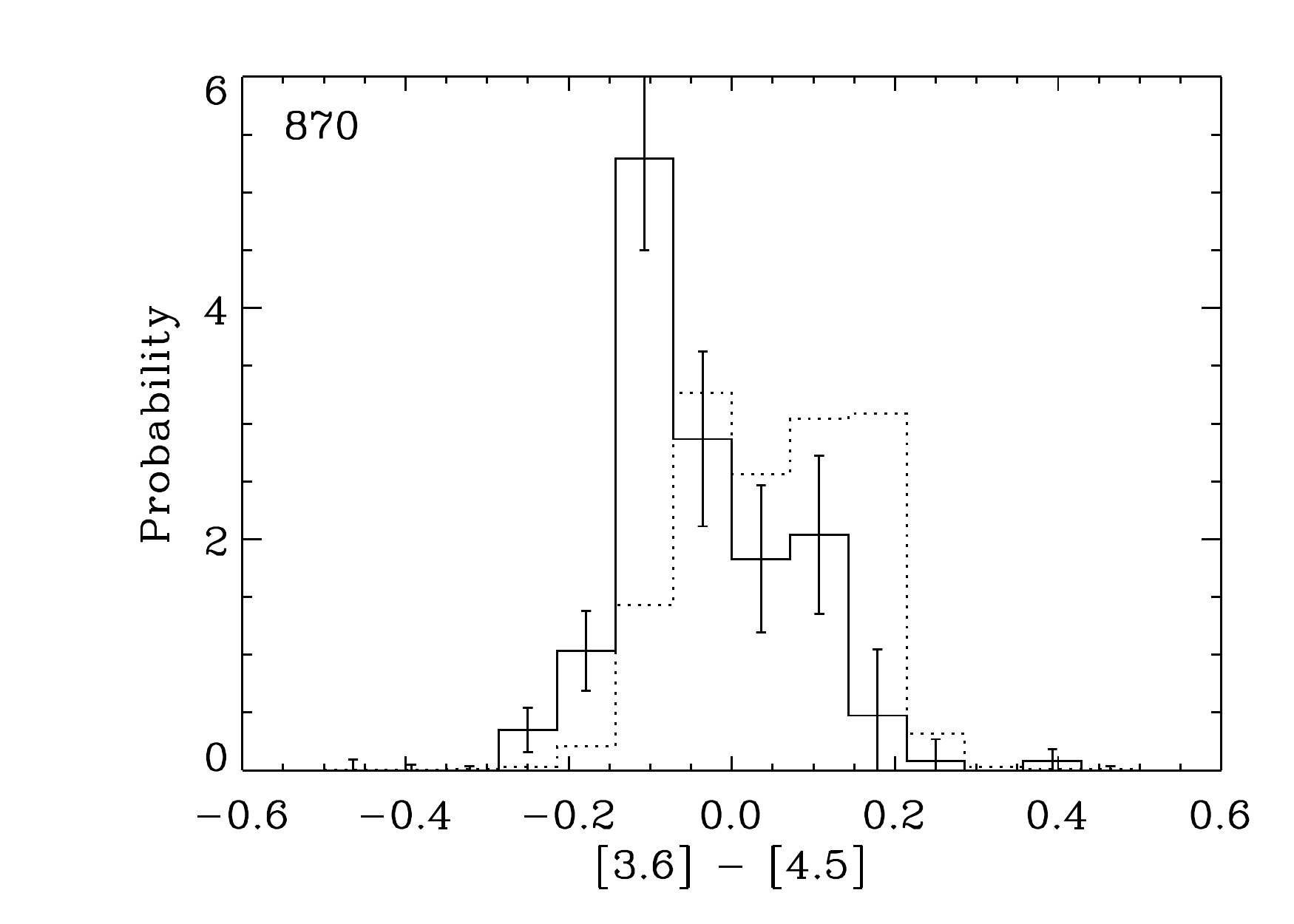}
\caption{IRAC $\log_{10}(S_{3.6}/S_{4.5})$ colour distributions of
  excess matching catalogue sources around submm positions in all four
  bands (solid histogram with Poisson uncertainties), our estimate of
  $p(c)$, compared to the general population (dotted histogram), our
  estimate of $\rho(c)$. There is a clear trend of redder colours in
  the IRAC bands for galaxies selected at longer submm wavelengths.}
\label{fig:chist}
\end{figure*}

In a similar way we compare the $\log_{10}(S_{3.6}/S_{4.5})$ colour of
the background population to the excess around submm positions. This
IRAC colour was chosen because it is the primary discriminator for
redshift used in \citet{devlin2009}, \citet{marsden2009}, and
\citet{pascale2009} -- with the trend that smaller ratios (i.e. redder
colours) correlate with higher redshifts. Earlier studies have also
used IRAC colours as a crude redshift estimator
\citep[e.g.,][]{pope2006,yun2008,wilson2008}.  The normalized excess
colour distribution, $p(c)$ (Fig.~\ref{fig:chist}) agrees with this
expectation. There is a trend from bluer colours starting at
250\,\micron\ (submm excess and background colours nearly
indistinguishable), to significantly redder colours at 500 and
870\,micron.

Individually, $p(S_{24})$, $p(S_\mathrm{r})$, and $p(c)$ give
normalized estimates of $q(S_{24},S_\mathrm{r},c)$ marginalized over
the remaining variables. If they were completely independent of one
another, we could estimate $q(S,c)$ as
\begin{equation}
q(S,c) \simeq E \times p(S_{24}) p(S_\mathrm{r}) p(c),
\end{equation}
where $E$ is the total excess measured in Fig.~\ref{fig:r_excess}.
Ideally we would like to bin the excess counts in cubes of $S_{24}$,
$S_\mathrm{r}$, and $c$ simultaneously. Unfortunately, given the
sample size, it is not possible to obtain a statistically significant
measurement of this distribution. Similarly, we approximate the
background counts
\begin{equation}
\rho(S,c) \simeq \rho(S_{24}) \rho(S_\mathrm{r}) \rho(c).
\end{equation}
We proceed under the assumption that these quantities are independent,
but in the next section use Monte Carlo simulations to establish a
reasonable threshold for this approximate LR to obtain counterparts.

\subsection{Normalized LR}

Since each submm peak appears to be produced by a blend of several
matching catalogue sources, we cannot use the \citet{sutherland1992}
reliability, $R$, as a normalized probability that a given candidate
is {\em the single} counterpart. Furthermore, its absolute
normalization depends on precise measurements of $\sigma_\mathrm{r}$,
$\rho(S,c)$ and $q(S,c)$, all of which have moderate uncertainties for
this sample (especially $q(S,c)$, as noted in the previous section).

\begin{figure*}
\centering
\includegraphics[width=0.49\linewidth]{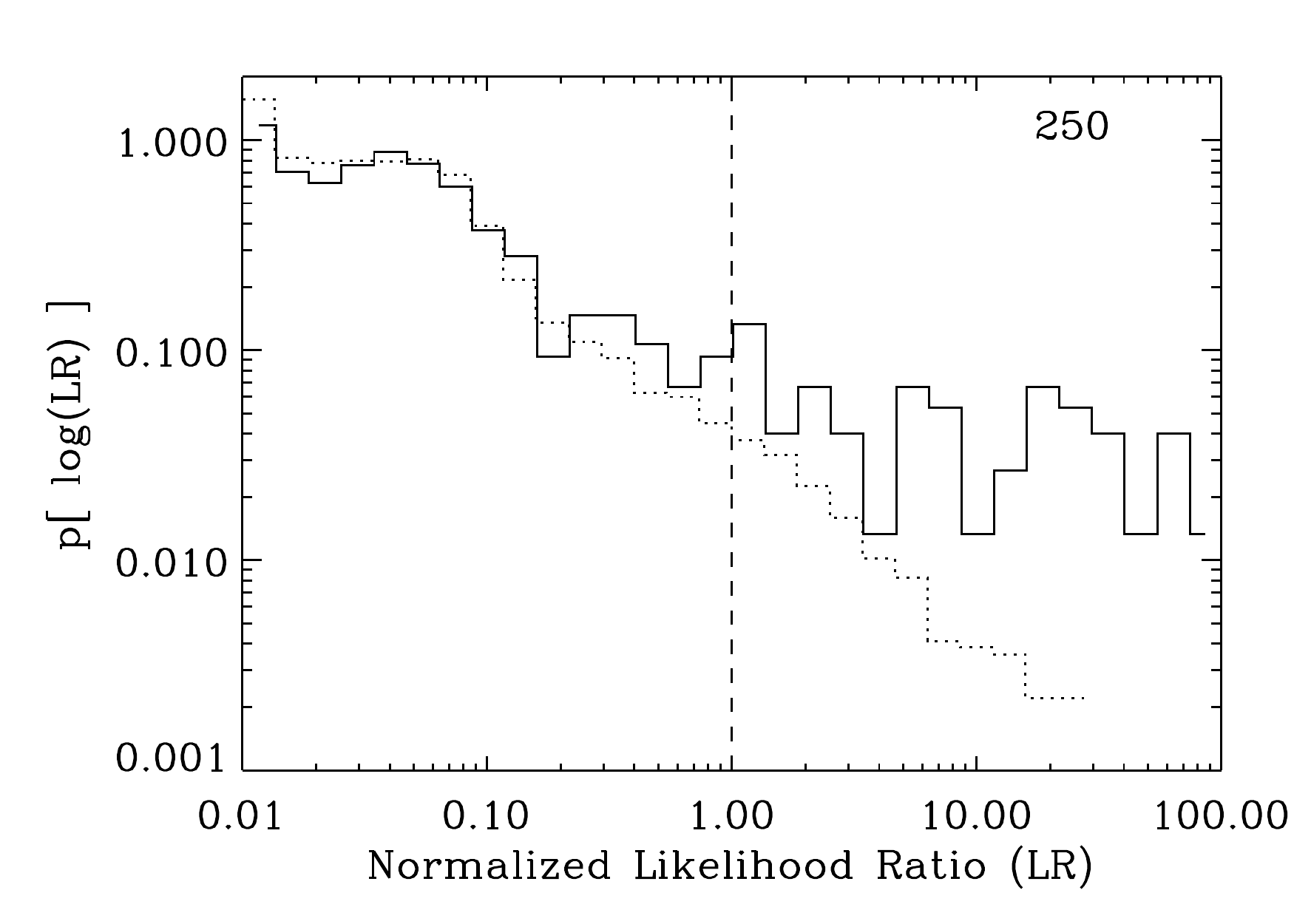}
\includegraphics[width=0.49\linewidth]{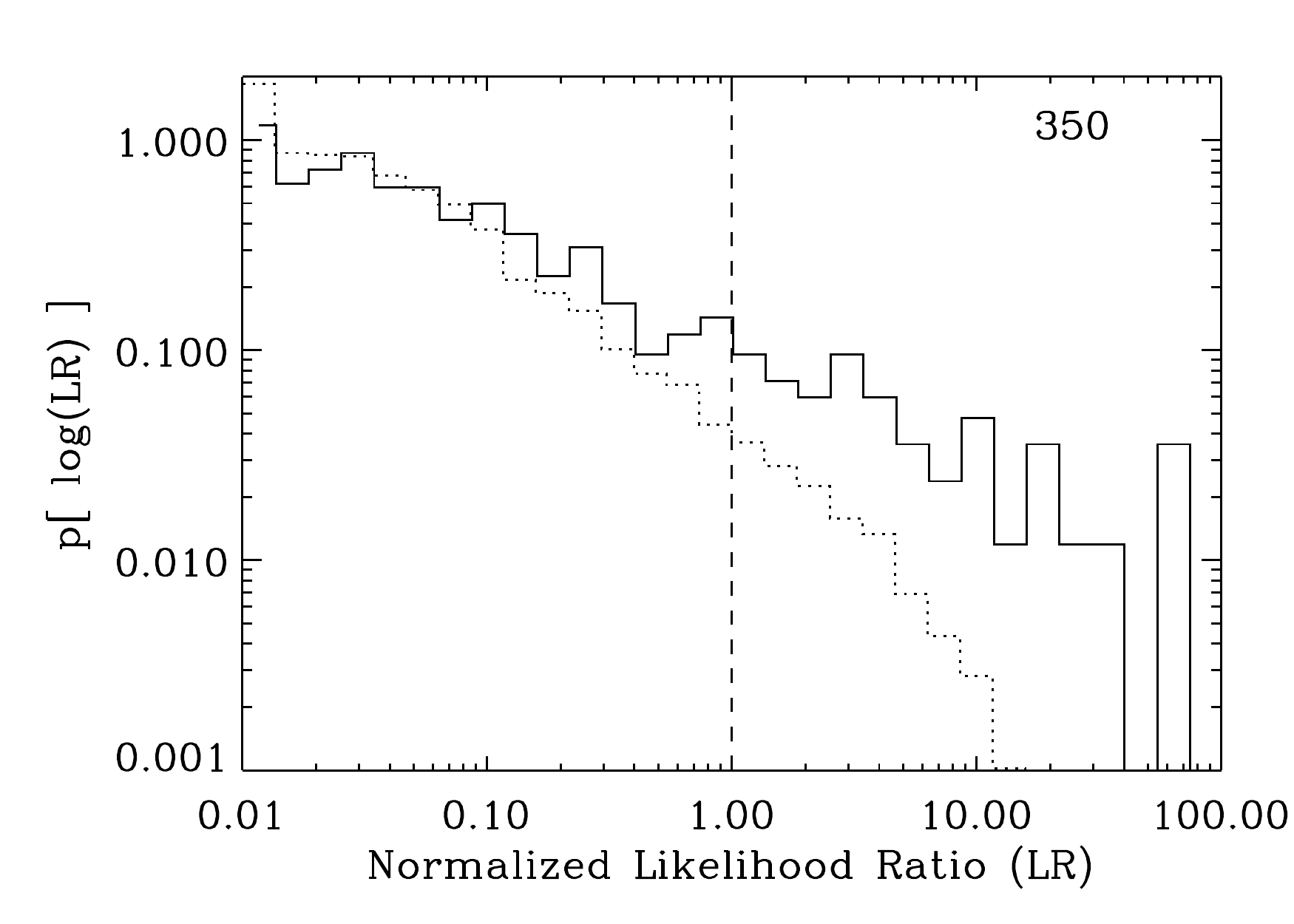}
\includegraphics[width=0.49\linewidth]{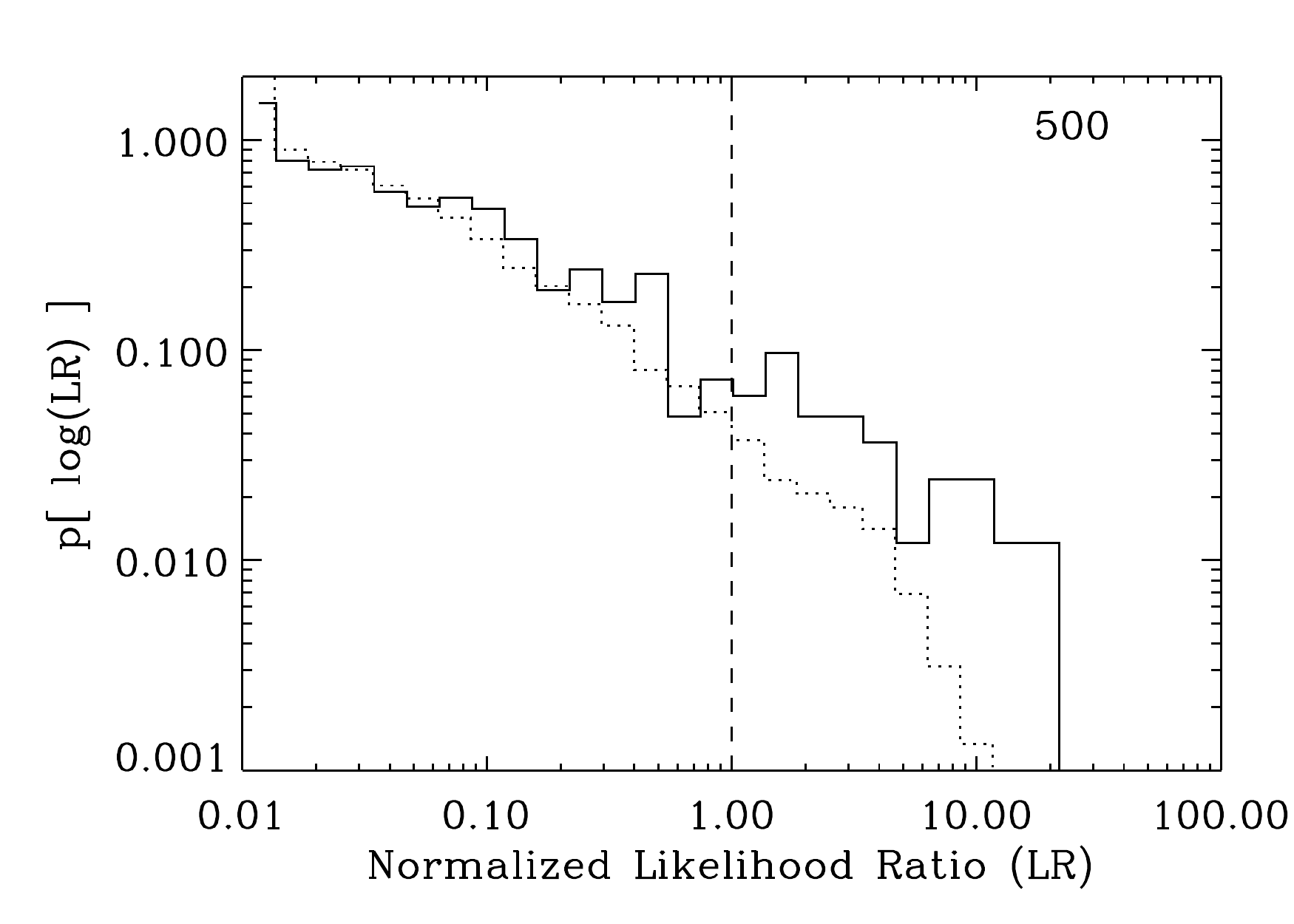}
\includegraphics[width=0.49\linewidth]{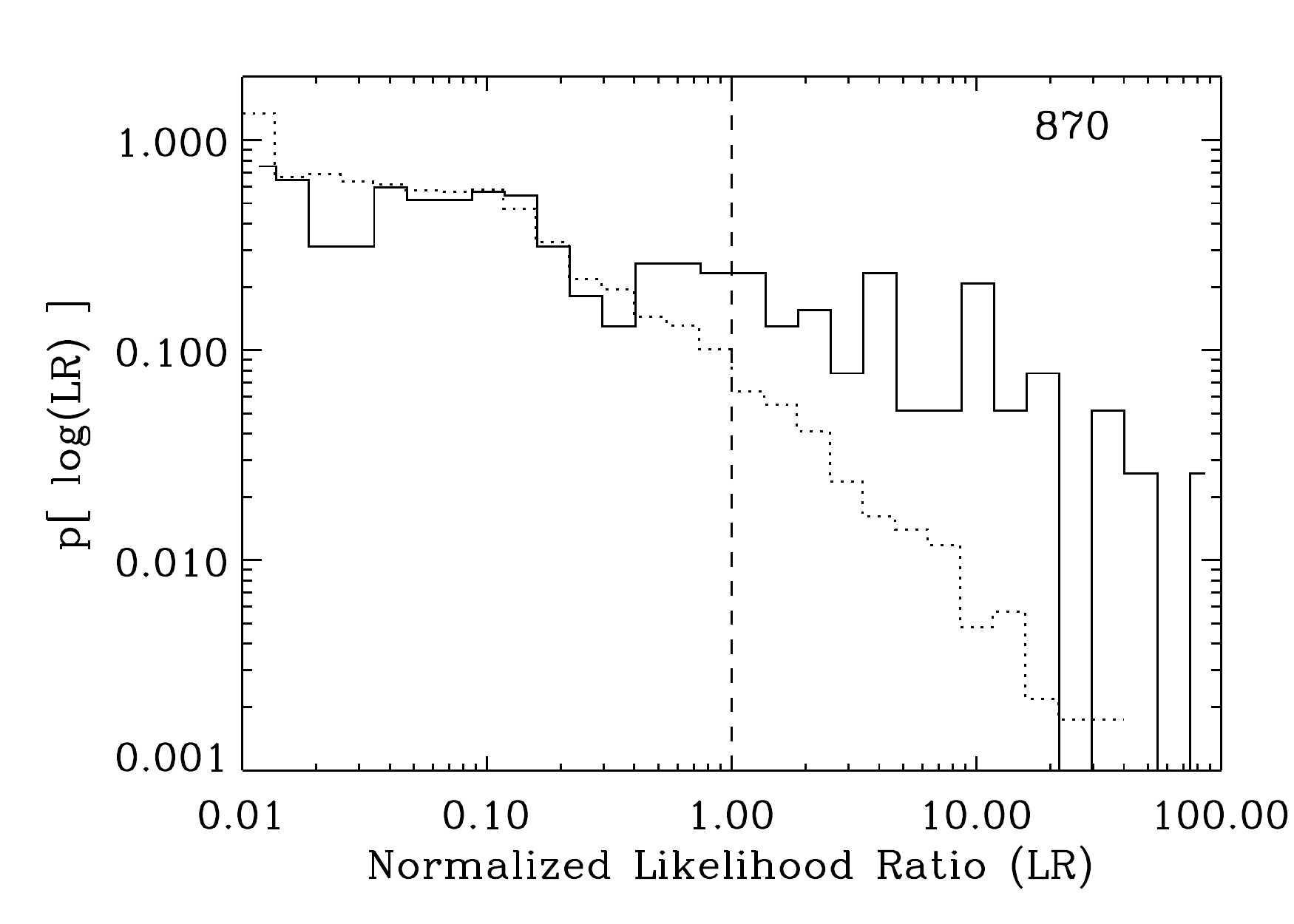}
\caption{Distribution of likelihood ratios (LR) for potential
  counterparts in the matching catalogue out to 60\,arcsec from submm
  peak positions (solid histograms) compared to random positions
  (dotted histograms). Thresholds for identifying potential
  counterparts are given by the vertical dashed lines, such that the
  integrated tail of the distribution for random positions gives a
  false positive rate of 10\%. All LRs are normalized to this value
  (i.e., potential IDs with $\mathrm{LR} > 1.00$ have less than a 10\%
  chance of being false).}
\label{fig:lr}
\end{figure*}

Instead, we compare the LRs for all potential candidate matches to
submm peaks, with the distribution of LRs for false matches to 10,000
random positions. This procedure enables us to select a threshold LR
that gives an acceptable rate of false-positives, which we set to
10\%. This approach to identifying the threshold LR is similar to that
of \citet{mann1997}. The results of our calculation are shown in
Fig.~\ref{fig:lr}.  The solid histogram is the distribution of LR for
all matching catalogue sources out to 60\,arcsec (this maximum search
radius easily contains $>99$\% of the true counterparts) from each
submm position. The dotted histograms show the `background'
distribution obtained by searching for counterparts around random
positions. As expected, for large values of the LR, there are more
sources around submm peaks than random positions. Also,
unsurprisingly, the distinction between the two curves is greatest at
870\,\micron\ (which has the smallest positional uncertainties), and
progressively worse across the BLAST bands to longer wavelengths.

\begin{table*}
  \caption{Excess count and ID statistics. The number of peaks
    are from the BLAST and LESS catalogues, using 3.75$\sigma$ and
    3.72$\sigma$ thresholds, respectively. The excess counts per source and
    $\sigma_\mathrm{r}$
    are the best-fit positional uncertainty model parameters
    (with $\chi_\mathrm{r}^2$
    giving the reduced chi-squared in each case) -- see Section~\ref{sec:pos}.
    The fact that the excesses are greater than 1 is an indication of how
    confused the data are, and a warning that finding single counterparts is
    challenging.
    Expected numbers of IDs are calculated from the product
    of the number of submm peaks with the
    average excesses per source. `Found IDs' are the total number of potential
    counterparts with false-identification rates of 10\% in the matching
    catalogue.
    The `Real ID' rate subtracts the expected number of false
    IDs for the list (10\% of the number of submm peaks; see discussion of
    threshold LR values in Section~\ref{sec:id}), and also expresses that
    quantity as a percentage of the total expected.}

  \vspace{0.2cm}
  \centering
  \begin{tabular}{lccccccc}
    \hline
    $\lambda$ & Peaks & Excess      & $\sigma_\mathrm{r}$ & $\chi^2_\mathrm{r}$ & Expected IDs & Found IDs & Real IDs \\
    (\micron) & (\#)    & (per peak)& (arcsec)   &            & (\#)         & (\#)      & (\#)  \\
    \hline
    250 & 64 & 3.2 & 10.4 & 1.8 & 204.8 & 52 & 45.6 (22\%) \\
    350 & 67 & 3.4 & 10.4 & 1.5 & 227.8 & 50 & 43.3 (19\%) \\
    500 & 55 & 3.7 & 15.4 & 1.0 & 203.5 & 31 & 25.5 (13\%) \\
    870 & 81$^\dagger$ & 2.2 &  6.9 & 1.8 & 178.2 & 66 & 57.9 (32\%) \\
    \hline \\
  \end{tabular} \\
  $^\dagger$There are 81 peaks from LESS that land in the region of coverage
  considered here, but only 42 peaks are cross-matched to the BLAST peak
  lists. The full list of 81 peaks is used to estimate priors even though we
  only discuss the properties of the smaller subset of cross-matched peaks
  in this paper.
  \label{tab:pos}
\end{table*}

A threshold in the LR is chosen such that the fraction of sources with
{\em larger} LRs for the simulation (random positions) is 10\%. This
cut is indicated as a vertical dashed line in Fig.~\ref{fig:lr}. For
convenience, we normalize the LRs by these threshold values, so that
potential counterparts have values that are greater than 1. Also,
noting the ratio of the density of sources around submm and random
positions, a candidate counterpart at this threshold is about three
times more likely to be real than spurious (this value does {\em not},
however, reflect on the {\em absolute} probability that the source is
either real or spurious).

The results of applying this cut to potential identifications around
submm positions are given in Table~\ref{tab:pos}. For each band the
total number of expected counterparts is the product of the number of
submm positions with the average excess of matching catalogue sources
around each spot. The estimated number of true IDs found is then the
number of sources detected above the cut on the LR, after subtracting
the 10\% spurious fraction. This test shows that as many as 32\% of
the individual matching catalogue sources contributing to the
870\,\micron\ peaks have been identified, and as few as 13\% at
500\,\micron. The trend of this result is certainly expected, due to
the increased confusion caused by larger beams. However, the extent of
the problem of identifying confused sources is larger than some would
have expected.

\section[]{Data Tables}
\label{sec:tables}

The following data tables are provided in this Appendix: the
match-filtered submm peak lists in the BLAST bands
(Tables~\ref{tab:250cat}-\ref{tab:500cat}), as well as the LESS
peaks from \citet{weiss2009} that land within the region that was
analyzed (Table~\ref{tab:870cat}); the correspondence between the
matching catalogue and each of the submm lists
(Tables~\ref{tab:match}); and the properties of the matched sources
(re-measured submm flux densities, SED fits and redshifts,
Table~\ref{tab:stuff}).


\begin{table*}
  \caption{250\,\micron\ peak list produced from matched-filtered maps that
    land within the coverage of the matching catalogue
    (Figure~\ref{fig:coverage}). On average, each of these peaks is a blend of
    3.2 sources from our matching catalogue. The
    `Previously Published Name' is taken from the supplement to
    \citet{devlin2009} when available (new peaks have no entry in this
    column), but the short submm `SID' is used throughout this paper.
    The `Match' column refers to IDs in Table~\ref{tab:match}.
    The positions are centroids of the submm peaks. The flux densities are
    raw values from the maps (with no correction for flux boosting), and the
    noises and \snr\ refer to instrumental noise only (excluding source
    confusion).}
  \centering
  \begin{tabular}{ccccccc}
    \hline
    SID & Previously Published & Match & R.A. & Dec. & Flux Density & \snr\ \\
    250 &  Name         & ID
    & ($^{\mathrm h}$\ \ \ $^{\mathrm m}$\ \ \ $^{\mathrm s}$)
    & ($^{\circ}$\ \ \ $^{\prime}$\ \ \ $^{\prime\prime}$)
    & (mJy) \\
    \hline
  3 & BLAST J033235$-$275530 (250\micron) & 66 & 03 32 34.99 & $-$27 55 31.1 &  196 $\pm$   14 & 13.93 \\
  5 & BLAST J033229$-$274414 (250\micron) & 61,62 & 03 32 29.66 & $-$27 44 16.3 &  162 $\pm$   13 & 11.62 \\
 16 & BLAST J033129$-$275722 (250\micron) & 10,11 & 03 31 29.94 & $-$27 57 24.1 &  112 $\pm$   14 &  7.93 \\
 17 & BLAST J033249$-$275842 (250\micron) & 86 & 03 32 49.32 & $-$27 58 41.5 &  104 $\pm$   14 &  7.41 \\
 24 & BLAST J033246$-$275744 (250\micron) & 81 & 03 32 46.05 & $-$27 57 45.0 &   99 $\pm$   14 &  7.11 \\
 27 & BLAST J033258$-$274322 (250\micron) & 92 & 03 32 58.31 & $-$27 43 24.9 &   97 $\pm$   14 &  6.79 \\
 32 & BLAST J033145$-$274635 (250\micron) & 26 & 03 31 45.37 & $-$27 46 36.6 &   87 $\pm$   14 &  6.21 \\
 36 & BLAST J033145$-$275730 (250\micron) & 27 & 03 31 45.86 & $-$27 57 29.0 &   84 $\pm$   14 &  5.96 \\
 38 & BLAST J033217$-$275905 (250\micron) & 52,53 & 03 32 17.07 & $-$27 59 06.9 &   83 $\pm$   14 &  5.91 \\
 45 & BLAST J033221$-$275630 (250\micron) & 57 & 03 32 21.77 & $-$27 56 26.0 &   79 $\pm$   14 &  5.65 \\
 51 & BLAST J033152$-$273931 (250\micron) & 34,35 & 03 31 52.57 & $-$27 39 33.1 &   76 $\pm$   14 &  5.40 \\
 52 & BLAST J033308$-$274805 (250\micron) & 100,101,102 & 03 33 08.62 & $-$27 48 04.5 &   74 $\pm$   13 &  5.37 \\
 54 & BLAST J033318$-$274610 (250\micron) & 111,112 & 03 33 17.83 & $-$27 46 08.5 &   74 $\pm$   14 &  5.32 \\
 57 & BLAST J033145$-$274205 (250\micron) & 25 & 03 31 44.99 & $-$27 42 06.4 &   74 $\pm$   14 &  5.19 \\
 58 & BLAST J033218$-$275216 (250\micron) & 55,56 & 03 32 18.01 & $-$27 52 18.7 &   72 $\pm$   14 &  5.16 \\
 59 & BLAST J033149$-$274335 (250\micron) & 30 & 03 31 49.74 & $-$27 43 34.7 &   72 $\pm$   14 &  5.14 \\
 60 & BLAST J033241$-$273818 (250\micron) & 74,75 & 03 32 41.87 & $-$27 38 17.9 &   72 $\pm$   14 &  5.06 \\
 67 & BLAST J033319$-$275423 (250\micron) & 116 & 03 33 19.03 & $-$27 54 23.7 &   70 $\pm$   14 &  4.98 \\
 69 & BLAST J033237$-$273527 (250\micron) & 70,71 & 03 32 37.78 & $-$27 35 43.6 &   69 $\pm$   14 &  4.95 \\
 70 & BLAST J033135$-$273933 (250\micron) & 12 & 03 31 35.75 & $-$27 39 44.2 &   70 $\pm$   14 &  4.94 \\
 72 & BLAST J033316$-$275043 (250\micron) & 109 & 03 33 16.20 & $-$27 50 39.6 &   68 $\pm$   13 &  4.94 \\
 75 & BLAST J033205$-$274645 (250\micron) & 43 & 03 32 05.14 & $-$27 46 45.6 &   68 $\pm$   14 &  4.91 \\
 78 &                 & 1 & 03 31 27.89 & $-$27 44 49.5 &   68 $\pm$   14 &  4.86 \\
 80 & BLAST J033128$-$273916 (250\micron) & 4,5 & 03 31 29.06 & $-$27 39 05.2 &   68 $\pm$   14 &  4.81 \\
 82 & BLAST J033243$-$273919 (250\micron) & 78 & 03 32 43.61 & $-$27 39 20.5 &   67 $\pm$   14 &  4.79 \\
 85 & BLAST J033141$-$274439 (250\micron) & 19,20 & 03 31 41.15 & $-$27 44 38.0 &   67 $\pm$   14 &  4.75 \\
 91 & BLAST J033223$-$273642 (250\micron) & 59 & 03 32 22.81 & $-$27 36 43.4 &   66 $\pm$   14 &  4.72 \\
 92 & BLAST J033140$-$275633 (250\micron) & 18 & 03 31 40.33 & $-$27 56 36.3 &   66 $\pm$   14 &  4.70 \\
 93 & BLAST J033211$-$275859 (250\micron) &  & 03 32 11.86 & $-$27 59 00.8 &   66 $\pm$   14 &  4.70 \\
 97 & BLAST J033222$-$280019 (250\micron) & 58 & 03 32 22.14 & $-$28 00 21.1 &   65 $\pm$   13 &  4.66 \\
 99 & BLAST J033129$-$275910 (250\micron) & 9 & 03 31 29.19 & $-$27 59 11.1 &   66 $\pm$   14 &  4.64 \\
111 & BLAST J033130$-$275604 (250\micron) & 7,8 & 03 31 29.99 & $-$27 56 02.1 &   63 $\pm$   14 &  4.53 \\
112 &                 & 63 & 03 32 29.95 & $-$27 43 14.4 &   64 $\pm$   14 &  4.52 \\
113 & BLAST J033235$-$274932 (250\micron) & 67 & 03 32 35.24 & $-$27 49 28.2 &   62 $\pm$   13 &  4.51 \\
121 & BLAST J033230$-$275905 (250\micron) & 64 & 03 32 30.17 & $-$27 59 05.9 &   63 $\pm$   14 &  4.44 \\
130 & BLAST J033259$-$273535 (250\micron) & 93 & 03 32 59.71 & $-$27 35 33.4 &   61 $\pm$   14 &  4.37 \\
146 & BLAST J033151$-$274431 (250\micron) & 31,32 & 03 31 51.13 & $-$27 44 38.1 &   60 $\pm$   13 &  4.29 \\
151 & BLAST J033147$-$274147 (250\micron) &  & 03 31 47.89 & $-$27 41 41.2 &   60 $\pm$   14 &  4.27 \\
154 &                 &  & 03 32 06.88 & $-$27 38 00.7 &   60 $\pm$   14 &  4.25 \\
178 & BLAST J033232$-$275304 (250\micron) &  & 03 32 33.03 & $-$27 53 04.8 &   59 $\pm$   14 &  4.18 \\
185 & BLAST J033317$-$274118 (250\micron) & 110 & 03 33 17.38 & $-$27 41 12.5 &   59 $\pm$   14 &  4.15 \\
188 & BLAST J033154$-$274406 (250\micron) & 37 & 03 31 54.28 & $-$27 44 04.2 &   58 $\pm$   14 &  4.15 \\
210 & BLAST J033135$-$274705 (250\micron) &  & 03 31 35.03 & $-$27 47 09.1 &   57 $\pm$   14 &  4.09 \\
222 & BLAST J033200$-$280234 (250\micron) &  & 03 32 00.67 & $-$28 02 32.1 &   57 $\pm$   14 &  4.06 \\
226 &                 &  & 03 31 29.94 & $-$27 39 41.9 &   58 $\pm$   14 &  4.05 \\
235 & BLAST J033156$-$280306 (250\micron) &  & 03 31 56.85 & $-$28 03 09.4 &   57 $\pm$   14 &  4.03 \\
244 &                 & 17 & 03 31 39.58 & $-$27 41 29.5 &   56 $\pm$   14 &  4.00 \\
246 & BLAST J033211$-$280242 (250\micron) & 45 & 03 32 11.14 & $-$28 02 42.6 &   56 $\pm$   14 &  4.00 \\
250 & BLAST J033251$-$274530 (250\micron) &  & 03 32 51.55 & $-$27 45 33.3 &   56 $\pm$   14 &  3.99 \\
255 & BLAST J033144$-$275521 (250\micron) & 24 & 03 31 44.76 & $-$27 55 19.5 &   55 $\pm$   13 &  3.98 \\
    \hline
  \end{tabular}
\label{tab:250cat}
\end{table*}

\begin{table*}
  \centering
  \begin{tabular}{ccccccc}
    \hline
    SID & Previously Published & Match & R.A. & Dec. & Flux Density & \snr\ \\
    250 &  Name         & ID
    & ($^{\mathrm h}$\ \ \ $^{\mathrm m}$\ \ \ $^{\mathrm s}$)
    & ($^{\circ}$\ \ \ $^{\prime}$\ \ \ $^{\prime\prime}$)
    & (mJy) \\
    \hline
279 & BLAST J033311$-$274313 (250\micron) &  & 03 33 11.13 & $-$27 43 12.0 &   56 $\pm$   14 &  3.93 \\
289 & BLAST J033217$-$275054 (250\micron) & 54 & 03 32 17.43 & $-$27 50 55.7 &   54 $\pm$   13 &  3.91 \\
298 & BLAST J033243$-$275146 (250\micron) & 79,80 & 03 32 43.50 & $-$27 51 45.6 &   54 $\pm$   14 &  3.89 \\
312 & BLAST J033228$-$273545 (250\micron) &  & 03 32 28.66 & $-$27 35 53.0 &   54 $\pm$   14 &  3.86 \\
316 & BLAST J033238$-$275651 (250\micron) & 72 & 03 32 37.99 & $-$27 56 57.6 &   53 $\pm$   13 &  3.86 \\
325 & BLAST J033225$-$273822 (250\micron) & 60 & 03 32 24.97 & $-$27 38 22.2 &   54 $\pm$   14 &  3.84 \\
328 & BLAST J033205$-$280054 (250\micron) & 42 & 03 32 05.54 & $-$28 00 54.2 &   53 $\pm$   13 &  3.84 \\
348 & BLAST J033150$-$275100 (250\micron) &  & 03 31 50.24 & $-$27 51 11.0 &   53 $\pm$   13 &  3.81 \\
354 & BLAST J033248$-$274443 (250\micron) &  & 03 32 47.55 & $-$27 44 58.3 &   53 $\pm$   14 &  3.80 \\
358 & BLAST J033210$-$274253 (250\micron) & 47 & 03 32 10.51 & $-$27 42 55.3 &   53 $\pm$   14 &  3.80 \\
369 &                 & 87 & 03 32 49.09 & $-$27 36 19.7 &   53 $\pm$   14 &  3.78 \\
378 & BLAST J033131$-$274601 (250\micron) &  & 03 31 31.77 & $-$27 46 05.1 &   52 $\pm$   14 &  3.76 \\
381 & BLAST J033311$-$275226 (250\micron) &  & 03 33 11.65 & $-$27 52 27.0 &   52 $\pm$   13 &  3.76 \\
389 & BLAST J033251$-$275936 (250\micron) & 88 & 03 32 51.85 & $-$27 59 40.9 &   53 $\pm$   14 &  3.75 \\
    \hline
  \end{tabular}
\end{table*}

\begin{table*}
  \caption{350\,\micron\ peak list produced from matched-filtered maps.
    Columns have the same meaning as in Table~\ref{tab:250cat}.
    On average, each of these peaks is a blend of 3.4 sources from our matching
    catalogue.}
  \centering
  \begin{tabular}{ccccccc}
    \hline
    SID & Previously Published & Match & R.A. & Dec. & Flux Density & \snr\ \\
    350 &  Name         & ID
    & ($^{\mathrm h}$\ \ \ $^{\mathrm m}$\ \ \ $^{\mathrm s}$)
    & ($^{\circ}$\ \ \ $^{\prime}$\ \ \ $^{\prime\prime}$)
    & (mJy) \\
    \hline
  3 & BLAST J033234$-$275531 (350\micron) & 66 & 03 32 34.88 & $-$27 55 31.3 &   89 $\pm$   10 &  8.24 \\
  7 & BLAST J033249$-$275833 (350\micron) & 86 & 03 32 49.49 & $-$27 58 33.8 &   79 $\pm$   10 &  7.36 \\
  8 & BLAST J033229$-$274414 (350\micron) & 61,62 & 03 32 29.37 & $-$27 44 16.4 &   74 $\pm$   10 &  6.95 \\
 10 & BLAST J033220$-$275647 (350\micron) &  & 03 32 20.15 & $-$27 56 45.4 &   73 $\pm$   10 &  6.82 \\
 11 & BLAST J033258$-$274328 (350\micron) & 92 & 03 32 58.39 & $-$27 43 25.0 &   72 $\pm$   10 &  6.73 \\
 12 & BLAST J033229$-$274305 (350\micron) & 63 & 03 32 29.49 & $-$27 43 03.4 &   71 $\pm$   10 &  6.72 \\
 14 & BLAST J033328$-$275700 (350\micron) & 118 & 03 33 27.65 & $-$27 57 03.4 &   72 $\pm$   10 &  6.66 \\
 16 & BLAST J033129$-$275722 (350\micron) & 10,11 & 03 31 29.61 & $-$27 57 22.6 &   70 $\pm$   11 &  6.41 \\
 19 & BLAST J033204$-$274650 (350\micron) & 43 & 03 32 04.59 & $-$27 46 51.7 &   67 $\pm$   10 &  6.35 \\
 20 & BLAST J033207$-$275815 (350\micron) & 44 & 03 32 07.34 & $-$27 58 18.1 &   68 $\pm$   10 &  6.30 \\
 22 & BLAST J033128$-$273927 (350\micron) & 2,3,4 & 03 31 28.02 & $-$27 39 32.2 &   71 $\pm$   11 &  6.25 \\
 31 & BLAST J033311$-$274135 (350\micron) & 105 & 03 33 11.16 & $-$27 41 32.5 &   64 $\pm$   10 &  5.95 \\
 34 & BLAST J033247$-$274224 (350\micron) & 83 & 03 32 47.24 & $-$27 42 31.9 &   62 $\pm$   10 &  5.86 \\
 35 & BLAST J033150$-$274333 (350\micron) & 30 & 03 31 49.77 & $-$27 43 28.2 &   63 $\pm$   10 &  5.84 \\
 37 & BLAST J033140$-$274435 (350\micron) & 19,20 & 03 31 40.71 & $-$27 44 36.9 &   62 $\pm$   10 &  5.76 \\
 42 & BLAST J033138$-$274122 (350\micron) & 17 & 03 31 39.04 & $-$27 41 22.3 &   60 $\pm$   10 &  5.67 \\
 43 & BLAST J033318$-$274606 (350\micron) & 111,112 & 03 33 18.39 & $-$27 46 13.4 &   60 $\pm$   10 &  5.64 \\
 45 & BLAST J033210$-$275206 (350\micron) & 46 & 03 32 11.53 & $-$27 52 03.8 &   61 $\pm$   10 &  5.62 \\
 47 & BLAST J033151$-$274428 (350\micron) & 31,32 & 03 31 51.20 & $-$27 44 33.4 &   59 $\pm$   10 &  5.56 \\
 49 & BLAST J033135$-$275448 (350\micron) & 13,14,15 & 03 31 35.44 & $-$27 54 47.5 &   60 $\pm$   10 &  5.56 \\
 52 & BLAST J033252$-$273756 (350\micron) &  & 03 32 52.12 & $-$27 37 47.9 &   58 $\pm$   10 &  5.36 \\
 58 & BLAST J033321$-$275513 (350\micron) & 117 & 03 33 21.74 & $-$27 55 13.1 &   56 $\pm$   10 &  5.19 \\
 64 & BLAST J033302$-$275640 (350\micron) & 95 & 03 33 02.20 & $-$27 56 43.7 &   54 $\pm$   10 &  5.05 \\
 65 & BLAST J033218$-$275207 (350\micron) & 55,56 & 03 32 18.12 & $-$27 52 12.5 &   54 $\pm$   10 &  5.03 \\
 69 & BLAST J033237$-$273541 (350\micron) & 70,71 & 03 32 37.62 & $-$27 35 42.4 &   54 $\pm$   11 &  4.96 \\
 79 & BLAST J033307$-$275833 (350\micron) &  & 03 33 06.98 & $-$27 58 35.5 &   52 $\pm$   10 &  4.84 \\
 80 & BLAST J033217$-$275906 (350\micron) & 53 & 03 32 17.78 & $-$27 59 20.2 &   52 $\pm$   10 &  4.83 \\
 84 & BLAST J033318$-$274131 (350\micron) &  & 03 33 19.72 & $-$27 41 38.7 &   52 $\pm$   10 &  4.79 \\
 85 & BLAST J033230$-$275905 (350\micron) & 64 & 03 32 30.00 & $-$27 59 02.4 &   51 $\pm$   10 &  4.79 \\
 88 & BLAST J033211$-$273731 (350\micron) & 49 & 03 32 11.34 & $-$27 37 24.1 &   51 $\pm$   10 &  4.76 \\
 90 & BLAST J033253$-$280107 (350\micron) & 89 & 03 32 53.05 & $-$28 01 11.3 &   51 $\pm$   10 &  4.76 \\
 91 & BLAST J033321$-$273908 (350\micron) &  & 03 33 21.00 & $-$27 39 10.0 &   51 $\pm$   10 &  4.75 \\
 92 & BLAST J033308$-$275130 (350\micron) & 104 & 03 33 08.61 & $-$27 51 29.3 &   50 $\pm$   10 &  4.74 \\
 93 & BLAST J033153$-$274952 (350\micron) & 36 & 03 31 54.84 & $-$27 49 37.7 &   51 $\pm$   10 &  4.73 \\
 95 & BLAST J033152$-$280325 (350\micron) & 33 & 03 31 53.14 & $-$28 03 35.7 &   51 $\pm$   10 &  4.72 \\
 98 &                 & 108 & 03 33 15.41 & $-$27 45 32.9 &   50 $\pm$   10 &  4.69 \\
108 & BLAST J033215$-$273929 (350\micron) & 51 & 03 32 15.25 & $-$27 39 31.8 &   50 $\pm$   10 &  4.62 \\
112 & BLAST J033316$-$275056 (350\micron) & 109 & 03 33 17.16 & $-$27 50 55.7 &   48 $\pm$   10 &  4.60 \\
115 & BLAST J033305$-$274414 (350\micron) & 96 & 03 33 06.26 & $-$27 44 14.1 &   49 $\pm$   10 &  4.59 \\
116 & BLAST J033245$-$275739 (350\micron) & 81 & 03 32 46.04 & $-$27 57 35.2 &   49 $\pm$   10 &  4.59 \\
117 & BLAST J033134$-$274629 (350\micron) &  & 03 31 34.56 & $-$27 46 25.9 &   49 $\pm$   10 &  4.57 \\
124 & BLAST J033153$-$273931 (350\micron) & 34,35 & 03 31 52.62 & $-$27 39 29.9 &   49 $\pm$   10 &  4.54 \\
128 & BLAST J033243$-$275509 (350\micron) & 76,77 & 03 32 42.86 & $-$27 55 13.1 &   48 $\pm$   10 &  4.51 \\
137 & BLAST J033217$-$275906 (350\micron) & 52 & 03 32 17.89 & $-$27 58 46.5 &   47 $\pm$   10 &  4.44 \\
144 & BLAST J033152$-$275035 (350\micron) &  & 03 31 51.22 & $-$27 50 41.8 &   46 $\pm$   10 &  4.40 \\
145 & BLAST J033308$-$274813 (350\micron) & 100,101,103 & 03 33 08.49 & $-$27 48 16.5 &   47 $\pm$   10 &  4.40 \\
152 & BLAST J033145$-$274153 (350\micron) &  & 03 31 44.98 & $-$27 41 50.5 &   46 $\pm$   10 &  4.32 \\
154 & BLAST J033238$-$274620 (350\micron) &  & 03 32 38.52 & $-$27 46 18.5 &   46 $\pm$   10 &  4.32 \\
161 & BLAST J033212$-$275459 (350\micron) &  & 03 32 12.74 & $-$27 54 59.1 &   45 $\pm$   10 &  4.29 \\
169 & BLAST J033225$-$273810 (350\micron) & 60 & 03 32 24.93 & $-$27 38 17.3 &   46 $\pm$   10 &  4.25 \\
    \hline
  \end{tabular}
\label{tab:350cat}
\end{table*}

\begin{table*}
  \centering
  \begin{tabular}{ccccccc}
    \hline
    SID & Previously Published & Match & R.A. & Dec. & Flux Density & \snr\ \\
    350 &  Name         & ID
    & ($^{\mathrm h}$\ \ \ $^{\mathrm m}$\ \ \ $^{\mathrm s}$)
    & ($^{\circ}$\ \ \ $^{\prime}$\ \ \ $^{\prime\prime}$)
    & (mJy) \\
    \hline
170 & BLAST J033201$-$274142 (350\micron) & 41 & 03 32 02.02 & $-$27 41 38.0 &   45 $\pm$   10 &  4.25 \\
177 & BLAST J033129$-$275911 (350\micron) & 9 & 03 31 29.63 & $-$27 59 09.1 &   46 $\pm$   10 &  4.22 \\
184 &                 &  & 03 32 10.66 & $-$27 58 56.9 &   45 $\pm$   10 &  4.21 \\
187 & BLAST J033141$-$275529 (350\micron) & 23 & 03 31 41.76 & $-$27 55 28.4 &   44 $\pm$   10 &  4.18 \\
188 & BLAST J033147$-$274755 (350\micron) & 28 & 03 31 47.60 & $-$27 47 52.9 &   45 $\pm$   10 &  4.18 \\
199 & BLAST J033217$-$275407 (350\micron) &  & 03 32 17.92 & $-$27 54 08.0 &   44 $\pm$   10 &  4.13 \\
235 & BLAST J033127$-$274430 (350\micron) & 1 & 03 31 27.90 & $-$27 44 32.8 &   43 $\pm$   10 &  4.04 \\
259 & BLAST J033235$-$280135 (350\micron) & 68,69 & 03 32 35.90 & $-$28 01 48.7 &   42 $\pm$   10 &  3.97 \\
264 & BLAST J033210$-$275612 (350\micron) &  & 03 32 11.27 & $-$27 56 12.1 &   42 $\pm$   10 &  3.96 \\
273 & BLAST J033234$-$280043 (350\micron) &  & 03 32 34.46 & $-$28 00 36.3 &   42 $\pm$   10 &  3.94 \\
276 & BLAST J033313$-$273709 (350\micron) &  & 03 33 13.26 & $-$27 37 14.9 &   43 $\pm$   10 &  3.93 \\
291 & BLAST J033136$-$274933 (350\micron) &  & 03 31 36.08 & $-$27 49 40.9 &   41 $\pm$   10 &  3.89 \\
298 & BLAST J033312$-$275611 (350\micron) & 106,107 & 03 33 13.37 & $-$27 56 04.2 &   41 $\pm$   10 &  3.88 \\
317 &                 &  & 03 32 29.08 & $-$28 01 53.8 &   41 $\pm$   10 &  3.84 \\
322 & BLAST J033148$-$280212 (350\micron) & 29 & 03 31 48.41 & $-$28 02 20.9 &   41 $\pm$   10 &  3.83 \\
340 & BLAST J033132$-$274314 (350\micron) &  & 03 31 32.37 & $-$27 43 16.5 &   41 $\pm$   10 &  3.80 \\
362 &                 &  & 03 33 04.96 & $-$27 37 26.0 &   40 $\pm$   10 &  3.76 \\
    \hline
  \end{tabular}
\end{table*}

\begin{table*}
  \caption{500\,\micron\ peak list produced from matched-filtered maps.
    Columns have the same meaning as in Table~\ref{tab:250cat}.
    On average, each of these peaks is a blend of 3.7 sources from our matching
    catalogue.}
  \centering
  \begin{tabular}{ccccccc}
    \hline
    SID & Previously Published & Match & R.A. & Dec. & Flux Density & \snr\ \\
    500 &  Name         & ID
    & ($^{\mathrm h}$\ \ \ $^{\mathrm m}$\ \ \ $^{\mathrm s}$)
    & ($^{\circ}$\ \ \ $^{\prime}$\ \ \ $^{\prime\prime}$)
    & (mJy) \\
    \hline
  5 & BLAST J033328$-$275659 (500\micron) & 118 & 03 33 28.17 & $-$27 56 57.0 &   62 $\pm$    8 &  7.43 \\
  7 & BLAST J033311$-$275610 (500\micron) & 106 & 03 33 11.77 & $-$27 56 11.2 &   56 $\pm$    8 &  6.81 \\
  8 & BLAST J033207$-$275811 (500\micron) & 44 & 03 32 08.12 & $-$27 58 05.0 &   53 $\pm$    8 &  6.51 \\
  9 & BLAST J033129$-$275545 (500\micron) & 6,7,8 & 03 31 29.24 & $-$27 55 49.0 &   53 $\pm$    8 &  6.41 \\
 11 & BLAST J033128$-$275713 (500\micron) & 10,11 & 03 31 29.51 & $-$27 57 22.0 &   53 $\pm$    8 &  6.28 \\
 12 & BLAST J033258$-$274325 (500\micron) & 92 & 03 32 58.68 & $-$27 43 27.9 &   52 $\pm$    8 &  6.26 \\
 13 & BLAST J033215$-$275027 (500\micron) &  & 03 32 15.44 & $-$27 50 30.6 &   49 $\pm$    8 &  6.08 \\
 14 & BLAST J033153$-$273921 (500\micron) & 34,35 & 03 31 52.05 & $-$27 39 21.8 &   50 $\pm$    8 &  6.08 \\
 17 & BLAST J033128$-$273942 (500\micron) & 2,3,4 & 03 31 28.18 & $-$27 39 42.3 &   53 $\pm$    8 &  6.01 \\
 20 & BLAST J033256$-$280102 (500\micron) & 91 & 03 32 56.40 & $-$28 01 05.4 &   48 $\pm$    8 &  5.84 \\
 21 & BLAST J033217$-$275212 (500\micron) & 55,56 & 03 32 18.29 & $-$27 52 12.7 &   47 $\pm$    8 &  5.74 \\
 23 & BLAST J033321$-$275510 (500\micron) & 117 & 03 33 22.24 & $-$27 55 14.4 &   47 $\pm$    8 &  5.70 \\
 24 & BLAST J033318$-$274926 (500\micron) & 113,114,115 & 03 33 17.67 & $-$27 49 26.6 &   46 $\pm$    8 &  5.67 \\
 26 &                 & 70,71 & 03 32 37.60 & $-$27 35 31.5 &   46 $\pm$    8 &  5.61 \\
 31 & BLAST J033220$-$275631 (500\micron) & 57 & 03 32 20.86 & $-$27 56 26.3 &   43 $\pm$    8 &  5.27 \\
 33 & BLAST J033212$-$275558 (500\micron) & 50 & 03 32 13.11 & $-$27 56 03.4 &   43 $\pm$    8 &  5.24 \\
 35 & BLAST J033156$-$274511 (500\micron) &  & 03 31 58.33 & $-$27 45 13.5 &   42 $\pm$    8 &  5.20 \\
 37 & BLAST J033249$-$274227 (500\micron) & 83,84,85 & 03 32 47.94 & $-$27 42 25.3 &   43 $\pm$    8 &  5.14 \\
 41 & BLAST J033253$-$273759 (500\micron) & 90 & 03 32 52.93 & $-$27 37 59.7 &   41 $\pm$    8 &  5.05 \\
 43 & BLAST J033301$-$275625 (500\micron) &  & 03 33 00.48 & $-$27 56 08.5 &   41 $\pm$    8 &  5.05 \\
 44 & BLAST J033309$-$275125 (500\micron) & 104 & 03 33 09.35 & $-$27 51 20.1 &   41 $\pm$    8 &  5.05 \\
 48 & BLAST J033135$-$275448 (500\micron) & 13,14,15,16 & 03 31 36.29 & $-$27 54 42.9 &   41 $\pm$    8 &  4.99 \\
 53 & BLAST J033229$-$274426 (500\micron) & 61,62 & 03 32 29.05 & $-$27 44 30.6 &   41 $\pm$    8 &  4.96 \\
 56 & BLAST J033235$-$275518 (500\micron) & 66 & 03 32 35.23 & $-$27 55 14.6 &   41 $\pm$    8 &  4.91 \\
 57 &                 & 108 & 03 33 14.92 & $-$27 45 32.0 &   40 $\pm$    8 &  4.90 \\
 58 & BLAST J033210$-$273725 (500\micron) & 49 & 03 32 10.34 & $-$27 37 20.0 &   40 $\pm$    8 &  4.87 \\
 64 & BLAST J033129$-$275918 (500\micron) & 9 & 03 31 29.49 & $-$27 59 14.7 &   41 $\pm$    8 &  4.77 \\
 68 & BLAST J033211$-$275210 (500\micron) & 46 & 03 32 10.34 & $-$27 51 54.0 &   39 $\pm$    8 &  4.74 \\
 84 & BLAST J033253$-$274459 (500\micron) &  & 03 32 54.41 & $-$27 44 52.3 &   37 $\pm$    8 &  4.58 \\
 88 & BLAST J033153$-$274943 (500\micron) &  & 03 31 52.34 & $-$27 49 27.5 &   38 $\pm$    8 &  4.57 \\
 91 & BLAST J033213$-$274302 (500\micron) & 48 & 03 32 13.32 & $-$27 43 01.3 &   38 $\pm$    8 &  4.55 \\
103 &                 &  & 03 32 34.40 & $-$27 50 17.3 &   36 $\pm$    8 &  4.43 \\
104 & BLAST J033229$-$274314 (500\micron) & 63 & 03 32 29.47 & $-$27 43 08.2 &   36 $\pm$    8 &  4.42 \\
109 & BLAST J033256$-$274539 (500\micron) &  & 03 32 56.26 & $-$27 45 49.6 &   35 $\pm$    8 &  4.37 \\
128 & BLAST J033318$-$274115 (500\micron) &  & 03 33 19.16 & $-$27 41 15.3 &   35 $\pm$    8 &  4.28 \\
133 & BLAST J033238$-$275638 (500\micron) & 72 & 03 32 38.87 & $-$27 56 44.9 &   35 $\pm$    8 &  4.26 \\
137 & BLAST J033323$-$274900 (500\micron) &  & 03 33 23.47 & $-$27 48 53.7 &   35 $\pm$    8 &  4.26 \\
144 &                 & 86 & 03 32 49.97 & $-$27 58 33.2 &   35 $\pm$    8 &  4.24 \\
156 & BLAST J033137$-$273743 (500\micron) &  & 03 31 37.89 & $-$27 37 30.7 &   36 $\pm$    8 &  4.16 \\
157 & BLAST J033157$-$275930 (500\micron) & 40 & 03 31 56.89 & $-$27 59 26.3 &   34 $\pm$    8 &  4.15 \\
162 & BLAST J033143$-$274817 (500\micron) & 22 & 03 31 43.92 & $-$27 48 19.2 &   33 $\pm$    8 &  4.14 \\
176 & BLAST J033300$-$274852 (500\micron) & 94 & 03 32 59.71 & $-$27 48 53.8 &   32 $\pm$    8 &  4.06 \\
177 &                 & 29 & 03 31 48.34 & $-$28 02 03.3 &   34 $\pm$    8 &  4.06 \\
191 &                 &  & 03 32 13.32 & $-$27 59 08.4 &   33 $\pm$    8 &  4.03 \\
199 & BLAST J033247$-$275415 (500\micron) & 82 & 03 32 47.54 & $-$27 54 14.0 &   33 $\pm$    8 &  4.01 \\
200 &                 & 98 & 03 33 06.56 & $-$28 00 46.8 &   33 $\pm$    8 &  4.01 \\
202 & BLAST J033154$-$275343 (500\micron) & 38,39 & 03 31 54.94 & $-$27 53 35.8 &   33 $\pm$    8 &  4.00 \\
203 & BLAST J033215$-$275901 (500\micron) & 52 & 03 32 17.57 & $-$27 58 46.5 &   33 $\pm$    8 &  4.00 \\
205 & BLAST J033153$-$274943 (500\micron) & 36 & 03 31 54.51 & $-$27 49 43.2 &   33 $\pm$    8 &  3.99 \\
210 &                 & 42 & 03 32 06.41 & $-$28 01 07.1 &   32 $\pm$    8 &  3.97 \\
222 & BLAST J033309$-$275347 (500\micron) & 97 & 03 33 09.20 & $-$27 54 00.5 &   32 $\pm$    8 &  3.94 \\
223 & BLAST J033222$-$274555 (500\micron) &  & 03 32 22.75 & $-$27 45 52.9 &   31 $\pm$    8 &  3.93 \\
248 & BLAST J033145$-$274148 (500\micron) &  & 03 31 46.11 & $-$27 41 44.2 &   31 $\pm$    8 &  3.87 \\
250 & BLAST J033235$-$274931 (500\micron) & 67 & 03 32 35.56 & $-$27 49 04.7 &   32 $\pm$    8 &  3.87 \\
253 & BLAST J033307$-$274001 (500\micron) &  & 03 33 07.81 & $-$27 39 36.2 &   32 $\pm$    8 &  3.86 \\
261 & BLAST J033148$-$280315 (500\micron) &  & 03 31 49.06 & $-$28 03 13.2 &   31 $\pm$    8 &  3.84 \\
267 & BLAST J033243$-$273914 (500\micron) & 78 & 03 32 43.67 & $-$27 39 38.1 &   31 $\pm$    8 &  3.83 \\
    \hline
  \end{tabular}
\label{tab:500cat}
\end{table*}

\begin{table*}
  \caption{870\,\micron\ peaks from \citet{weiss2009} that land within
    the coverage of the matching catalogue. The columns have the same meaning
    as in Table~\ref{tab:250cat}, although the \snr\ is measured with respect
    to the combined instrumental and confusion noise.
    On average, each of these peaks is a blend of 2.2 sources in our matching
    catalogue.
    In addition, the 33
    peaks marked with an asterisk in the `Match ID' column have no matches
    in any of the BLAST peak catalogues
    (Tables~\ref{tab:250cat}-\ref{tab:500cat}). These peaks are used
    to measure priors, but are not otherwise analyzed in this paper.}
  \centering
  \begin{tabular}{ccccccc}
    \hline
    SID & Previously Published & Match & R.A. & Dec. & Flux Density & \snr\ \\
    870 &  Name         & ID
    & ($^{\mathrm h}$\ \ \ $^{\mathrm m}$\ \ \ $^{\mathrm s}$)
    & ($^{\circ}$\ \ \ $^{\prime}$\ \ \ $^{\prime\prime}$)
    & (mJy) \\
    \hline
  1 & LESS J033314.3$-$275611 & 107 & 03 33 14.26 & $-$27 56 11.2 & 14.7 $\pm$  1.2 & 12.48 \\
  2 & LESS J033302.5$-$275643 & 95 & 03 33 02.50 & $-$27 56 43.6 & 12.2 $\pm$  1.2 & 10.31 \\
  3 & LESS J033321.5$-$275520 & 117 & 03 33 21.51 & $-$27 55 20.2 & 11.9 $\pm$  1.2 & 10.12 \\
  4 & LESS J033136.0$-$275439 & 13,14,15 & 03 31 36.01 & $-$27 54 39.2 & 11.2 $\pm$  1.1 &  9.72 \\
  5 & LESS J033129.5$-$275907 & 9 & 03 31 29.46 & $-$27 59 07.3 & 10.1 $\pm$  1.2 &  8.45 \\
  6 & LESS J033257.1$-$280102 & 91 & 03 32 57.14 & $-$28 01 02.1 &  9.8 $\pm$  1.2 &  8.16 \\
  7 & LESS J033315.6$-$274523 & 108 & 03 33 15.55 & $-$27 45 23.6 &  9.4 $\pm$  1.2 &  7.89 \\
  9 & LESS J033211.3$-$275210 & 46 & 03 32 11.29 & $-$27 52 10.4 &  9.4 $\pm$  1.2 &  7.71 \\
 10 & LESS J033219.0$-$275219 & 56 & 03 32 19.02 & $-$27 52 19.4 &  9.3 $\pm$  1.2 &  7.60 \\
 11 & LESS J033213.6$-$275602 & 50 & 03 32 13.58 & $-$27 56 02.5 &  9.2 $\pm$  1.2 &  7.58 \\
 12 & LESS J033248.1$-$275414 & 82 & 03 32 48.12 & $-$27 54 14.7 &  8.9 $\pm$  1.2 &  7.24 \\
 13 & LESS J033249.2$-$274246 & 84,85 & 03 32 49.23 & $-$27 42 46.6 &  8.9 $\pm$  1.2 &  7.21 \\
 14 & LESS J033152.6$-$280320 & 33 & 03 31 52.64 & $-$28 03 20.4 &  9.5 $\pm$  1.3 &  7.19 \\
 16 & LESS J033218.9$-$273738 & * & 03 32 18.89 & $-$27 37 38.7 &  8.2 $\pm$  1.2 &  6.87 \\
 17 & LESS J033207.6$-$275123 & * & 03 32 07.59 & $-$27 51 23.0 &  7.8 $\pm$  1.2 &  6.36 \\
 18 & LESS J033205.1$-$274652 & 43 & 03 32 05.12 & $-$27 46 52.1 &  7.7 $\pm$  1.2 &  6.30 \\
 19 & LESS J033208.1$-$275818 & 44 & 03 32 08.10 & $-$27 58 18.7 &  7.5 $\pm$  1.2 &  6.19 \\
 20 & LESS J033316.6$-$280018 & * & 03 33 16.56 & $-$28 00 18.8 &  7.5 $\pm$  1.2 &  6.15 \\
 25 & LESS J033157.1$-$275940 & 40 & 03 31 57.05 & $-$27 59 40.8 &  7.0 $\pm$  1.2 &  5.83 \\
 26 & LESS J033136.9$-$275456 & 15,16 & 03 31 36.90 & $-$27 54 56.1 &  6.8 $\pm$  1.2 &  5.80 \\
 28 & LESS J033302.9$-$274432 & * & 03 33 02.92 & $-$27 44 32.6 &  7.0 $\pm$  1.2 &  5.62 \\
 31 & LESS J033150.0$-$275743 & * & 03 31 49.96 & $-$27 57 43.9 &  6.7 $\pm$  1.2 &  5.51 \\
 32 & LESS J033243.6$-$274644 & * & 03 32 43.57 & $-$27 46 44.0 &  6.8 $\pm$  1.2 &  5.46 \\
 33 & LESS J033149.8$-$275332 & * & 03 31 49.78 & $-$27 53 32.9 &  6.8 $\pm$  1.2 &  5.46 \\
 34 & LESS J033217.6$-$275230 & 55 & 03 32 17.64 & $-$27 52 30.3 &  6.8 $\pm$  1.2 &  5.40 \\
 36 & LESS J033149.2$-$280208 & 29 & 03 31 49.15 & $-$28 02 08.7 &  6.9 $\pm$  1.3 &  5.36 \\
 38 & LESS J033310.2$-$275641 & * & 03 33 10.20 & $-$27 56 41.5 &  6.4 $\pm$  1.2 &  5.22 \\
 40 & LESS J033246.7$-$275120 & * & 03 32 46.74 & $-$27 51 20.9 &  6.4 $\pm$  1.2 &  5.18 \\
 42 & LESS J033231.0$-$275858 & 65 & 03 32 31.02 & $-$27 58 58.1 &  6.4 $\pm$  1.2 &  5.13 \\
 43 & LESS J033307.0$-$274801 & 99,100 & 03 33 07.00 & $-$27 48 01.0 &  6.4 $\pm$  1.3 &  5.12 \\
 45 & LESS J033225.7$-$275228 & * & 03 32 25.71 & $-$27 52 28.5 &  6.3 $\pm$  1.2 &  5.10 \\
 49 & LESS J033124.5$-$275040 & * & 03 31 24.45 & $-$27 50 40.9 &  6.6 $\pm$  1.3 &  5.05 \\
 50 & LESS J033141.2$-$274441 & 19,20 & 03 31 41.15 & $-$27 44 41.5 &  6.1 $\pm$  1.2 &  5.02 \\
 51 & LESS J033144.8$-$274425 & * & 03 31 44.81 & $-$27 44 25.1 &  6.2 $\pm$  1.2 &  5.01 \\
 52 & LESS J033128.5$-$275601 & 7 & 03 31 28.51 & $-$27 56 01.3 &  6.2 $\pm$  1.2 &  4.94 \\
 53 & LESS J033159.1$-$275435 & * & 03 31 59.12 & $-$27 54 35.5 &  6.2 $\pm$  1.2 &  4.93 \\
 55 & LESS J033302.2$-$274033 & * & 03 33 02.20 & $-$27 40 33.6 &  6.1 $\pm$  1.2 &  4.90 \\
 56 & LESS J033153.2$-$273936 & 35 & 03 31 53.17 & $-$27 39 36.1 &  6.0 $\pm$  1.2 &  4.89 \\
 57 & LESS J033152.0$-$275329 & * & 03 31 51.97 & $-$27 53 29.7 &  6.1 $\pm$  1.3 &  4.87 \\
 59 & LESS J033303.9$-$274412 & * & 03 33 03.87 & $-$27 44 12.2 &  6.0 $\pm$  1.2 &  4.77 \\
 60 & LESS J033317.5$-$275121 & * & 03 33 17.47 & $-$27 51 21.5 &  5.8 $\pm$  1.2 &  4.75 \\
 61 & LESS J033245.6$-$280025 & * & 03 32 45.63 & $-$28 00 25.3 &  5.9 $\pm$  1.2 &  4.73 \\
 63 & LESS J033308.5$-$280044 & 98 & 03 33 08.46 & $-$28 00 44.3 &  6.0 $\pm$  1.3 &  4.71 \\
 64 & LESS J033201.0$-$280025 & * & 03 32 01.00 & $-$28 00 25.6 &  5.8 $\pm$  1.2 &  4.70 \\
 65 & LESS J033252.4$-$273527 & * & 03 32 52.40 & $-$27 35 27.7 &  5.9 $\pm$  1.3 &  4.67 \\
 67 & LESS J033243.3$-$275517 & 76,77 & 03 32 43.28 & $-$27 55 17.9 &  5.9 $\pm$  1.3 &  4.67 \\
 68 & LESS J033233.4$-$273918 & * & 03 32 33.44 & $-$27 39 18.5 &  5.8 $\pm$  1.2 &  4.65 \\
 69 & LESS J033134.3$-$275934 & * & 03 31 34.26 & $-$27 59 34.3 &  5.7 $\pm$  1.2 &  4.65 \\
 70 & LESS J033144.0$-$273832 & * & 03 31 43.97 & $-$27 38 32.5 &  5.7 $\pm$  1.2 &  4.64 \\
 72 & LESS J033240.4$-$273802 & 73 & 03 32 40.40 & $-$27 38 02.5 &  5.7 $\pm$  1.2 &  4.63 \\
    \hline
  \end{tabular}
\label{tab:870cat}
\end{table*}

\begin{table*}
  \centering
  \begin{tabular}{ccccccc}
    \hline
    SID & Previously Published & Match & R.A. & Dec. & Flux Density & \snr\ \\
    870 &  Name         & ID
    & ($^{\mathrm h}$\ \ \ $^{\mathrm m}$\ \ \ $^{\mathrm s}$)
    & ($^{\circ}$\ \ \ $^{\prime}$\ \ \ $^{\prime\prime}$)
    & (mJy) \\
    \hline
 73 & LESS J033229.3$-$275619 & * & 03 32 29.33 & $-$27 56 19.3 &  5.8 $\pm$  1.2 &  4.63 \\
 74 & LESS J033309.3$-$274809 & 101,102,103 & 03 33 09.34 & $-$27 48 09.9 &  5.8 $\pm$  1.3 &  4.62 \\
 75 & LESS J033126.8$-$275554 & 6 & 03 31 26.83 & $-$27 55 54.6 &  5.8 $\pm$  1.3 &  4.61 \\
 77 & LESS J033157.2$-$275633 & * & 03 31 57.23 & $-$27 56 33.2 &  5.5 $\pm$  1.2 &  4.42 \\
 79 & LESS J033221.3$-$275623 & 57 & 03 32 21.25 & $-$27 56 23.5 &  5.5 $\pm$  1.2 &  4.40 \\
 80 & LESS J033142.2$-$274834 & 21,22 & 03 31 42.23 & $-$27 48 34.4 &  5.4 $\pm$  1.2 &  4.38 \\
 81 & LESS J033127.5$-$274440 & 1 & 03 31 27.45 & $-$27 44 40.4 &  5.7 $\pm$  1.3 &  4.38 \\
 82 & LESS J033253.8$-$273810 & 90 & 03 32 53.77 & $-$27 38 10.9 &  5.3 $\pm$  1.2 &  4.35 \\
 84 & LESS J033154.2$-$275109 & * & 03 31 54.22 & $-$27 51 09.8 &  5.5 $\pm$  1.3 &  4.33 \\
 88 & LESS J033155.2$-$275345 & 38,39 & 03 31 55.19 & $-$27 53 45.3 &  5.4 $\pm$  1.3 &  4.28 \\
 89 & LESS J033248.4$-$280023 & * & 03 32 48.44 & $-$28 00 23.8 &  5.3 $\pm$  1.2 &  4.25 \\
 90 & LESS J033243.7$-$273554 & * & 03 32 43.65 & $-$27 35 54.1 &  5.4 $\pm$  1.3 &  4.23 \\
 91 & LESS J033135.3$-$274033 & * & 03 31 35.25 & $-$27 40 33.7 &  5.3 $\pm$  1.3 &  4.22 \\
 92 & LESS J033138.4$-$274336 & * & 03 31 38.36 & $-$27 43 36.0 &  5.2 $\pm$  1.2 &  4.22 \\
 94 & LESS J033307.3$-$275805 & * & 03 33 07.27 & $-$27 58 05.0 &  5.3 $\pm$  1.2 &  4.20 \\
 95 & LESS J033241.7$-$275846 & * & 03 32 41.74 & $-$27 58 46.1 &  5.2 $\pm$  1.2 &  4.18 \\
 96 & LESS J033313.0$-$275556 & 106 & 03 33 13.03 & $-$27 55 56.8 &  5.2 $\pm$  1.2 &  4.18 \\
 97 & LESS J033313.7$-$273803 & * & 03 33 13.65 & $-$27 38 03.4 &  5.1 $\pm$  1.2 &  4.16 \\
 98 & LESS J033130.2$-$275726 & 10,11 & 03 31 30.22 & $-$27 57 26.0 &  5.1 $\pm$  1.2 &  4.11 \\
 99 & LESS J033251.5$-$275536 & * & 03 32 51.45 & $-$27 55 36.0 &  5.3 $\pm$  1.3 &  4.11 \\
101 & LESS J033151.5$-$274552 & * & 03 31 51.47 & $-$27 45 52.1 &  5.1 $\pm$  1.3 &  4.08 \\
104 & LESS J033258.5$-$273803 & * & 03 32 58.46 & $-$27 38 03.0 &  4.9 $\pm$  1.2 &  4.05 \\
106 & LESS J033140.1$-$275631 & 18 & 03 31 40.09 & $-$27 56 31.4 &  4.9 $\pm$  1.2 &  4.03 \\
107 & LESS J033130.9$-$275150 & * & 03 31 30.85 & $-$27 51 50.9 &  5.0 $\pm$  1.2 &  4.02 \\
108 & LESS J033316.4$-$275033 & 109 & 03 33 16.42 & $-$27 50 33.1 &  5.0 $\pm$  1.2 &  4.02 \\
113 & LESS J033236.4$-$275845 & * & 03 32 36.42 & $-$27 58 45.9 &  5.0 $\pm$  1.3 &  3.94 \\
114 & LESS J033150.8$-$274438 & 31,32 & 03 31 50.81 & $-$27 44 38.5 &  4.9 $\pm$  1.3 &  3.90 \\
116 & LESS J033154.4$-$274525 & * & 03 31 54.42 & $-$27 45 25.5 &  4.9 $\pm$  1.3 &  3.84 \\
117 & LESS J033128.0$-$273925 & 2,3,4 & 03 31 28.02 & $-$27 39 25.2 &  5.0 $\pm$  1.3 &  3.83 \\
120 & LESS J033328.5$-$275655 & 118 & 03 33 28.45 & $-$27 56 55.9 &  4.9 $\pm$  1.3 &  3.79 \\
122 & LESS J033139.6$-$274120 & 17 & 03 31 39.62 & $-$27 41 20.4 &  4.7 $\pm$  1.2 &  3.77 \\
    \hline
  \end{tabular}
\end{table*}


\begin{table*}
  \setlength{\tabcolsep}{0.09cm}
  \caption{Matches between the external catalogue with each of the 4
    submm peak lists.  The first column gives a unique short ID that
    is used throughout this paper. Since all radio sources and the FIDEL
    catalogue use the SIMPLE survey \citep[{\em Spitzer} IRAC/MUSYC
    Public Legacy in ECDF-S;][]{gawiser2006} as a positional prior,
    the SIMPLE IDs for each source are given in column two. The only
    exceptions are 870\,\micron\ peaks with no identification in the
    matching catalogue; these peaks have simply been added to the
    catalogue with the SIMPLE ID replaced by the 870\,\micron\ SID
    referring to Table~\ref{tab:870cat} (in boldface brackets).  The
    third and fourth columns give the IRAC J2000 coordinates of each
    source (or the 870\,\micron\ position when unavailable). There are
    then 4 blocks of 5 columns for each submm band giving: the submm
    SID for each respective submm peak list to which the source was
    matched (Tables~\ref{tab:250cat}-\ref{tab:870cat}); the radial
    separation in arcsec; the likelihood ratio; and finally the
    24\,\micron\ and 1.4\,GHz $P$ values when available.}
  \tiny
  \centering
  \begin{tabular}{cccc|ccccc|ccccc|ccccc|ccccc}
    \hline
    ID & SIMPLE & R.A. & Dec.
    & \multicolumn{5}{c}{250\,\micron}
    & \multicolumn{5}{c}{350\,\micron}
    & \multicolumn{5}{c}{500\,\micron}
    & \multicolumn{5}{c}{870\,\micron} \\
    & & &
    & SID & $\Delta r$ & LR & $P_{24}$ & $P_\mathrm{r}$
    & SID & $\Delta r$ & LR & $P_{24}$ & $P_\mathrm{r}$
    & SID & $\Delta r$ & LR & $P_{24}$ & $P_\mathrm{r}$
    & SID & $\Delta r$ & LR & $P_{24}$ & $P_\mathrm{r}$ \\

    & & ($^{\mathrm h}$\ \ \ $^{\mathrm m}$\ \ \ $^{\mathrm s}$) &
    ($^{\circ}$\ \ \ $^{\prime}$\ \ \ $^{\prime\prime}$)
    & 250 & ($^{\prime\prime}$) & & &
    & 350 & ($^{\prime\prime}$) & & &
    & 500 & ($^{\prime\prime}$) & & &
    & 870 & ($^{\prime\prime}$) & & & \\
  \hline
  1 &     37075 & 03 31 27.53 & $-$27 44 39.3 &  78 & 11.2 &  14.4 & 0.098 & 0.058 & 235 &  8.2 &  11.2 & 0.061 & 0.035 &\ldots&\ldots&\ldots& \ldots&\ldots &  81 &  1.5 &  43.1 & 0.003 & 0.002 \\
  2 &     45028 & 03 31 27.58 & $-$27 39 27.6 &\ldots&\ldots&\ldots& \ldots&\ldots &  22 &  7.4 &   1.7 & 0.215 & 0.066 &\ldots&\ldots&\ldots& \ldots&\ldots & 117 &  6.3 &   2.2 & 0.168 & 0.048 \\
  3 &     45037 & 03 31 27.95 & $-$27 39 36.4 &\ldots&\ldots&\ldots& \ldots&\ldots &  22 &  4.4 &   3.7 & 0.076 & 0.038 &  17 &  6.6 &   1.6 & 0.141 & 0.075 & 117 & 11.3 &   1.2 & \ldots & \ldots \\
  4 &     45195 & 03 31 28.81 & $-$27 39 17.0 &\ldots&\ldots&\ldots& \ldots&\ldots &\ldots&\ldots&\ldots& \ldots&\ldots &\ldots&\ldots&\ldots& \ldots&\ldots & 117 & 13.3 &   1.0 & \ldots & \ldots \\
  5 &     45836 & 03 31 28.86 & $-$27 39 04.1 &  80 &  2.9 &   1.3 & 0.013 & 0.051 &\ldots&\ldots&\ldots& \ldots&\ldots &\ldots&\ldots&\ldots& \ldots&\ldots &\ldots&\ldots&\ldots& \ldots&\ldots \\
  6 &     18957 & 03 31 27.18 & $-$27 55 51.0 &\ldots&\ldots&\ldots& \ldots&\ldots &\ldots&\ldots&\ldots& \ldots&\ldots &   9 & 27.4 &   2.5 & \ldots & \ldots &\ldots&\ldots&\ldots& \ldots&\ldots \\
  7 &     18523 & 03 31 28.87 & $-$27 56 07.8 &\ldots&\ldots&\ldots& \ldots&\ldots &\ldots&\ldots&\ldots& \ldots&\ldots &\ldots&\ldots&\ldots& \ldots&\ldots &  52 &  8.1 &   1.7 & 0.206 & 0.492 \\
  8 &     17966 & 03 31 30.06 & $-$27 56 02.2 & 111 &  0.8 &  16.9 & 0.001 & 0.001 &\ldots&\ldots&\ldots& \ldots&\ldots &\ldots&\ldots&\ldots& \ldots&\ldots &\ldots&\ldots&\ldots& \ldots&\ldots \\
  9 & {\bf (5)} & 03 31 29.46 & $-$27 59 07.3 &  99 &  5.2 & \ldots & \ldots & \ldots & 177 &  2.8 & \ldots & \ldots & \ldots &  64 &  7.4 & \ldots & \ldots & \ldots &\ldots&\ldots&\ldots& \ldots&\ldots \\
 10 &     16359 & 03 31 29.89 & $-$27 57 22.4 &  16 &  1.8 &   8.0 & 0.014 & 0.003 &  16 &  3.8 &  13.7 & 0.052 & 0.012 &  11 &  5.1 &   9.9 & 0.082 & 0.020 &  98 &  5.6 &  16.9 & 0.092 & 0.022 \\
 11 &     15753 & 03 31 30.74 & $-$27 57 33.8 &  16 & 14.4 &   1.2 & 0.615 & 0.056 &\ldots&\ldots&\ldots& \ldots&\ldots &\ldots&\ldots&\ldots& \ldots&\ldots &\ldots&\ldots&\ldots& \ldots&\ldots \\
 12 &     44643 & 03 31 36.10 & $-$27 39 40.6 &  70 &  5.8 &   3.0 & 0.031 & 0.029 &\ldots&\ldots&\ldots& \ldots&\ldots &\ldots&\ldots&\ldots& \ldots&\ldots &\ldots&\ldots&\ldots& \ldots&\ldots \\
 13 &     21125 & 03 31 35.50 & $-$27 54 35.4 &\ldots&\ldots&\ldots& \ldots&\ldots &  49 & 12.1 &   5.0 & 0.336 & 0.146 &  48 & 12.9 &   3.3 & 0.359 & 0.163 &   4 &  7.8 &  10.2 & 0.198 & 0.073 \\
 14 &     20735 & 03 31 35.91 & $-$27 54 43.3 &\ldots&\ldots&\ldots& \ldots&\ldots &  49 &  7.5 &   1.1 & 0.201 & 0.549 &\ldots&\ldots&\ldots& \ldots&\ldots &   4 &  4.3 &   4.6 & 0.090 & 0.355 \\
 15 &     20560 & 03 31 36.15 & $-$27 54 49.6 &\ldots&\ldots&\ldots& \ldots&\ldots &  49 &  9.7 &   2.3 & 0.197 & 0.162 &  48 &  7.0 &   4.2 & 0.128 & 0.105 &  26 & 11.8 &   2.7 & \ldots & \ldots \\
 16 &     19870 & 03 31 36.94 & $-$27 55 10.6 &\ldots&\ldots&\ldots& \ldots&\ldots &\ldots&\ldots&\ldots& \ldots&\ldots &\ldots&\ldots&\ldots& \ldots&\ldots &  26 & 14.5 &   1.3 & \ldots & \ldots \\
 17 &     42263 & 03 31 39.53 & $-$27 41 19.4 & 244 & 10.1 &  41.3 & 0.023 & 0.050 &  42 &  7.1 &  70.1 & 0.013 & 0.028 &\ldots&\ldots&\ldots& \ldots&\ldots & 122 &  1.5 &  79.8 & 0.001 & 0.002 \\
 18 &     18083 & 03 31 40.17 & $-$27 56 22.4 &  92 & 14.1 &   1.7 & 0.221 & 0.198 &\ldots&\ldots&\ldots& \ldots&\ldots &\ldots&\ldots&\ldots& \ldots&\ldots & 106 &  9.1 &   8.1 & 0.120 & 0.102 \\
 19 &     37304 & 03 31 40.97 & $-$27 44 34.9 &  85 &  3.9 &   5.2 & 0.038 & 0.031 &  37 &  4.0 &   9.7 & 0.040 & 0.033 &\ldots&\ldots&\ldots& \ldots&\ldots &  50 &  7.0 &  15.2 & 0.094 & 0.076 \\
 20 &     36814 & 03 31 41.37 & $-$27 44 46.7 &\ldots&\ldots&\ldots& \ldots&\ldots &\ldots&\ldots&\ldots& \ldots&\ldots &\ldots&\ldots&\ldots& \ldots&\ldots &  50 &  6.0 &   1.6 & 0.129 & 0.078 \\
 21 &     31008 & 03 31 41.63 & $-$27 48 30.3 &\ldots&\ldots&\ldots& \ldots&\ldots &\ldots&\ldots&\ldots& \ldots&\ldots &\ldots&\ldots&\ldots& \ldots&\ldots &  80 &  9.0 &   1.1 & 0.235 & 0.501 \\
 22 &     30829 & 03 31 42.80 & $-$27 48 36.7 &\ldots&\ldots&\ldots& \ldots&\ldots &\ldots&\ldots&\ldots& \ldots&\ldots &\ldots&\ldots&\ldots& \ldots&\ldots &  80 &  7.9 &   1.1 & 0.399 & 0.165 \\
 23 &     19031 & 03 31 41.77 & $-$27 55 37.8 &\ldots&\ldots&\ldots& \ldots&\ldots & 187 &  9.3 &   1.4 & 0.280 & 0.105 &\ldots&\ldots&\ldots& \ldots&\ldots &\ldots&\ldots&\ldots& \ldots&\ldots \\
 24 &     18659 & 03 31 43.58 & $-$27 55 28.1 & 255 & 17.9 &   1.1 & \ldots & \ldots &\ldots&\ldots&\ldots& \ldots&\ldots &\ldots&\ldots&\ldots& \ldots&\ldots &\ldots&\ldots&\ldots& \ldots&\ldots \\
 25 &     40970 & 03 31 44.46 & $-$27 42 12.0 &  57 &  9.0 &   1.7 & 0.098 & 0.061 &\ldots&\ldots&\ldots& \ldots&\ldots &\ldots&\ldots&\ldots& \ldots&\ldots &\ldots&\ldots&\ldots& \ldots&\ldots \\
 26 & {\bf (125)} & 03 31 46.02 & $-$27 46 21.2 &  32 & 17.6 & \ldots & \ldots & \ldots &\ldots&\ldots&\ldots& \ldots&\ldots &\ldots&\ldots&\ldots& \ldots&\ldots & 125 &  0.0 & \ldots & \ldots & \ldots \\
 27 &     16076 & 03 31 46.56 & $-$27 57 34.8 &  36 & 10.9 &  21.2 & 0.033 & 0.035 &\ldots&\ldots&\ldots& \ldots&\ldots &\ldots&\ldots&\ldots& \ldots&\ldots &\ldots&\ldots&\ldots& \ldots&\ldots \\
 28 &     31651 & 03 31 48.03 & $-$27 48 01.7 &\ldots&\ldots&\ldots& \ldots&\ldots & 188 & 10.5 &   6.2 & 0.113 & 0.117 &\ldots&\ldots&\ldots& \ldots&\ldots &\ldots&\ldots&\ldots& \ldots&\ldots \\
 29 &     10254 & 03 31 48.95 & $-$28 02 13.6 &\ldots&\ldots&\ldots& \ldots&\ldots & 322 & 10.2 &   2.7 & 0.211 & 0.183 & 177 & 13.0 &   3.7 & 0.283 & 0.252 &  36 &  5.6 &  11.1 & 0.090 & 0.074 \\
 30 &     38757 & 03 31 49.73 & $-$27 43 26.1 &  59 &  8.6 &  10.5 & 0.030 & 0.042 &  35 &  2.2 &   4.4 & 0.003 & 0.004 &\ldots&\ldots&\ldots& \ldots&\ldots &\ldots&\ldots&\ldots& \ldots&\ldots \\
 31 &     37046 & 03 31 50.30 & $-$27 44 46.8 &\ldots&\ldots&\ldots& \ldots&\ldots &\ldots&\ldots&\ldots& \ldots&\ldots &\ldots&\ldots&\ldots& \ldots&\ldots & 114 & 10.7 &   1.5 & \ldots & \ldots \\
 32 &     36827 & 03 31 51.09 & $-$27 44 37.0 & 146 &  1.3 &   8.1 & 0.003 & 0.003 &  47 &  4.0 &   9.7 & 0.019 & 0.023 &\ldots&\ldots&\ldots& \ldots&\ldots & 114 &  3.9 &  30.6 & 0.019 & 0.021 \\
 33 &      8829 & 03 31 52.48 & $-$28 03 18.6 &\ldots&\ldots&\ldots& \ldots&\ldots &  95 & 19.2 &   1.0 & \ldots & \ldots &\ldots&\ldots&\ldots& \ldots&\ldots &  14 &  2.8 &  17.4 & 0.112 & 0.012 \\
 34 &     45260 & 03 31 52.07 & $-$27 39 26.6 &  51 &  9.3 &   5.9 & 0.109 & 0.013 & 124 &  8.1 &  16.4 & 0.087 & 0.010 &  14 &  4.8 &  14.1 & 0.038 & 0.004 &\ldots&\ldots&\ldots& \ldots&\ldots \\
 35 &     45144 & 03 31 53.11 & $-$27 39 37.3 &  51 &  8.3 &   1.1 & 0.166 & 0.169 & 124 &  9.8 &   2.3 & 0.208 & 0.211 &  14 & 21.0 &   1.9 & 0.451 & 0.458 &  56 &  1.4 &  11.5 & 0.010 & 0.010 \\
 36 &     28865 & 03 31 54.67 & $-$27 49 44.4 &\ldots&\ldots&\ldots& \ldots&\ldots &  93 &  7.0 &   1.5 & 0.193 & 0.144 & 205 &  2.5 &   1.9 & 0.040 & 0.029 &\ldots&\ldots&\ldots& \ldots&\ldots \\
 37 &     37463 & 03 31 54.98 & $-$27 44 10.5 & 188 & 11.2 &   1.2 & 0.431 & 0.052 &\ldots&\ldots&\ldots& \ldots&\ldots &\ldots&\ldots&\ldots& \ldots&\ldots &\ldots&\ldots&\ldots& \ldots&\ldots \\
 38 &     22221 & 03 31 54.82 & $-$27 53 41.0 &\ldots&\ldots&\ldots& \ldots&\ldots &\ldots&\ldots&\ldots& \ldots&\ldots &\ldots&\ldots&\ldots& \ldots&\ldots &  88 &  6.6 &   5.7 & 0.109 & 0.042 \\
 39 &     22204 & 03 31 55.76 & $-$27 53 47.5 &\ldots&\ldots&\ldots& \ldots&\ldots &\ldots&\ldots&\ldots& \ldots&\ldots &\ldots&\ldots&\ldots& \ldots&\ldots &  88 &  7.9 &   2.5 & 0.207 & 0.224 \\
 40 &     13515 & 03 31 56.86 & $-$27 59 39.0 &\ldots&\ldots&\ldots& \ldots&\ldots &\ldots&\ldots&\ldots& \ldots&\ldots & 157 & 12.7 &   3.4 & 0.317 & 0.196 &  25 &  3.1 &  17.3 & 0.044 & 0.022 \\
 41 &     41664 & 03 32 02.14 & $-$27 41 28.5 &\ldots&\ldots&\ldots& \ldots&\ldots & 170 &  9.6 &   2.6 & 0.220 & 0.132 &\ldots&\ldots&\ldots& \ldots&\ldots &\ldots&\ldots&\ldots& \ldots&\ldots \\
 42 &     11649 & 03 32 04.66 & $-$28 00 57.6 & 328 & 12.1 &   2.0 & 0.375 & 0.071 &\ldots&\ldots&\ldots& \ldots&\ldots &\ldots&\ldots&\ldots& \ldots&\ldots &\ldots&\ldots&\ldots& \ldots&\ldots \\
 43 &     33160 & 03 32 04.87 & $-$27 46 47.3 &  75 &  3.9 &   4.7 & 0.016 & 0.015 &  19 &  5.8 &   3.2 & 0.032 & 0.028 &\ldots&\ldots&\ldots& \ldots&\ldots &  18 &  5.8 &  11.8 & 0.032 & 0.028 \\
 44 &     15657 & 03 32 08.24 & $-$27 58 13.9 &\ldots&\ldots&\ldots& \ldots&\ldots &\ldots&\ldots&\ldots& \ldots&\ldots &   8 &  9.0 &   1.2 & 0.447 & 0.181 &  19 &  5.1 &   3.9 & 0.251 & 0.075 \\
 45 &      9589 & 03 32 10.99 & $-$28 02 36.0 & 246 &  6.8 &   1.2 & 0.162 & 0.134 &\ldots&\ldots&\ldots& \ldots&\ldots &\ldots&\ldots&\ldots& \ldots&\ldots &\ldots&\ldots&\ldots& \ldots&\ldots \\
 46 &     24758 & 03 32 11.35 & $-$27 52 13.0 &\ldots&\ldots&\ldots& \ldots&\ldots &\ldots&\ldots&\ldots& \ldots&\ldots &\ldots&\ldots&\ldots& \ldots&\ldots &   9 &  2.7 &   3.5 & 0.091 & 0.041 \\
 47 &     40039 & 03 32 10.49 & $-$27 43 08.8 & 358 & 13.5 &   2.0 & 0.146 & 0.313 &\ldots&\ldots&\ldots& \ldots&\ldots &\ldots&\ldots&\ldots& \ldots&\ldots &\ldots&\ldots&\ldots& \ldots&\ldots \\
 48 &     39690 & 03 32 12.54 & $-$27 43 06.0 &\ldots&\ldots&\ldots& \ldots&\ldots &\ldots&\ldots&\ldots& \ldots&\ldots &  91 & 11.4 &   3.6 & 0.280 & 0.248 &\ldots&\ldots&\ldots& \ldots&\ldots \\
 49 &     47577 & 03 32 11.63 & $-$27 37 25.9 &\ldots&\ldots&\ldots& \ldots&\ldots &  88 &  4.3 &  64.1 & 0.005 & 0.001 &  58 & 18.2 &  20.1 & 0.055 & 0.012 &\ldots&\ldots&\ldots& \ldots&\ldots \\
 50 &     18807 & 03 32 13.85 & $-$27 55 59.9 &\ldots&\ldots&\ldots& \ldots&\ldots &\ldots&\ldots&\ldots& \ldots&\ldots &  33 & 10.4 &   1.8 & 0.455 & 0.183 &  11 &  4.4 &   7.4 & 0.181 & 0.048 \\
 51 &     44585 & 03 32 16.19 & $-$27 39 30.3 &\ldots&\ldots&\ldots& \ldots&\ldots & 108 & 12.6 &  35.7 & 0.032 & 0.190 &\ldots&\ldots&\ldots& \ldots&\ldots &\ldots&\ldots&\ldots& \ldots&\ldots \\
 52 &     14672 & 03 32 17.04 & $-$27 58 46.5 &\ldots&\ldots&\ldots& \ldots&\ldots &\ldots&\ldots&\ldots& \ldots&\ldots & 203 &  7.0 &   1.8 & 0.398 & 0.004 &\ldots&\ldots&\ldots& \ldots&\ldots \\
 53 &     13333 & 03 32 17.05 & $-$27 59 16.6 &  38 &  9.7 &  30.3 & 0.019 & 0.054 &  80 & 10.2 &  24.3 & 0.021 & 0.059 &\ldots&\ldots&\ldots& \ldots&\ldots &\ldots&\ldots&\ldots& \ldots&\ldots \\
 54 &     25701 & 03 32 17.87 & $-$27 50 59.0 & 289 &  6.7 &   5.1 & 0.017 & 0.060 &\ldots&\ldots&\ldots& \ldots&\ldots &\ldots&\ldots&\ldots& \ldots&\ldots &\ldots&\ldots&\ldots& \ldots&\ldots \\
 55 &     24159 & 03 32 17.18 & $-$27 52 20.5 &  58 & 11.2 &   1.3 & 0.259 & 0.113 &  65 & 14.8 &   1.5 & 0.351 & 0.168 &\ldots&\ldots&\ldots& \ldots&\ldots &\ldots&\ldots&\ldots& \ldots&\ldots \\
 56 &     24739 & 03 32 19.05 & $-$27 52 14.5 &\ldots&\ldots&\ldots& \ldots&\ldots &  65 & 12.6 &   2.6 & 0.450 & 0.139 &  21 & 10.3 &   1.6 & 0.384 & 0.106 &  10 &  5.0 &  10.9 & 0.161 & 0.032 \\
 57 &     17994 & 03 32 21.60 & $-$27 56 23.3 &  45 &  3.5 &   2.9 & 0.014 & 0.048 &\ldots&\ldots&\ldots& \ldots&\ldots &  31 & 10.2 &   1.0 & 0.085 & 0.232 &  79 &  4.7 &   4.1 & 0.023 & 0.072 \\
 58 &     11555 & 03 32 22.59 & $-$28 00 23.4 &  97 &  6.3 &   8.4 & 0.006 & 0.011 &\ldots&\ldots&\ldots& \ldots&\ldots &\ldots&\ldots&\ldots& \ldots&\ldots &\ldots&\ldots&\ldots& \ldots&\ldots \\
 59 &     48158 & 03 32 23.67 & $-$27 36 48.2 &  91 & 12.4 &  64.5 & 0.006 & 0.056 &\ldots&\ldots&\ldots& \ldots&\ldots &\ldots&\ldots&\ldots& \ldots&\ldots &\ldots&\ldots&\ldots& \ldots&\ldots \\
 60 &     46695 & 03 32 25.04 & $-$27 38 22.4 & 325 &  1.0 &   1.6 & 0.003 & 0.001 & 169 &  5.3 &   1.1 & 0.044 & 0.024 &\ldots&\ldots&\ldots& \ldots&\ldots &\ldots&\ldots&\ldots& \ldots&\ldots \\
 61 &     35317 & 03 32 29.86 & $-$27 44 24.2 &   5 &  8.4 &  87.8 & 0.002 & 0.010 &   8 & 10.1 &   4.2 & 0.003 & 0.014 &  53 & 12.6 &   1.3 & 0.005 & 0.021 &\ldots&\ldots&\ldots& \ldots&\ldots \\
 62 &     37400 & 03 32 29.98 & $-$27 44 04.8 &   5 & 12.2 &  60.8 & 0.010 & 0.038 &   8 & 14.2 &  19.9 & 0.013 & 0.048 &\ldots&\ldots&\ldots& \ldots&\ldots &\ldots&\ldots&\ldots& \ldots&\ldots \\
 63 &     39003 & 03 32 29.93 & $-$27 43 01.0 &\ldots&\ldots&\ldots& \ldots&\ldots &  12 &  6.4 &   1.2 & 0.142 & 0.091 & 104 &  9.5 &   1.4 & 0.240 & 0.164 &\ldots&\ldots&\ldots& \ldots&\ldots \\
 64 &     13126 & 03 32 30.55 & $-$27 59 11.3 & 121 &  7.5 & 124.9 & 0.004 & 0.017 &  85 & 11.5 &  69.4 & 0.008 & 0.036 &\ldots&\ldots&\ldots& \ldots&\ldots &\ldots&\ldots&\ldots& \ldots&\ldots \\
 65 &     14440 & 03 32 31.45 & $-$27 58 51.5 &\ldots&\ldots&\ldots& \ldots&\ldots &  85 & 22.1 &   1.0 & \ldots & \ldots &\ldots&\ldots&\ldots& \ldots&\ldots &  42 &  8.7 &   8.6 & 0.176 & 0.103 \\
 66 &     17925 & 03 32 35.07 & $-$27 55 32.6 &   3 &  1.9 &  28.1 & 0.000 & 0.001 &   3 &  2.9 &  17.8 & 0.001 & 0.002 &\ldots&\ldots&\ldots& \ldots&\ldots &\ldots&\ldots&\ldots& \ldots&\ldots \\
 67 &     29372 & 03 32 35.71 & $-$27 49 15.9 & 113 & 13.8 &   3.4 & 0.131 & 0.186 &\ldots&\ldots&\ldots& \ldots&\ldots & 250 & 11.5 &   2.7 & 0.100 & 0.147 &\ldots&\ldots&\ldots& \ldots&\ldots \\
 68 &     10608 & 03 32 36.05 & $-$28 01 37.9 &\ldots&\ldots&\ldots& \ldots&\ldots & 259 & 10.9 &   1.4 & 0.320 & 0.132 &\ldots&\ldots&\ldots& \ldots&\ldots &\ldots&\ldots&\ldots& \ldots&\ldots \\
 69 &     10758 & 03 32 36.43 & $-$28 01 51.0 &\ldots&\ldots&\ldots& \ldots&\ldots & 259 &  7.4 &   2.0 & 0.294 & 0.092 &\ldots&\ldots&\ldots& \ldots&\ldots &\ldots&\ldots&\ldots& \ldots&\ldots \\
 70 &     50164 & 03 32 37.47 & $-$27 35 47.8 &  69 &  5.9 &   4.3 & 0.038 & 0.067 &  69 &  5.7 &   3.6 & 0.036 & 0.064 &\ldots&\ldots&\ldots& \ldots&\ldots &\ldots&\ldots&\ldots& \ldots&\ldots \\
 71 &     50104 & 03 32 37.89 & $-$27 35 49.9 &  69 &  6.5 &   4.7 & 0.046 & 0.050 &  69 &  8.2 &   3.4 & 0.068 & 0.074 &\ldots&\ldots&\ldots& \ldots&\ldots &\ldots&\ldots&\ldots& \ldots&\ldots \\
 72 &     16640 & 03 32 38.92 & $-$27 57 00.3 & 316 & 12.5 &  27.7 & 0.016 & 0.082 &\ldots&\ldots&\ldots& \ldots&\ldots & 133 & 15.5 &   6.9 & 0.023 & 0.114 &\ldots&\ldots&\ldots& \ldots&\ldots \\
 73 &     46989 & 03 32 40.05 & $-$27 38 08.6 &\ldots&\ldots&\ldots& \ldots&\ldots &\ldots&\ldots&\ldots& \ldots&\ldots &\ldots&\ldots&\ldots& \ldots&\ldots &  72 &  7.6 &   1.2 & 0.057 & 0.199 \\
 74 &     46485 & 03 32 42.58 & $-$27 38 25.9 &  60 & 12.3 &  20.6 & 0.030 & 0.072 &\ldots&\ldots&\ldots& \ldots&\ldots &\ldots&\ldots&\ldots& \ldots&\ldots &\ldots&\ldots&\ldots& \ldots&\ldots \\
 75 &     46947 & 03 32 42.87 & $-$27 38 17.1 &  60 & 13.2 &   1.1 & 0.516 & 0.021 &\ldots&\ldots&\ldots& \ldots&\ldots &\ldots&\ldots&\ldots& \ldots&\ldots &\ldots&\ldots&\ldots& \ldots&\ldots \\
 76 &     19784 & 03 32 43.19 & $-$27 55 14.4 &\ldots&\ldots&\ldots& \ldots&\ldots & 128 &  4.5 &   9.5 & 0.024 & 0.030 &\ldots&\ldots&\ldots& \ldots&\ldots &  67 &  3.7 &  31.1 & 0.017 & 0.022 \\
 77 &     19900 & 03 32 43.75 & $-$27 55 16.2 &\ldots&\ldots&\ldots& \ldots&\ldots &\ldots&\ldots&\ldots& \ldots&\ldots &\ldots&\ldots&\ldots& \ldots&\ldots &  67 &  6.4 &   2.5 & 0.277 & 0.154 \\
 78 &     44806 & 03 32 43.49 & $-$27 39 29.1 &  82 &  8.7 &   1.9 & 0.087 & 0.190 &\ldots&\ldots&\ldots& \ldots&\ldots &\ldots&\ldots&\ldots& \ldots&\ldots &\ldots&\ldots&\ldots& \ldots&\ldots \\
 79 &     25546 & 03 32 44.04 & $-$27 51 43.3 & 298 &  7.6 &  21.0 & 0.048 & 0.025 &\ldots&\ldots&\ldots& \ldots&\ldots &\ldots&\ldots&\ldots& \ldots&\ldots &\ldots&\ldots&\ldots& \ldots&\ldots \\
 80 &     25673 & 03 32 44.25 & $-$27 51 41.1 & 298 & 10.9 &  67.0 & 0.003 & 0.040 &\ldots&\ldots&\ldots& \ldots&\ldots &\ldots&\ldots&\ldots& \ldots&\ldots &\ldots&\ldots&\ldots& \ldots&\ldots \\
 81 &     15411 & 03 32 45.96 & $-$27 57 45.3 &  24 &  1.3 &  30.2 & 0.000 & 0.001 & 116 & 10.1 &   3.7 & 0.017 & 0.050 &\ldots&\ldots&\ldots& \ldots&\ldots &\ldots&\ldots&\ldots& \ldots&\ldots \\
 82 &     21056 & 03 32 47.96 & $-$27 54 16.3 &\ldots&\ldots&\ldots& \ldots&\ldots &\ldots&\ldots&\ldots& \ldots&\ldots &\ldots&\ldots&\ldots& \ldots&\ldots &  12 &  2.6 &   1.9 & 0.137 & 0.035 \\
 83 &     40193 & 03 32 47.88 & $-$27 42 32.9 &\ldots&\ldots&\ldots& \ldots&\ldots &\ldots&\ldots&\ldots& \ldots&\ldots &  37 &  7.6 &   1.7 & 0.023 & 0.028 &\ldots&\ldots&\ldots& \ldots&\ldots \\
 84 &     40147 & 03 32 48.93 & $-$27 42 51.3 &\ldots&\ldots&\ldots& \ldots&\ldots &\ldots&\ldots&\ldots& \ldots&\ldots &\ldots&\ldots&\ldots& \ldots&\ldots &  13 &  6.1 &   3.9 & 0.478 & 0.210 \\
 85 &     40166 & 03 32 49.42 & $-$27 42 35.3 &\ldots&\ldots&\ldots& \ldots&\ldots &\ldots&\ldots&\ldots& \ldots&\ldots &\ldots&\ldots&\ldots& \ldots&\ldots &  13 & 11.6 &   1.7 & \ldots & \ldots \\
 86 &     14647 & 03 32 49.33 & $-$27 58 45.0 &  17 &  3.5 &  12.8 & 0.033 & 0.008 &   7 & 11.4 &   6.8 & 0.195 & 0.055 & 144 & 14.6 &   7.6 & 0.263 & 0.082 &\ldots&\ldots&\ldots& \ldots&\ldots \\
 87 &     48957 & 03 32 49.40 & $-$27 36 36.1 & 369 & 17.0 &   1.1 & \ldots & \ldots &\ldots&\ldots&\ldots& \ldots&\ldots &\ldots&\ldots&\ldots& \ldots&\ldots &\ldots&\ldots&\ldots& \ldots&\ldots \\
 88 &     12980 & 03 32 51.79 & $-$27 59 56.2 & 389 & 15.2 &   1.2 & 0.145 & 0.166 &\ldots&\ldots&\ldots& \ldots&\ldots &\ldots&\ldots&\ldots& \ldots&\ldots &\ldots&\ldots&\ldots& \ldots&\ldots \\
 89 &     11316 & 03 32 53.05 & $-$28 01 17.8 &\ldots&\ldots&\ldots& \ldots&\ldots &  90 &  6.5 &   2.9 & 0.102 & 0.106 &\ldots&\ldots&\ldots& \ldots&\ldots &\ldots&\ldots&\ldots& \ldots&\ldots \\
 90 & {\bf (82)} & 03 32 53.77 & $-$27 38 10.9 &\ldots&\ldots&\ldots& \ldots&\ldots &  52 & 31.8 & \ldots & \ldots & \ldots &  41 & 15.8 & \ldots & \ldots & \ldots &   5 &  0.0 & \ldots & \ldots & \ldots \\
  \hline
  \end{tabular}
  \normalsize
\label{tab:match}
\end{table*}


\begin{table*}
  \setlength{\tabcolsep}{0.09cm}
  \tiny
  \centering
  \begin{tabular}{cccc|ccccc|ccccc|ccccc|ccccc}
    \hline
    ID & SIMPLE & R.A. & Dec.
    & \multicolumn{5}{c}{250\,\micron}
    & \multicolumn{5}{c}{350\,\micron}
    & \multicolumn{5}{c}{500\,\micron}
    & \multicolumn{5}{c}{870\,\micron} \\
    & & &
    & SID & $\Delta r$ & LR & $P_{24}$ & $P_\mathrm{r}$
    & SID & $\Delta r$ & LR & $P_{24}$ & $P_\mathrm{r}$
    & SID & $\Delta r$ & LR & $P_{24}$ & $P_\mathrm{r}$
    & SID & $\Delta r$ & LR & $P_{24}$ & $P_\mathrm{r}$ \\

    & & ($^{\mathrm h}$\ \ \ $^{\mathrm m}$\ \ \ $^{\mathrm s}$) &
    ($^{\circ}$\ \ \ $^{\prime}$\ \ \ $^{\prime\prime}$)
    & & ($^{\prime\prime}$) & & &
    & & ($^{\prime\prime}$) & & &
    & & ($^{\prime\prime}$) & & &
    & & ($^{\prime\prime}$) & & & \\
  \hline
 91 & {\bf (6)} & 03 32 57.14 & $-$28 01 02.1 &\ldots&\ldots&\ldots& \ldots&\ldots &\ldots&\ldots&\ldots& \ldots&\ldots &  20 & 10.3 & \ldots & \ldots & \ldots &\ldots&\ldots&\ldots& \ldots&\ldots \\
 92 &     38977 & 03 32 57.63 & $-$27 43 18.0 &  27 & 11.4 &   2.4 & 0.100 & 0.136 &\ldots&\ldots&\ldots& \ldots&\ldots &\ldots&\ldots&\ldots& \ldots&\ldots &\ldots&\ldots&\ldots& \ldots&\ldots \\
 93 &     50259 & 03 32 59.32 & $-$27 35 34.1 & 130 &  5.3 &  21.0 & 0.030 & 0.019 &\ldots&\ldots&\ldots& \ldots&\ldots &\ldots&\ldots&\ldots& \ldots&\ldots &\ldots&\ldots&\ldots& \ldots&\ldots \\
 94 &     29432 & 03 32 59.32 & $-$27 48 58.6 &\ldots&\ldots&\ldots& \ldots&\ldots &\ldots&\ldots&\ldots& \ldots&\ldots & 176 &  7.0 &  10.2 & 0.029 & 0.094 &\ldots&\ldots&\ldots& \ldots&\ldots \\
 95 &     17004 & 03 33 01.59 & $-$27 56 49.4 &\ldots&\ldots&\ldots& \ldots&\ldots &  64 &  9.9 &   2.5 & 0.240 & 0.142 &\ldots&\ldots&\ldots& \ldots&\ldots &\ldots&\ldots&\ldots& \ldots&\ldots \\
 96 &     37597 & 03 33 06.16 & $-$27 44 15.3 &\ldots&\ldots&\ldots& \ldots&\ldots & 115 &  1.8 &   1.3 & 0.011 & 0.016 &\ldots&\ldots&\ldots& \ldots&\ldots &\ldots&\ldots&\ldots& \ldots&\ldots \\
 97 &     21942 & 03 33 07.78 & $-$27 53 51.3 &\ldots&\ldots&\ldots& \ldots&\ldots &\ldots&\ldots&\ldots& \ldots&\ldots & 222 & 20.9 &   1.4 & 0.226 & 0.335 &\ldots&\ldots&\ldots& \ldots&\ldots \\
 98 & {\bf (63)} & 03 33 08.46 & $-$28 00 44.3 &\ldots&\ldots&\ldots& \ldots&\ldots &\ldots&\ldots&\ldots& \ldots&\ldots & 200 & 25.2 & \ldots & \ldots & \ldots &\ldots&\ldots&\ldots& \ldots&\ldots \\
 99 &     31482 & 03 33 06.63 & $-$27 48 02.0 &\ldots&\ldots&\ldots& \ldots&\ldots &\ldots&\ldots&\ldots& \ldots&\ldots &\ldots&\ldots&\ldots& \ldots&\ldots &  43 &  5.0 &   2.6 & 0.099 & 0.396 \\
100 &     30926 & 03 33 07.25 & $-$27 48 08.4 &\ldots&\ldots&\ldots& \ldots&\ldots &\ldots&\ldots&\ldots& \ldots&\ldots &\ldots&\ldots&\ldots& \ldots&\ldots &  43 &  8.1 &   1.1 & 0.074 & 0.189 \\
101 &     31537 & 03 33 09.15 & $-$27 48 16.8 &\ldots&\ldots&\ldots& \ldots&\ldots & 145 &  8.7 &   1.4 & 0.278 & 0.169 &\ldots&\ldots&\ldots& \ldots&\ldots &  74 &  7.4 &   4.9 & 0.222 & 0.126 \\
102 &     30714 & 03 33 09.71 & $-$27 48 01.6 &  52 & 14.7 &  27.5 & 0.031 & 0.052 & 145 & 22.0 &   2.6 & \ldots & \ldots &\ldots&\ldots&\ldots& \ldots&\ldots &  74 &  9.7 &  11.3 & 0.015 & 0.025 \\
103 &     30905 & 03 33 09.77 & $-$27 48 21.0 &\ldots&\ldots&\ldots& \ldots&\ldots &\ldots&\ldots&\ldots& \ldots&\ldots &\ldots&\ldots&\ldots& \ldots&\ldots &  74 & 12.5 &   2.2 & \ldots & \ldots \\
104 &     26260 & 03 33 10.12 & $-$27 51 24.8 &\ldots&\ldots&\ldots& \ldots&\ldots &\ldots&\ldots&\ldots& \ldots&\ldots &  44 & 11.3 &   1.3 & 0.263 & 0.237 &\ldots&\ldots&\ldots& \ldots&\ldots \\
105 &     41860 & 03 33 11.79 & $-$27 41 38.4 &\ldots&\ldots&\ldots& \ldots&\ldots &  31 & 10.2 &   6.5 & 0.256 & 0.035 &\ldots&\ldots&\ldots& \ldots&\ldots &\ldots&\ldots&\ldots& \ldots&\ldots \\
106 &     18562 & 03 33 12.62 & $-$27 55 51.7 &\ldots&\ldots&\ldots& \ldots&\ldots & 298 & 15.9 &   2.1 & \ldots & \ldots &\ldots&\ldots&\ldots& \ldots&\ldots &  96 &  7.5 &   1.2 & 0.024 & 0.073 \\
107 & {\bf (1)} & 03 33 14.26 & $-$27 56 11.2 &\ldots&\ldots&\ldots& \ldots&\ldots & 298 & 13.8 & \ldots & \ldots & \ldots &   7 & 32.9 & \ldots & \ldots & \ldots &   1 &  0.0 & \ldots & \ldots & \ldots \\
108 &     35704 & 03 33 15.42 & $-$27 45 24.0 &\ldots&\ldots&\ldots& \ldots&\ldots &  98 &  8.8 &   2.4 & 0.113 & 0.113 &  57 & 10.3 &   1.6 & 0.142 & 0.144 &   7 &  1.8 &   4.4 & 0.009 & 0.009 \\
109 &     26283 & 03 33 16.51 & $-$27 50 39.4 &  72 &  4.2 &  39.5 & 0.002 & 0.007 &\ldots&\ldots&\ldots& \ldots&\ldots &\ldots&\ldots&\ldots& \ldots&\ldots & 108 &  6.4 &   4.2 & 0.004 & 0.015 \\
110 &     41704 & 03 33 16.93 & $-$27 41 21.4 & 185 & 10.7 &  27.3 & 0.017 & 0.039 &\ldots&\ldots&\ldots& \ldots&\ldots &\ldots&\ldots&\ldots& \ldots&\ldots &\ldots&\ldots&\ldots& \ldots&\ldots \\
111 &     34567 & 03 33 17.78 & $-$27 46 05.9 &  54 &  2.6 &   7.9 & 0.012 & 0.011 &  43 & 11.1 &   5.9 & 0.122 & 0.115 &\ldots&\ldots&\ldots& \ldots&\ldots &\ldots&\ldots&\ldots& \ldots&\ldots \\
112 &     34104 & 03 33 17.78 & $-$27 46 23.6 &\ldots&\ldots&\ldots& \ldots&\ldots &  43 & 12.9 &   1.8 & 0.294 & 0.150 &\ldots&\ldots&\ldots& \ldots&\ldots &\ldots&\ldots&\ldots& \ldots&\ldots \\
113 &     28590 & 03 33 17.43 & $-$27 49 48.7 &\ldots&\ldots&\ldots& \ldots&\ldots &\ldots&\ldots&\ldots& \ldots&\ldots &  24 & 22.3 &   2.7 & 0.300 & 0.177 &\ldots&\ldots&\ldots& \ldots&\ldots \\
114 &     29761 & 03 33 17.81 & $-$27 49 10.4 &\ldots&\ldots&\ldots& \ldots&\ldots &\ldots&\ldots&\ldots& \ldots&\ldots &  24 & 16.3 &   1.1 & 0.538 & 0.375 &\ldots&\ldots&\ldots& \ldots&\ldots \\
115 &     28465 & 03 33 18.69 & $-$27 49 40.1 &\ldots&\ldots&\ldots& \ldots&\ldots &\ldots&\ldots&\ldots& \ldots&\ldots &  24 & 19.2 &   2.0 & 0.549 & 0.128 &\ldots&\ldots&\ldots& \ldots&\ldots \\
116 &     20681 & 03 33 18.90 & $-$27 54 33.5 &  67 &  9.9 &   1.9 & 0.129 & 0.141 &\ldots&\ldots&\ldots& \ldots&\ldots &\ldots&\ldots&\ldots& \ldots&\ldots &\ldots&\ldots&\ldots& \ldots&\ldots \\
117 &     19773 & 03 33 21.48 & $-$27 55 20.3 &\ldots&\ldots&\ldots& \ldots&\ldots &\ldots&\ldots&\ldots& \ldots&\ldots &\ldots&\ldots&\ldots& \ldots&\ldots &   3 &  0.4 &   2.4 & 0.010 & 0.015 \\
118 &     17028 & 03 33 28.56 & $-$27 56 54.1 &\ldots&\ldots&\ldots& \ldots&\ldots &  14 & 15.2 &   1.0 & 0.273 & 0.313 &   5 &  5.9 &   5.0 & 0.075 & 0.093 & 120 &  2.3 &  11.2 & 0.015 & 0.019 \\
  \hline
  \end{tabular}
\normalsize
\end{table*}

\begin{table*}
  \caption{Maximum-likelihood submm photometry, SED model fits and
    redshifts. Uncertainties in flux densities have the estimated
    confusion noise (Tables~\ref{tab:matched}) added in quadrature to
    the instrumental noise.  The temperatures, $T_\mathrm{obs}$ and
    10-1000\,\micron\ total infrared (TIR) fluxes, $S_\mathrm{TIR}$
    are for {\em observed-frame} modified blackbody fits of the form
    $S_\nu \propto \nu^{2.0}B_\nu(T_\mathrm{obs})$ (rest-frame values
    shown in Fig.~\ref{fig:seds}). The fractional uncertainties in the
    TIR fluxes, $\Delta S_\mathrm{TIR}/S_\mathrm{TIR}$, are produced
    from the same Monte Carlo simulation used to measure uncertainties
    in the temperatures.  SED fits are not provided for confused
    sources (see SEDs with `C' indicated in Fig.~\ref{fig:seds}). When
    redshifts are available, the TIR luminosity, $L_\mathrm{TIR}$, is
    also calculated, but integrating in the {\em rest-frame}, and
    using a more realistic SED fit from the library of
    \citet{dale2001} in order to estimate emission in the mid/far-IR
    (fits shown in Fig.~\ref{fig:seds}). Redshifts in boldface are
    optical spectroscopic measurements, redshifts in regular face are
    optical photometric estimates, and redshifts in brackets are
    IRAC-based photometric estimates. The superscripts indicate where
    the redshifts were found: `P' from the composite catalogue of
    \citet{pascale2009}; `I' from \citet{ivison2009}; `D' from
    \citet{dunlop2010}; and `C' from \citet{casey2010}.}
  \small
  \centering
  \begin{tabular}{cr@{$\pm$}lr@{$\pm$}lr@{$\pm$}lr@{$\pm$}lr@{$\pm$}lcccc}
    \hline
    ID &
    \multicolumn{2}{c}{$S_{250}$} &
    \multicolumn{2}{c}{$S_{350}$} &
    \multicolumn{2}{c}{$S_{500}$} &
    \multicolumn{2}{c}{$S_{870}$} &
    \multicolumn{2}{c}{$T_\mathrm{obs}$} & $S_\mathrm{TIR}$ &
    $\Delta S_\mathrm{TIR}/S_\mathrm{TIR}$ & $L_\mathrm{TIR}$ &
    Redshift \\
       &
    \multicolumn{2}{c}{(mJy)} &
    \multicolumn{2}{c}{(mJy)} &
    \multicolumn{2}{c}{(mJy)} &
    \multicolumn{2}{c}{(mJy)} &
    \multicolumn{2}{c}{(K)} &
       ($\log_{10}$ W\,m$^{-2}$) & & ($\log_{10}$ L\,$_\odot$) &
    $(z)$ \\
    \hline
  1 &   38&  18  &   41&  15  &    9&  13  &   5.1&  1.2 &  10.5&  1.1 & -15.3 & 0.36& 12.4 & (1.9)$^\mathrm{P}$\\
  2 &   17&  33  &  -40&  30  &  -92&  31  &   1.2&  2.2 &  \multicolumn{2}{c}{\ldots} &\ldots  & \ldots & \ldots  & 1.1$^\mathrm{P}$\\
  3 &   40&  31  &   95&  29  &  131&  30  &   2.3&  2.1 &  \multicolumn{2}{c}{\ldots} &\ldots  & \ldots & \ldots  & 0.8$^\mathrm{P}$\\
  4 &   26&  32  &   -2&  30  &    0&  31  &   2.1&  1.5 &  \multicolumn{2}{c}{\ldots} &\ldots  & \ldots & \ldots  & 1.7$^\mathrm{P}$\\
  5 &   62&  30  &   56&  29  &   29&  30  &   2.2&  1.4 &  \multicolumn{2}{c}{\ldots} &\ldots  & \ldots & \ldots  & 2.6$^\mathrm{P}$\\
  6 &   13&  21  &   30&  19  &   42&  18  &   3.6&  1.2 &   9.7&  1.0 & -15.5 & 0.35& 12.6 & 2.7$^\mathrm{P}$\\
  7 &  -20&  34  &   -7&  33  &  -28&  39  &   4.1&  1.3 &  \multicolumn{2}{c}{\ldots} &\ldots  & \ldots & \ldots  & (1.9)$^\mathrm{P}$\\
  8 &   93&  30  &   29&  29  &   49&  31  &  -0.2&  1.3 &  \multicolumn{2}{c}{\ldots} &\ldots  & \ldots & \ldots  & {\bf 0.677}$^\mathrm{I}$\\
  9 &   69&  18  &   47&  15  &   51&  13  &   9.6&  1.2 &   9.9&  0.7 & -15.1 & 0.24&  \ldots   & \ldots\\
 10 &  109&  24  &   85&  20  &   60&  18  &   3.3&  1.2 &  \multicolumn{2}{c}{\ldots} &\ldots  & \ldots & \ldots  & {\bf 1.482}$^\mathrm{C}$\\
 11 &  -14&  24  &  -25&  20  &  -14&  18  &   2.8&  1.2 &  \multicolumn{2}{c}{\ldots} &\ldots  & \ldots & \ldots  & 0.4$^\mathrm{P}$\\
 12 &   68&  18  &   40&  15  &   22&  13  &  -0.4&  1.2 &  17.7&  0.5 & -14.8 & 0.15& 12.4 & 1.2$^\mathrm{P}$\\
 13 &  -39&  47  &  -19&  52  &   29&  50  &   4.1&  1.8 &  \multicolumn{2}{c}{\ldots} &\ldots  & \ldots & \ldots  & \ldots\\
 14 &   31&  85  &   57& 101  &   -5& 101  &   2.9&  2.5 &  \multicolumn{2}{c}{\ldots} &\ldots  & \ldots & \ldots  & (1.2)$^\mathrm{P}$\\
 15 &   16&  58  &   20&  68  &   18&  67  &   6.3&  1.8 &  \multicolumn{2}{c}{\ldots} &\ldots  & \ldots & \ldots  & 0.1$^\mathrm{P}$\\
 16 &   24&  22  &   -8&  20  &   14&  19  &   4.4&  1.2 &   6.1&  1.9 & -16.2 & 0.72& 11.6 & 2.1$^\mathrm{P}$\\
 17 &   41&  18  &   43&  15  &   24&  13  &   4.2&  1.2 &  11.4&  1.8 & -15.2 & 0.48& 12.6 & {\bf 2.024}$^\mathrm{P}$\\
 18 &   55&  18  &   20&  15  &   13&  13  &   3.6&  1.2 &  13.0&  4.5 & -15.1 & 0.90& 13.1 & 3.0$^\mathrm{P}$\\
 19 &   32&  29  &   51&  26  &   51&  23  &   2.5&  1.3 &  \multicolumn{2}{c}{\ldots} &\ldots  & \ldots & \ldots  & (0.6)$^\mathrm{P}$\\
 20 &   28&  29  &   -2&  26  &  -23&  23  &   3.9&  1.3 &  \multicolumn{2}{c}{\ldots} &\ldots  & \ldots & \ldots  & 0.4$^\mathrm{P}$\\
 21 &   21&  27  &   13&  23  &   24&  22  &   2.9&  1.3 &  \multicolumn{2}{c}{\ldots} &\ldots  & \ldots & \ldots  & 1.2$^\mathrm{P}$\\
 22 &   -9&  27  &    4&  23  &    3&  22  &   3.2&  1.3 &  \multicolumn{2}{c}{\ldots} &\ldots  & \ldots & \ldots  & (2.2)$^\mathrm{P}$\\
 23 &   21&  19  &   37&  16  &   28&  15  &   2.5&  1.2 &  11.4&  2.1 & -15.4 & 0.67& 12.1 & (1.5)$^\mathrm{P}$\\
 24 &   28&  19  &   25&  16  &    0&  15  &   1.7&  1.2 &  21.6&  1.3 & -14.7 & 0.19&  9.9 & {\bf 0.095}$^\mathrm{P}$\\
 25 &   65&  18  &   33&  15  &   30&  13  &   1.4&  1.2 &  15.7&  1.0 & -14.9 & 0.24& 12.2 & 1.1$^\mathrm{I}$\\
 26 &   77&  18  &   22&  15  &   22&  13  &   3.9&  1.2 &  14.4&  3.7 & -14.9 & 0.90&  \ldots   & \ldots\\
 27 &   78&  18  &   45&  15  &   22&  13  &   0.4&  1.2 &  23.2&  0.9 & -14.3 & 0.09& 11.6 & {\bf 0.364}$^\mathrm{P}$\\
 28 &   40&  18  &   38&  15  &   13&  13  &   3.6&  1.2 &  11.9&  1.8 & -15.2 & 0.63& 11.3 & 0.6$^\mathrm{P}$\\
 29 &   35&  18  &   39&  15  &   30&  13  &   6.5&  1.2 &   9.6&  1.1 & -15.3 & 0.41& 12.5 & (2.1)$^\mathrm{P}$\\
 30 &   84&  18  &   53&  15  &   33&  13  &   3.2&  1.2 &  18.2&  0.4 & -14.5 & 0.08& 12.0 & {\bf 0.620}$^\mathrm{P}$\\
 31 &   18&  22  &  -11&  20  &   25&  19  &   1.7&  1.2 &   9.2&  2.9 & -16.0 & 0.90& 12.2 & (3.0)$^\mathrm{P}$\\
 32 &   57&  23  &   79&  20  &   15&  19  &   3.4&  1.2 &  13.5&  1.8 & -14.9 & 0.42& 12.6 & {\bf 1.605}$^\mathrm{C}$\\
 33 &   29&  18  &   33&  15  &   22&  13  &   8.1&  1.3 &   8.4&  1.0 & -15.4 & 0.33& 11.7 & (1.2)$^\mathrm{P}$\\
 34 &   42&  19  &   29&  17  &   25&  15  &   1.1&  1.2 &  14.5&  1.4 & -15.1 & 0.45& 12.8 & {\bf 2.342}$^\mathrm{C}$\\
 35 &   74&  20  &   50&  17  &   38&  15  &   5.2&  1.2 &  12.3&  1.2 & -14.9 & 0.37& 13.2 & 2.9$^\mathrm{P}$\\
 36 &   60&  18  &   60&  15  &   47&  13  &   3.2&  1.2 &  \multicolumn{2}{c}{\ldots} &\ldots  & \ldots & \ldots  & (2.1)$^\mathrm{P}$\\
 37 &   54&  18  &   43&  15  &   24&  13  &   2.6&  1.2 &  \multicolumn{2}{c}{\ldots} &\ldots  & \ldots & \ldots  & 0.6$^\mathrm{P}$\\
 38 &   30&  27  &   35&  24  &   49&  22  &   1.7&  1.3 &  \multicolumn{2}{c}{\ldots} &\ldots  & \ldots & \ldots  & \ldots\\
 39 &   14&  27  &    5&  24  &  -15&  22  &   3.4&  1.3 &  \multicolumn{2}{c}{\ldots} &\ldots  & \ldots & \ldots  & 1.8$^\mathrm{P}$\\
 40 &   11&  18  &   31&  15  &   26&  13  &   6.1&  1.2 &   8.3&  0.9 & -15.6 & 0.33& 11.9 & (1.6)$^\mathrm{P}$\\
 41 &   17&  18  &   31&  15  &    0&  13  &   1.3&  1.2 &  18.5&  1.4 & -15.0 & 0.30&  9.9 & 0.1$^\mathrm{P}$\\
 42 &   30&  18  &   12&  15  &   11&  13  &   2.0&  1.2 &  12.8&  6.0 & -15.4 & 0.90& 10.1 & 0.2$^\mathrm{P}$\\
 43 &   68&  18  &   58&  15  &   20&  13  &   6.3&  1.2 &  11.7&  1.0 & -15.0 & 0.32& 12.9 & {\bf 2.252}$^\mathrm{C}$\\
 44 &   55&  18  &   60&  15  &   65&  13  &   6.0&  1.2 &  10.8&  0.7 & -15.1 & 0.29&  9.7 & (0.1)$^\mathrm{P}$\\
 45 &   37&  18  &    5&  15  &    5&  13  &  -0.1&  1.2 &  32.6&  7.8 & -14.1 & 0.59& 12.5 & 0.7$^\mathrm{P}$\\
 46 &   22&  18  &   55&  15  &   43&  13  &   8.5&  1.2 &   8.8&  0.6 & -15.3 & 0.32& 12.2 & (1.6)$^\mathrm{P}$\\
 47 &   42&  18  &  -15&  16  &  -10&  14  &   1.7&  1.2 &  17.5&  1.3 & -15.2 & 0.24& 11.2 & 0.6$^\mathrm{P}$\\
 48 &   15&  18  &   38&  16  &   43&  14  &   2.0&  1.2 &  11.1&  3.0 & -15.4 & 0.90& 12.1 & 1.7$^\mathrm{D}$\\
 49 &   30&  18  &   52&  15  &   32&  13  &   2.1&  1.2 &  12.5&  1.4 & -15.2 & 0.53& 12.3 & {\bf 1.565}$^\mathrm{P}$\\
 50 &   16&  18  &   10&  15  &   35&  13  &   8.1&  1.2 &   7.1&  1.3 & -15.7 & 0.52& 10.7 & (0.6)$^\mathrm{P}$\\
    \hline
  \end{tabular}
  \normalsize
\label{tab:stuff}
\end{table*}

\begin{table*}
  \small
  \centering
  \begin{tabular}{cr@{$\pm$}lr@{$\pm$}lr@{$\pm$}lr@{$\pm$}lr@{$\pm$}lcccc}
    \hline
    ID &
    \multicolumn{2}{c}{$S_{250}$} &
    \multicolumn{2}{c}{$S_{350}$} &
    \multicolumn{2}{c}{$S_{500}$} &
    \multicolumn{2}{c}{$S_{870}$} &
    \multicolumn{2}{c}{$T_\mathrm{obs}$} & $S_\mathrm{TIR}$ &
    $\Delta S_\mathrm{TIR}/S_\mathrm{TIR}$ & $L_\mathrm{TIR}$ & Redshift \\
       &
    \multicolumn{2}{c}{(mJy)} &
    \multicolumn{2}{c}{(mJy)} &
    \multicolumn{2}{c}{(mJy)} &
    \multicolumn{2}{c}{(mJy)} &
    \multicolumn{2}{c}{(K)} &
       ($\log_{10}$ W\,m$^{-2}$) & & ($\log_{10}$ L\,$_\odot$) & $(z)$\\
    \hline
 51 &   48&  18  &   55&  15  &   12&  13  &  -0.6&  1.2 &  17.0&  0.5 & -14.9 & 0.21& 12.4 & {\bf 1.324}$^\mathrm{P}$\\
 52 &   33&  19  &   33&  16  &   21&  14  &   0.2&  1.2 &  18.2&  5.1 & -14.9 & 0.90& 11.6 & 0.6$^\mathrm{P}$\\
 53 &   74&  19  &   51&  16  &   17&  14  &   2.1&  1.2 &  21.0&  0.7 & -14.4 & 0.11& 10.5 & 0.1$^\mathrm{P}$\\
 54 &   56&  18  &   26&  15  &   17&  13  &   1.5&  1.2 &  20.5&  0.9 & -14.6 & 0.15& 10.3 & {\bf 0.124}$^\mathrm{D}$\\
 55 &   46&  19  &   24&  16  &   30&  14  &   3.0&  1.2 &  12.5&  2.1 & -15.2 & 0.56& 11.9 & {\bf 1.097}$^\mathrm{D}$\\
 56 &   35&  19  &   50&  17  &   25&  15  &   8.2&  1.2 &   9.2&  1.1 & -15.3 & 0.40& 12.6 & 2.3$^\mathrm{D}$\\
 57 &   69&  18  &   32&  15  &   37&  13  &   4.4&  1.2 &  12.6&  2.9 & -15.0 & 0.89& 12.9 & {\bf 2.277}$^\mathrm{C}$\\
 58 &   51&  18  &   18&  15  &   -5&  13  &  -0.3&  1.2 &  25.2&  1.4 & -14.4 & 0.14& 10.1 & 0.1$^\mathrm{P}$\\
 59 &   48&  18  &   26&  15  &   19&  13  &   1.0&  1.2 &  24.4&  1.2 & -14.4 & 0.13& 10.3 & 0.1$^\mathrm{P}$\\
 60 &   69&  18  &   42&  15  &   34&  13  &   1.8&  1.2 &  15.7&  2.1 & -14.8 & 0.38& 12.3 & 1.1$^\mathrm{P}$\\
 61 &   97&  20  &   42&  18  &   33&  16  &   1.9&  1.2 &  24.8&  0.8 & -14.1 & 0.08& 10.4 & {\bf 0.077}$^\mathrm{I}$\\
 62 &   65&  21  &   36&  18  &    1&  16  &   0.1&  1.2 &  24.8&  1.1 & -14.3 & 0.12& 10.1 & {\bf 0.076}$^\mathrm{P}$\\
 63 &   64&  18  &   73&  15  &   43&  13  &   4.0&  1.2 &  12.7&  2.0 & -14.9 & 0.56& 12.4 & 1.4$^\mathrm{P}$\\
 64 &   51&  19  &   26&  16  &    6&  14  &   2.5&  1.2 &  \multicolumn{2}{c}{\ldots} &\ldots  & \ldots & \ldots  & {\bf 0.125}$^\mathrm{P}$\\
 65 &   27&  18  &    4&  16  &  -10&  14  &   4.2&  1.2 &  \multicolumn{2}{c}{\ldots} &\ldots  & \ldots & \ldots  & (2.3)$^\mathrm{P}$\\
 66 &  185&  18  &   78&  15  &   35&  13  &   0.1&  1.2 &  22.1&  0.4 & -14.1 & 0.05&  9.7 & {\bf 0.038}$^\mathrm{D}$\\
 67 &   54&  18  &   35&  15  &   40&  13  &   2.7&  1.2 &  13.2&  1.4 & -15.1 & 0.25& 13.1 & 2.8$^\mathrm{P}$\\
 68 &   20&  20  &   27&  18  &   13&  16  &  -1.3&  1.2 &  \multicolumn{2}{c}{\ldots} &\ldots  & \ldots & \ldots  & 0.6$^\mathrm{P}$\\
 69 &   26&  20  &   10&  18  &   -3&  16  &   1.6&  1.2 &  \multicolumn{2}{c}{\ldots} &\ldots  & \ldots & \ldots  & (2.2)$^\mathrm{P}$\\
 70 &  102&  29  &   67&  26  &   95&  23  &  -1.4&  1.8 &  \multicolumn{2}{c}{\ldots} &\ldots  & \ldots & \ldots  & (1.8)$^\mathrm{P}$\\
 71 &  -31&  29  &  -11&  26  &  -53&  23  &   2.7&  1.8 &  \multicolumn{2}{c}{\ldots} &\ldots  & \ldots & \ldots  & 0.2$^\mathrm{P}$\\
 72 &   47&  18  &   29&  15  &   28&  13  &  -1.6&  1.2 &  22.3&  0.8 & -14.6 & 0.10& 11.1 & {\bf 0.297}$^\mathrm{P}$\\
 73 &   30&  18  &   35&  15  &   21&  13  &   4.4&  1.2 &  10.3&  2.0 & -15.3 & 0.66& 11.4 & {\bf 0.830}$^\mathrm{P}$\\
 74 &    6&  29  &   58&  26  &   49&  25  &   3.1&  1.8 &  \multicolumn{2}{c}{\ldots} &\ldots  & \ldots & \ldots  & {\bf 0.250}$^\mathrm{P}$\\
 75 &   43&  29  &  -42&  26  &  -42&  25  &  -1.1&  1.8 &  \multicolumn{2}{c}{\ldots} &\ldots  & \ldots & \ldots  & 0.1$^\mathrm{P}$\\
 76 &   79&  27  &   41&  24  &   24&  22  &   1.8&  1.5 &  \multicolumn{2}{c}{\ldots} &\ldots  & \ldots & \ldots  & {\bf 2.123}$^\mathrm{C}$\\
 77 &  -24&  27  &   13&  24  &    1&  22  &   3.8&  1.5 &  \multicolumn{2}{c}{\ldots} &\ldots  & \ldots & \ldots  & 2.8$^\mathrm{P}$\\
 78 &   65&  18  &   34&  15  &   21&  13  &   1.3&  1.2 &  19.2&  6.7 & -14.6 & 0.90& 12.1 & {\bf 0.733}$^\mathrm{P}$\\
 79 &   55&  27  &   33&  23  &    8&  22  &   0.4&  1.9 &  \multicolumn{2}{c}{\ldots} &\ldots  & \ldots & \ldots  & {\bf 0.279}$^\mathrm{P}$\\
 80 &  -10&  27  &    0&  23  &   -1&  22  &   0.4&  1.9 &  \multicolumn{2}{c}{\ldots} &\ldots  & \ldots & \ldots  & {\bf 0.279}$^\mathrm{P}$\\
 81 &   94&  18  &   52&  15  &   21&  13  &   2.3&  1.2 &  21.2&  0.7 & -14.3 & 0.11& 10.4 & {\bf 0.104}$^\mathrm{P}$\\
 82 &   45&  18  &   24&  15  &   27&  13  &   7.7&  1.2 &   9.0&  2.1 & -15.4 & 0.69& 12.8 & 2.8$^\mathrm{D}$$^\dagger$\\
 83 &   67&  21  &   70&  18  &   32&  16  &   2.4&  1.2 &  \multicolumn{2}{c}{\ldots} &\ldots  & \ldots & \ldots  & {\bf 0.981}$^\mathrm{P}$\\
 84 &  -28&  21  &  -38&  18  &  -22&  16  &   7.3&  1.2 &  \multicolumn{2}{c}{\ldots} &\ldots  & \ldots & \ldots  & (1.9)$^\mathrm{P}$\\
 85 &   10&  20  &    7&  18  &   34&  16  &   2.8&  1.2 &  \multicolumn{2}{c}{\ldots} &\ldots  & \ldots & \ldots  & {\bf 0.981}$^\mathrm{P}$\\
 86 &   98&  18  &   68&  15  &   21&  13  &   2.3&  1.2 &  16.4&  0.8 & -14.6 & 0.22& 13.3 & {\bf 2.326}$^\mathrm{C}$\\
 87 &   10&  18  &  -22&  15  &    3&  13  &   0.7&  1.2 &  29.7& 11.3 & -15.6 & 0.34& 10.3 & 0.4$^\mathrm{P}$\\
 88 &   33&  18  &   16&  15  &   16&  13  &   1.6&  1.2 &  18.6&  1.3 & -14.9 & 0.24& 11.6 & {\bf 0.620}$^\mathrm{P}$\\
 89 &   -5&  18  &   43&  15  &    8&  13  &   1.8&  1.2 &  10.0&  1.8 & -15.7 & 0.41& 11.8 & (1.6)$^\mathrm{P}$\\
 90 &   52&  18  &   51&  15  &   42&  13  &   4.6&  1.2 &  11.6&  1.8 & -15.1 & 0.35&  \ldots   & \ldots\\
 91 &   18&  18  &   40&  15  &   39&  13  &   8.9&  1.2 &   8.2&  0.8 & -15.4 & 0.30&  \ldots   & \ldots\\
 92 &   90&  18  &   54&  15  &   38&  13  &   2.8&  1.2 &  15.7&  1.7 & -14.7 & 0.33& 13.0 & 1.9$^\mathrm{P}$\\
 93 &   58&  18  &   25&  15  &    7&  13  &   0.9&  1.2 &  18.4&  0.8 & -14.8 & 0.17& 10.6 & 0.2$^\mathrm{P}$\\
 94 &   43&  18  &   13&  15  &   29&  13  &   1.3&  1.2 &  16.0&  6.0 & -15.0 & 0.90& 11.2 & 0.5$^\mathrm{P}$\\
 95 &   25&  18  &   39&  15  &   32&  13  &   4.7&  1.2 &   9.9&  1.4 & -15.4 & 0.46& 11.9 & 1.3$^\mathrm{P}$\\
 96 &   51&  18  &   52&  15  &   15&  13  &   0.2&  1.2 &  19.7&  4.8 & -14.6 & 0.90& 11.4 & 0.4$^\mathrm{P}$\\
 97 &   36&  18  &    9&  15  &   29&  13  &   1.9&  1.2 &  12.9&  2.2 & -15.3 & 0.50& 12.1 & 1.5$^\mathrm{P}$\\
 98 &  -35&  18  &   -7&  15  &   13&  13  &   5.2&  1.2 &   4.4&  1.3 & -16.5 & 0.66&  \ldots   & \ldots\\
 99 &  -30&  29  &   48&  28  &    4&  33  &   3.0&  1.3 &  \multicolumn{2}{c}{\ldots} &\ldots  & \ldots & \ldots  & (3.0)$^\mathrm{P}$\\
100 &   70&  31  &  -27&  30  &   35&  41  &   3.0&  1.3 &  \multicolumn{2}{c}{\ldots} &\ldots  & \ldots & \ldots  & {\bf 0.531}$^\mathrm{P}$\\
    \hline
  \end{tabular}
  \normalsize

  \flushleft
  $^\dagger$Source 82 has an optical photometric redshift limit
  $z>2.8$ from \citet{dunlop2010}. Since this limit is significantly
  larger than its IRAC-based photometric redshift $z_\mathrm{i}=1.6$
  we use the limit as the best estimate for this source.
\end{table*}

\begin{table*}
  \small
  \begin{center}
  \begin{tabular}{cr@{$\pm$}lr@{$\pm$}lr@{$\pm$}lr@{$\pm$}lr@{$\pm$}lcccc}
    \hline
    ID &
    \multicolumn{2}{c}{$S_{250}$} &
    \multicolumn{2}{c}{$S_{350}$} &
    \multicolumn{2}{c}{$S_{500}$} &
    \multicolumn{2}{c}{$S_{870}$} &
    \multicolumn{2}{c}{$T_\mathrm{obs}$} & $S_\mathrm{TIR}$ &
    $\Delta S_\mathrm{TIR}/S_\mathrm{TIR}$ & $L_\mathrm{TIR}$ & Redshift \\
       &
    \multicolumn{2}{c}{(mJy)} &
    \multicolumn{2}{c}{(mJy)} &
    \multicolumn{2}{c}{(mJy)} &
    \multicolumn{2}{c}{(mJy)} &
    \multicolumn{2}{c}{(K)} &
       ($\log_{10}$ W\,m$^{-2}$) & & ($\log_{10}$ L\,$_\odot$) & $(z)$\\
    \hline
101 &   24&  43  &  100&  42  &  -16&  49  &   3.6&  1.9 &  \multicolumn{2}{c}{\ldots} &\ldots  & \ldots & \ldots  & (2.2)$^\mathrm{P}$\\
102 &   38&  32  &  -14&  30  &  -11&  28  &   2.0&  1.2 &  \multicolumn{2}{c}{\ldots} &\ldots  & \ldots & \ldots  & {\bf 0.180}$^\mathrm{I}$\\
103 &    0&  27  &  -51&  25  &   45&  27  &  -0.5&  2.0 &  \multicolumn{2}{c}{\ldots} &\ldots  & \ldots & \ldots  & (1.4)$^\mathrm{P}$\\
104 &   53&  18  &   36&  15  &   35&  13  &   0.9&  1.2 &  16.9&  4.7 & -14.8 & 0.90& 12.6 & (1.6)$^\mathrm{P}$\\
105 &   21&  18  &   44&  15  &    9&  13  &   0.9&  1.2 &  18.3&  1.4 & -14.9 & 0.22& 10.8 & 0.3$^\mathrm{P}$\\
106 &   34&  19  &   28&  16  &   41&  14  &   3.9&  1.2 &  16.4&  1.0 & -14.8 & 0.17& 11.0 & 0.3$^\mathrm{P}$\\
107 &   27&  19  &   34&  16  &    9&  14  &  14.1&  1.2 &   6.4&  1.8 & -15.6 & 0.55&  \ldots   & \ldots\\
108 &   19&  18  &   46&  15  &   31&  13  &   8.8&  1.2 &   8.4&  0.9 & -15.4 & 0.33& 11.5 & 0.9$^\mathrm{P}$\\
109 &   76&  18  &   25&  15  &    0&  13  &   4.0&  1.2 &  22.8&  0.8 & -14.3 & 0.09& 10.3 & {\bf 0.087}$^\mathrm{P}$\\
110 &   48&  18  &   49&  15  &   23&  13  &   1.9&  1.2 &  22.5&  0.9 & -14.4 & 0.11& 10.6 & {\bf 0.148}$^\mathrm{P}$\\
111 &   63&  21  &   44&  18  &   13&  16  &   3.9&  1.2 &  13.3&  1.6 & -15.0 & 0.36& 12.8 & 2.1$^\mathrm{C}$\\
112 &   26&  21  &   29&  18  &   31&  16  &  -0.1&  1.2 &  18.2&  3.2 & -15.0 & 0.72& 12.4 & 1.4$^\mathrm{P}$\\
113 &   -6&  23  &   18&  20  &   -7&  19  &   3.3&  1.2 &  \multicolumn{2}{c}{\ldots} &\ldots  & \ldots & \ldots  & (1.9)$^\mathrm{P}$\\
114 &   26&  19  &    9&  16  &   20&  14  &  -0.7&  1.2 &  \multicolumn{2}{c}{\ldots} &\ldots  & \ldots & \ldots  & (2.0)$^\mathrm{P}$\\
115 &   22&  24  &   -8&  21  &   22&  21  &   1.8&  1.2 &  \multicolumn{2}{c}{\ldots} &\ldots  & \ldots & \ldots  & 0.9$^\mathrm{P}$\\
116 &   80&  18  &   25&  15  &   11&  13  &   0.6&  1.2 &  16.3&  0.6 & -14.8 & 0.14& 11.4 & 0.5$^\mathrm{I}$\\
117 &   48&  18  &   48&  15  &   33&  13  &  11.4&  1.2 &   8.7&  1.0 & -15.2 & 0.38& 12.2 & (1.5)$^\mathrm{P}$\\
118 &   53&  18  &   73&  15  &   52&  13  &   4.4&  1.2 &  11.8&  2.8 & -15.1 & 0.90&  \ldots   & \ldots\\
    \hline
  \end{tabular}\end{center}
  \normalsize
\end{table*}

\section[]{Postage Stamps}
\label{sec:stamps}

\begin{figure*}
\centering
\includegraphics[width=0.49\linewidth]{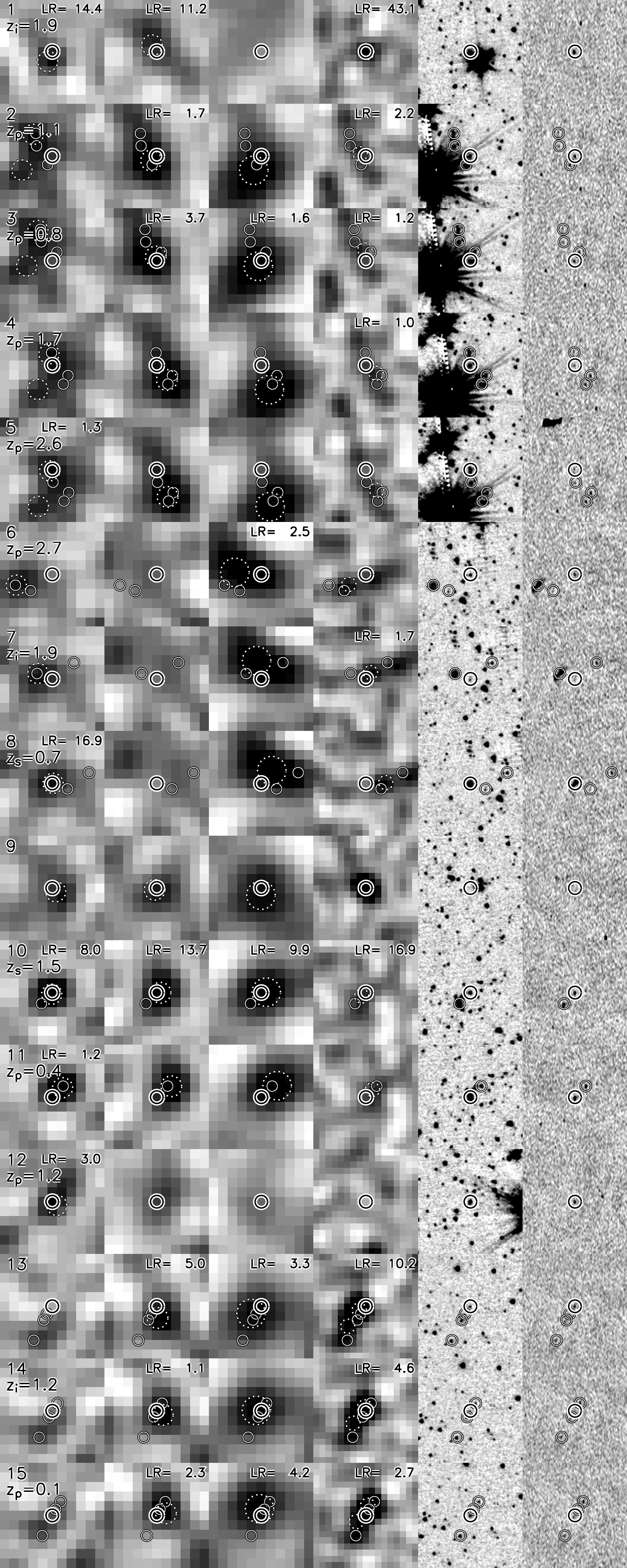}
\includegraphics[width=0.49\linewidth]{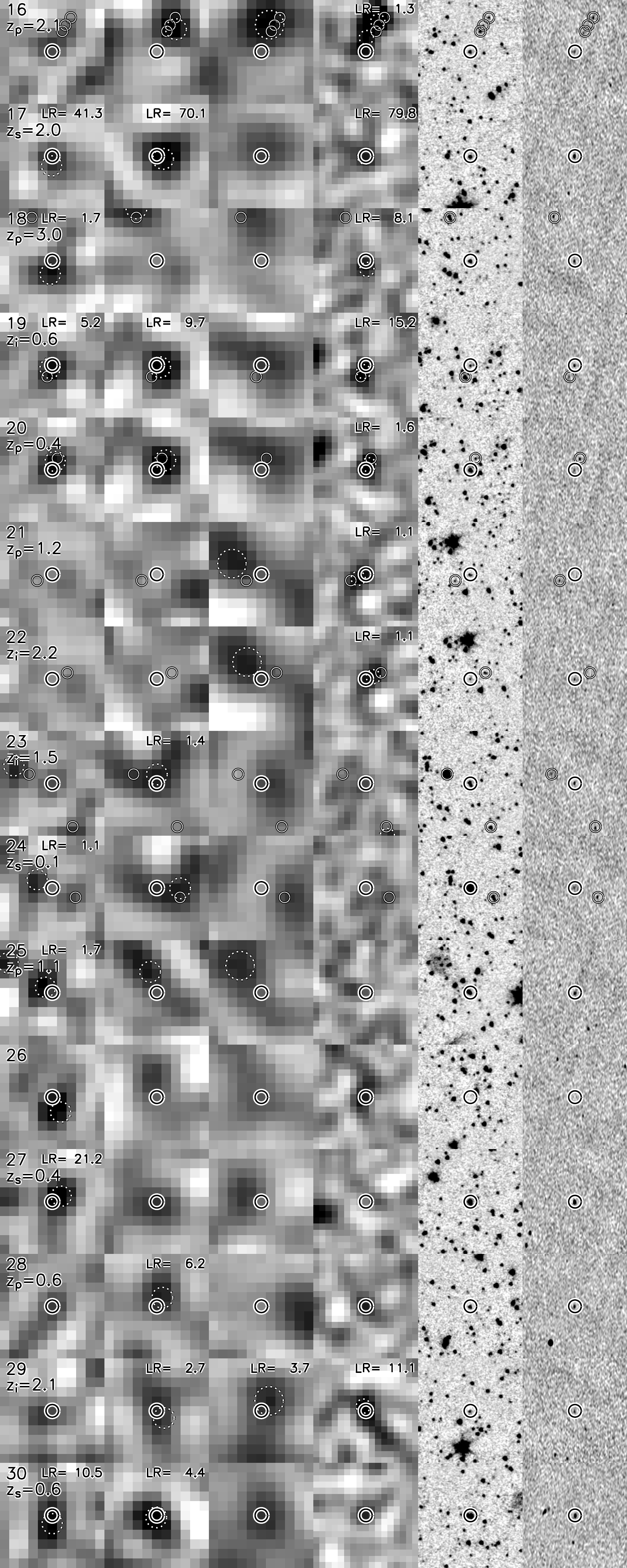}
\caption{$1.8' \times 1.8'$ Postage stamps for each of the fitted
  sources. Columns from left to right: 250\,\micron, 350\,\micron,
  500\,\micron, 870\,\micron (\snr\ maps, scaled between -$3\sigma$
  (white) and $+5\sigma$ (black), 3.6\,\micron, and 1.4\,GHz\
  (-2\,$\mu$Jy (white), +5\,$\mu$Jy (black)). Small heavy circles
  indicate the source under consideration. Light smaller circles
  indicate other nearby sources that have been simultaneously fit in
  the submm bands. Large dashed circles on the submm images indicate
  input positions obtained from local maxima in the match-filtered
  images. Sources were matched to the external catalogue, and the
  reported redshifts are those that were used in \citet{pascale2009}
  when available (subscript `s' indicates spectroscopic redshifts, `p'
  optical photometric redshifts, and `i' IRAC-based photometric
  redshifts). The likelihood ratios (LR) of the matches to each submm
  band are indicated in the top-right corners when available.}
\label{fig:stamps1}
\end{figure*}

This Appendix contains submm, IRAC and radio postage stamps for all
118 proposed identifications to the submm peaks (Fig.~\ref{fig:stamps1}).

\clearpage

\begin{figure*}
\centering
\includegraphics[width=0.49\linewidth]{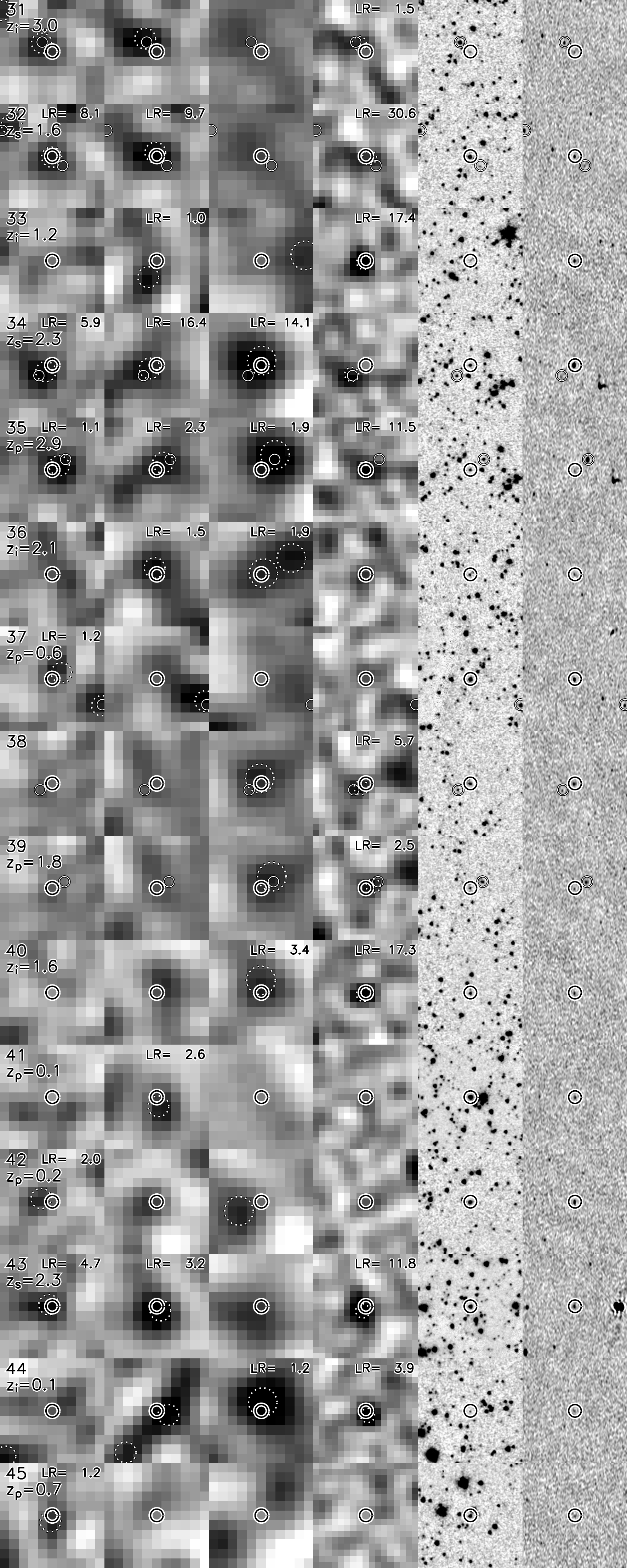}
\includegraphics[width=0.49\linewidth]{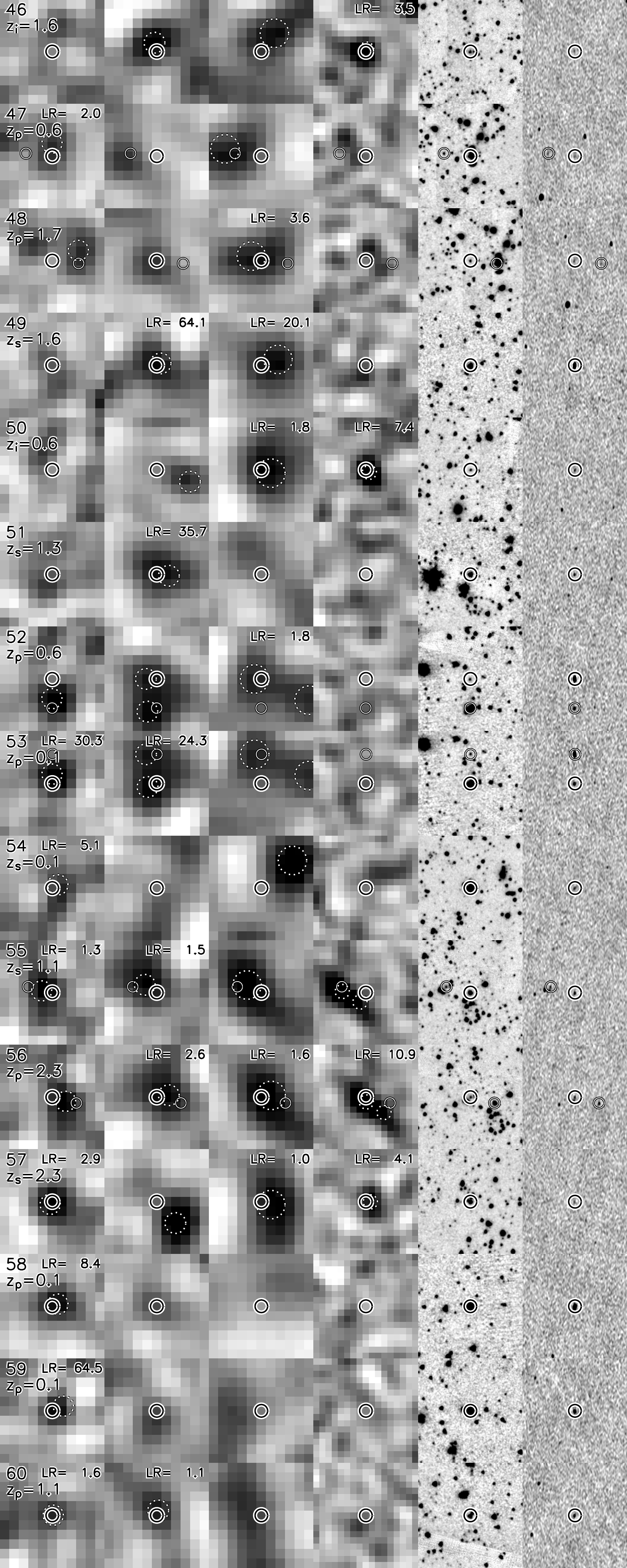}
\end{figure*}


\begin{figure*}
\centering
\includegraphics[width=0.49\linewidth]{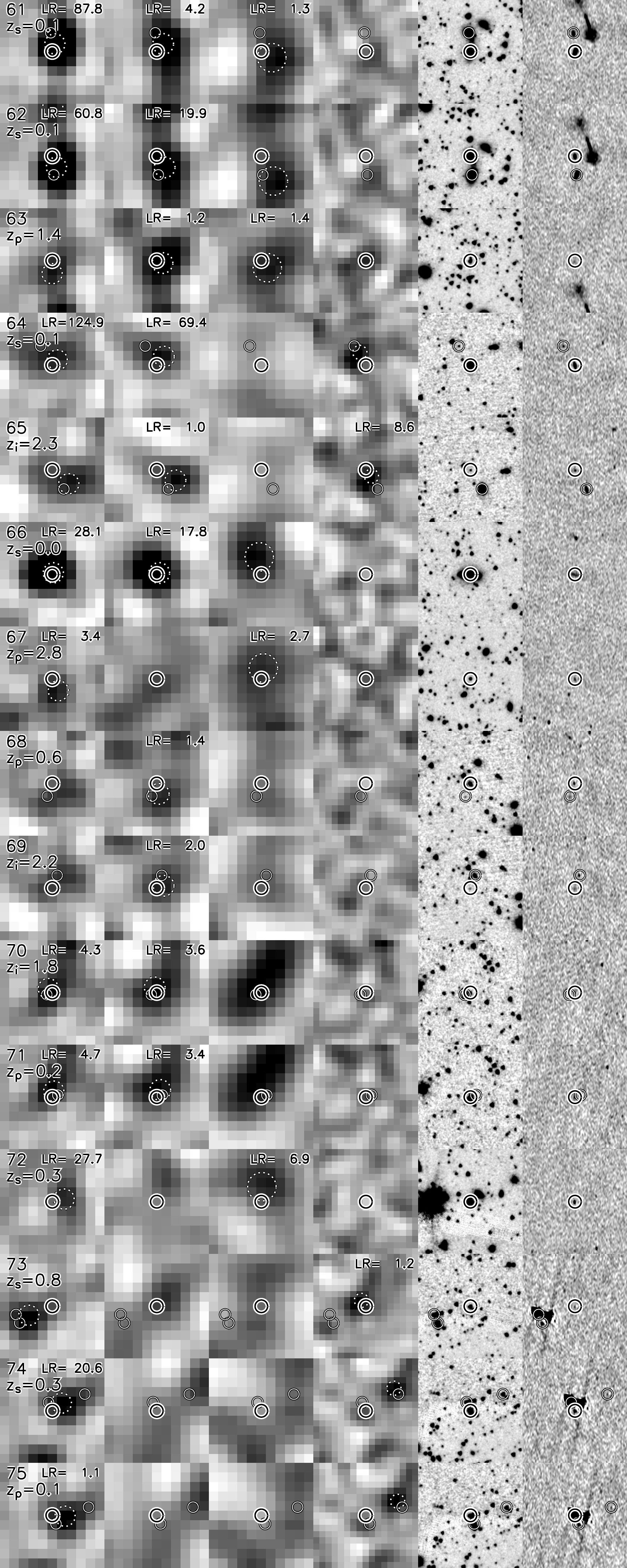}
\includegraphics[width=0.49\linewidth]{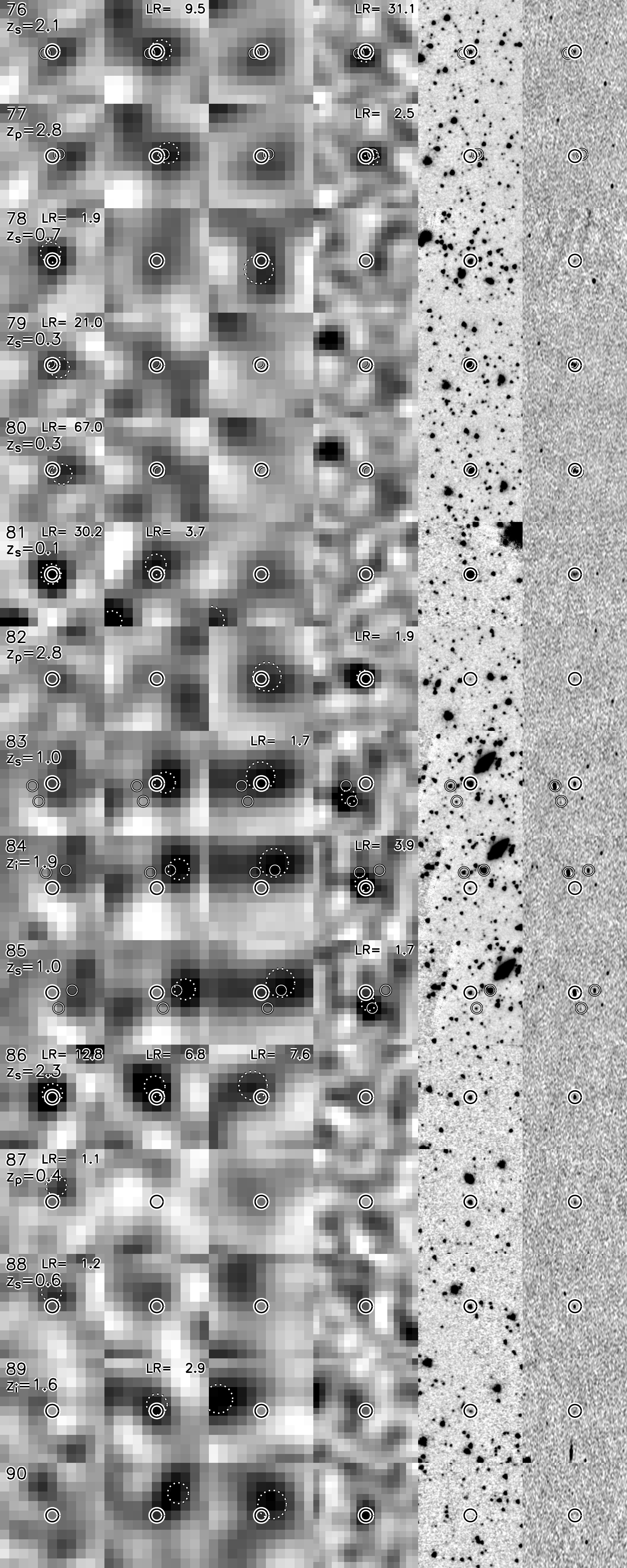}
\end{figure*}


\begin{figure*}
\centering
\includegraphics[width=0.49\linewidth]{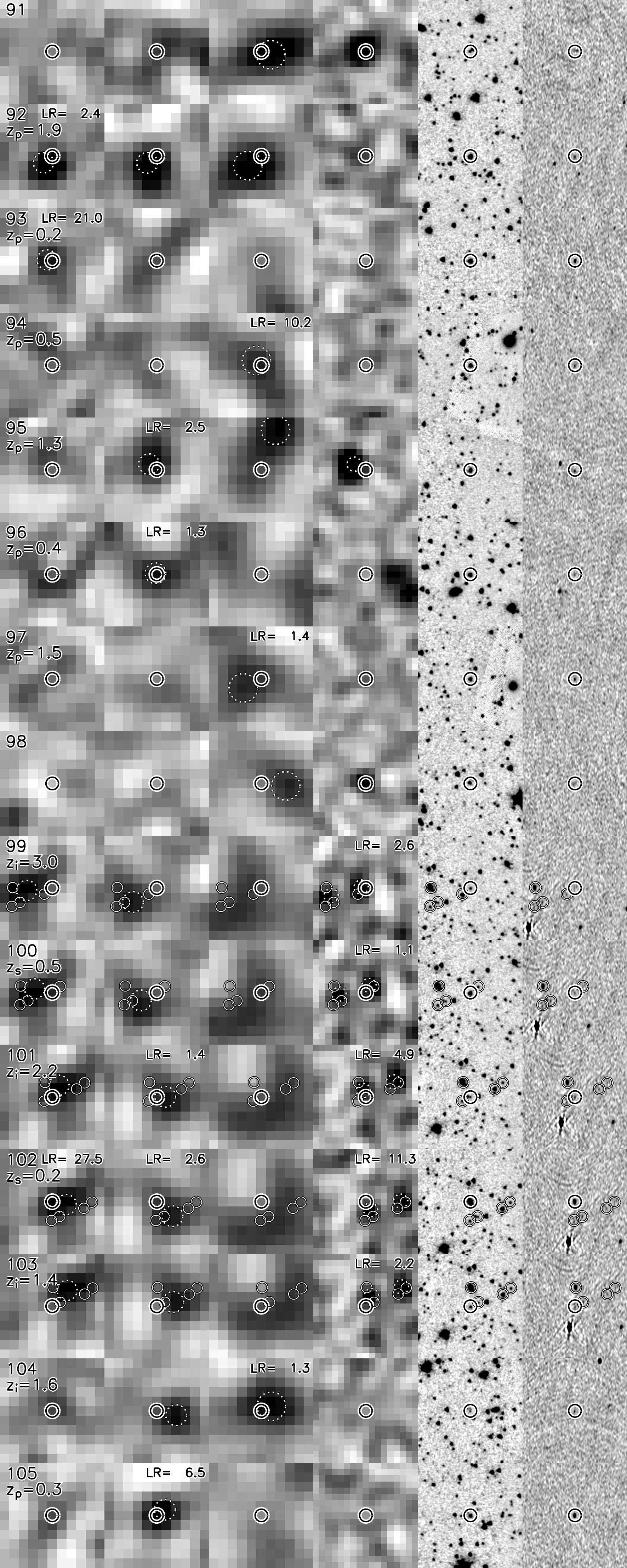}
\includegraphics[width=0.49\linewidth]{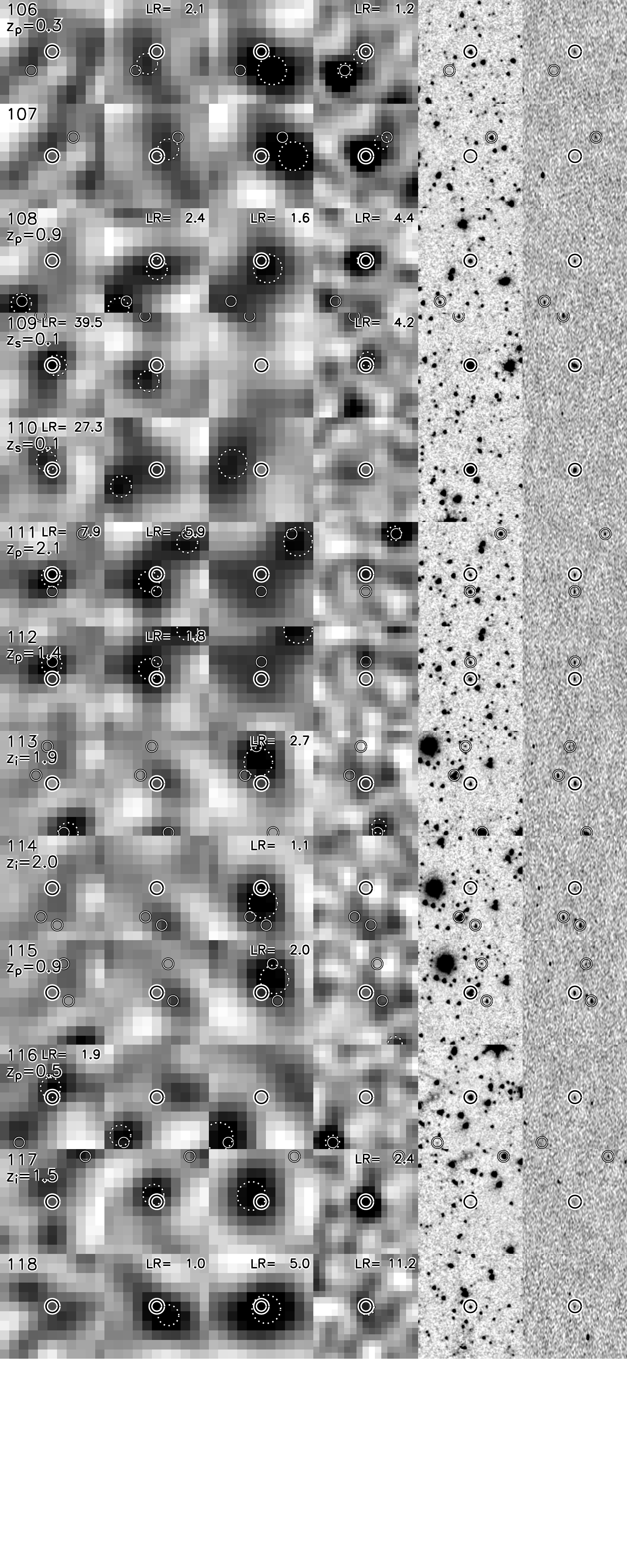}
\end{figure*}

\section[]{Spectral Energy Distributions}
\label{sec:seds}

\begin{figure*}
\centering
\includegraphics[width=0.95\linewidth]{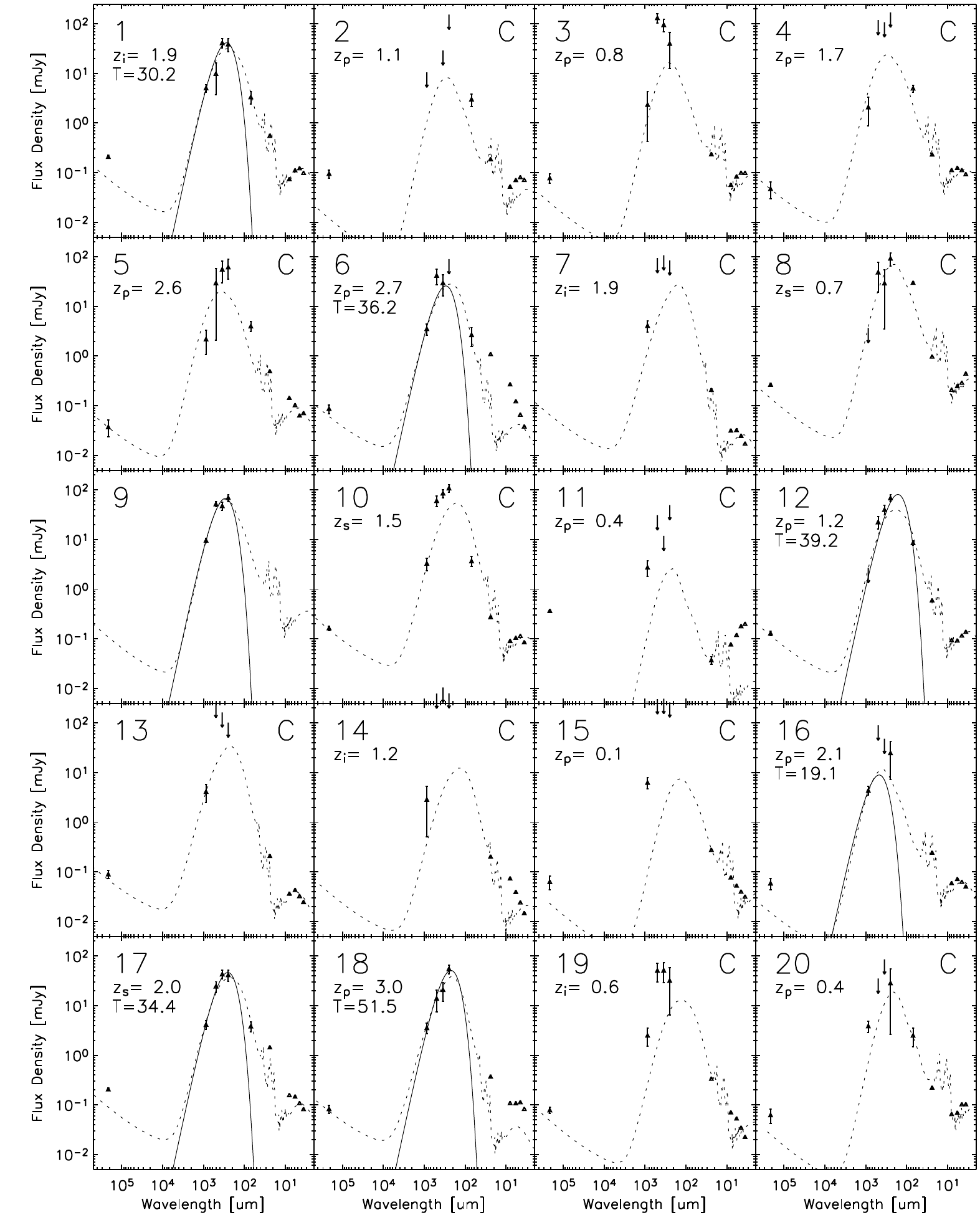}
\caption{Observed-frame spectral energy distributions. The flux
  densities at all 118 positions are fit simultaneously in each submm
  map. In this way flux densities in blends are divided up. All
  matched sources have IRAC photometry from SIMPLE, and most sources
  also have 24\,\micron\ (and occasional 70\,\micron) flux densities
  from FIDEL. Most sources also have 1.4\,GHz radio measurements. The
  only exceptions are sources for which the BLAST data have only been
  matched to an 870\,\micron\ peak from LESS, whose position has
  been used to re-measure flux densities in the BLAST bands (no SED
  information given). Redshift labels have the same meaning as in
  Fig.~\ref{fig:seds}. Dotted lines are SED fits to the submm and FIR
  photometry using the library of \citet{dale2001}. The `C' indicates
  42 sources that are confused to the point that the submm photometry
  is unusable, and only the \citet{dale2001} SEDs are fit (to the
  radio and mid/near-IR photometry). For the remaining 83 sources a
  modified blackbody with emissivity $\beta=2.0$ has been fit, with
  the maximum-likelihood {\em rest-frame} temperatures indicated for
  the 73 sources with redshifts.}
\label{fig:seds}
\end{figure*}

This Appendix contains observed-frame radio--submm--(mid/near)-IR SEDs
for all 118 proposed identifications to the submm peaks
(Fig.~\ref{fig:seds}).

\begin{figure*}
\centering
\includegraphics[width=0.95\linewidth]{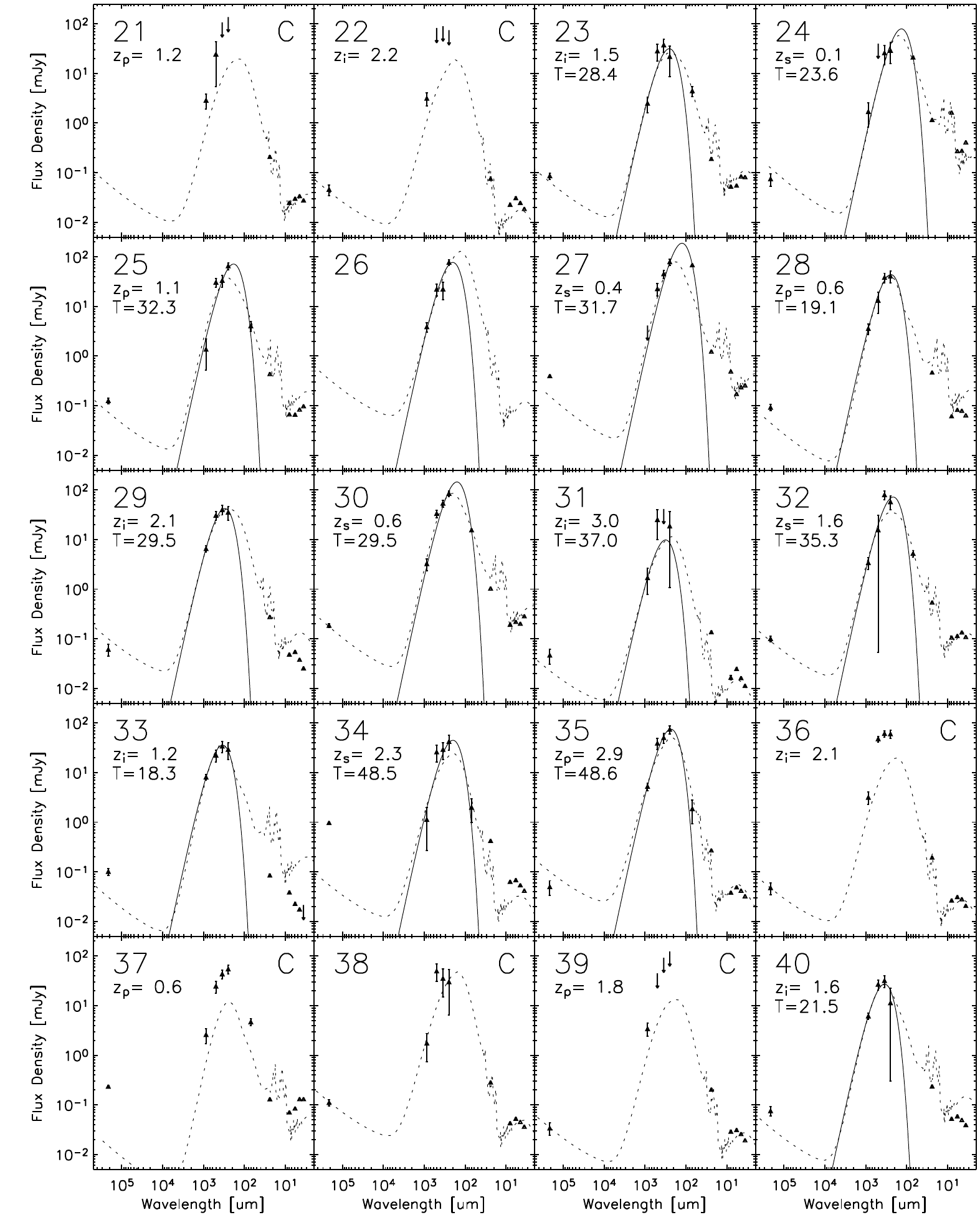}
\end{figure*}


\begin{figure*}
\centering
\includegraphics[width=0.95\linewidth]{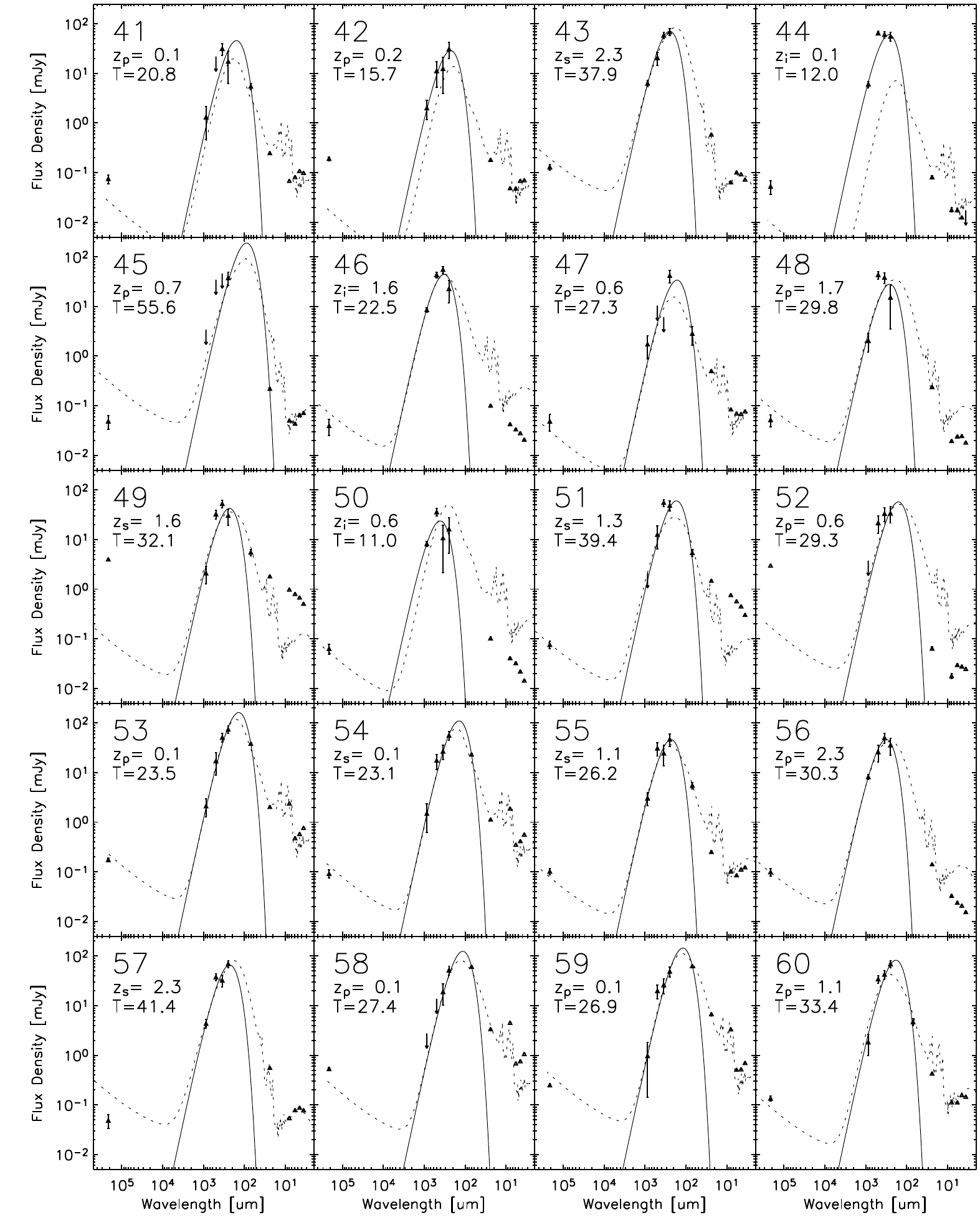}
\end{figure*}


\begin{figure*}
\centering
\includegraphics[width=0.95\linewidth]{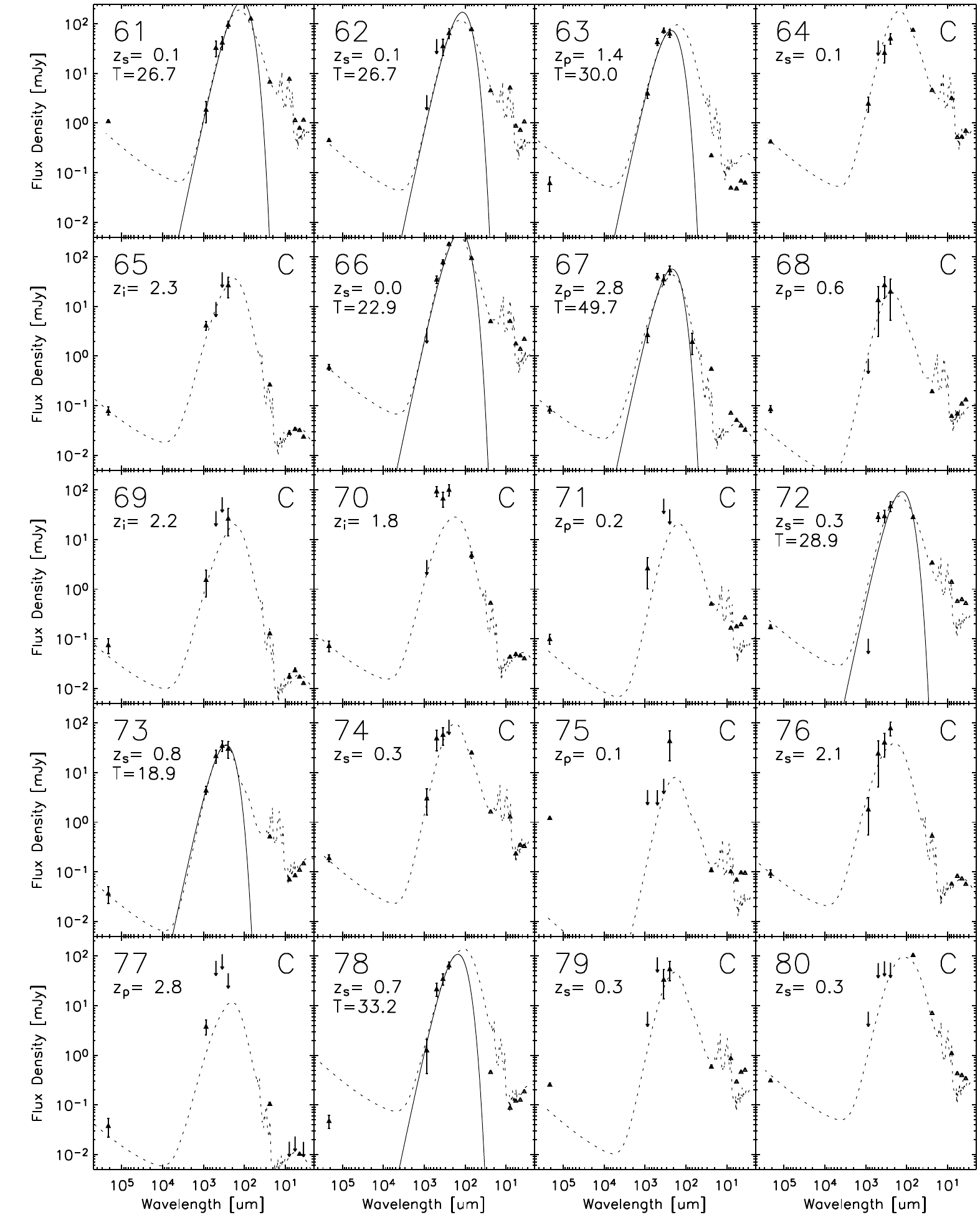}
\end{figure*}


\begin{figure*}
\centering
\includegraphics[width=0.95\linewidth]{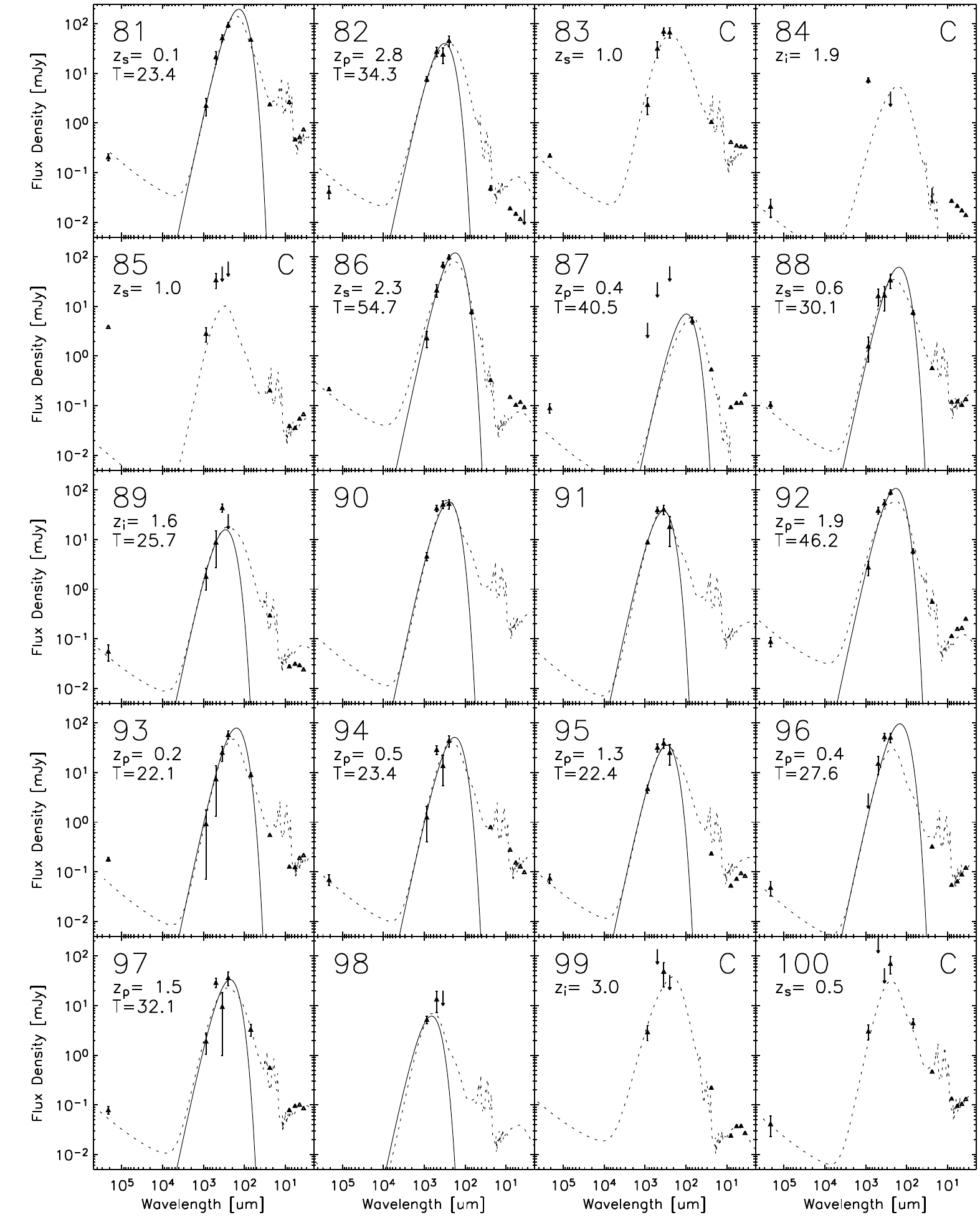}
\end{figure*}


\begin{figure*}
\centering
\includegraphics[width=0.95\linewidth]{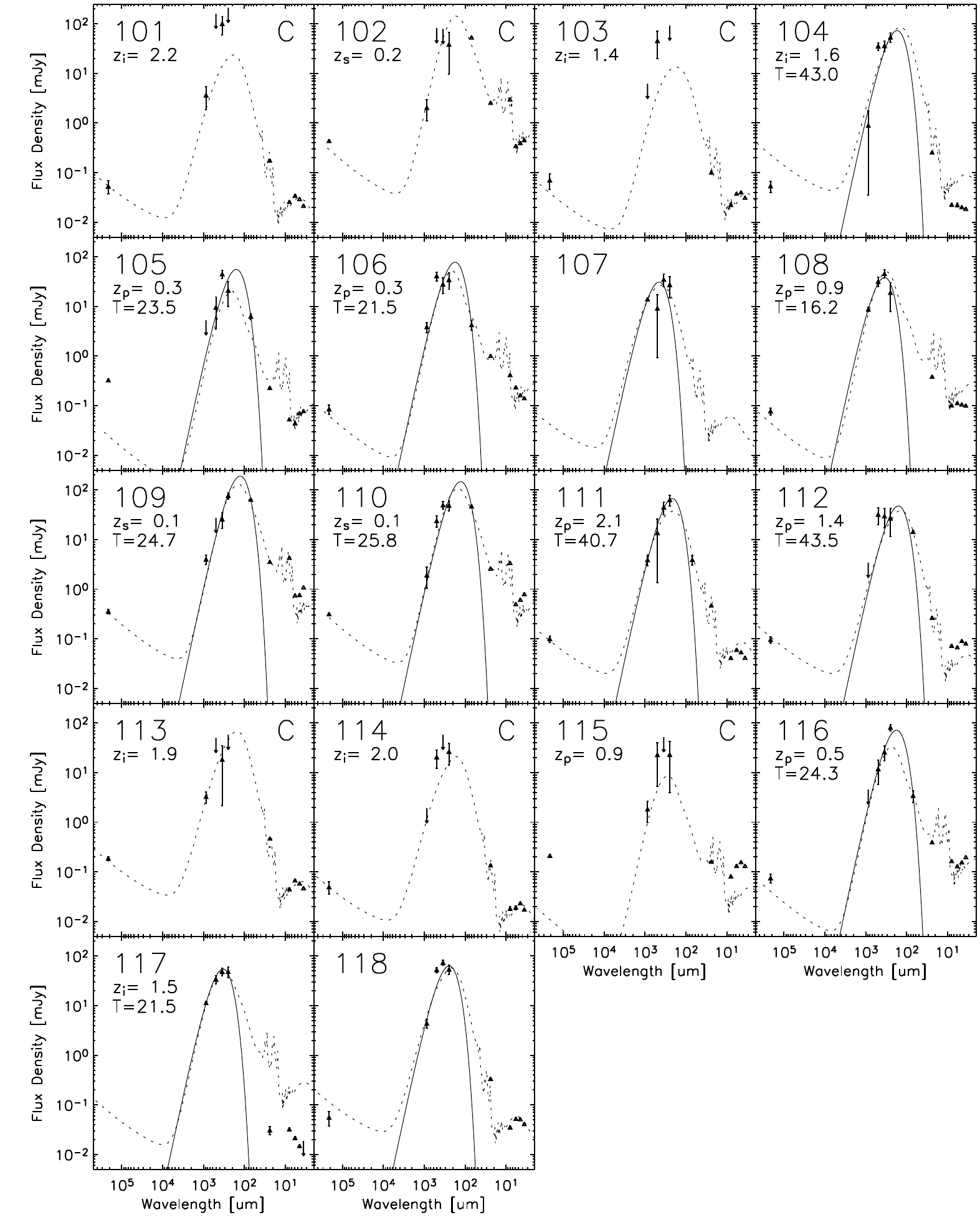}
\end{figure*}

\end{document}